\documentclass[superscriptaddress,
               twocolumn,
               twoside,longbibliography,floatfix,
               nofootinbib,a4paper,aps,pra,
               reprint]{revtex4-2}
\usepackage{epsfig}
\usepackage{amsmath}
\usepackage{amssymb}
\usepackage{amsthm}
\usepackage{color}
\usepackage[T1]{fontenc}
\usepackage{bbm}
\usepackage{bbold}
\usepackage{times}
\usepackage{amsfonts}
\usepackage{multirow}
\usepackage{enumitem}
\usepackage[normalem]{ulem}
\usepackage{latexsym}
\usepackage{mathrsfs}
\usepackage{verbatim}
\usepackage{float}
\usepackage[usenames,dvipsnames]{xcolor}
  \definecolor{carmine}{RGB}{150,0,24}
  \definecolor{c1}{RGB}{214,39,40}
  \definecolor{c2}{RGB}{31,119,180}
  \definecolor{c3}{RGB}{255,127,14}
\usepackage[colorlinks=true,
            linkcolor=blue,
            citecolor=carmine,
            urlcolor=blue
            ]{hyperref}
\usepackage{fontawesome}
  \renewcommand{\bell}{\text{\faBellO}}
  \renewcommand{\triangle}{\bigtriangleup}
  \newcommand{\at}{{\scriptsize\faAt}}
\usepackage{mathtools}
\usepackage{overpic}
\usepackage{graphicx}
  \graphicspath{{figs/}}
\usepackage[caption=false]{subfig}
\usepackage{kbordermatrix}
  \renewcommand{\kbldelim}{(}
  \renewcommand{\kbrdelim}{)}

\DeclareMathOperator{\Tr}{\text{Tr}}
\DeclarePairedDelimiter{\expec}{\langle}{\rangle}

\newcommand{\ket}[1]{|#1\rangle}

\newcommand{\ketbra}[2]{|#1\rangle\langle#2|}
\newcommand{\expect}[1]{\langle#1\rangle}

\newcommand{\bv}{\mathbf{v}}

\newcommand{\Cov}{\mathrm{Cov}}
\newcommand{\st}{\mathrm{~s.t.~}}
\newcommand{\token}{\mathrm{{count}}}
\newcommand{\match}{\mathrm{{match}}}
\newcommand{\RGB}{\mathrm{{RGB}}}
\newcommand\rout{\bgroup\markoverwith{\textcolor{red}{\rule[0.5ex]{2pt}{1pt}}}\ULon}

\newcommand{\id}{\mathbbm{1}}

\usepackage{pifont}
  \newcommand{\vmark}{\text{\ding{51}}}
  \newcommand{\xmark}{\text{\ding{55}}}
  \newcommand{\smark}{\text{\ding{103}}}

\newcommand{\squ}[1][black]{\textcolor{#1}{\rule{8pt}{8pt}}}

\begin{document}


\title{Bell nonlocality in networks}

\author{\underline{Armin Tavakoli}}
\altaffiliation{\href{mailto:armin.tavakoli@oeaw.ac.at}{armin.tavakoli\at oeaw.ac.at}}
\affiliation{Institute for Quantum Optics and Quantum Information - IQOQI Vienna, Austrian Academy of Sciences, Boltzmanngasse 3, 1090 Vienna, Austria}
\affiliation{Institute for Atomic and Subatomic Physics, Vienna University of Technology, 1020 Vienna, Austria}

\author{\underline{\smash{Alejandro Pozas-Kerstjens}}}
\altaffiliation{\href{mailto:physics@alexpozas.com}{physics\at alexpozas.com}}
\affiliation{Departamento de An\'alisis Matem\'atico, Universidad Complutense de Madrid, 28040 Madrid, Spain}
\affiliation{Instituto de Ciencias Matem\'aticas (CSIC-UAM-UC3M-UCM), Madrid, Spain}

\author{Ming-Xing Luo}
\altaffiliation{\href{mxluo@swjtu.edu.cn}{mxluo\at swjtu.edu.cn}}
\affiliation{School of Information Science and Technology, Southwest Jiaotong University, Chengdu 610031, China}

\author{Marc-Olivier Renou}
\altaffiliation{\href{mailto:marc-olivier.renou@icfo.eu}{marc-olivier.renou\at icfo.eu}}
\affiliation{ICFO - Institut de Ci\`encies Fot\`oniques, The Barcelona Institute of Science and Technology, 08860 Castelldefels (Barcelona), Spain}
\begin{abstract}
Bell's theorem proves that quantum theory is inconsistent with local physical models. It has propelled research in the foundations of quantum theory and quantum information science. As a fundamental feature of quantum theory, it impacts predictions far beyond the traditional scenario of the Einstein-Podolsky-Rosen paradox. In the last decade, the investigation of nonlocality has moved beyond Bell's theorem to consider more sophisticated experiments that involve several independent sources which distribute shares of physical systems among many parties in a network. Network scenarios, and the nonlocal correlations that they give rise to, lead to phenomena that have no counterpart in traditional Bell experiments, thus presenting a formidable conceptual and practical challenge. This review discusses the main concepts, methods, results and future challenges in the emerging topic of Bell nonlocality in networks.
\end{abstract}


{\hypersetup{linkcolor=black,urlcolor=black}\maketitle}
\def\thefootnote{}\footnotetext{The underlined authors share the first authorship}
\setcounter{footnote}{0}
\def\thefootnote{\arabic{footnote}}

{\hypersetup{linkcolor=black}\tableofcontents}

\newpage

\section{Introduction}
\label{secIntroduction}
Bell's theorem \cite{BellTheorem}, derived in 1964, constitutes a milestone in the understanding of quantum theory. The theorem states that no physical theory based only on local variables can account for all the predictions of quantum theory. Motivated both by foundational interest and the development of quantum information science, the last thirty years have witnessed intensive research on the topic of Bell nonlocality (see Refs.~\cite{Genovese2005, BrunnerRMP} for reviews, and also Ref.~\cite{ScaraniBook}). For a long time, research focus remained on simple experiments in the spirit of the Einstein-Podolsky-Rosen paradox \cite{EPR1935}, where two parties perform measurements on their respective shares of a two-particle state. However, recent years have seen a major development in the field: concepts in Bell nonlocality are applied to many-party scenarios featuring several independent sources that each distribute a physical state. The independence of the sources reflects a network structure over which parties are connected. This endeavour, referred to as Bell nonlocality in networks, concerns a substantially broader manifestation of nonlocality as compared to that arising in the intensively researched  Einstein-Podolsky-Rosen scenario.

The motivation behind the research program on Bell nonlocality in networks is two-fold.
Firstly, from the foundational point of view, it is interesting to understand quantum theory and its relationship to other physical models in more sophisticated and qualitatively new scenarios that naturally continue the spirit of Bell's theorem. Secondly, from the point of view of quantum information science, it is becoming increasingly relevant to apply quantum phenomena in networks towards information processing tasks. This is due the technological developments towards scalable quantum networks (see e.g.~Refs.~\cite{Kimble2008, Wehner2018, Kozlowski2019} for an overview of recent progress on the development of a quantum internet). Moreover, Bell nonlocality in networks also brings novel forms of entanglement to the forefront of the physical process. When many \textit{a priori} independent shares of entangled states are distributed in a quantum network, a party that holds multiple shares originating from different sources can perform entangled measurements  (i.e.~projections of distinct quantum systems in an entangled basis) to \textit{a posteriori} distribute entanglement between distant, initially independent, systems in the network. This is fundamentally based on the process of entanglement swapping \cite{Zukowski1993}, which has no counterpart in traditional tests of Bell inequalities. However, as it has turned out, the analysis of Bell nonlocality in networks cannot fall back on the long-established ideas and tools for standard Bell inequality experiments. Instead, the topic has turned out a formidable challenge which demands new lines of thought.

This article presents a general review of the present knowledge of Bell nonlocality in networks. It is also intended to serve as a self-contained introduction to the topic, as a highlight of its most notable progress, and as pointer to its main open problems. In the remainder of this introduction, we first briefly present standard Bell scenarios, then introduce network Bell scenarios and finally give an overview of the main methods used to analyse correlations in networks. In Section \ref{section2}, we discuss in depth the two smallest, and seminal, examples of networks, which are known as the bilocal scenario and the triangle scenario. In Section \ref{section3}, we present general methodology for analysing classical and quantum correlations in networks. In Section \ref{section4}, we discuss how the network configuration imposes ultimate limits of physical predictions. In Section \ref{section5}, we review experiments demonstrating network nonlocality. In Section \ref{section6}, we outline a number of additional topics in network nonlocality and closely related subjects. Finally, in Section \ref{section8} we conclude the review and present a list of interesting open problems and research directions.

\subsection{Standard Bell scenarios}
\label{secIntroBell}
\begin{figure*}[t!]
	\centering
		\begin{tabular}{ccl}
			\subfloat[\label{fig:Bell}]{
				\centering
	      \begin{minipage}[t]{0.39\textwidth}
	      	\centering
					\includegraphics[scale=0.42]{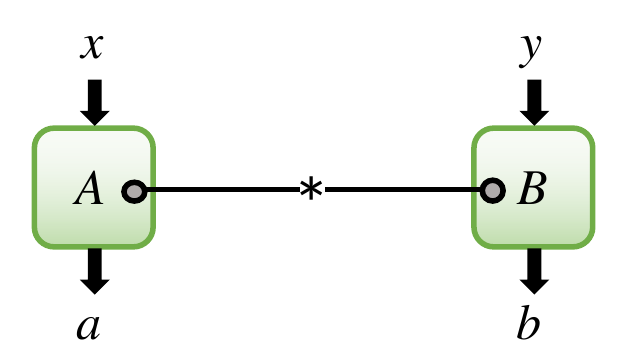}
		 \end{minipage}}
			&
			\qquad\qquad
			&
		\subfloat[\label{fig:bilocal}]{
				\centering
	      \begin{minipage}[t]{0.49\textwidth}
	      	\centering
					\includegraphics[scale=0.42]{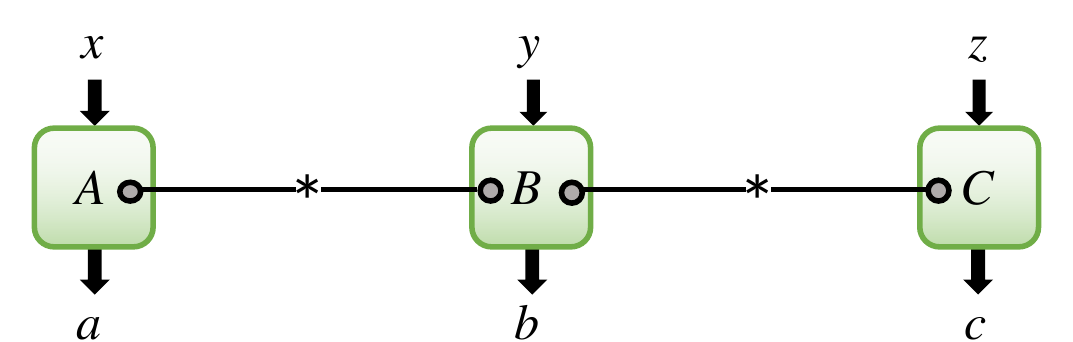}
			\end{minipage}
			}
			\\
			\subfloat[\label{fig:triangle}]{
				\centering
	      \begin{minipage}[t]{0.39\textwidth}
	      	\centering
					\includegraphics[scale=0.42]{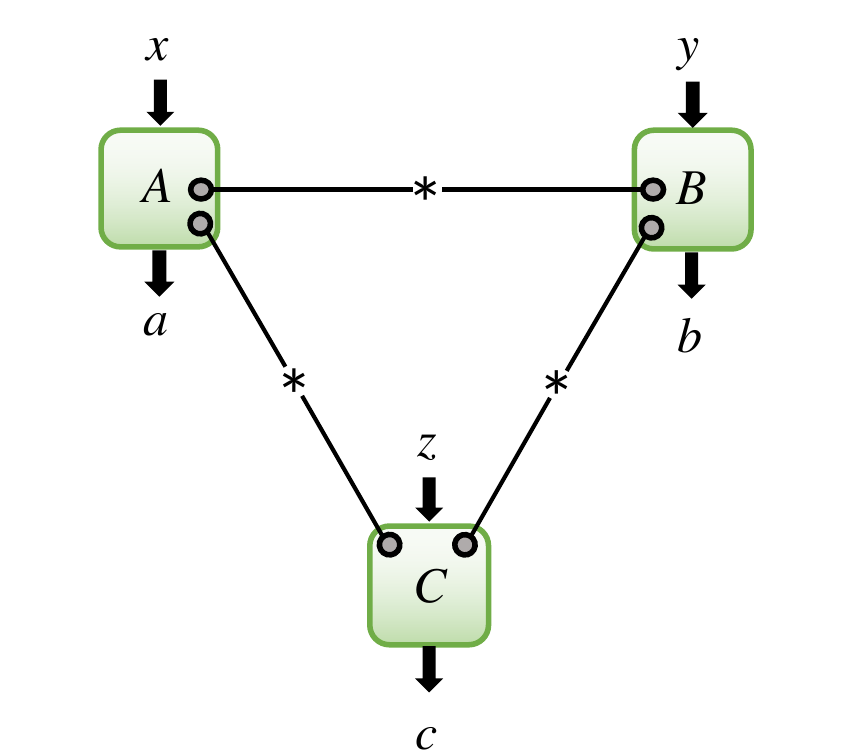}
				\end{minipage}
			}
			&
			\qquad\qquad
			&
			\subfloat[\label{fig:complex}]{
				\centering
	      \begin{minipage}[t]{0.49\textwidth}
	      	\centering
					\includegraphics[scale=0.42]{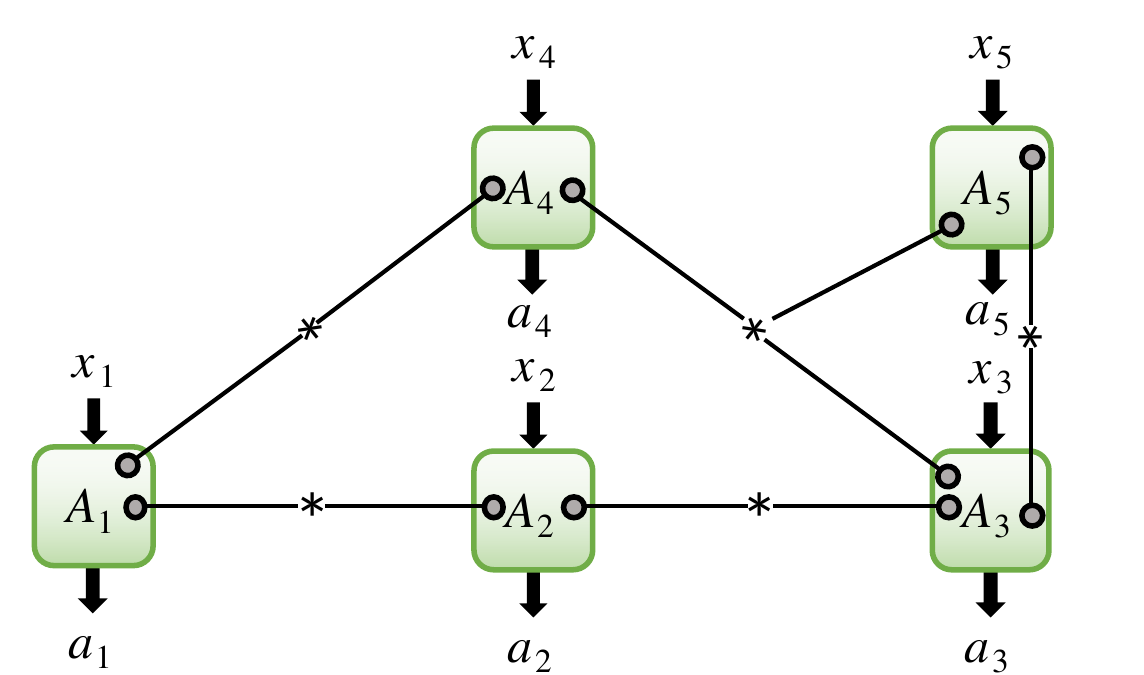}
				\end{minipage}
			}
		\end{tabular}
	\caption{Illustrations of network Bell scenarios. (a) The standard Bell scenario, with two parties that share a common source. This is the simplest scenario for studying nonlocality and it is the focus of previous reviews~\cite{BrunnerRMP, Genovese2005}. (b) The bilocal scenario, where two extremal parties each share a source with a central party. This is the scenario that underlies quantum repeaters. (c) The triangle scenario, with three parties are connected to each other through bipartite sources. (d) A more complex network with cycles, multipartite sources, and conditionally independent parties.}
	\label{FigureNetworks}
\end{figure*}

A standard Bell scenario features several parties who perform measurements on the respective shares of a system originally emitted from a single physical source.
For simplicity, let us focus on the bipartite scenario, although the definitions and conclusions naturally generalise to more parties.
Our parties,  call them Alice and Bob, freely and independently choose a measurement setting each, denoted $x$ and $y$ respectively, and obtain outcomes labelled $a$ and $b$, in spacelike separated events (see Figure~\ref{fig:Bell}). When repeated a large number of times, the relative frequencies correspond to a conditional probability distribution, $p(a,b|x,y)$, which we simply refer to as $P$ or alternatively as the correlations.

\textit{A priori}, the only assumption made on the correlations stems from relativity, namely that the output of either party reveals no information about the input chosen by the other party.
This so-called no-signaling principle is formalised as
\begin{equation}
	\begin{split}
		&\forall b,x,y: \hspace{3mm}\sum_{a}p(a,b|x,y)=p(b|x,y)\stackrel{\text{NS}}{=}p(b|y), \\
		&\forall a,x,y: \hspace{3mm} \sum_{b}p(a,b|x,y)=p(a|x,y)\stackrel{\text{NS}}{=}p(a|x).
	\end{split}
\label{eq:nonsignaling}
\end{equation}

A local-variable theory is an attempt at giving a physical explanation for the correlations $P$ based on the principle that the outcome of Alice (resp. Bob)  only depends on her (resp. his) input and some stochastic property that is carried with the shares of the system due to their common past. This stochastic property, historically called a local hidden variable\footnote{Since $\lambda$ does not actually need to be hidden, we will simply refer to it as a local variable.}, is denoted $\lambda$ and subject to some unknown probability density $\mu(\lambda)$. Therefore, we say that the correlations admit a local model (i.e.~they satisfy Bell's local causality assumption) if they can be written as
\begin{equation}
 \label{BellLocal}
p(a,b|x,y)=\int d\lambda \mu(\lambda) p(a|x,\lambda)p(b|y,\lambda),
\end{equation}
where, importantly, Alice's and Bob's choices of settings are assumed to be independent from the local variable [i.e.~$\mu(\lambda|x,y)\,{=}\,\mu(\lambda)$]. In this sense, any correlation consistent with the no-signaling principle that does not admit a local model is said to be nonlocal.

\begin{figure}
	\centering
	\includegraphics[width=0.9\columnwidth]{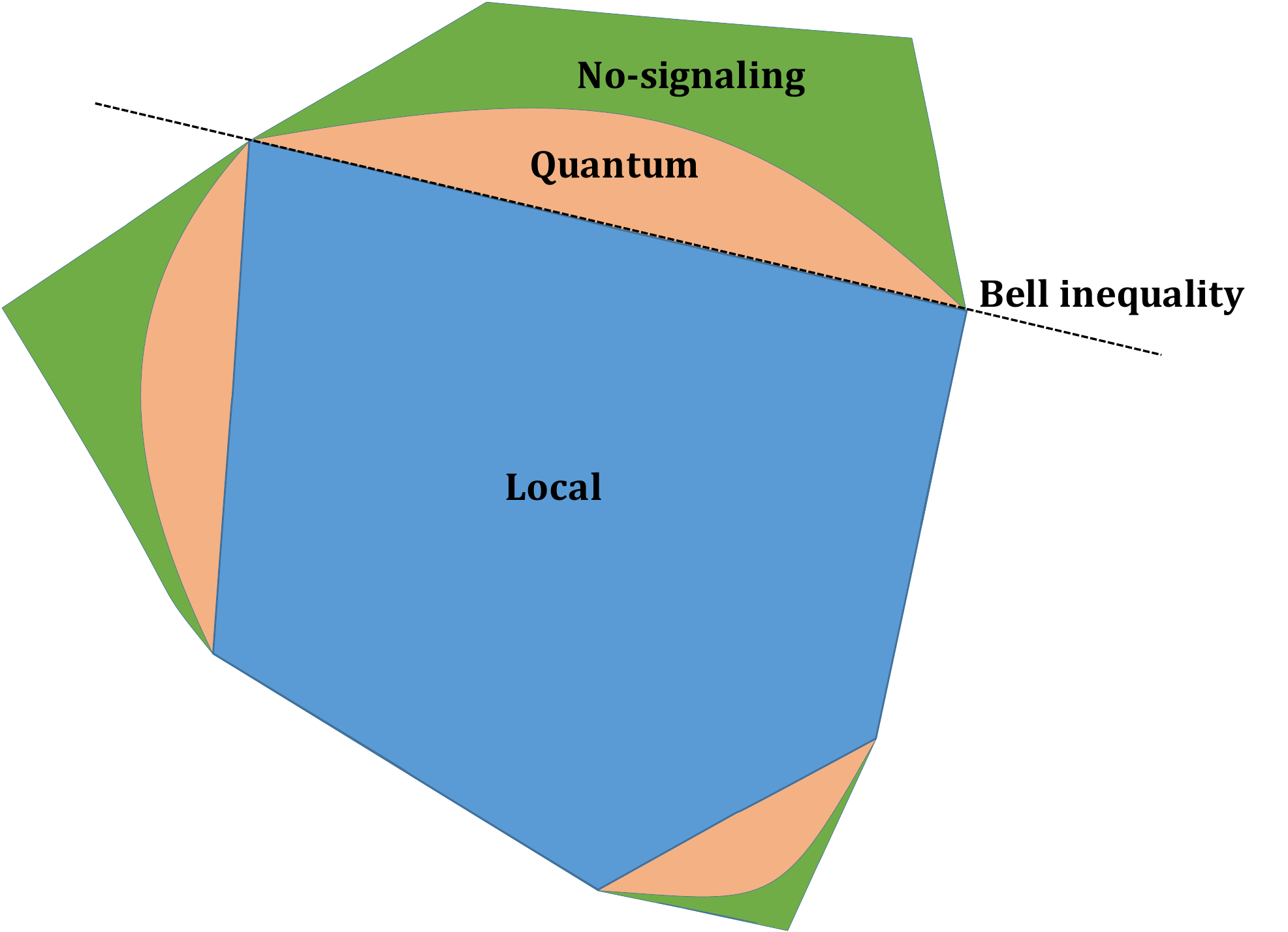}
\caption{Illustration of the local polytope, the quantum set of correlations (convex) and the no-signaling polytope. Optimal Bell inequalities are hyperplanes corresponding to a facet of the local polytope. More generally, any hyperplane delimiting the local polytope constitutes a valid Bell inequality.}
\label{FigBellScenario}
\end{figure}

When the number of inputs and outputs is finite, the set of correlations with a local model admits a simple characterisation in terms of a polytope \cite{Fine1982}, i.e.~a high-dimensional convex polyhedron, often referred to as the local polytope.
The polytope structure follows from the fact that since any randomness in the response functions of Alice and Bob can be absorbed in $\mu(\lambda)$ one may, without loss of generality, assume that the response functions are deterministic, i.e.~that $p(a|x,\lambda)\,{\in}\,\{0,1\}$ and $p(b|y,\lambda)\,{\in}\,\{0,1\}$. Therefore, each choice of deterministic response functions constitutes a vertex of a polytope in the space of correlations.
Since these are finitely many, one can also take $\lambda$ from an equally sized set and replace the integral in Eq.~\eqref{BellLocal} by a summation.
Due to the freedom of varying $\mu(\lambda)$, the achievable correlations are the convex hull of these vertices, which is the local polytope (see the blue polytope in Figure~\ref{FigBellScenario}).
This allows one to decide whether a given $p(a,b|x,y)$ admits a local model either by determining the existence of a suitable $\mu(\lambda)$ (a problem which can be cast as a linear program \cite{Kaszlikowski2000,Zukowski1999}) or, alternatively, by determining all the linear inequalities that characterise the facets of the local polytope and checking that they are all satisfied.
The latter inequalities, or more generally any hyperplane which separates the local polytope, are known as Bell inequalities (see Figure~\ref{FigBellScenario} for an illustration). Proving that at least one Bell inequality is violated is sufficient for asserting that correlations are nonlocal.

The simplest and most popular Bell inequality  is the so-called Clauser-Horne-Shimony-Holt (CHSH) Bell inequality \cite{CHSH1969}. The CHSH inequality concerns the simplest nontrivial Bell scenario, namely that in which Alice and Bob each choose between just two measurements, $x\,{\in}\,\{0,1\}$ and $y\,{\in}\,\{0,1\}$, and produce binary outcomes $a\,{\in}\,\{0,1\}$ and $b\,{\in}\,\{0,1\}$. In this scenario, the nontrivial faces of the local polytope are, up to re-labelings, given by a single inequality:
\begin{equation}\label{CHSH}
\mathcal{S}_\text{CHSH}\coloneqq \langle A_0B_0\rangle+\langle A_0B_1\rangle+\langle A_1B_0\rangle-\langle A_1B_1\rangle\leq 2,
\end{equation}
where $\langle A_xB_y\rangle\,{=}\,\sum_{a,b}(-1)^{a+b}p(a,b|x,y)$ is the joint expectation value of the observations of Alice and Bob when their settings are $x$ and $y$, respectively. Using the short-hand notation  $\hat{A}_\pm\,{\coloneqq}\, A_0 \pm A_1$, the CHSH expression becomes $\mathcal{S}_\text{CHSH} \,{=}\, \langle\hat{A}_+ B_0+\hat{A}_- B_1\rangle$. For local models, the expectation value is taken over $\lambda$. Then, using the triangle inequality, Eq.~\eqref{CHSH} follows from $|B_y|\leq 1$ and $|\hat{A}_+| + |\hat{A}_-|\leq 2$.

Remarkably, Eq.~\eqref{CHSH} can be violated in quantum theory.
This highlights the fact that quantum theory does not satisfy local causality.
In quantum theory, correlations are given by the Born rule
\begin{equation}
  p(a,b|x,y)=\Tr\left(A_{a|x}\otimes B_{b|y}\rho\right),
\end{equation}
for some measurements $\{A_{a|x}\}$ and $\{B_{b|y}\}$\footnote{These are in general positive operator-valued measures, i.e.~the measurement operators are non-negative and sum to identity.} and a quantum state $\rho$. The general statement that quantum theory eludes local models is known as Bell's theorem \cite{BellTheorem}. In order to find a quantum violation of the CHSH inequality, let Alice and Bob share the maximally entangled two-qubit state $\ket{\phi^+}\,{=}\,\frac{1}{\sqrt{2}}\left(\ket{00}+\ket{11}\right)$ and let them perform locally anticommuting measurements, corresponding to the observables $A_0\,{=}\,\sigma_1$, $A_1\,{=}\,\sigma_3$, $B_0\,{=}\,\frac{\sigma_1+\sigma_3}{\sqrt{2}}$ and $B_1\,{=}\,\frac{\sigma_1-\sigma_3}{\sqrt{2}}$, where by $\vec{\sigma}\,{=}\,(\sigma_1,\sigma_2,\sigma_3)$ we denote the Pauli matrices.
The expectation values, computed from the Born rule, are  $\expect{A_xB_y}\,{=}\,\frac{(-1)^{xy}}{\sqrt{2}}$.
This gives a violation of the CHSH inequality as $\mathcal{S}_\text{CHSH}\,{=}\,2\sqrt{2}>2$, which proves the nonlocal nature of correlations in quantum theory.
In fact, quantum theory does not allow for violations larger than $\mathcal{S}_\text{CHSH}=2\sqrt{2}$ \cite{Cirelson1980}.

The existence of quantum models, which  enable stronger correlations than local-variable ones, raises the question of the ultimate limits on correlations in Bell scenarios.
The no-signaling principle is considered as the minimal requirement that correlations must respect in Bell scenarios.
Therefore, by only demanding that correlations are no-signaling, one obtains the most general notion of nonlocal correlations (see Figure~\ref{FigBellScenario}).
Their relevance is showcased in the celebrated Popescu-Rohrlich distribution, which reads $p(a,b|x,y)=\frac{1}{4}\left[1+(-1)^{a+b+xy}\right]$ and is consistent with the no-signaling principle described in Eq.~\eqref{eq:nonsignaling}.
It gives the algebrically maximal violation of the CHSH inequality, $\mathcal{S}_\text{CHSH}=4$ \cite{Popescu1994}.

The first experimental violation of the CHSH inequality was reported already in 1972 \cite{Freedman1972}. In 1982, it was followed by a more faithful demonstration \cite{Aspect1982}. In recent years, quantum violations of  CHSH-type inequalities were reported in rigorous experiments that closed the various experimental loopholes that otherwise could have been exploited in order to maintain a local-variable model \cite{ZeilingerLoopholeFree2015,HansonLoopholeFree2015,ShalmLoopholefree,WeinfurterLoopholeFree}.

\subsection{Networks: Beyond the standard Bell scenario}
\label{secIntroNetwork}
Standard Bell scenarios have inspired the development of network Bell scenarios.
Network Bell scenarios come with a crucial conceptual difference: they feature more than one source.
Each source in a network, labeled $i\,{=}\,1,\ldots,m$, is assumed to be independent from the rest.
Although most of the works surveyed in this review consider networks with only bipartite sources, it is equally permissible to consider sources that emit more than two subsystems.
Notice that this assumption of source independence is irrelevant in standard Bell scenarios since these only involve a single source.
This has direct implications in the interpretation of what the local variables represent. In standard Bell scenarios one may either view the local variable $\lambda$ as a physical influence carried with particles due to a shared past (this is Bell's original interpretation~\cite{BellTheorem}) or as a free and unlimited resource of shared randomness made accessible to the parties.
However, in network Bell scenarios, due to the presence of independent sources, one can no longer consider the two interpretations equivalent: global shared randomness is no longer a free resource.
In this sense, network Bell scenarios are more restricted than their traditional counterparts.

While the independence of the sources in general constitutes an additional assumption, it is arguably a natural one when considering distant sources that are built and operated separately.
Each source distributes shares of a physical system to a given subset of the parties present in the network.
Therefore, some parties can hold several, initially independent, shares originating from different sources.
The manner in which sources connect the parties can be described as a set of relations $i\,{\rightarrow}\, j$ indicating that source $i$ emits a share to party $j$.
The parties, labeled $j\,{=}\,1,\ldots, n$, choose, freely and independently, inputs $x_j$ based on which they apply a measurement to their system share(s) and obtain an outcome $a_j$.
In the limit of many rounds, this gives rise to correlations in the network that are written $p(a_1,\ldots, a_n|x_1,\ldots,x_n)$\footnote{For networks with three parties (as in Figures~\ref{fig:bilocal} and \ref{fig:triangle}) we typically use the handier notations $x,y,z$ to denote inputs and $a,b,c$ to denote outputs. The probability distribution is then written $p(a,b,c|x,y,z)$.}.
In Figure~\ref{FigureNetworks} we illustrate a few examples of network Bell scenarios.

\begin{figure*}
	\hfill
	\subfloat[\label{fig:NetworkL_L}]{
		\includegraphics[width=.85\columnwidth]{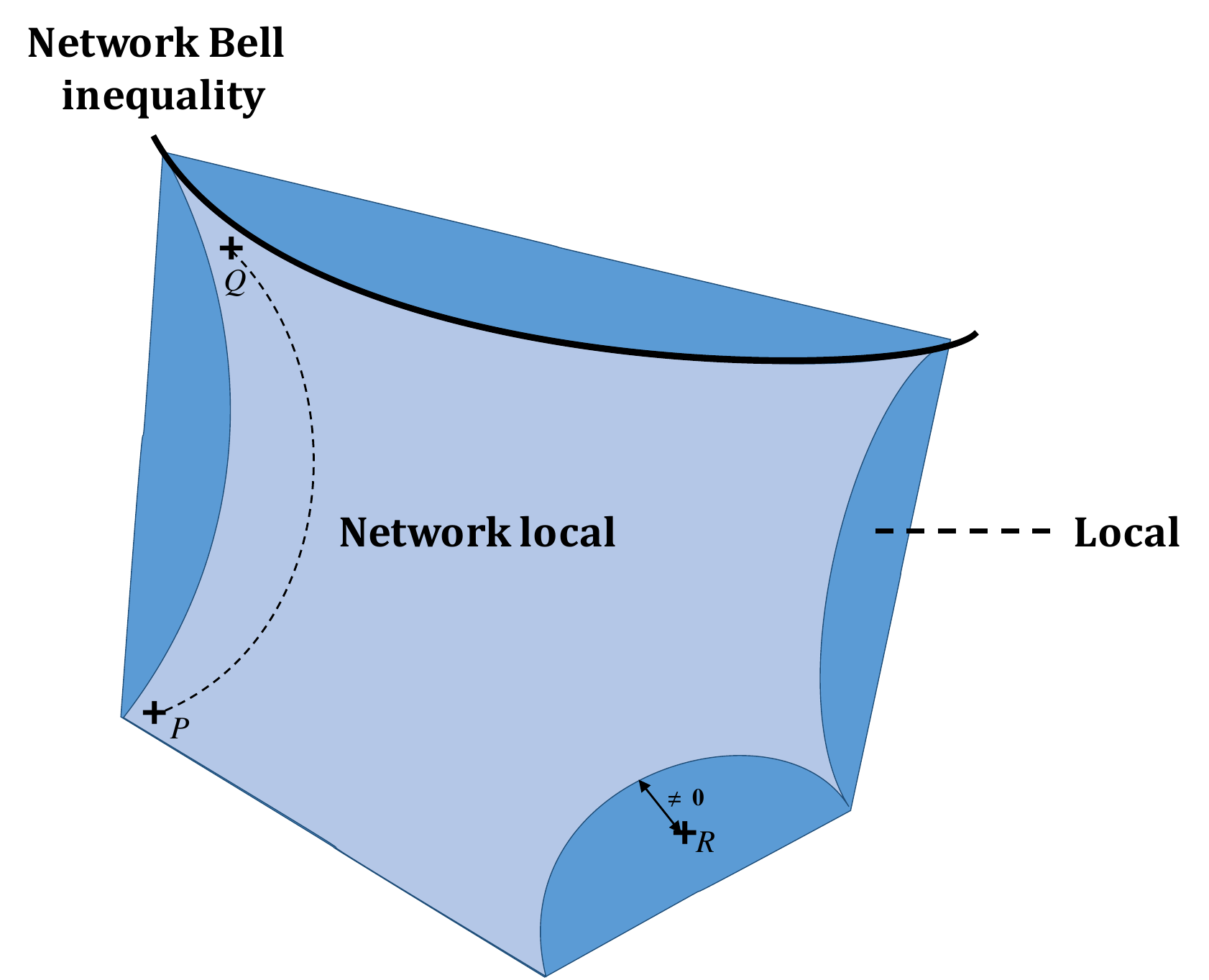}
	}
	\hfill
	\subfloat[\label{fig:NetworkL_NetworkQ}]{
		\includegraphics[width=.9\columnwidth]{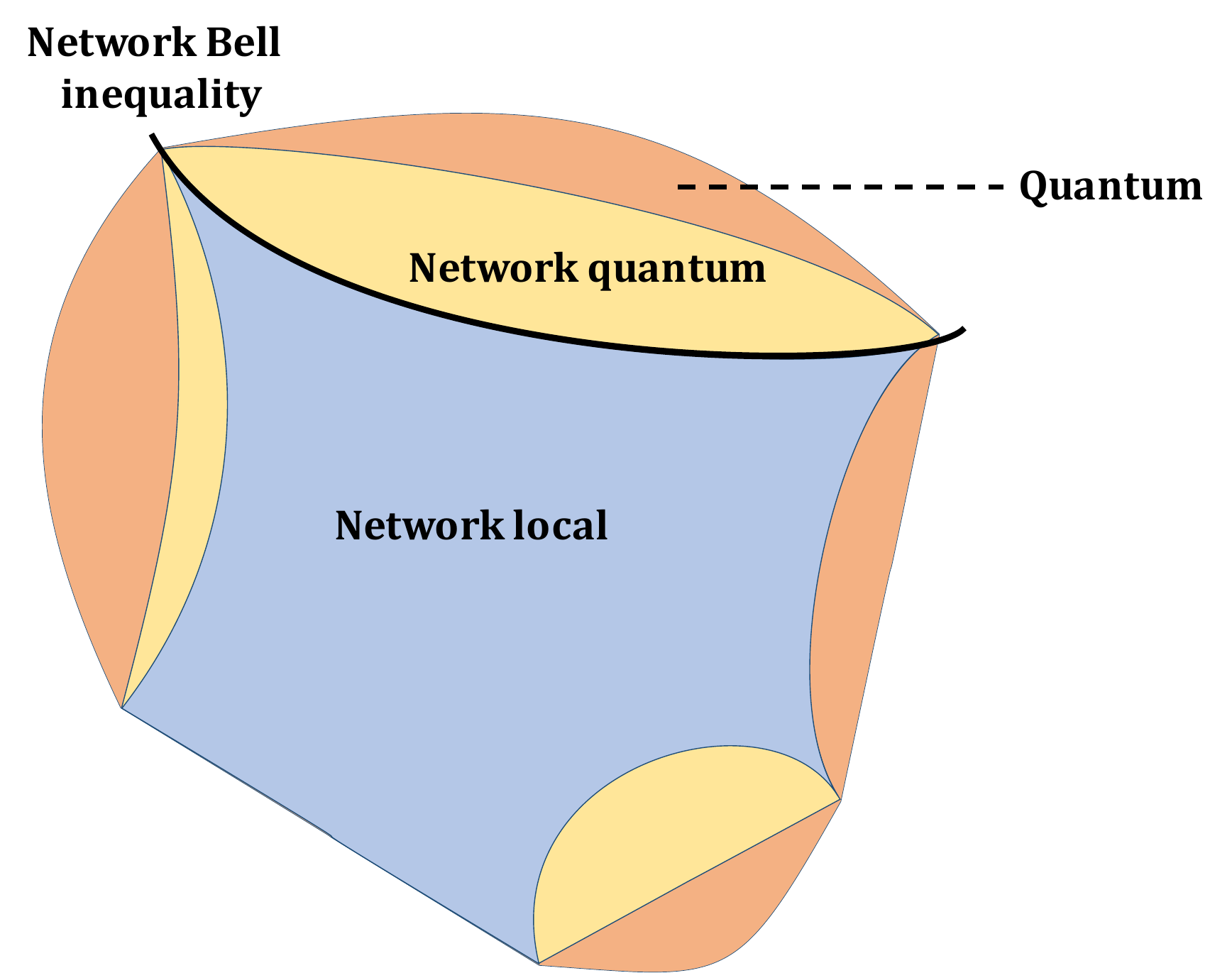}
	}
	\hfill
	\caption{Illustration of sets of network correlations.
	(a) The set of network local correlations is contained in the set of local correlations, so there exist Bell-local correlations that violate network Bell inequalities. Thus, when restricting to correlations generated by local systems, a violation of a network Bell inequality can be interpreted as a certification that the correlations were not generated according to the network structure. (b) The set of network local correlations is also contained in the set of network quantum correlations, which is itself contained in the set of quantum correlations without independent sources. Therefore, network Bell inequalities, once the underlying network configuration is fixed, have a meaning analogous to that of standard Bell inequalities (i.e.~identifying nonlocal correlations).}
	\label{FigureNetworkSets}
\end{figure*}

In local-variable models for the correlations in the network, the independence of the $m$ sources is manifested by associating a stochastic local variable, $\lambda_i$, to each of them. The outcome of party $j$ is modelled as a stochastic response conditioned on their input $x_j$ and the local variables that they have received.
Naturally generalising Eq.~(\ref{BellLocal}), the local correlations achievable in a network take the form
\begin{equation}\label{Eqlocal}
	\begin{split}
		p(\bar{a}|\bar{x})=&\int d\lambda_1\mu_1(\lambda_1)\ldots \int d\lambda_m \mu_m(\lambda_m) \\
		& \times p(a_1|x_1,\bar{\lambda}_1)\ldots p(a_n|x_n,\bar{\lambda}_n),
	\end{split}
\end{equation}
where $\bar{a}\,{=}\,(a_1,\ldots,a_n)$, $\bar{x}\,{=}\,(x_1,\ldots,x_n)$, $\bar{\lambda}_j$ denotes the collection of local variables received by party $j$ and $\mu_i(\lambda_i)$ denotes the probability density of $\lambda_i$.
These models are called network local models.
The independence of the sources is the key feature of the above definition, which distinguishes it from standard multipartite Bell scenarios~\cite{Svetlichny1987}.
In analogy with standard Bell scenarios, the network local models can without loss of generality be restricted to a finite alphabet, i.e.~each $\lambda_i$ can be taken to have a finite cardinality and the corresponding $\mu_i(\lambda_i)$ can be restricted to a discrete probability distribution \cite{Fritz2016}.
In Ref.~\cite{Rosset2018} it was shown how to determine an upper bound on the alphabet size of the local variables for any given network based on the number of inputs and outputs and the network structure.

The independence of the sources causes the set of correlations attainable in a network to acquire a more complex structure than the local polytope (see Figure~\ref{fig:NetworkL_L}).
A few qualitative properties of the set of local correlations in a network are
\begin{enumerate}[label=(\roman*)]
	\item It is contained in the local polytope.
	\item It is not convex.
	\item It is closed, connected and characterised by a finite number of polynomial inequalities.
\end{enumerate}
The first property is self-evident, because the local polytope corresponds to relaxing the independence of the sources in Eq.~\eqref{Eqlocal}. However, property (ii) highlights a crucial difference between the network local set and the local polytope. The non-convexity of the set of local correlations in networks \cite{Branciard2010} has important consequences. It makes the characterisation of correlations in networks a much more difficult task than in standard Bell scenarios, for both analytical and numerical considerations. A particular aspect of this is that optimal Bell inequalities are no longer separating hyperplanes. The reason is that the deterministic strategies that constitute the vertices of the local set are also the vertices of the network local set, and the facets that connect these vertices are the linear Bell inequalities of the local polytope (see Figure~\ref{FigureNetworkSets}).
Consequently, in network scenarios, one must consider nonlinear Bell inequalities.
However, property (iii) stipulates that these nonlinear Bell inequalities can be restricted to polynomial expressions  \cite{Fritz2012}.
Furthermore, since the network local set is closed, every point outside the network local set upholds some (perhaps very small) degree of tolerance to noise.

Characterising network local correlations is a central problem in the field of network nonlocality, in the same way that characterising Bell-local correlations is the central problem in the field of Bell nonlocality~\cite{BrunnerRMP}.
However, in contrast to standard Bell nonlocality, in the case of nonlocality in networks the violation of a Bell-like inequality can signal two different phenomena: either it implies the use of non-classical resources, provided that the network structure is assumed (see Figure~\ref{fig:NetworkL_NetworkQ}) or it implies that the hypothesised network structure is false, provided that classical resources are assumed (see Figure~\ref{fig:NetworkL_L}).
Thus the former is an interpretation analogous to that of standard nonlocality while the latter signifies a witness of the underlying network structure.

In quantum models of network correlations, the independence of the sources leads one to associate an independent quantum state, $\rho_i$, to each of the sources.
Each party performs a measurement $\{A^{(j)}_{a_j|x_j}\}$ on their respective shares.
The correlations are then given by the Born rule
\begin{equation}
	p(\bar{a}|\bar{x})=\Tr\left[\left(A^{(1)}_{a_1|x_1}\otimes \ldots \otimes A^{(n)}_{a_n|x_n}\right)\left(\rho_1\otimes \ldots \otimes \rho_m\right)\right],
	\label{eq:quantummodel}
\end{equation}
where the Hilbert spaces are suitably ordered as given by the network configuration.
Note that each state $\rho_i$ can be entangled and that the tensor product structure $\rho_1\otimes \ldots \otimes \rho_m$ indicates that there is neither entanglement nor classical correlation between the various states, as implied by the independence of the sources.
Alike the network local set, the set of network quantum correlations is non-convex and closed (see Figure~\ref{fig:NetworkL_NetworkQ}).
It naturally contains the network local set but it does not contain the local polytope (or vice versa).
There exist correlations that, while being possible to generate in a network with quantum sources, cannot be generated by the same number of parties if they all share a global local variable, and vice versa\footnote{Examples of both cases will be discussed throughout the review. Eqs.~\eqref{eq:mermin}-\eqref{eq:svet} represent illustrations of quantum network correlations that cannot be generated with a global local variable, while the GHZ distribution of Eq.~\eqref{eq:GHZ} is an example of correlations with a local model that cannot be generated in a quantum network.}.

Moreover, in analogy with no-signaling correlations generalising quantum correlations in standard Bell scenarios, one may consider what are the most general correlations that can arise in a given network.
However, it turns out that the correlations implied by a theory-independent treatment of networks featuring independent sources are far less straightforward to characterise and require the introduction of concepts outside the scope of this review (see, for instance, Refs.~\cite{Henson2014,chiribella2009framework}).
Despite this, in Section~\ref{section4} we discuss available methods for deriving constraints that follow exclusively from the structure of the network, irrespective of the nature of the sources.

\subsection{Brief overview on network methods}
\label{sec:methods}
While methods for analysing local, quantum and no-signaling correlations in standard Bell nonlocality greatly benefit from the convexity of these correlation sets \cite{BrunnerRMP}, methods for analysing their counterparts in networks in general require significantly different approaches. To these purposes a variety of methods have been developed and investigated.
They all come with different advantages and drawbacks, most notably concerning generality (ideally applying to any correlation in any network), strength (ideally recovering the boundaries of the sets) and practicality (ideally being computable with commonly accessible resources).
While these methods will later be reviewed individually, we here provide a brief overview of some selected methods accompanied with pointers to the specific sections where they are further discussed.

\begin{table}[ht!]
    \begin{tabular}{c|c|c|c|c}
      & \begin{tabular}{@{}c@{}}Disguised \\ Bell nonlocality\end{tabular} & \begin{tabular}{@{}c@{}}Algebraic \\ inequalities\end{tabular} &
      Inflation & \begin{tabular}{@{}c@{}} Explicit \\ decompositions\end{tabular} \\ \hline \hline
     \begin{tabular}{@{}c@{}} Type of  \\ networks \end{tabular} & Any & \begin{tabular}{@{}c@{}} With condit. \\ independence \end{tabular} & Any & \begin{tabular}{@{}c@{}} With condit. \\ independence \end{tabular} \\ \hline
     Example  & Triangle & Star & Triangle & Bilocal \\  \hline
     Section & \ref{subsec:OverviewReplaceInputs} & \ref{subsubsec:BilocCor}, \ref{subsec:CondIndep} &
     \ref{Inflation} & \ref{subsubsec:NonlocEJM}, \ref{sec:Simulation}
  \end{tabular}
  \caption{Selection of methods for characterising correlations that admit local variable models in networks discussed in this review, along with the features that identify the networks they are applicable to, and examples of such families.}
  \label{tab:AllMethodsForNetwNonloc}
\end{table}

Firstly, Table~\ref{tab:AllMethodsForNetwNonloc} contains a selection of methods dedicated to studying network local correlations. Many approaches exist, which range from the explicit parametrisation of network local models~\cite{Branciard2012,Tavakoli2020} to the use of hierarchies of relaxations of the sets of compatible correlations~\cite{wolfe2019inflation}, or the analytical construction of network Bell inequalities \cite{Branciard2010, Branciard2012, Tavakoli2014}. The methods contained in the table, and additional ones, are collected in Sections~\ref{section2} and \ref{section3}.

\begin{table*}
  \begin{tabular}{c|c|c|c|c|c|c}
    & Scalar extension & Classical inflation & Quantum inflation & Entropy cones & Covariance matrices & Finner inequalities \\ \hline \hline
    Local        & \vmark & \vmark & \vmark & \vmark & \xmark & \xmark \\ \hline
    Quantum      & \vmark & \xmark & \vmark & \vmark & \vmark & \xmark \\ \hline
    Theory-independent & \xmark & \vmark & \xmark & \vmark & \smark & \vmark \\ \hline
    Section & \ref{sec:scalarextension} & \ref{sec:cinflation}, \ref{sec:nonsignaling} & \ref{sec:qinflation} & \ref{Entropy}, \ref{sec:nonsignaling} & \ref{subsec:CovMatrix} & \ref{subsec:FinnerIneq}
  \end{tabular}
  \caption{Methods for characterising network correlations of different natures and the types of correlations they can constrain. Given that the sets of local, quantum and  theory-independent correlations compatible with a same network follow a strict inclusion order, a method designed for constraining correlations of one type can naturally constrain correlations of all the corresponding sub-types (i.e., higher up in the table). The \vmark{} signs in the table denote instead the existence of variants of the method that distinguish the selected correlations from more general ones (i.e., lower in the table).
Analogously, the \xmark{} sign denotes that the method is not capable of singling out the selected correlations from more general ones.
The \smark{} sign denotes the fact that the ability of the method to constrain network correlations has only been proven in certain networks, and its applicability to the general case remains unknown.
  }
  \label{tab:GeneralMethods}
\end{table*}

Secondly, Table~\ref{tab:GeneralMethods} collects techniques that allow the characterisation not only of network local models, but also of quantum models in networks of the form of Eq.~\eqref{eq:quantummodel} and theory-independent network correlations.
On one hand, network quantum correlations are particularly interesting since they encompass the network nonlocal correlations that are expected to be attainable in experiments.
As in the case of network local correlations, there exist characterisations of very different nature, with examples being the use of functional analysis or infinite hierarchies of relaxations.
These methods are contained, mainly, in Section~\ref{section3}.
On the other hand, theory-independent correlations are the most general correlations that can be created in the network with space-like separation, and thus their study can be understood as the analysis of the limits that the network structure imposes in the correlations that can be produced in it.
The techniques that characterise these correlations are reviewed mostly in Section~\ref{section4}.

\section{Elementary network scenarios}
\label{section2}

In this section we focus on the two quintessential examples of network Bell scenarios that arise in the simplest multipartite setting, namely that involving only three parties. There are only three qualitatively different network configurations based on three parties. (i) All parties are connected to a single source. This is the standard tripartite Bell scenario, which is not the focus of this review (see instead, e.g.,~Ref.~\cite{BrunnerRMP}). (ii) One party is separately connected to each of the two other parties. This three-on-a-line scenario is illustrated in Figure~\ref{fig:bilocal} and it is commonly referred to as the bilocal scenario\footnote{The term ``bilocal scenario'' does not mean that the scenario only can be analysed in terms of network local models. Instead, it is common in the literature to use the term bilocal scenario to refer to any realisation of the network in Figure~\ref{fig:bilocal}, while bilocal correlations maintains its standard meaning as correlations of the form given by Eq.~\eqref{bilocal}.}. (iii) All three parties are pairwise connected to each other. This network is illustrated in  Figure~\ref{fig:triangle} and is known as the triangle scenario. The bilocal scenario and the triangle scenario are the focus of this section. However, before discussing them in detail, we use these two scenarios in order to discuss the relationship between standard Bell nonlocality and network nonlocality.

\subsection{Bell nonlocality disguised as network nonlocality}
\label{subsec:OverviewReplaceInputs}
There are many, qualitatively different, manifestations of network nonlocality (see, for instance, the discussions in Refs.~\cite{Fritz2012,Branciard2012,Fraser2018,Renou2019}).
This is reflected in one of the central questions of the field, namely: which forms of network nonlocality ought to be viewed as genuine to the network structure?
While all forms of network nonlocality elude the model in Eq.~\eqref{Eqlocal}, some manifestations are more contrived in the sense that they trace back to standard Bell nonlocality.
Here, following Ref.~\cite{Fritz2012}, we discuss a few typical examples of how standard Bell nonlocality may disguise itself as network nonlocality.

One manifestation of disguised Bell nonlocality is that only a part of the network demonstrates Bell nonlocality. As a simple example, consider the bilocal scenario (recall, Figure~\ref{fig:bilocal}), in which parties Alice, Bob and Charlie obtain the correlations $p(a,b,c|x,y,z)$.
Imagine that Alice and Bob decide to test the CHSH inequality using only the marginal distribution $p(a,b|x,y)$.
In a network local model \eqref{Eqlocal}, in which the correlations take the form in Eq.~\eqref{bilocal}, this distribution becomes
\begin{align}
  p(a,b|x,y)&=\sum_c \Big[\int d\lambda_1d\lambda_2 \mu_1(\lambda_1)\mu_2(\lambda_2) \notag\\
  & \qquad \qquad \times p(a|x,\lambda_1)p(b|y,\lambda_1,\lambda_2)p(c|z,\lambda_2)\Big] \notag\\
  =&\int\!\!d\lambda_1 \mu_1(\lambda_1)p(a|x,\lambda_1)\!\!\int\!\!d\lambda_2 \mu_2(\lambda_2) p(b|y,\lambda_1,\lambda_2) \notag\\
  =&\int d\lambda_1\mu(\lambda_1)p(a|x,\lambda_1)p(b|y,\lambda_1),
\end{align}
where in the second line we have used that $\sum_c p(c|z,\lambda_2)\,{=}\,1$ since it is a probability distribution, and in the third line we have used that $\int d\lambda_2 \mu(\lambda_2) p(b|y,\lambda_1,\lambda_2)\,{=}\,p(b|y,\lambda_1)$. The result is precisely the standard local model in Eq.~\eqref{BellLocal}.
Therefore, if $p(a,b|x,y)$ is nonlocal in the standard sense, then the distribution $p(a,b,c|x,y,z)$ has no network local model~\cite{Branciard2012}.
Thus, even extreme cases like the distribution obtained when Alice and Bob violate a Bell inequality and Charlie outputs a random number does not admit a network local model.
This is an instance of standard Bell nonlocality disguised as network nonlocality.
Notably, this is not restricted to our example of the bilocal scenario but readily extends to any network.

\begin{figure*}
  \centering
  \hfill
  \begin{minipage}[c]{0.4\textwidth}
    \subfloat[\label{fig:BellRNG}]{
    \begin{overpic}[width=0.7\textwidth]{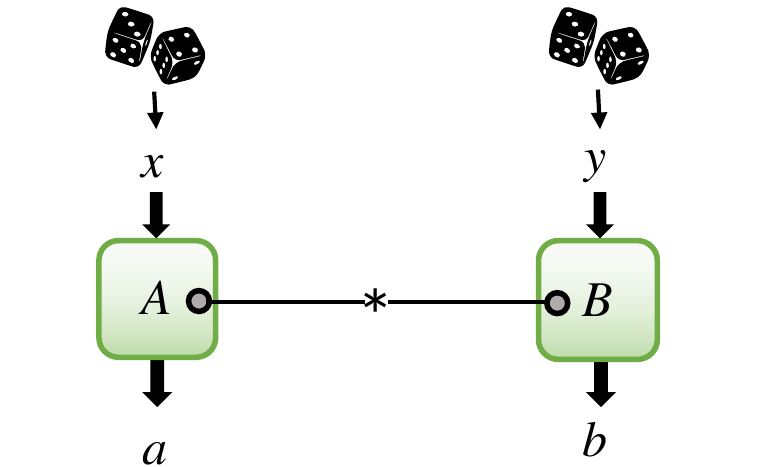}
      \put(105,50){\rotatebox{30}{\huge\faArrowRight}}
      \put(105,20){\rotatebox{-30}{\huge\faArrowRight}}
    \end{overpic}
  }
  \end{minipage}
  \renewcommand\thesubfigure{b\arabic{subfigure}}
  \setcounter{subfigure}{0}
  \begin{tabular}{l}
    \subfloat[\label{fig:ChainRNG}]{
      \includegraphics[width=0.32\textwidth]{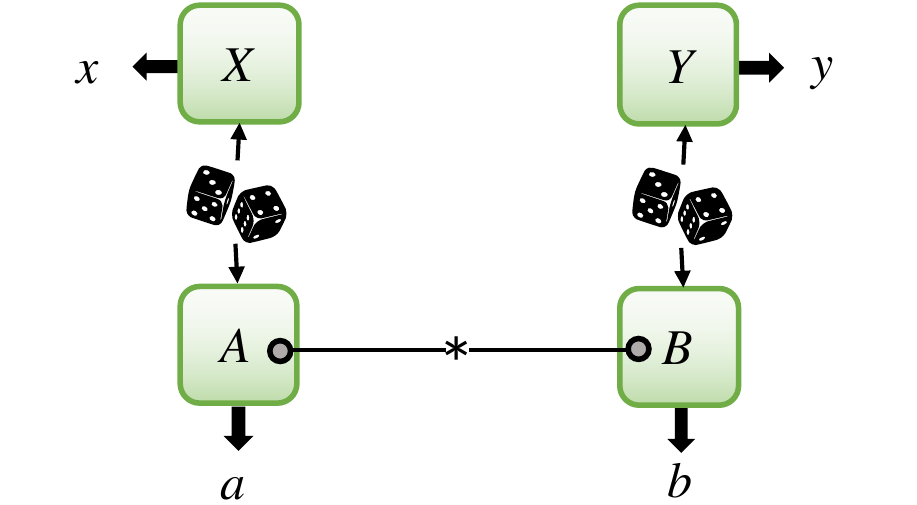}
    }
    \\
    \subfloat[\label{fig:TriangleRNG}]{
      \hspace*{0.47cm}
      \includegraphics[width=0.265\textwidth]{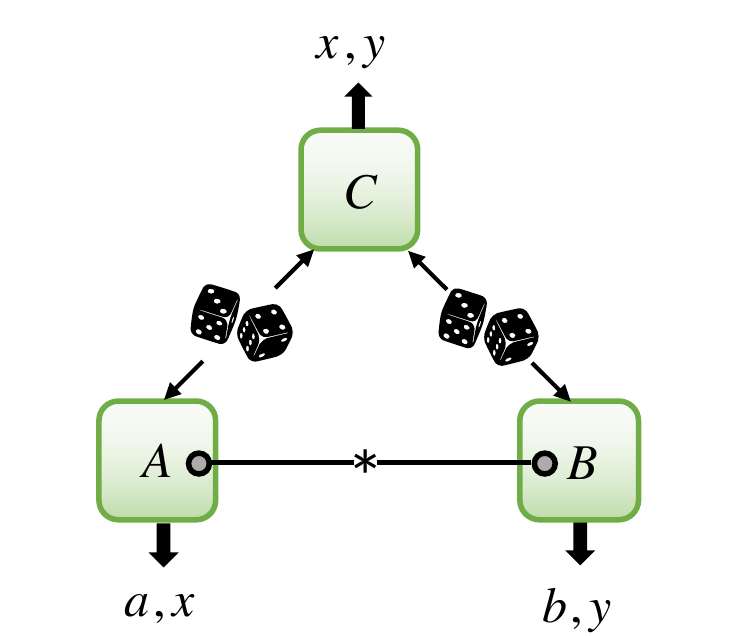}
    }
  \end{tabular}
  \hspace*{3cm}
  \caption{Bell nonlocality disguised as network nonlocality. (a) The standard bipartite Bell scenario, in which the inputs are obtained from two independent random number generators. (b1) Four-party line network without inputs, isomorphic to the standard Bell scenario by reinterpreting the outputs of the random number generators as outputs of new parties. Note that this reinterpretation does not require any physical modification of a setup. (b2) Triangle network strategy to disguise the standard Bell scenario.}
 \label{fig:IsomorphismBellNonLocNetwNonloc}
\end{figure*}

Disguised Bell nonlocality is not always as explicit as in the above example. Sometimes, it may be concealed and distributed among all the parties in the network. A prime example is the possibility to establish a correspondence between the correlations arising in a standard Bell scenario and correlations arising in a four-party network in the shape of a line \cite{Branciard2012,Fritz2012}. In the Bell scenario, Alice and Bob each use a random number generator to decide their inputs $x$ and $y$ (see Figure~\ref{fig:BellRNG}). Then, they perform their measurements and obtain the correlations $p(a,b|x,y)$. These correlations can always be mapped to a network scenario in which the inputs are replaced by new parties, connected respectively to Alice and Bob via new bipartite sources. This yields a four-party line network in which parties have no inputs and $x$ and $y$ now act as the outputs of the new parties (see Figure~\ref{fig:ChainRNG}). It is now possible to generate network correlations $p(x,a,b,y)$ that are network local (resp. network quantum) if and only if $p(a,b|x,y)$ is local (resp. quantum). This follows from re-interpreting Alice's and Bob's private random number generators as sources of shared randomness from which the new parties in the network deterministically produce their output.
Thanks to the shared randomness, Alice and Bob can then recover $x$ and $y$ respectively and proceed to perform their standard Bell inequality test (see Ref.~\cite{Fritz2012} for a proof).
More generally, such an argument applies to any experiment with inputs in which the input $x_j$ of one party $A_j$ is replaced by a new party $X_j$ and a new source that connects only to parties $X_j$ and $A_j$.
Interestingly, this also highlights an important difference between network nonlocality and standard Bell nonlocality: the latter always requires the parties to have inputs whereas the above examples show that, with the source independence assumption, the former is possible even when no party has an input.

In a similar spirit, another type of map is possible between the Bell scenario and a network Bell scenario.
This time, the inputs of both Alice and Bob are interpreted as the outputs of a single new party who is independently connected to Alice and Bob via two new sources.
This leads to a triangle-shaped network (see Figure~\ref{fig:TriangleRNG}).
Again, the argument relies on interpreting Alice's and Bob's random number generators as classical sources that distribute, say, one bit of shared randomness.
This allows the new party, call it Charlie, to learn both the settings of Alice and Bob, which form his output, $\tilde{c}\,{=}\,(x,y)$.
Similary, since Alice and Bob have now received $x$ and $y$ from the respective new sources, they perform e.g.~a test of the CHSH inequality and output $\tilde{a}\,{=}\,(a,x)$ and $\tilde{b}\,{=}\,(b,y)$.
If the correlations $p(a,b|x,y)$ violate the CHSH inequality, then the correlations $p(\tilde{a},\tilde{b},\tilde{c})$ do not admit a local model in the triangle network \cite{Fritz2012}.
In the case of maximal quantum violation of the CHSH inequality, the network nonlocal distribution $p(\tilde{a},\tilde{b},\tilde{c})$ is known as the Fritz distribution, which is given explicitly in Eq.~\eqref{eq:fritz}.
However, all the nonlocality stems from the entanglement distributed between Alice and Bob and therefore is, in essence, that of a CHSH inequality test.
In fact, such disguised Bell nonlocality can elude network local models even when the three sources are subject to classical correlations~\cite{Supic2020independence}.
Arguments of the type seen here extend to more general network scenarios as well.

Notice that these different examples of disguised network nonlocality can be combined and therefore apply to a vast variety of network scenarios. This further emphasises the importance of distinguishing more genuine forms of network nonlocality. Finally, it is interesting to point out that, within quantum theory, entangled measurements may not be necessary in order go beyond disguised network nonlocality. This plausibility is due to the fact that quantum theory features a phenomenon known as ``nonlocality without entanglement'', whereby there exist measurements whose eigenstates are product states but nevertheless cannot be reliably distinguished by two separated parties, even when classical communication is allowed \cite{Bennett1999}.

\subsection{The bilocal scenario}
\label{sec:bilocality}
The bilocal scenario is the simplest network scenario.
It corresponds to the scenario underlying entanglement swapping experiments (see e.g.~Refs.~\cite{Zukowski1993, Zukowski1995, Bose1999}).
It features three parties who receive random and independent inputs $x$, $y$ and $z$, respectively, and produce outcomes $a$, $b$ and $c$.
The experiment gives rise to correlations contained in a conditional probability distribution of the form $p(a,b,c|x,y,z)$.
The network features two independent sources: one emits a system shared between Alice and Bob, and the other emits a system shared between Bob and Charlie (see Figure~\ref{fig:bilocal}).
Therefore, Alice and Charlie \textit{a priori} share no correlations. Due to the lack of a shared source between Alice and Charlie, it is said that they are conditionally independent.
Conditional independence refers to independence with respect to the inputs of a party and the systems received from the sources\footnote{In the bilocal scenario, Alice and Charlie are conditionally independent because, even when their respective inputs and the systems sent by the sources are known, the joint distribution of their outcomes factorises. Note that, in contrast, Alice's and Charlie's outcomes are not independent if one conditions on Bob's outcome.}.
This means that regardless of the physical model considered, the marginal distribution between Alice and Charlie must factor into $p(a,c|x,z)\,{=}\,p(a|x)p(c|z)$.
Conditional independence is the key property exploited in many of the methods that analyse correlations in networks.

While nonlocality in the bilocal scenario has been studied earlier in the context of entanglement swapping, the concept of network local correlations in the bilocal scenario was first proposed in Ref.~\cite{Branciard2010}. The probability distribution $p(a,b,c|x,y,z)$ admits a network local model (or, for simplicity, a bilocal model) if it can be written in the form
\begin{multline}\label{bilocal}
  p(a,b,c|x,y,z)=\int d\lambda_1\mu_1(\lambda_1)\int d\lambda_2 \mu_2(\lambda_2) \\
  \times p(a|x,\lambda_1)p(b|y,\lambda_1,\lambda_2)p(c|z,\lambda_2).
\end{multline}
This is Eq.~\eqref{Eqlocal} in the context of the bilocal scenario.
To show that $p(a,b,c|x,y,z)$ does not admit a bilocal model, one must show that no decomposition of the form \eqref{bilocal} exists.
This is often a nontrivial matter.
Nevertheless, there are to date several approaches to detect non-bilocal correlations.
On the one hand, one may opt for methods tailored for more general networks, such as the inflation technique (detailed in Section~\ref{Inflation}) or scalar extension (detailed in Section~\ref{sec:scalarextension}).
On the other hand, one can attempt to construct explicit bilocal Bell inequalities, i.e.~inequalities defined for probability distributions in the bilocal scenario that are satisfied by all bilocal correlations.
Here, we focus on bilocal Bell inequalities and save the discussion of more general methodology, of which the bilocal scenario is a special case, for later.

Before proceeding, it is relevant to note that standard entanglement swapping protocols can enable a restricted form of nonlocality in the bilocal scenario.
In such a protocol, Bob performs a standard entanglement swapping measurement \cite{Pan1998} and then Alice and Charlie test the CHSH inequality conditioned on just one of Bob's outcomes \cite{Jennewein2001}.
Motivated by benchmarking the success of entanglement swapping, rather than by nonlocality in networks, many such experiments, which we detail in Section~\ref{section5}, have been performed before Bell nonlocality in networks emerged as a topic of its own.
The focus of this section will, however, be on more general approaches to the bilocal scenario, which do not target direct tests of standard Bell inequalities between Alice and Charlie via entanglement swapping protocols.

\subsubsection{Nonlocality and the Bell state measurement} \label{subsubsec:BilocCor}
The first example of a bilocal Bell inequality that was conceptually different from testing a standard Bell inequality after entanglement swapping, was reported in Ref.~\cite{Branciard2010} and later shown to be a special instance of a more general bilocal Bell inequality derived in Ref.~\cite{Branciard2012}. Here we present the latter bilocal Bell inequality.
Consider the bilocal scenario with Alice and Charlie each having two possible inputs, $x,z\,{\in}\,\{0,1\}$ and each producing one of two possible outputs  $a,c\,{\in}\,\{0,1\}$.
Bob has a fixed input and produces one of four possible outputs, which for simplicity are represented as two bits $b\,{=}\,(b_0,b_1)\,{\in}\,\{0,1\}^2$.
The correlations can therefore be written  as $p(a,b,c|x,z)$, where the label $y$ has been omitted. Next, define linear combinations of the probabilities as
\begin{align}
  \label{bilocexpectation}\nonumber
  & I_+\coloneqq  \frac{1}{4}\expect{\hat{A}_+B^0\hat{C}_+}=\frac{1}{4}\sum_{a,b,c,x,z}(-1)^{a+b_0+c}p(a,b,c|x,z),\\
  & I_-\coloneqq \frac{1}{4}\expect{\hat{A}_-B^1\hat{C}_-}\!=\!\frac{1}{4}\!\!\!\sum_{a,b,c,x,z}\!\!\!\!(-1)^{a+b_1+c+x+z}p(a,b,c|x,z),
\end{align}
with shorthand notations $\hat{A}_\pm\,{\coloneqq}\,A_0\pm A_1$ and $\hat{C}_\pm\,{\coloneqq}\,C_0\pm C_1$, where $A_x$ and $C_z$ are the observables of Alice and Charlie. In any physical model respecting the independence of the sources, one can use that $|\expect{B^0}|,|\expect{B^1}|\,{\leq}\, 1$ to obtain  $|I_\pm|\leq \frac{1}{4}\expect{|\hat{A}_\pm|}\expect{|\hat{C}_\pm|}$, where the expectation values are over $\lambda_1$ and $\lambda_2$ respectively. Then, the Cauchy-Schwarz inequality gives
\begin{equation}
  \label{brgpstep}
  \sqrt{|I_+|}+\sqrt{|I_-|}\leq \frac{1}{2}\sqrt{\expect{|\hat{A}_+|+|\hat{A}_-|}}\sqrt{\expect{|\hat{C}_+|+|\hat{C}_-|}}.
\end{equation}
The expressions under the square roots are similar to those encountered previously in our discussion of the CHSH inequality. In a network local model, it holds that  $\expect{|\hat{A}_+|+|\hat{A}_-|}\leq 2$  and similarly for the second square root. This leads to the bilocal Bell inequality of BRGP \cite{Branciard2012}, which reads
\begin{equation}
  \label{BRGP}
  \mathcal{S}_{\text{BRGP}}\coloneqq \sqrt{|I_+|}+\sqrt{|I_-|}\leq 1.
\end{equation}
Observing $\mathcal{S}_{\text{BRGP}}\,{>}\,1$ in an experiment implies that no bilocal model of the generated correlation is possible\footnote{As in the case of inequalities in Bell scenarios, in reality Eq.~\eqref{BRGP} represents a whole family of inequalities, generated by relabeling inputs and outputs in the parties' measurements. A violation of any of them is a sufficient condition for incompatibility with a bilocal model.}. Importantly, however, a violation of the BRGP inequality is not to be considered a necessary condition for non-bilocal correlations; it is highly plausible that there exist correlations that do not admit a bilocal model that nevertheless do not violate the  BRGP inequality.

Moreover, Ref.~\cite{Branciard2012} presents two more bilocal Bell inequalities, for different values of Bob's available inputs and outputs.
While these are mathematically similar to the above BRGP inequality, they are tailored for quantum protocols based on separable measurements and partial entanglement swapping measurements, respectively.

The most interesting quantum protocols that violate the BRGP inequality are based on entanglement swapping, i.e.~a procedure in which Bob measures his two independent systems in an entangled basis in such a way that the state shared between Alice and Charlie becomes entangled. Let each source emit  a copy of the maximally entangled state, i.e.~Alice and Bob share $\ket{\phi^+}_{\text{AB}_1}$ and Bob and Charlie share $\ket{\phi^+}_{\text{B}_2\text{C}}$. Bob performs a complete Bell state measurement of his two particles. This paradigmatic measurement (famous from e.g.~quantum teleportation \cite{Teleportation} and dense coding \cite{Bennett1992}) is a projection in a basis $\{\ket{\psi_{b_0b_1}}\}$ of maximally entangled two-qubit states, namely
\begin{equation}\label{BSM}
  \begin{split}
    & \ket{\psi_{00}}=\frac{\ket{00}+\ket{11}}{\sqrt{2}}, \qquad \qquad\ket{\psi_{01}}=\frac{\ket{01}+\ket{10}}{\sqrt{2}},\\
    & \ket{\psi_{10}}=\frac{\ket{00}-\ket{11}}{\sqrt{2}}, \qquad \qquad \ket{\psi_{11}}=\frac{\ket{01}-\ket{10}}{\sqrt{2}}.
  \end{split}
\end{equation}
These states are also interchangably denoted $\{\ket{\phi^+},\ket{\psi^+},\ket{\phi^-},\ket{\psi^-}\}$ respectively.
Depending on his output, Bob renders Alice's and Charlie's initially independent qubits in one of these maximally entangled states, $\ket{\psi_{b_0b_1}}_{AC}$; but they cannot know which of the four states they hold due to the randomness in Bob's outcome.
If Alice and Charlie each measure the anticommuting observables $A_0\,{=}\,C_0\,{=}\,\frac{\sigma_1+\sigma_3}{\sqrt{2}}$ and  $A_1\,{=}\,C_1\,{=}\,\frac{\sigma_1-\sigma_3}{\sqrt{2}}$, one obtains the violation $\mathcal{S}_\text{BRGP}\,{=}\,\sqrt{2}\,{>}\,1$.
Hence, no bilocal model can account for the predictions of quantum theory.
Interestingly, it is also possible to violate the BRGP inequality (but to the lesser magnitude of $\mathcal{S}_{\text{BRGP}}\,{=}\,2^{1/4}\,{>}\,1$) using only separable measurements for Bob \cite{Saunders2017}.
Thus, perhaps surprisingly, a violation does not imply entanglement swapping in the quantum network.
The reason for this traces back to disguised Bell nonlocality; separable measurements can be used to achieve standard Bell nonlocality between Alice-Bob and Bob-Charlie respectively, which to some extent enables violations of the BRGP inequality.

The Bell state measurement is a natural component of a quantum protocol for violating the BRGP inequality.
However, it was shown in Refs.~\cite{Gisin2017,Andreoli2017} that any such protocol can fundamentally be understood in terms of the CHSH inequality.
The key observation is that while the Bell state measurement itself indeed enables entanglement swapping, it acts in a separable manner on the level of the quantities $I_\pm$.
Specifically $\sum_{b_0,b_1} (-1)^{b_0} \ketbra{\psi_{b_0b_1}}{\psi_{b_0b_1}}\,{=}\,\sigma_1\,{\otimes}\,\sigma_1$ and $\sum_{b_0,b_1} (-1)^{b_1} \ketbra{\psi_{b_0b_1}}{\psi_{b_0b_1}}\,{=}\,\sigma_3\,{\otimes}\,\sigma_3$.
If the sources emit arbitrary two-qubit states $\rho_{AB_1}$ and $\rho_{B_2C}$ respectively, one finds that
\begin{equation}
  \mathcal{S}_{\text{BRGP}}\leq \frac{1}{2}\sqrt{\mathcal{S}_\text{CHSH}(A_0,A_1)_{\rho_{AB_1}}}
  \sqrt{\mathcal{S}_\text{CHSH}(C_0,C_1)_{\rho_{B_2C}}},
\end{equation}
where the quantities under the square roots are the CHSH parameters \eqref{CHSH} obtained when Alice or Charlie tests the CHSH inequality with Bob and the latter performs the measurements $\sigma_1$ and $\sigma_3$. Refs.~\cite{Gisin2017, Andreoli2017} used this reduction to the CHSH scenario in order to prove that all pure entangled states can violate the BRGP inequality. Moreover, these works gave a general criterion (similar to the Horodecki criterion for the CHSH inequality \cite{Horodecki1995}) for the magnitude of the quantum violation obtainable from a given pair of two-qubit states. Such an analysis, restricted to a particular class of two-qubit states, was derived earlier in Ref.~\cite{Mukherjee2016}.

An important aspect of Bell-type experiments is the quantification of nonlocal correlations, which can be achieved in several ways.
In standard Bell scenarios two operational approaches are particularly notable: (i) determining the amount of isotropic noise that the entangled state can be exposed to before the correlations admit a local model (see e.g.~Refs.~\cite{Kaszlikowski2000,Acin2002}), and (ii) determining the amount of superluminal communication with which a local model must be supplemented in order to simulate quantum correlations (see e.g.~Refs.~\cite{Maudlin1992,Brassard1999,Steiner2000,Bacon2003,Toner2003,Barrett2011}).
Exploration of the second approach in quantum networks is an interesting problem which has so far only been explored for particular protocols based on entanglement swapping in the bilocal scenario \cite{BranciardBrunner2012}.
The former approach, however, straightforwardly extends to network scenarios. For the violation of the BRGP inequality, consider that each source emits a mixture of the maximally entangled state and isotropic noise,
\begin{equation}\label{isotropic}
  \rho_v^{\text{iso}}\coloneqq v\ketbra{\phi^+}{\phi^+}+\frac{1-v}{4}\id,
\end{equation}
where $v\,{\in}\,[0,1]$ is the visibility of the state in each source.
The critical visibility per source required to violate the inequality is $v_{\text{crit}}\,{=}\,\frac{1}{\sqrt{2}}$.
This follows immediately from the connection with the CHSH inequality, for which the critical visibility also is $v_{\text{crit}}\,{=}\,\frac{1}{\sqrt{2}}$.
Importantly, however, the BRGP inequality offers an advantage over a conventional entanglement swapping protocol (i.e.~when Alice and Charlie test the CHSH inequality post-selected on one of Bob's outputs): in such a protocol, the critical visibility per source required to violate the CHSH inequality between Alice and Charlie is $v_{\text{crit}}\,{=}\,2^{-\frac{1}{4}}$.

\subsubsection{Nonlocality and the Elegant Joint Measurement}\label{subsubsec:NonlocEJM}
Whereas quantum nonlocality in the bilocal scenario is expected to be qualitatively different from standard Bell nonlocality, we have seen that quantum protocols for the BRGP inequality essentially boil down to separate tests of the CHSH inequality. This raises the question of whether non-bilocal correlations that bear no immediate resemblance to standard Bell nonlocality can be generated using entanglement swapping measurements different from the Bell state measurement.

One such entanglement swapping measurement was proposed in Ref.~\cite{Gisin2019}.
This, so-called Elegant Joint Measurement, is a projection of two qubits in a basis of equally entangled states.
These states have the symmetry that if either qubit is lost, the four reduced states form a regular tetrahedron inside the Bloch sphere.
The Elegant Joint Measurement, $\{\ket{e_b}\}_{b=1}^4$, is a projection onto the following basis:
\begin{equation}\label{EJM}
  \begin{split}
    & \ket{e_1}\coloneqq \frac{1}{\sqrt{8}}(-1-i,0,-2i,-1+i)^\dagger,
    \\
    & \ket{e_2}\coloneqq\frac{1}{\sqrt{8}}(1-i,2i,0,1+i)^\dagger,
    \\
    &\ket{e_3}\coloneqq\frac{1}{\sqrt{8}}(-1+i,2i,0,-1-i)^\dagger,
    \\
    & \ket{e_4}\coloneqq\frac{1}{\sqrt{8}}(1+i,0,-2i,1-i)^\dagger.
 \end{split}
\end{equation}
This measurement was first introduced in Ref.~\cite{Gisin1999} in the context of encoding quantum information in antiparallel spins. It also admits another operational interpretation as the optimal projective two-qubit measurement for discriminating the reduced states of either qubit \cite{Czartowski2020}. This, together with its connection to symmetric informationally complete measurements \cite{Renes2004}, has enabled its generalisation to higher-dimensional systems \cite{Czartowski2020}. Another generalisation  extends it to a one-parameter family of two-qubit measurements in which the radius of the reduced-state tetrahedron can be continuously tuned. This allows one to construct measurements intermediate between the Elegant Joint Measurement and the Bell state measurement \cite{Tavakoli2020}. A relevant open problem consists in generalising the Elegant Joint Measurement to systems of many qubits.

The Elegant Joint Measurement enables quantum correlations in the bilocal scenario that are qualitatively different from those obtained through the Bell state measurement, and can be detected through the violation of bilocal Bell inequalities that do not, in any apparent manner, reduce to a known standard Bell inequality. Such bilocal Bell inequalities were presented in Ref.~\cite{Tavakoli2020}. Consider a scenario in which Alice and Charlie have three inputs each and binary outcomes, while Bob has a single input and four outcomes. Bilocal correlations in this scenario are restricted by the following bilocal Bell inequality:
\begin{multline}
  \label{TGB1}
  \mathcal{S}_{\text{TGB}}\coloneqq \frac{1}{3}\left(\sum_{y=z}\expect{B^yC_z}-\sum_{x=y}\expect{A_xB^y}\right)\\
  -\sum_{x\neq y\neq z\neq x}\expect{A_xB^yC_z}\leq 3,
\end{multline}
which is valid when all the (one-, two- and three-party) expectation values that do not appear in the inequality are zero\footnote{The index $y\,{=}\,1,2,3$ labels the three bits $b\,{=}\,(b_1,b_2,b_3)\in\{-1,+1\}^3$ satisfying the constraint \mbox{$b_1b_2b_3\,{=}\,1$}, used to denote Bob's four-valued outcome. The vanishing auxiliary expectation values are the expressions $\expect{A_x}$, $\expect{B^y}$, $\expect{C_z}$, $\expect{A_xB^y}$, $\expect{A_xC_z}$, $\expect{B^yC_z}$, $\expect{A_xB^yC_z}$ not appearing in the inequality.}.
This inequality has been proven to hold for bilocal models satisfying additional symmetries, and numerical evidence supports that it holds also for general bilocal models.
Nevertheless, a fully general analytical proof is still lacking.
Note that while the inequality is linear, it is nevertheless a nontrivial constraint on bilocal correlations due to the restrictions imposed on the auxiliary expectation values.
Notably Ref.~\cite{Tavakoli2020} also proposed other, nonlinear, inequalities that can discriminate both quantum and post-quantum correlations.

The bilocal Bell inequality \eqref{TGB1} is tailored to a quantum protocol based on Bob implementing the Elegant Joint Measurement.
If the sources each distribute a singlet state and Alice and Charlie each measure the three Pauli observables, one obtains the violation $\mathcal{S}_{\text{TGB}}\,{=}\,4$.
In contrast, no protocol based on the Bell state measurement achieving a violation of the inequality is currently known.
Assessing the strength of the inequality via the visibility of two-qubit Werner states\footnote{Note that these are equivalent to the states of Eq.~\eqref{isotropic} up to a local unitary.} \cite{Werner1989}, the critical visibility for a violation is found to be $v_\text{crit}\,{=}\,\frac{\sqrt{37}-1}{6}\,{\approx}\,0.847$, but the best known bilocal model for the full distribution has a critical visibility of $v_\text{crit}\,{\approx}\, 0.791$.
This indicates that tighter bilocal Bell inequalities could still be possible.
Furthermore, the optimality of the quantum violation remains an open problem.
Finally, the visibility per source needed for a violation is lower than that of the swapped state to violate the CHSH inequality.

\subsubsection{Explicit bilocal decompositions}\label{sec:Simulation}
How can one detect correlations that admit a bilocal model?
One answer is to find all bilocal Bell inequalities, thus characterising the boundary of the bilocal set of correlations, and checking that they are all satisfied.
This is, however, not a practical approach since no method is known for systematically finding these inequalities.
A more straightforward answer is to construct an explicit bilocal decomposition \eqref{bilocal}.
To this end, one may exploit that the general bilocal model \eqref{bilocal} can without loss of generality be taken as a deterministic bilocal model, i.e.~
\begin{align}\nonumber
  p(a,b,c|x,y,z)=&\sum_{\lambda_1,\lambda_2}p_1(\lambda_1)p_2(\lambda_2) \\\label{detbiloc}
  & \times D^{\text{A}}_{\lambda_1}(a|x) D^{\text{B}}_{\lambda_{1}\lambda_2}(b|y)D^{\text{C}}_{\lambda_2}(c|z),
\end{align}
where $p_1$ and $p_2$ are arbitrary probability distributions and $(D^{\text{A}},D^{\text{B}},D^{\text{C}})$ are deterministic response functions, indexed for each of the parties by their relevant local variable. Denote the number of inputs and outputs for Alice by $s_\text{A}$ and $o_\text{A}$: there are only $s_\text{A}^{o_\text{A}}$ deterministic response functions for Alice, and similarly for Bob and Charlie. Therefore, one can limit the cardinality of the relevant local variable accordingly. Consequently, the matter of simulating a given correlation is reduced to finding a suitable pair of probability distributions $p_1$ and $p_2$. Nonetheless, this can still not be straightforwardly evaluated due to the non-convexity of the problem.

It is possible to re-formulate the deterministic bilocal model \eqref{detbiloc} such that it becomes handier to search for an explicit bilocal decomposition.
This possibility is explored in Ref.~\cite{Branciard2012}, where bilocal models are analysed on the level of correlators, as opposed to probabilities.
Although this approach readily extends to some larger networks and also to other input/output scenarios, we exemplify it here for a simple case and note that its generalisation is conceptually straightforward.

Let Alice, Bob and Charlie have binary outputs and associate the bit strings $\alpha\,{=}\,\alpha_1\ldots \alpha_{s_\text{A}}$,  $\beta\,{=}\,\beta_1\ldots \beta_{s_\text{B}}$ and $\gamma\,{=}\,\gamma_1\ldots \gamma_{s_\text{C}}$ to the deterministic responses of each party for each measurement.
We write $q_{\alpha\beta\gamma}$ for the probability of adopting this deterministic strategy.
From these probabilities, define correlators through the discrete Fourier transform
\begin{equation}\label{FT}
  e_{ijk}=\sum_{\alpha,\beta,\gamma}(-1)^{\alpha\cdot i+\beta\cdot j+\gamma\cdot k}q_{\alpha\beta\gamma},
\end{equation}
where $i$, $j$ and $k$ also are bit strings. The correlators are now one-to-one with the probabilities. Following common practice, one may define $A_x\,{=}\,(-1)^{\alpha_x}$, $B_y\,{=}\,(-1)^{\beta_y}$ and $C_z\,{=}\,(-1)^{\gamma_z}$ and therefore write the correlators as  expectation values
\begin{equation}
  e_{ijk}=\expect{A_1^{i_1}\ldots A_{s_\text{A}}^{i_{s_\text{A}}}B_1^{j_1}\ldots B_{s_\text{B}}^{j_{s_\text{B}}}C_1^{k_1}\ldots C_{s_\text{C}}^{k_{s_\text{C}}}}.
\end{equation}
The correlators can now be divided into two sets: those that are fixed by the given $p(a,b,c|x,y,z)$ and those that are not. This handy distinction can be made because some correlators are physically relevant (e.g.~$\expec{A_1B_1C_2}$) whereas others are not (e.g.~$\expec{A_1B_1C_2C_1}$). The formers correspond to the cases when $i$, $j$ and $k$ each contain at most one element ``$1$''  and the latters correspond to all other cases, which may now be viewed as degrees of freedom in the bilocal model. Restricting to these degrees of freedom, the conditions for a bilocal simulation can be expressed straightforwardly. Firstly, since the probabilities must remain positive, one reverses the Fourier transform in \eqref{FT} and imposes that for all $\alpha$, $\beta$, $\gamma$
\begin{equation}
  q_{\alpha\beta\gamma}=\frac{1}{2^{s_\text{A}+s_\text{B}+s_\text{C}}}\sum_{i,j,k}(-1)^{\alpha\cdot i+\beta\cdot j+\gamma\cdot k}e_{ijk}\geq 0.
\end{equation}
Secondly, to respect the independence of the sources, one imposes that Alice and Charlie are conditionally independent. This amounts to the nonlinear constraint
\begin{equation}\label{quad}
  \forall i,k: \quad e_{i\bar{0}k}=e_{i\bar{0}\bar{0}}e_{\bar{0}\bar{0}k},
\end{equation}
where $\bar{0}=0\ldots 0$. This method has been employed in Refs.~\cite{Branciard2012, Tavakoli2020} to efficiently construct bilocal simulations of noisy quantum correlations. The main reason that the method is more efficient than brute-force search over $p_1$ and $p_2$ in Eq.~\eqref{detbiloc} is that the degrees of freedom in the bilocal model are separated from the constraints implied by $p(a,b,c|x,y,z)$ and that the independence of the sources now appers as a quadratic constraint \eqref{quad} on only some of the correlators. Its practical usefulness in larger networks is yet to be explored.

\subsection{The triangle network}
\label{sec:triangleintro}
Going beyond the bilocal scenario, one may add a third, independent, source connecting Alice and Charlie, resulting in the triangle network captured in Figure~\ref{fig:triangle}. This is the simplest nontrivial scenario that does not feature conditional independence between any pair of parties. Consequently, the main ideas leveraged in Section~\ref{sec:bilocality} to analyse the bilocal scenario do not apply to the triangle scenario, and finding nonlocal correlations in this network (beyond Fritz's example, discussed in Section~\ref{subsec:OverviewReplaceInputs}) is a difficult task.
Following directly from the generic definition in Eq.~\eqref{Eqlocal}, local models in the triangle network admit the form
\begin{align}
	p(a,b,c|x,y,z)&\!=\!\int d\lambda_1\mu_1(\lambda_1)\int d\lambda_2\mu_2(\lambda_2)\int d\lambda_3\mu_3(\lambda_3) \notag\\
	&\times p(a|x,\lambda_1,\lambda_2)p(b|y,\lambda_2,\lambda_3)p(c|z,\lambda_1,\lambda_3).
	\label{eq:localtriangle}
\end{align}
While we already have seen in Section~\ref{subsec:OverviewReplaceInputs} that quantum theory can elude such models, we here focus on qualitatively different approaches to analyzing nonlocality the triangle network.

\subsubsection{Two interesting distributions}
\label{subsec:OverviewTC}
An important example of quantum nonlocality in the triangle scenario was reported in Ref.~\cite{Renou2019}.
There, the authors propose a quantum probability distribution incompatible with any triangle local model \eqref{eq:localtriangle}, which in no apparent way reduces to disguised Bell nonlocality.
Such nonlocality is therefore, arguably, inherently different.
However, the arguments suggesting that the distribution cannot be reduced to a form of disguised Bell nonlocality are only qualitative and intuitive.
This stresses the need for formalising a rigorous concept of genuine network nonlocality that goes beyond  failure to admit the network local model of Eq.~\eqref{Eqlocal}.

The distribution is obtained from the following quantum experiment, where the parties always perform the same measurement (so one can omit the inputs $x$, $y$ and $z$) with four possible outcomes, labeled by $a,b,c\,{\in}\,\{\bar{0}, \bar{1}_0, \bar{1}_1, \bar{2}\}$.
Each source distributes the state $\frac{\ket{10}+\ket{01}}{\sqrt{2}}$, and every party measures in the same basis, given by the vectors
\begin{equation}
  \begin{split}
      \ket{\bar{0}}&\coloneqq\ket{00},\\ \ket{\bar{1}_0}\coloneqq c\ket{01}+s\ket{10},&\qquad \ket{\bar{1}_1}\coloneqq s\ket{01}-c\ket{10},\\ \ket{\bar{2}}&\coloneqq\ket{11},
  \end{split}
\end{equation}
with $c$, $s$ being real parameters that satisfy $c^2\,{+}\,s^2\,{=}\,1$.
The resulting probability distribution, $P_{\mathrm{RGB4}}$, is explicitly given in Eq.~\eqref{eq:genuine} in Appendix~\ref{AppendixDistributions}.
Notice that the measurement can be implemented in two steps.
In a first step, the parties measure the operators associated to $\bar{0}$, $\bar{1}$, $\bar{2}$, where \mbox{$\bar{1}\,{\coloneqq}\, \ketbra{01}{01}\,{+}\, \ketbra{10}{10}$}.
Second, whenever $\bar{1}$ is obtained, the parties measure $\bar{1}_0$, $\bar{1}_1$, which are the only entangled measurement operators.
This reformulation allows to interpret this experiment in terms of token-counting strategies, following Refs.~\cite{Renou2020short,Renou2020long}.
The states $\ket{0}$, $\ket{1}$ can be interpreted as the absence or the presence of a token.
Thus, the sources can be thought of as sending one token each, to the right ($\ket{01}$) or to the left ($\ket{10}$), in superposition.
A party first measures the total number of tokens received, obtaining $\bar{0}$, $\bar{1}$ or $\bar{2}$ tokens, and then performs an additional entangled measurement if needed.
Since each source distributes one token, the sum of token numbers obtained by all the parties is always three.
As we further explain in Section~\ref{subsec:TCandCM}, the rigidity property of classical token-counting strategies in the triangle, which demonstrates that there is essentially a unique way to classically simulate the distribution of the token counts, allows to prove that the resulting probability distribution is nonlocal for $0.785\,{\lesssim}\, c^2\,{<}\,1$.

The key reason that allows the distribution $P_{\RGB4}$ to be proven nonlocal is that it is tailored to uphold convenient properties.
While this is not always desirable, especially when considering experimental implementations, there exist proposals for experiments that demonstrate single-photon nonlocality based on the nonlocality of $P_{\RGB4}$~\cite{abiuso2021singlephoton}.
However, more often than not, one wants to assert the nonlocality of a probability distribution of interest, perhaps arising from more intuitive measurements in the triangle network.
The price of adopting this perspective, i.e.~to investigate the nonlocality of quantum correlations arising from more natural measurements (including fully entangled ones) is that rigorous proofs of nonlocality are difficult.
Interestingly, the distribution generated when the three parties perform a Bell state measurement (naively the most natural candidate for obtaining nonlocality) on shared maximally entangled states, admits a triangle local model\footnote{The distribution can be realised when the sources distribute a four-valued classical variable each, two of the parties output the value of the variable received that is shared with the remaining party, and this remaining party generates its output according to a deterministic strategy defined in advance.} \cite{Gisin2019}.
Therefore, the quest for quantum triangle nonlocality motivates the use of other natural entanglement swapping measurements.

Refs.~\cite{gisin2017elegant,Gisin2019} investigated the case of all three parties performing the Elegant Joint Measurement \eqref{EJM} on maximally entangled states emitted by all three sources. This leads to a very symmetric distribution~\cite{gisin2017elegant}, defined by
\begin{equation}\label{elegantdist}
  p_3=\frac{25}{256}, \qquad p_2=\frac{1}{256}, \qquad p_1=\frac{5}{256},
\end{equation}
where the subindex denotes the number of parties that obtain the same outcome. Based on its symmetry, the fact that its realisation requires complete entanglement swapping measurements and the inability of current approaches at simulating it in a network local model \cite{gisin2017elegant}, this distribution is conjectured to be network nonlocal in a manner that cannot be reduced to disguised Bell nonlocality. Current techniques for identifying network nonlocal correlations~\cite{wolfe2019inflation} have, however, failed to prove the conjecture so far. Also, sophisticated numerics for constructing network local models (further discussed in Section~\ref{sec:neural}) have thus far failed to disprove the conjecture \cite{krivachy2019neural}. Therefore, the question remains open.

\subsubsection{Inflation in the triangle network}
\label{subsec:OverviewInfl}
Inflation, introduced by Wolfe et al. \cite{wolfe2019inflation}\footnote{While Ref.~\cite{wolfe2019inflation} formally introduces and formalises the concept, there is record of the use of inflation-like arguments in earlier works \cite{Henson2014}.}, is a powerful concept that enables the analysis of correlations in any network and, as will be discussed in Section~\ref{sec:causalinference}, also in more general causal structures.
The triangle network is the simplest network for which the methods developed for characterising bilocal correlations, which mostly relied on the conditional independence of Alice and Charlie, fail.
This makes the triangle network the natural scenario to apply inflation methods.
In fact, many notable results in the triangle network, described below, have been obtained via inflation.
Before summarising these results, we briefly illustrate the general principles behind inflation, which are applied in different ways for particular types of correlations throughout the review.

\paragraph*{Inflation through example:} Inflation is an attempt at proof through contradiction.
If a distribution can be generated with some sources and measurement devices in a particular network scenario, then one can consider which kinds of distributions one would be able to generate if given access to multiple copies of such sources and measurement devices.
The networks that are generated by arranging those copies are called inflations of the original network.
Correlations compatible with the inflated networks satisfy two important properties.
First, they are highly symmetric, since they are invariant under suitable permutations of copies of a same original element.
Second, when marginalising over sources and measurement devices that reproduce (collections of) parts of the original network, the marginal distribution coincides with (the product of) the corresponding marginals of the original distribution.
Inflation susbstitutes the question of whether a probability distribution can be generated in a given network to finding distributions compatible with the inflation networks.
The original distribution cannot be generated in the original network if a compatible distribution in any of its inflations is proved not to exist.
While it seems rather unnatural to analyse large inflated networks in order to determine compatibility on simpler networks, it turns out that the symmetries and independences present in inflated networks make its characterisation much easier.

\begin{figure}
  \hfill
  \begin{minipage}[b]{0.45\columnwidth}
    \hspace*{-2.5cm}
    \subfloat[\label{fig:TriangleCInf}]{
    			\begin{minipage}{0.45\columnwidth}
    				\vspace*{0.37cm}
    				\hspace*{-.25cm}
    				\includegraphics[scale=0.32]{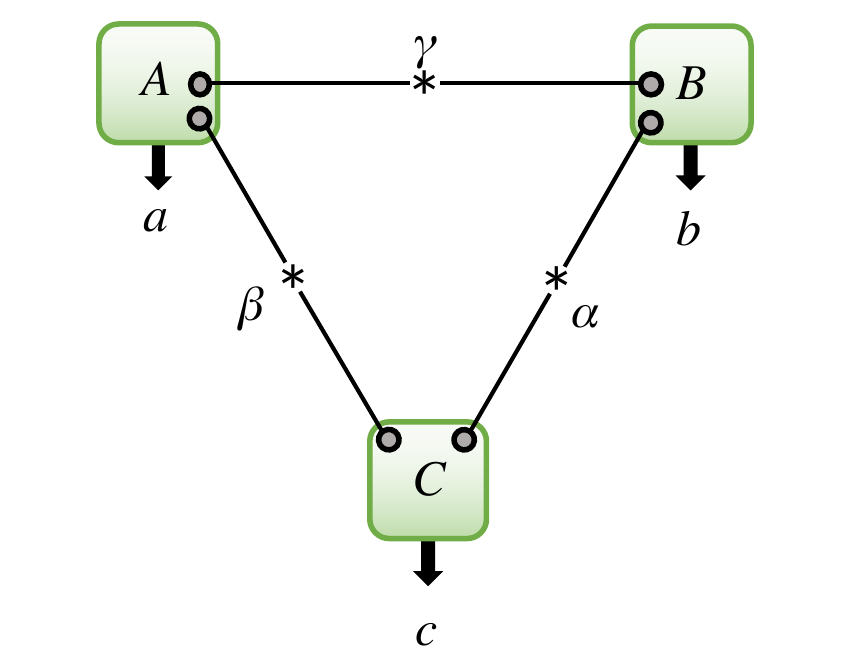}
    			\end{minipage}
    	}
  \end{minipage}
  	\hfill
  	\begin{minipage}[b]{0.49\columnwidth}
    	\subfloat[\label{fig:TriangleCInfInflated}]{
    			\includegraphics[scale=0.32]{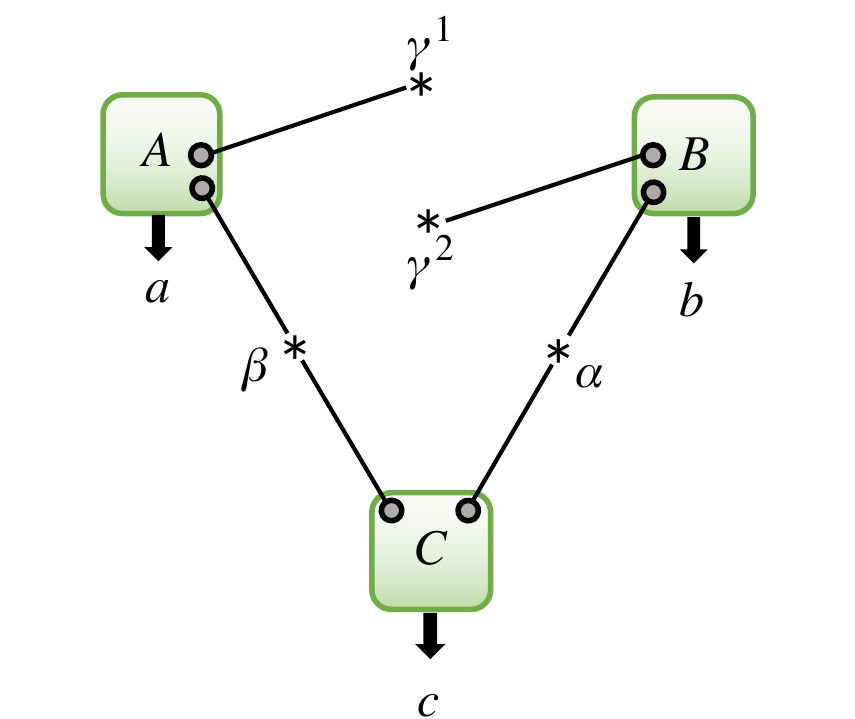}
    	}
  \end{minipage}
  \hfill
  \caption{Illustration of inflation in the triangle scenario. (a) is the triangle network, formed by three parties and three bipartite sources. By measuring the systems that are sent from their respective sources, the parties produce outcomes $a$, $b$ and $c$ according to the probability distribution $P^\text{ABC}_\text{obs}$. (b) A possible inflation considers two copies of the source $\gamma$. One sends a system only to Alice, and the other one sends a system only to Bob. By measuring their respective systems the parties produce outcomes according to the distribution $P^\text{ABC}_\text{inf}$. In both cases, Alice and Charlie receive the same systems from the same type of sources, so $P^\text{AC}_\text{obs}\,{=}\,P^\text{AC}_\text{inf}$. This is not the case for Alice and Bob, since $P^\text{AB}_\text{inf}\,{=}\,P^\text{A}_\text{inf}P^\text{B}_\text{inf}$ (which also equates to $P^\text{A}_\text{obs}P^\text{B}_\text{obs}$), while $P^\text{AB}_\text{obs}\not=P^\text{A}_\text{obs}P^\text{B}_\text{obs}$ in general.
  }
  \label{fig:InflationGHZ}
\end{figure}

In order to illustrate this reasoning, let us show, following Ref.~\cite{wolfe2019inflation}, that a shared random bit (also called the GHZ distribution), given by
\begin{equation}
	P_\text{GHZ}\coloneqq\frac{1}{2}([000]+[111]),\footnote{We use the shorthand $[abc]$ for denoting the product of Kronecker deltas $\delta_{A,a}\delta_{B,b}\delta_{C,c}$, where $A$, $B$ and $C$ are the random variables that represent the outcomes of Alice's, Bob's, and Charlie's measurements, respectively.}
	\label{eq:GHZ}
\end{equation}
cannot be generated in the triangle scenario. Assume by contradiction that the distribution is realisable in the triangle network. This means that there exist sources $\alpha$, $\beta$, $\gamma$ and measurement devices $A$, $B$, $C$, arranged as illustrated in Figure~\ref{fig:TriangleCInf}, that generate $P_\text{GHZ}$. Consider next the inflation in Figure~\ref{fig:TriangleCInfInflated}, where two copies of the source $\gamma$ are available: one sends a system to Alice, and the other sends a system to Bob. The sub-network composed of Alice, Charlie, and the sources that connect to them, is the same in the original network and in the inflation since $\gamma^1$ is an exact copy of $\gamma$, and thus a distribution compatible with the inflation must satisfy $P^\text{AC}_{\text{inf}}\,{=}\,P^\text{AC}_\text{GHZ}\,{=}\,\frac{1}{2}\left([00]+[11]\right)$. The same reasoning applies to the sub-network composed of Bob, Charlie and their sources, so $P^\text{BC}_{\text{inf}}\,{=}\,P^\text{BC}_\text{GHZ}\,{=}\,\frac{1}{2}\left([00] + [11]\right)$. Moreover, given that, in the inflation, Alice and Bob are not connected by a common source, distributions that can be generated in the inflated scenario satisfy $P^\text{AB}_\text{inf}\,{=}\,P^\text{A}_\text{inf}P^\text{B}_\text{inf}$ (which, furthermore, can be identified with $P^\text{A}_\text{GHZ} P^\text{B}_\text{GHZ}$ because the individual sub-networks composed of each party and their respective sources are the same in the inflation and in the original scenario). These three conditions are incompatible: if there are no occurrences of $a\,{\not=}\,c$ nor of $b\,{\not=}\,c$ in the inflated network because of the first two conditions, it must be the case that $a\,{=}\,b$ every time the three parties measure their shares. However, this is impossible if $P^\text{AB}_\text{inf}$ is of a product form as per the third condition. One has reached a contradiction, which proves that the premise ``$P_\text{GHZ}$ can be generated in the triangle scenario'' is false.

Note that the proof above has made no assumption about the nature of the sources $\alpha$, $\beta$ and $\gamma$. Therefore, the conclusion reached is independent of whether they distribute local variables, quantum states or states of more general probabilistic theories. In fact, the concept of inflation is applied in different ways depending on whether one wants to characterise local, quantum, or theory-independent correlations in networks. All these constructions are detailed later in the review, in Sections \ref{sec:cinflation}, \ref{sec:qinflation} and \ref{sec:nonsignaling}, respectively.

\paragraph*{Results in the triangle from inflation:} The original idea of inflation appears in Ref.~\cite{wolfe2019inflation}, followed by a series of applications leading to novel results. Not only Ref.~\cite{wolfe2019inflation} recovers the known result that the GHZ distribution in Eq.~\eqref{eq:GHZ} does not admit a local model in the triangle~\cite{SteudelGHZ}, but also provides the first proof that the so-called W distribution, or $P_\text{W}$, defined as\footnote{This distribution corresponds to the coordinated action of all parties outputting $0$ except one, which outputs $1$.}
\begin{equation}
  P_\text{W}\coloneqq\frac{1}{3}\left([100]+[010]+[001]\right),
  \label{eq:W}
\end{equation}
does not admit such a model. Moreover, Ref.~\cite{wolfe2019inflation} explains how inflation can be used to transform consistency inequalities satisfied by all probability distributions into compatibility inequalities in networks, presenting the complete characterisation of the set of binary-outcome probability distributions that admit a particular inflation of the triangle scenario. This is, Ref.~\cite{wolfe2019inflation} provides a first complete description of a nontrivial relaxation\footnote{Let us stress the fact that the complete characterisation is of a relaxation of the set of triangle local correlations. There may exist binary-outcome correlations that do not admit a model of the form of Eq.~\eqref{eq:localtriangle} but that satisfy all the inequalities of the characterisation provided in Ref.~\cite{wolfe2019inflation}.} of the set of binary-outcome correlations that admit a triangle local model.

The characterisation above was later found in Ref.~\cite{Fraser2018} not to admit quantum violations.
This lead to the conjecture that, for probability distributions with no inputs and binary outcomes, the sets of distributions generated in the triangle network by classical variables and by quantum systems coincide.
Furthermore, Ref.~\cite{Fraser2018} provides, using larger inflations, the first inequalities for the triangle network (with no inputs and four outcomes per party) that admit quantum violations.
While the original argument of Ref.~\cite{Fritz2012} only established network nonlocality in the triangle for a small set of individual distributions, the inequalities in Ref.~\cite{Fraser2018} allow one to witness network nonlocality in a region around these paradigmatic distributions, thus providing noise-tolerant witnesses of nonlocality.

Ref.~\cite{wolfe2019qinflation} presents a modification of the original inflation argument that allows the analysis not only of classical network models, but also quantum network models of the type of Eq.~\eqref{eq:quantummodel}. This ``quantum inflation'' argument is employed to prove that the W distribution can neither be generated by measuring quantum states in the triangle scenario, and to find the amount of noise required in order for it to admit a quantum model. When considering distributions with inputs, the authors prove that the Mermin-GHZ distribution (see \cite[Eq. (12)]{wolfe2019qinflation} for its definition) does not admit a quantum model in the triangle. For the two cases above, and exploiting duality theory of semidefinite programming, they also find witnesses of incompatibility with triangle quantum models in the form of polynomial expressions. Finally, Ref.~\cite{wolfe2019qinflation} finds bounds on well-known multipartite Bell inequalities in the classical and quantum triangle networks, illustrating the existence of gaps between different sets of correlations. A reproduction of the results for two famous multipartite Bell inequalities is given below:

\begin{widetext}
\begin{align}
	\label{eq:mermin}
	&\begin{array}[b]{l}
		\textbf{Mermin's Inequality~\cite{Mermin1990}} \\ \\
		\quad\expec*{A_1 B_0 C_0}+\expec*{A_0 B_1 C_0}+\expec*{A_0 B_0 C_1}-\expec*{A_1 B_1 C_1}
	\end{array}
	\leq \begin{cases}
		2 				& \mathcal{L}^\triangle,\;\mathcal{L}^\bell \\
		3.085^*		& \mathcal{Q}^\triangle \\
		4 & \mathcal{Q}^\bell,\;\mathcal{NS}^\triangle,\;\mathcal{NS}^\bell
	\end{cases}
\\
	\label{eq:svet}
	&\begin{array}[b]{l}
		\textbf{Svetlichny's Inequality~\cite{Svetlichny1987}} \\ \\
		\qquad\expec*{A_1 B_0 C_0}+\expec*{A_0 B_1 C_0}+\expec*{A_0 B_0 C_1}-\expec*{A_1 B_1 C_1}\\
		\qquad-\expec*{A_0 B_1 C_1}-\expec*{A_1 B_0 C_1}-\expec*{A_1 B_1 C_0}+\expec*{A_0 B_0 C_0}
	\end{array}
	\leq \begin{cases}
		4 				& \mathcal{L}^\triangle,\;\mathcal{L}^\bell\\
		4.405^* 	& \mathcal{Q}^\triangle \\
		4\sqrt{2} & \mathcal{Q}^\bell\\
		8 				& \mathcal{NS}^\triangle,\;\mathcal{NS}^\bell
	\end{cases}
\end{align}
\end{widetext}
where $\triangle$ denotes the triangle scenario, $\bell$ refers to the standard tripartite Bell scenario, and $\mathcal{L}$, $\mathcal{Q}$ and $\mathcal{NS}$ represent the sets of local, quantum and theory-independent correlations that can be generated in each of the scenarios, respectively.

The bounds with the asterisk are not necessarily tight, and thus it remains open whether they can be lowered by means of more computational power.
In particular, a very important problem is having a clear understanding of the relations between correlations that do not admit network local models and correlations that are nonlocal according to other, standard definitions of multipartite nonlocality~\cite{Svetlichny1987}.

It is also possible to employ inflation arguments to establish theory-independent limits on the correlations that are achievable in networks, providing a notion of no-signaling in networks.
In a nutshell, no-signaling in networks amounts to the inability of one party to learn the local operations of another (we defer a more detailed discussion on this concept to Section~\ref{sec:nsinflation}). By inflating the network, necessary conditions on this inability can be established. Ref.~\cite{Gisin2020} applies such ideas to the triangle scenario without inputs and binary outcomes, obtaining simple analytical constraints on no-signaling correlations in the network.

\section{Methods for general networks}
\label{section3}
The fauna of networks rapidly grows more diverse as one ventures beyond the three-party scenarios considered in the previous section. Here we review methods of considerable generality that apply to the characterisation of correlations in broad, or even fully general, classes of networks. Throughout the discussion, we exemplify the various methods in concrete network scenarios.

\subsection{Conditional independence}
\label{subsec:CondIndep}
\begin{figure}
	\subfloat[\label{fig:star}]{
		\begin{minipage}[b]{0.49\textwidth}
			\includegraphics[scale=0.35]{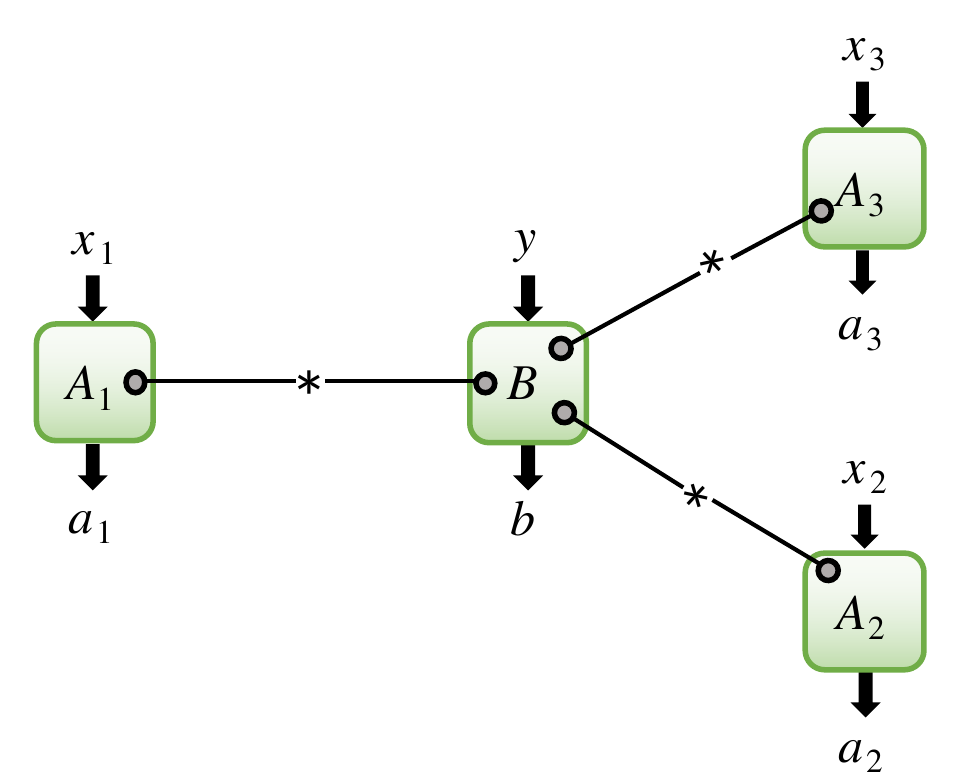}
		\end{minipage}
	}
	\hfill
	\\
	\subfloat[\label{fig:chain}]{
		\begin{minipage}{0.49\textwidth}
			\includegraphics[scale=0.35]{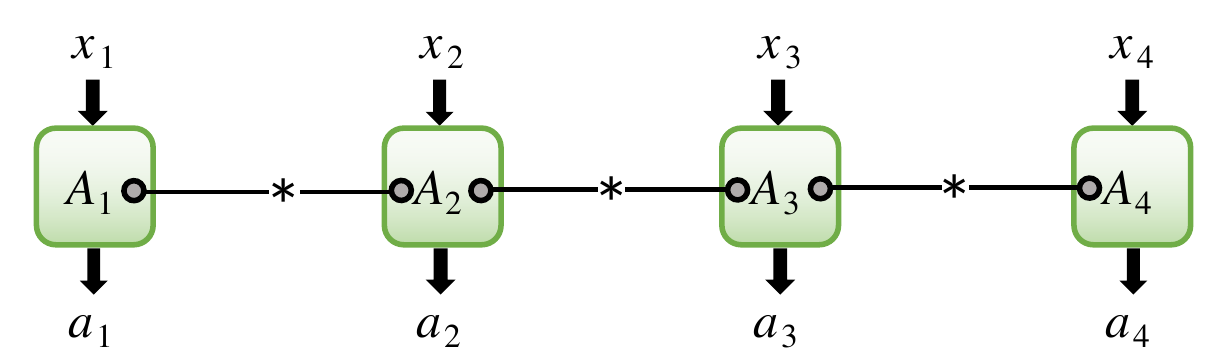}
		\end{minipage}
	}
	\caption{Families of networks with conditional independences. (a) Star network with $m\,{=}\,3$ sources. Upon tracing out the central party, the remaining become independent of each other. (b) Chain network with $m\,{=}\,3$ sources. Tracing any non-extremal party separates the remaining in two sets that are independent of each other.}\label{FigureStarAndChain}
\end{figure}

Any network in which at least two parties do not receive shares from the same source contains a degree of conditional independence.
This property is independent from the physical model: by marginalising over some parties, other parties become independent.
Evidently, most networks are of this type; notable examples are parties that are pairwise connected in a chain configuration or parties that are pairwise connected in a star configuration (see Figure~\ref{FigureStarAndChain}).
In this section, we discuss methods for constructing Bell inequalities for networks with conditional independence.

\subsubsection{Variable elimination methods}
\label{secExtensionMethod}
We have already seen how to exploit conditional independence in order to construct a Bell inequality for the bilocal scenario.
Specifically, the construction of the BRGP inequality \eqref{BRGP} hinges on the ability of exploiting the conditional independence of Alice and Charlie by eliminating Bob through a suitable ansatz for a Bell inequality.
Methods relying on this approach apply to many different networks and input/output scenarios.

A generalisation of the BRGP inequality, using methods that closely parallel the original work, was presented in Ref.~\cite{Luo2018}.
In this work, explicit Bell inequalities were derived for networks with an arbitrary number of parties under the restriction that they all have binary inputs and outputs.
The method applies to any network that features conditional independence and is centered on identifying the largest set of mutually unconnected parties, i.e.~a set of $k$ parties such that no two of them receive a share from the same source.
The bilocal scenario is a special case corresponding to $k\,{=}\,2$.
Another example is the network illustrated Figure~\ref{fig:complex} in which e.g.~parties $\{A_2,A_5\}$ form a set of $k\,{=}\,2$ conditionally independent parties.
Once $k$ has been determined, one can construct correlation quantities and a Bell inequality ansatz analogous to that encountered in the BRGP inequality (i.e.~$I_\pm$ in Eq.~\eqref{bilocexpectation} and the left-hand side of Eq.~\eqref{BRGP}, respectively).
In further analogy, one can then eliminate all but the $k$ conditionally independent parties and exploit their independence to establish a network Bell inequality.
The maximal quantum violation of these inequalities is based on maximally entangled states and Greenberger-Horne-Zeilinger (GHZ) states \cite{Greenberger1989}, and is found to also parallel that encountered for the BRGP inequality.
Later, more generalisations were found by using the stabiliser formalism~\cite{Hsu2021}, which allows to find nonlinear Bell inequalities in the bilocal scenario that are tailored for non-maximally entangled states.

Another avenue for constructing network Bell inequalities by exploiting conditional independence is through the network extension method developed in Ref.~\cite{Rosset2016}. The premise of this method is that one has an $n$-party network, in which all parties produce binary outputs, for which one already has a Bell inequality. Then, one can extend the network by adding one source that connects (say) party number $n$ to a new party, $n\,{+}\,1$, who has binary inputs and outputs. The added party is therefore conditionally independent of the first $n\,{-}\,1$ parties. This fact can be leveraged to construct a Bell inequality for the extended network by means of solving a quantifier elimination problem. While quantifier elimination is in general challenging, several concrete examples of solutions in particular networks are presented in Ref.~\cite{Rosset2016}. The practical advantage of the network extension method is that it can be applied iteratively. In particular, one may start from a standard Bell scenario, for which the construction of a Bell inequality is straightforward, and iteratively expand it to a variety of different  network configurations. Notably, this method is limited to expanding networks through the addition of bipartite sources. This limitation was, however, overcome in a subsequent work \cite{Tavakoli2016b} which showed how the extension method could in principle be used to derive Bell inequalities for any network that does not feature a cycle. Notably, however, the network Bell inequalities obtained by this method are, typically, tailored for violation using quantum protocols based on separable measurements.

A third approach to network Bell inequalities is based on variable elimination, and is notably not restricted to binary outputs. This approach is based on interpreting network Bell inequalities as instances of a marginal problem \cite{Chaves2016}. It is known that also standard Bell inequalities can be viewed in this framework \cite{Budroni2012, Fritz2013}. To see how this interpretation arises, consider that Alice and Bob each perform measurements $x$ and $y$ with $a_x$ and $b_y$ denoting the outcomes. Is it possible to explain the probability distribution observed in the Bell experiment as the marginal distribution of a larger, unphysical, joint probability distribution $\vec{p}\,{=}\,p(a_1,\dots,a_{|x|},b_1,\dots,b_{|y|})$ corresponding to a scenario in which Alice and Bob are able to perform all their respective measurements in each round? Notice that $\vec{p}$ contains several unobservable events because Alice and Bob each only perform one measurement per round in a Bell experiment. The answer to this question is equivalent to the existence of a local model and can be decided by eliminating the unobservable events from $\vec{p}$ using e.g.~Fourier-Motzkin elimination \cite{Williams1986}, which can be evaluated as a linear program. The resulting constraints on the observable events in $\vec{p}$ (compatible with normalisation, $\sum_j p_j\,{=}\,1$, and positivity, $\vec{p}\geq 0$) are precisely the Bell inequalities. Now, considering network Bell scenarios, one encounters a similar situation but with the key difference that it now features also nonlinear constraints on the larger probability distribution $\vec{p}$. These nonlinear constraints stem from conditional independence, i.e.~that the marginal distribution of two unconnected parties factorises. As shown in Ref.~\cite{Chaves2016}, one can consider the unobservable events in $\vec{p}$ that feature nonlinearities and find the smallest number of free variables required to linearise the problem. Then it can, once again, be solved by means of Fourier-Motzkin elimination. In this manner, one can systematically construct Bell inequalities for networks with conditional independence.

\subsubsection{Stars, chains, and beyond}
\label{StarAndChainedNet}
An important example of a multiparty network is a star-shaped configuration.
Such a star network is composed of a central party that is separately connected, via $m$ bipartite sources, to $m$ branch parties (see Figure~\ref{fig:star}).
Correlations in the network arise through the central party jointly measuring the $m$ independent shares received from the sources and the branch parties locally measuring the single shares received from the corresponding sources.
Nonlocality in star networks was first investigated in Ref.~\cite{Tavakoli2014} in a scenario where the central party has a fixed input and produces an output consisting of $m$ bits while each branch party has binary inputs and outputs.
Naturally extending the ideas behind the BRGP inequality, this construction of network Bell inequalities is based on defining linear combinations $I_j$ (where $j\,{=}\,1,\ldots,2^{m-1}$) of the observed probabilities in a manner analogous to that encountered in Eq.~\eqref{bilocexpectation} for the bilocal scenario.
This leads to the following Bell inequality for star networks
\begin{equation}
  \mathcal{S}_{\text{TSCA}}\coloneqq \sum_{j}|I_j|^{1/m}\leq 2^{m-2}.
  \label{StarIneq}
\end{equation}
In particular, for  $m\,{=}\,2$, this reduces to the BRGP inequality of Eq.~\eqref{BRGP}.
By considering complete, $m$-partite entanglement swapping, quantum theory can violate these inequalities regardless of the number of branches in the network.
The violation emerges by each source distributing a maximally entangled two-qubit state and the central party jointly measuring the $m$ independent qubits in a basis of GHZ states, i.e.~a basis composed of states equivalent to $\frac{\ket{0}^{\otimes m}+\ket{1}^{\otimes m}}{\sqrt{2}}$ up to local unitary transformations.
Conditioned on the output, this renders the $m$, initially independent, branch parties in a globally entangled state. Suitable local measurements then lead to the quantum violation $\mathcal{S}_{\text{TSCA}}\,{=}\,2^{m-2}\sqrt{2}$ \cite{Tavakoli2014}.
However, it is not known whether these are the maximal violations.
Notably, if the sources instead emit isotropic states of the form of Eq.~\eqref{isotropic}, a violation is possible when the visibility per source is at least $v\,{=}\,\frac{1}{\sqrt{2}}$, independently of $m$.
Moreover, a modification of the inequality \eqref{StarIneq} has also been developed \cite{Tavakoli2014} which admits an identical noise tolerance per source but is tailored to be violated by a quantum protocol based on separable measurements; this modification may be thus viewed as wirings of many independent standard Bell inequality tests.

The star inequalities \eqref{StarIneq} have themselves been generalised in several ways.
Ref.~\cite{Tavakoli2016} extended the analysis to star networks that feature multipartite sources and showed that networks featuring such sources exhibit nonlocality with an exponentially improving tolerance to noise, in analogy with standard multipartite Bell inequalities \cite{Mermin1990}.
Another generalisation is one in which a standard Bell inequality for binary-outcome scenarios is mapped to a Bell inequality for a star network.
This map has the property that the local bound of the original Bell inequality also appears as the local bound of the associated network Bell inequality.
Moreover, every quantum strategy in the original Bell scenario can be adapted into a quantum strategy in the network scenario which returns the same degree of violation \cite{Tavakoli2017}.
This method allows one to first investigate a standard Bell inequality and then map the results to the more challenging network scenario. Such constructions may rightfully be viewed as examples of disguised Bell nonlocality, as in Section~\ref{subsec:OverviewReplaceInputs}.

Star networks are also interesting from the point of view of activation of nonlocality. It is known that some entangled states that do not violate any standard Bell inequality (see e.g.~Ref.~\cite{Augusiak2014} for a discussion of such states) can have their nonlocality activated when distributed in a star network \cite{Sende2005,Cavalcanti2011}. This situation is further discussed in Section~\ref{withoutsource-independent}, but notably the analysis is framed in the context of standard multipartite Bell inequalities, i.e.~it does not invoke the independence of the sources. It is therefore expected that star networks, that actively exploit the independence of the sources, should be able to reveal the nonlocality of even noisier states. In spite of this general belief, there presently exists no nontrivial example of quantum correlations in a network with independent sources that reveals the nonlocality of a Bell-local state.

Another natural type of network is shaped like a chain. These networks notably represent quantum repeaters \cite{Briegel1998,Sangouard2011}, on which most long distance quantum communication protocols are based.
A chain network is composed of $m$ sources and $n\,{=}\,m\,{+}\,1$ parties in such a way that the source number $i$ connects to parties number $i$ and $i\,{+}\,1$.
An example is illustrated in Figure~\ref{fig:chain}. Beyond the  bilocal scenario, nonlocality in chain networks was first investigated in Ref.~\cite{Branciard2012} for the special case of $n\,{=}\,4$ parties and further discussed in Ref.~\cite{Fritz2012}.
This was then extended to arbitrarily long chains in Ref.~\cite{Mukherjee2015}, where, in particular, the authors consider scenarios in which only the left-most and right-most party have binary inputs and outputs while all middle parties have a fixed input and four outputs. A natural realisation of this scenario is through complete entanglement swapping in a sequence for a two-qubit state, i.e.~the entanglement initially shared between parties $1$ and $2$ is swapped sequentially throughout the chain so that it ends up being shared between party $1$ and party $n$ (see also Ref.~\cite{Goebel2008}).
The nonlocality of this scenario can naturally be detected via an end-to-end CHSH violation.
However, for sources of visibility $v$ and perfect swappings, the end-to-end visibility of the swapped state decreases exponentially as $v^m$, which makes this CHSH violation (or any protocol based on it, such as device independent quantum key distribution) hard for large $m$: the critical single source visibility, $v\,{=}\,2^{-\frac{1}{m+1}}$, scales exponentially.
This analysis, based on standard Bell nonlocality, does not include the source independence assumption.
Including it, Ref.~\cite{Mukherjee2015} introduces linear network Bell inequalities which obtain the same visibility threshold.
In contrast to this scaling, the network extension method (see Section~\ref{secExtensionMethod}) was applied to the chain network with binary inputs and outputs, obtaining a network inequality violated with a constant critical visibility per source of $\frac{1}{\sqrt{2}}$ \cite{Rosset2016}. The quantum protocol, however, does not exploit entanglement swapping, so the same features can be obtained via parallel violations of the CHSH inequality for each of the sources.
The importance of this protocol raises the question of the existence of a network inequality exploiting the source independence assumption which (i) cannot be violated with separable measurements and (ii) would obtain a better asymptotic scaling in the visibility of the sources required.

Other, related, types of networks have also been considered in the literature. Ref.~\cite{Mukherjee2017} developed Bell inequalities for a particular network structure that features $m$ sources that each emit an $m$-partite state. In a subsequent work \cite{Mukherjee2020} it was shown that these networks can support quantum nonlocality when all the observers who hold more than one independent share of a system have no input. Ref.~\cite{Chaves2016} derived an explicit Bell inequality and its quantum violation for a four-party network featuring two sources that each emits three-partite states. Finally, Ref.~\cite{Luo2018} presented Bell inequalities and quantum violations for polygon-shaped networks and a number of other network configurations.

\subsubsection{Semidefinite relaxations}
\label{sec:scalarextension}
Semidefinite programming~\cite{Boyd,Vandenberghe} is a powerful tool for studying quantum correlations.
Many different types of quantum correlations, arising in different scenarios with different assumptions, can be bounded by means of semidefinite programming relaxations (see e.g.~Refs.~\cite{navascues2007npa,navascues2008npa2,Navascues2015, Kogias2015,Wang2019,bowles2019sequential,Tavakoli2020b,Tavakoli2020c,Chaturvedi2020, Jee2020}).
In particular, the so-called Navascu\'es-Pironio-Ac\'in (NPA) relaxation hierarchy allows to compute increasingly strong necessary conditions for a given probability distribution to admit a quantum model in a standard Bell experiment~\cite{navascues2007npa,navascues2008npa2}.
In essence, the NPA hierarchy relaxes the problem
\begin{equation}
  \begin{split}
    \min &\quad f(p) \\
    \st &\quad p\in\mathcal{Q},
  \end{split}
  \label{eq:optimizeQ}
\end{equation}
where $f(p)$ is a linear functional of the elements of a quantum probability distribution \mbox{$p=p(\bar{a}|\bar{x})\in\mathcal{Q}$} [e.g.~a Bell operator like those in Eq.~\eqref{CHSH} or Eq.~\eqref{eq:mermin}], by the hierarchy of problems
\begin{equation}
  \begin{split}
    \min &\quad f(p) \\
    \st &\quad \Gamma^{(k)}(p)\succeq 0,
  \end{split}
  \label{eq:optimizeNPA}
\end{equation}
where $k=1,2,\dots,\infty$, and $\Gamma^{(k)}(p)$ are called moment matrices, built from expectation values of the operators characterising the parties' measurements.
A rigorous definition of these moment matrices and examples of their use in the context of Bell nonlocality can be found in Refs.~\cite{navascues2007npa,navascues2008npa2,npo}.
The sets of correlations $\mathcal{Q}^{(k)}\,{=}\,\{p:\Gamma^{(k)}(p)\succeq 0\}$ are relaxations of the set of quantum correlations, $\mathcal{Q}$, that satisfy $\mathcal{Q}^{(1)}\supseteq\mathcal{Q}^{(2)}\supseteq\dots\supseteq\mathcal{Q}^{(\infty)}\supset\mathcal{Q}$~\cite{navascues2008npa2}.
Thus, in practice, instead of solving the problem \eqref{eq:optimizeQ} which is difficult and, in general, undecidable~\cite{Slofstra,Ji2020Connes}, one provides increasingly tight bounds on the actual solution by solving Eq.~\eqref{eq:optimizeNPA} for increasing $k$.
Each of these problems, in contrast with Eq.~\eqref{eq:optimizeQ}, can be cast as a semidefinite program, which can be solved efficiently.
In fact, these problems are common in a wide variety of disciplines~\cite{parrilo,LasserreBook}, and multiple software packages exist that generate and solve them~\cite{cvx1,cvx2,yalmip,ncpol2sdpa,picos,mosek,sdpa,gurobi}.

While the theoretical arguments of the NPA construction are also satisfied by network quantum correlations, in general the additional constraints on the moment matrices that are imposed by the network structure break the connection between Eq.~\eqref{eq:optimizeNPA} and semidefinite programming.
A simple example is the already studied bilocal scenario (recall Figure~\ref{fig:bilocal} and Section~\ref{sec:bilocality}).
The fact that there is no source connecting Alice and Charlie implies that in the bilocal network any expectation value that does not involve operators of Bob factorises into the separate expectation values of Alice's and Charlie's operators.
This property translates into nonlinear constraints between the elements of the $\Gamma^{(k)}(p)$, which are not admitted in semidefinite programming.
Therefore, while Eq.~\eqref{eq:optimizeNPA} can be adapted to correlations in quantum networks by adding the necessary constraints, an efficient method for solving the problem does not exist in general.

Semidefinite programs can be used to address quantum nonlocality in any network, by combining the NPA hierarchy with inflation arguments \cite{wolfe2019qinflation} (we expand on this in Section~\ref{sec:qinflation}).
Here, however, we focus on an adaptation of the NPA hierarchy that applies to networks which feature conditional independence, in order to bound the corresponding set of quantum correlations \cite{Pozas2019}.
This adaptation has the form of Eq.~\eqref{eq:optimizeNPA}, but it uses extended moment matrices that generalise those appearing in the NPA hierarchy.
Typically, the rows and columns of moment matrices are indexed by a product of the operators associated to each party, and each cell in the matrix is associated to the expectation value of the product of the operator indexing the column and the (adjoint of the) operator indexing the row.
In Ref.~\cite{Pozas2019}, not only products of operators are used to index the rows and columns of moment matrices, but also products of operators and scalar quantities associated to expectation values.
For this reason, the method is called scalar extension.
The fact that the rows and columns are indexed by nonlinear quantities results in cells inside the moment matrix also representing nonlinear quantities, so polynomial constraints in the probability distribution can be encoded in linear constraints between the corresponding matrix elements.
These constraints are compatible with semidefinite programming, so one can again use efficient solving algorithms.

As an illustration, consider the bilocal scenario of Figure~\ref{fig:bilocal}, with Alice and Charlie performing two dichotomic measurements each, $\{A_0,A_1\}$ and $\{C_0,C_1\}$, respectively.
A possible extended moment matrix is that generated by the set of operators $\{\mathbb{1},A_0A_1,C_0C_1,\expect{A_0A_1}\mathbb{1}\}$, reading
\begin{equation}
  \Gamma = \kbordermatrix{
    & \mathbb{1} & A_0A_1 & C_0C_1 & \expect{A_0A_1}\mathbb{1} \\
    \mathbb{1} & 1 & v_1 & v_2 & v_3 \\
    (A_0A_1)^\dagger & & 1 & v_4 & v_5 \\
    (C_0C_1)^\dagger & & & 1 & v_6 \\
    \expect{A_0A_1}^*\mathbb{1} & & & & v_7
    }.
    \label{eq:scalarextension}
\end{equation}
Per the original NPA prescription~\cite{navascues2007npa}, every cell in the moment matrix above contains the expectation value of the operator characterising the row and the operator characterising the column under an unknown quantum state, i.e., $\Gamma_{R,C}\,{=}\,\Tr\left(\rho R^\dagger C\right)\,{=}\,\expect{R^\dagger C}_\rho$.
Note that the last row and column are the only ones different from standard NPA, as they are indexed by the operator $\expect{A_0A_1}\mathbb{1}$, the identity operator multiplied by the scalar $\expect{A_0A_1}_\rho$.

It is easy to see that, in the matrix above, \mbox{$v_1\,{=}\,\expect{A_0A_1}_\rho$} and \mbox{$v_3\,{=}\,\expect{\expect{A_0A_1}_\rho\mathbb{1}}_\rho\,{=}\,\expect{A_0A_1}_\rho$}, and that \mbox{$v_5\,{=}\,\expect{\expect{A_0A_1}_\rho(A_0A_1)^\dagger}_\rho\,{=}\,|\expect{A_0A_1}_\rho|^2$} and \mbox{$v_7\,{=}\,\expect{\expect{A_0A_1}_\rho\expect{A_0A_1}_\rho^*\mathbb{1}}_\rho\,{=}\,|\expect{A_0A_1}_\rho|^2$}.
Therefore, two immediate linear constraints, that are not imposed by the network structure but rather by scalar extension, are $v_1\,{=}\,v_3$ and $v_5\,{=}\,v_7$.
More importantly, the structure underlying the bilocal scenario imposes that expectations involving only operators of Alice and Charlie factorise.
In Eq.~\eqref{eq:scalarextension}, the factorisation $\expect{A_0A_1C_0C_1}_\rho\,{=}\,\expect{A_0A_1}_\rho\expect{C_0C_1}_\rho$ is realised by the linear equality $v_4\,{=}\,v_6$.
Therefore, the factorisation can be imposed, via scalar extension, in a way amenable to semidefinite programming.
Note that the example above is just an illustration.
When applying scalar extension to a given problem, a general procedure consists in extending the corresponding moment matrix with as many columns as the smallest number of factors that allow to set equality constraints for all elements in the moment matrix that must factorise (see, e.g., Refs.~\cite{Pozas2019,AlexThesis}).

While conceptually simple, scalar extension has proved itself useful in the analysis of correlations in networks with conditional independences.
Indeed, small levels of scalar extension are capable (see Refs.~\cite[Chapter 5]{AlexThesis} and \cite{scalarextension}) of identifying as incompatible with a bilocal model mostly all the distributions that are known to be incompatible due to the version of the BRGP inequality of Eq.~\eqref{BRGP} based on separable measurements.
Yet, it remains unknown whether, upon the choice of a sufficiently large extension, one is able to discard all distributions incompatible with a network.
Recent results in noncommutative polynomial optimisation~\cite{klep2020optimization} hint to a positive answer.

Scalar extension is employed in Ref.~\cite{Pozas2019}, in combination with semidefinite relaxations of sets of local-variable correlations~\cite{Steeg2011}, to demonstrate that the ability of measurement devices to create nonlocal correlations can be activated.
In a tripartite grey-box scenario where the devices held by the parties are known to have non-unity detection efficiencies, it is known that detection efficiencies below $\eta\,{=}\,2/3$ do not allow to certify the presence of standard Bell nonlocality in the two-input, two-output scenario\footnote{The scenario exemplified in Ref.~\cite{Pozas2019} contains one party that performs a single measurement, which effectively acts as a projection of the state shared by the remaining parties.
Consequently, the detection efficiency required to demonstrate nonlocality is that required to demonstrate it with the projected states, which is given in Refs.~\cite{eberhard1993detection,massar2003detection}.}.
However, by using scalar extension, Ref.~\cite{Pozas2019} proves that violations of bilocality can be certified from the observed correlations for detection efficiencies down to $\eta\,{=}\,0.5291$.
Moreover, scalar extension has been employed in Ref.~\cite{Tavakoli2020} for providing quantum bounds to bilocal Bell inequalities tailored for the Elegant Joint Measurement.

\subsection{Inflation of networks}
\label{Inflation}
Inflation is a powerful concept that allows to set constraints on the correlations that can arise in any network.
Its primary idea is to analyse the hypothetical situation where one has access to multiple copies of the sources and measurement devices that conform the network, and can arrange them in arbitrary configurations.
By analysing the correlations that are generated in these new configurations, the fact that the network elements are copies of those in the original network allows to translate conclusions obtained in the inflation networks back to the original one, providing a means of studying the correlations that are generated there.
While this concept was briefly illustrated in Section~\ref{subsec:OverviewInfl}, here we provide more details on how the concept of inflation materialises into systematic methods that constrain the classical and quantum correlations in networks.

\subsubsection{Classical inflation}
\label{sec:cinflation}
Inflation was originally developed in Ref.~\cite{wolfe2019inflation} with the aim of solving the problem of causal compatibility: given a causal structure (of which the network scenarios discussed in this review are concrete instances) and a probability distribution over its visible nodes, determine whether the distribution can be generated in the structure.
In the context of Bell nonlocality in networks, where the visible nodes of the structure represent the parties' outcomes after measuring their respective systems, this problem is equivalent to that of discerning whether an observed probability distribution admits a local model in the network, this is, if the given distribution admits a model of the type \eqref{Eqlocal}.

The method that Ref.~\cite{wolfe2019inflation} proposes for solving the causal compatibility problem was exemplified in Section~\ref{subsec:OverviewInfl} for the triangle network.
We now discuss the general situation.
One begins by assuming that the distribution under scrutiny, $P_\text{obs}$, can be generated in the network, i.e.~that there exist random variables (corresponding to the sources) and deterministic functions of such variables (which correspond to the parties) that give rise to the target distribution.
Then, one considers inflations of the network by adding copies of the sources and parties available, and connecting them in such a way that the each copy of a party is only connected to copies of the sources that connect to the original party in the original network.

\begin{figure}
  \centering
  \includegraphics[width=\columnwidth]{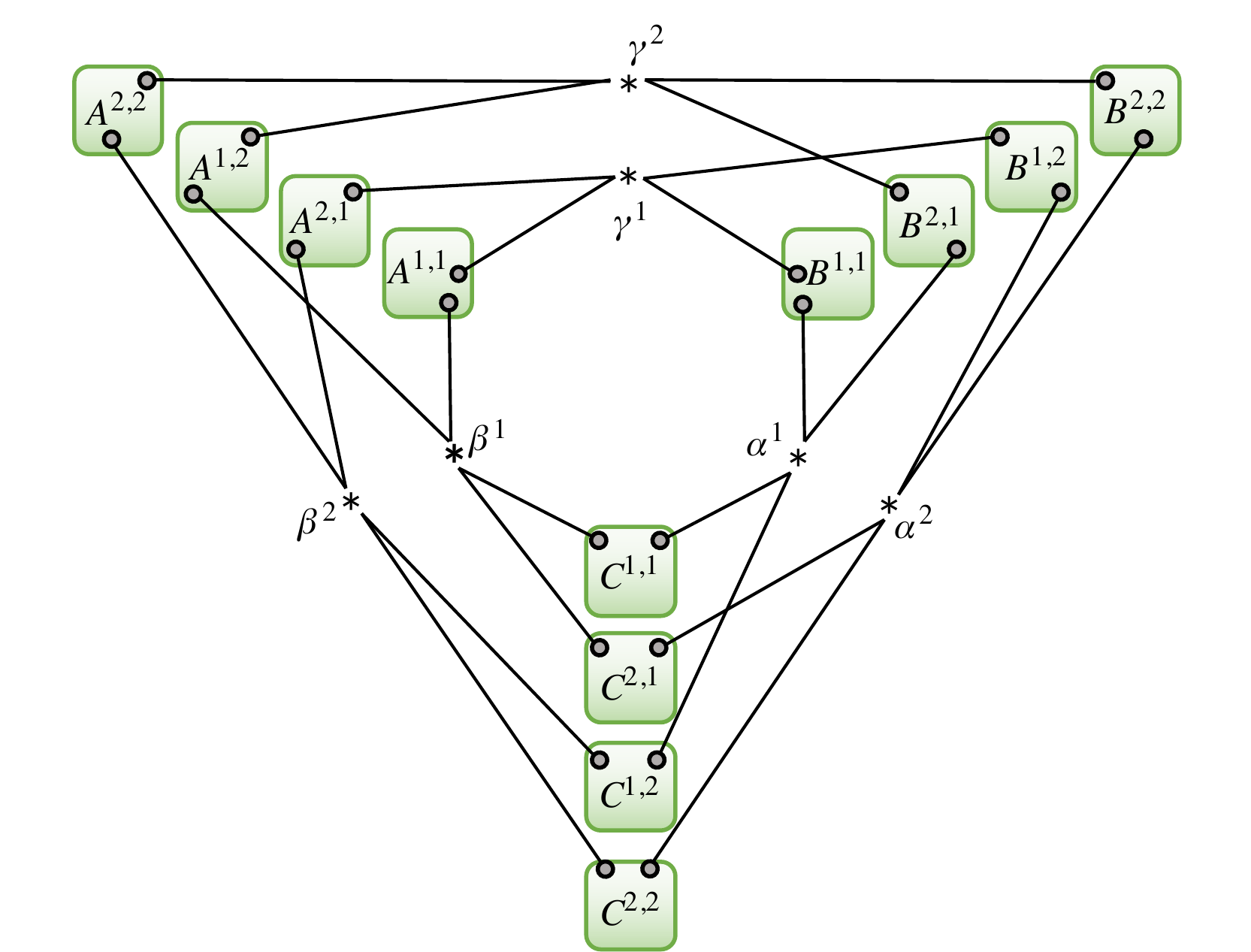}
  \caption{The web inflation of the triangle network (for simplicity, the arrows that denote inputs and outputs are omitted). It is also regarded as the second step in the hierarchy of inflations defined in Ref.~\cite{navascues2017convergence}. Note that copies of the individual random variables sent to the original parties are sent to various parties in the inflation. This is impossible in quantum mechanics due to the no-cloning theorem~\cite{cloning}, and thus this inflation can be used to identify correlations that cannot be written as Eq.~\eqref{eq:localtriangle}.}
  \label{fig:webinflation}
\end{figure}

The inflated network is analysed, and necessary constraints on compatible probability distributions are derived, which are later ``translated'' into constraints in the original network.
This translation is possible because certain marginals of the distributions compatible with the inflated scenario, namely those which exactly reproduce (collections of) sub-networks of the original network, must exactly match (products of) marginals of the original distribution.
For instance, in the so-called web inflation of the triangle scenario, illustrated in Figure~\ref{fig:webinflation}, any marginal of the form $p_\text{inf}(a^{i,j}, b^{j,k}, c^{k,i})$ must match the original distribution under scrutiny, $p_\text{obs}(a,b,c)$, since the sub-network formed by parties $\{A^{i,j},B^{j,k},C^{k,i}\}$ and the local variables that connect to them, $\{\alpha^k,\beta^i,\gamma^j\}$, reproduce the original triangle scenario.
Moreover, in addition to identifications with $P_\text{obs}$, the distributions that can be generated in the inflated network satisfy many other constraints that are derived from the intrinsic symmetries of the inflated network.
This is the case of, for example, $p_\text{inf}(a^{1,1},b^{1,1},c^{1,2})$ in Figure \ref{fig:webinflation}.
This marginal does not correspond to any marginal of the original distribution  because the copies of the source $\beta$ that send systems to $A^{1,1}$ and $C^{1,2}$ are different.
Nevertheless, it must coincide with all of the form $p_\text{inf}(a^{i,j},b^{j,k},c^{k,l\not=i})$ because all are created by the same sources and measurement devices, wired in the exact same way.

Finally, one needs to determine whether a $P_\text{inf}$ satisfying all possible constraints of the type outlined above exists.
Determining the existence of such a probability distribution can be formulated as a linear program, which can always be evaluated in an efficient manner, for instance by using the algorithms \cite{cvx1,cvx2,yalmip,mosek,gurobi}.
This constitutes a great simplification of the original, non-convex problem of finding a model of the form of Eq.~\eqref{Eqlocal}.

\begin{figure*}
  \centering
  \hfill
  \begin{tabular}{ll}
    \subfloat[\label{fig:TriangleSubsystems}]{
      \centering
      \begin{minipage}[t]{0.49\textwidth}
        \centering
        \begin{overpic}[scale=0.40]{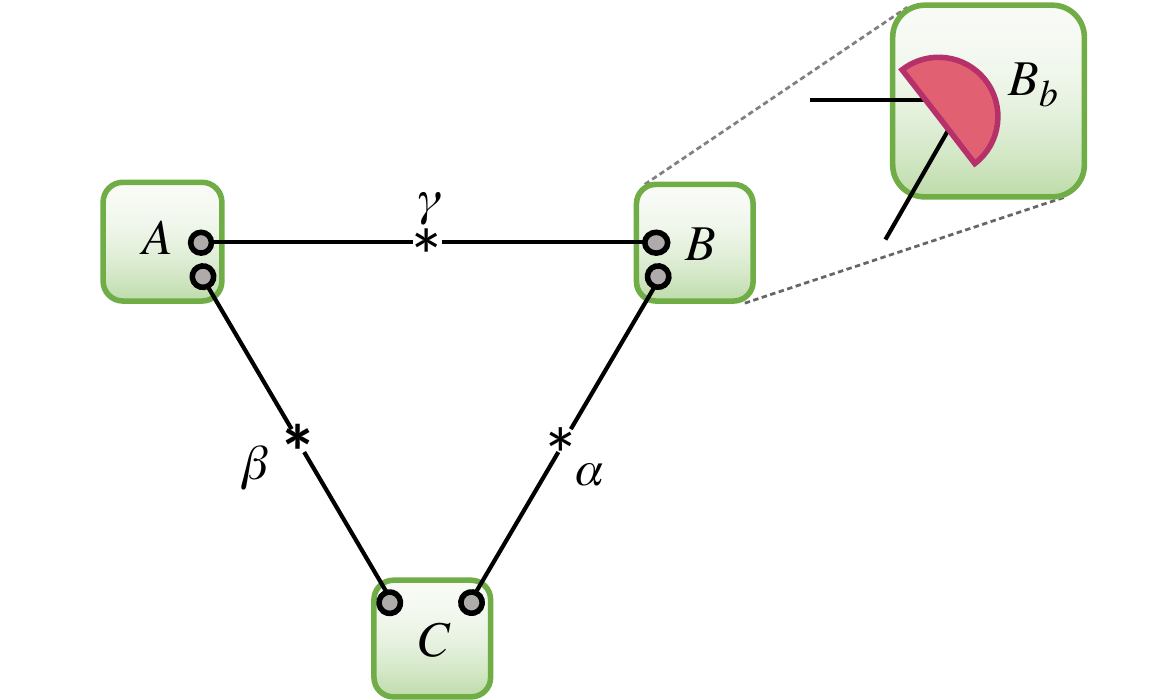}
           \put(90,22){\huge\faArrowRight}
        \end{overpic}
      \end{minipage}
    }
    &
    \subfloat[\label{fig:TriangleQInf}]{
      \centering
      \begin{minipage}[t]{0.49\textwidth}
        \centering
        \includegraphics[scale=0.40]{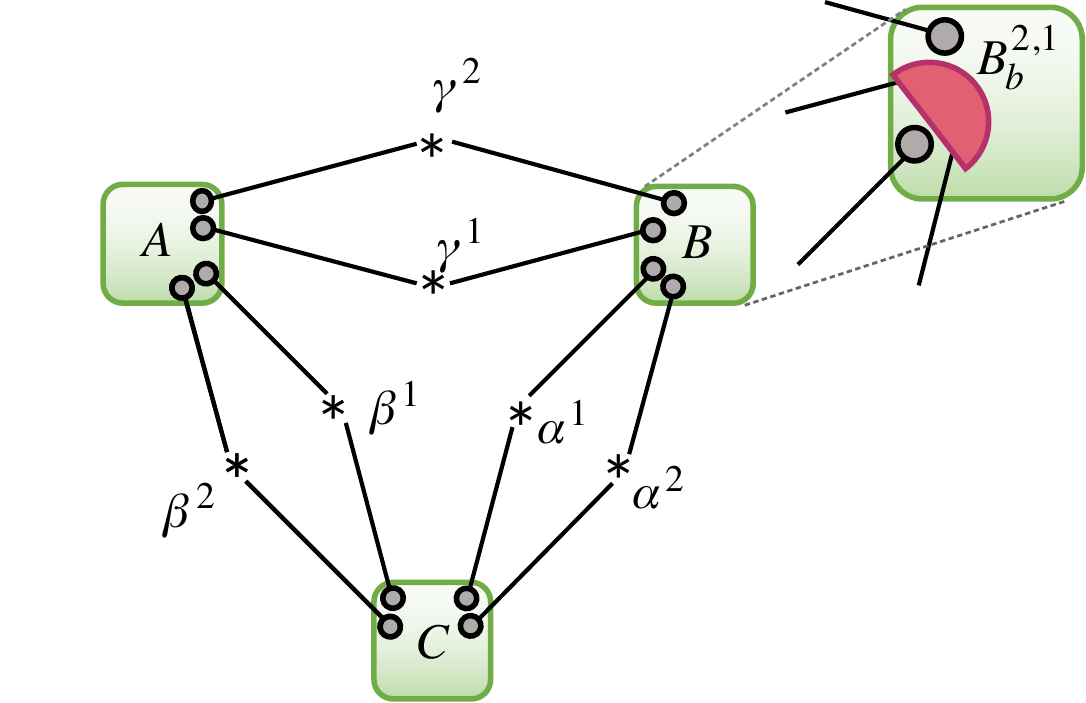}
      \end{minipage}
    }
  \end{tabular}

  \hfill
  \caption{Exemplification of quantum inflation in the triangle scenario (for simplicity, the arrows that denote inputs and outputs are omitted). (a) In the original scenario, the different outcomes $b$ for Bob's measurement are produced by applying $\{B_b\}_b$, depicted as a semi-circle,
  on the shares $\alpha_\text{B}$ and $\gamma_\text{B}$. The outcomes of Alice's and Charlie's measurements are produced similarly.
  (b) In quantum inflation, one distributes $n$ (in the case shown, $n\,{=}\,2$) independent copies of the same states to the parties, which apply their measurements on different pairs of copies of the states they receive. In the case shown, the operators $\{B_b^{2,1}\}_b$ are applied on the shares $\alpha_\text{B}^2$ and $\gamma_\text{B}^1$.
  In contrast with Figure~\ref{fig:webinflation}, this inflation is not affected by no-cloning, and thus can be used to identify correlations that cannot be written as Eq.~\eqref{eq:quantummodel} in the triangle network.
  }
  \label{fig:qinflation}
\end{figure*}

There are infinitely many ways to inflate a network, and finding just a single inflation for which there exists no distribution satisfying all the above constraints proves that the original distribution does not have a network local model.
Moreover, it was later proved~\cite{navascues2017convergence} that one can define a hierarchy of inflation tests that converges to the conclusive answer: if all the tests in such hierarchy are passed, the distribution is guaranteed to have a local model in the network.
In addition, passing a particular test guarantees a maximum total variation distance between the distribution and the set of network local models.
Thus, inflation is a powerful tool for determining that an observed probability distribution does not have a local model in a network.

\subsubsection{Quantum inflation}
\label{sec:qinflation}
It is not straightforward to generalise the inflation technique for the case when the sources distribute shares of joint quantum states instead of local variables.
The reason for this is the fact that classical inflation uses cloning of classical information, and this is impossible when the information is encoded in quantum systems~\cite{cloning}.
In Ref.~\cite{wolfe2019qinflation}, this problem with no-cloning is circumvented by not copying the subsystems that the sources send to the parties as classical inflation does, but by considering copies of the sources themselves.
In such a situation, the parties receive multiple copies of quantum states, and can choose on which copies they will perform their measurements.
This scheme is illustrated in Figure~\ref{fig:qinflation}.
If a probability distribution can be realised in a given quantum network, then there exists a global quantum state on an inflated scenario and measurements whose statistics (i) are invariant under the permutation of Hilbert spaces corresponding to the different copies of a same source, and (ii) reproduce the original distribution when the parties make a choice of copies that reproduces the original network.
The problem of finding a quantum state and measurements whose statistics are invariant under suitable permutations can be solved via non-commuting polynomial optimisation using hierarchies of semidefinite programs~\cite{npo}.
The non-existence of a solution of one of the problems in the hierarchy constitutes a proof that the original distribution cannot be realised in the original quantum network.

Ref.~\cite{wolfe2019qinflation} outlines with examples various families of problems for which quantum inflation can be used to provide a solution.
These include the certification of a distribution being impossible to generate in a quantum network, the optimisation over the set of distributions that can be generated in a quantum network, the extraction of polynomial witnesses of incompatibility with quantum network models and, as a more practical application, the bounding of the information that a malicious party can obtain when eavesdropping on a quantum repeater.
Moreover, quantum inflation is, at the time of writing, the only method for characterising quantum correlations in networks that is capable of analysing continuous-valued distributions.

The implementation of quantum inflation synthesises two different hierarchies: the one of inflations, and for each inflation, the non-commuting polynomial optimisation hierarchy used to determine whether a distribution admits such an inflation.
The latter hierarchy is well known to be convergent in a monotonic way, so each level in the hierarchy is not weaker than the previous.
As for the former, slight variations allow to define a non-monotonic albeit converging hierarchy~\cite{ligthart2021}.
There is evidence \cite[Chapter 5]{AlexThesis} supporting that each hierarchy enforces constraints of different nature.
Thus, in situations of limited computational resources, it is possible to prioritise one hierarchy over the other if prior knowledge or intuition about the distribution under scrutiny is available.

Quantum inflation, in the same way as the scalar extension method described in Section~\ref{sec:scalarextension}, can be modified~\cite{Steeg2011} to characterise the sets of correlations that admit network local models.
Indeed, by doing so quantum inflation generates semidefinite relaxations of the linear program that is defined by classical inflation at the same inflation level.
This gives rise to an interesting characterization that, by combining and trading-off various relaxations, is capable of alleviating to some extent the computational load of analyzing network local models.

\subsection{Rigidity-based arguments}
\label{subsec:TCandCM}
Network nonlocality can also be proven from making use of the rigidity property of some classical strategies. In a first step, this property is employed to restrict the possible local explanations of a quantum distribution to a tiny family of local strategies. In a second step, this family is ruled out, exploiting its characteristics.

In Refs.~\cite{Renou2020short,Renou2020long}, given a fixed network, a classical strategy associated to a correlation $P_\mathrm{R}$ is said to be rigid if it is the only classical strategy which obtains this correlation.
In other words, rigidity can be thought of as a self-testing of classical strategies: the observation of the correlation $P_\mathrm{R}$ is sufficient to characterise the classical strategy which obtained it.
This characterisation is defined up to a sole relabelling of the local variables and parties' outcome functions.
Two families of strategies are shown to be rigid in some restricted classes of networks:
token-counting and color-matching strategies.
As will be seen in the following, this rigidity property can be used to show that certain quantum network correlations $P$ admitting a rigid coarse-graining $P_\mathrm{R}$ are nonlocal.

\begin{figure*}
  \begin{minipage}[c]{0.35\textwidth}
    \subfloat[\label{fig:TCTriangle}]{
    \begin{overpic}[width=0.85\textwidth]{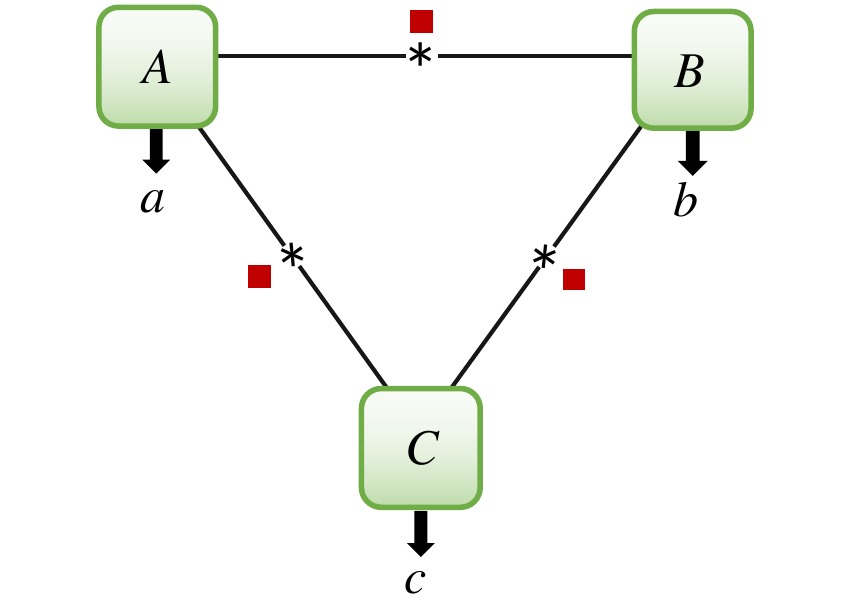}
      \put(100,40){\huge\faArrowRight}
    \end{overpic}
    }
  \end{minipage}
    \begin{minipage}[c]{0.6\textwidth}
    \subfloat[\label{fig:TCCounts}]{
      \includegraphics[width=0.9\textwidth]{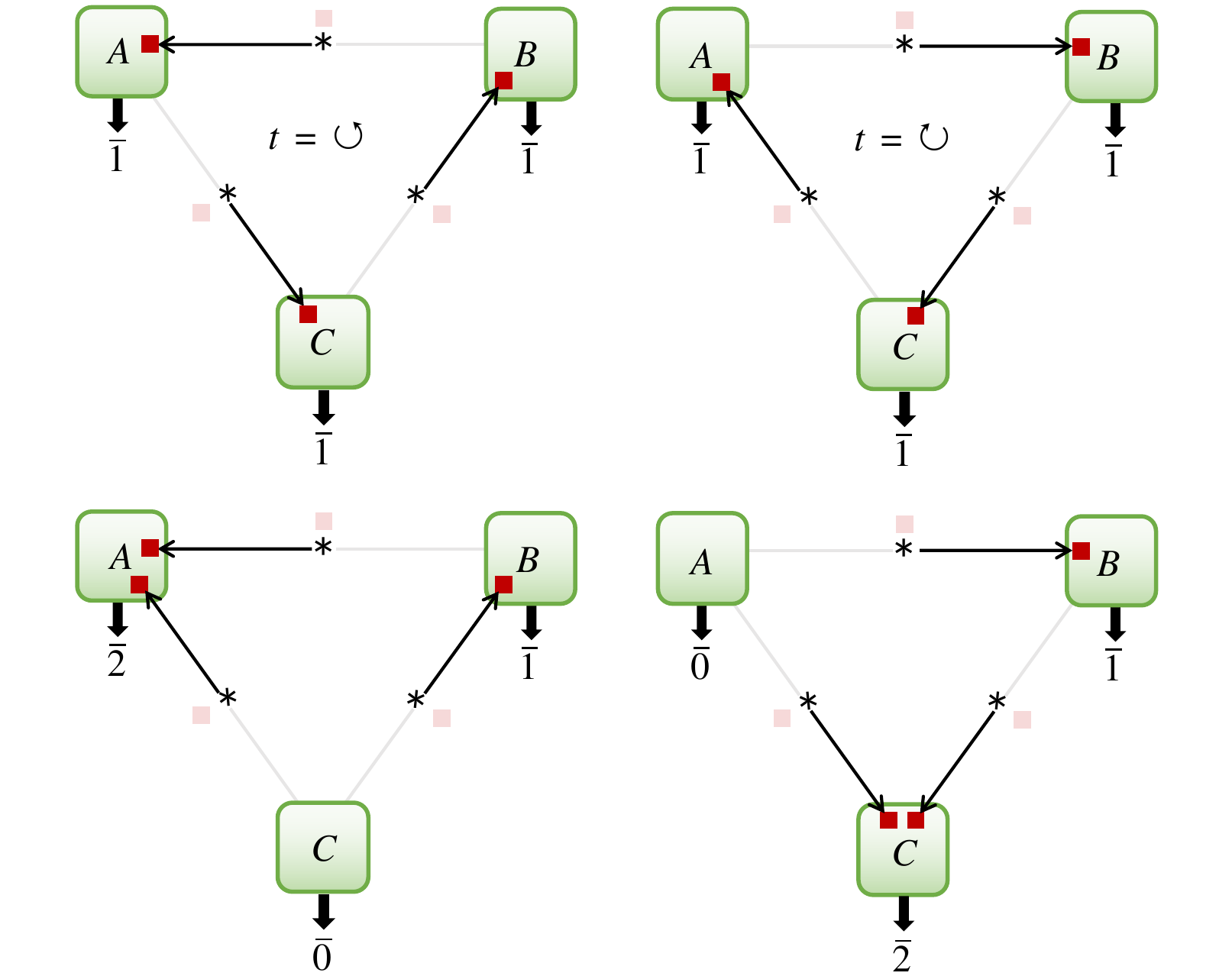}
    }
  \end{minipage}
  \caption{Token counts in the uniform token-counting strategy in the triangle scenario.
    (a) Each source uniformly distributes a unique token (red square) to one of the two parties it is connected to, either in a quantum way, using states of the form $\ket{\psi^+}\,{=}\,\frac{1}{\sqrt{2}}(\ket{01}+\ket{10})$, or a classical way, distributing states of the form $\frac{1}{2} (\ketbra{10}{10}+\ketbra{01}{01})$.
    (b) Each party first measures the total number $\bar{0}, \bar{1}, \bar{2}$ of tokens received, giving rise to the distribution $P_\token$, which is the same for classical and quantum strategies. $P_\token$ is a rigid distribution, so this is the unique classical strategy that is capable of generating it in the triangle network. In the quantum counting strategy, whenever $\bar{1}$ token is detected by a party, that party further measures $\{\bar{1}_0, \bar{1}_1\}$ to measure where the token comes from, obtaining the full correlation $P$.
    When all parties obtain $\bar{1}$, the token trajectories are undetermined between $t\,{=}\circlearrowleft$ and $t\,{=}\circlearrowright$ (top row).
    In all other cases, the token trajectories are fully determined by the token counts (bottom row).
  }
  \label{fig:TriangleTC}
\end{figure*}

\paragraph*{Triangle nonlocality from three-token-counting rigidity:}
A primal example of proofs of nonlocality from rigidity arguments is the proof that the quantum-triangle distribution $P_{\mathrm{RGB4}}$, introduced in Section~\ref{subsec:OverviewTC} and explicitly given in Eq.~\eqref{eq:genuine}, does not admit a local model in the triangle network. To obtain it, each source distributes maximally entangled states $\ket{\psi^+}$, which can be interpreted as a unique token sent in superposition to the right ($\ket{01}$) and to the left ($\ket{10}$) parties it is connected to. Each party first measures in the same basis given by $\{\bar{0}\,{\coloneqq}\,\ketbra{00}{00}$, $\bar{1}\,{\coloneqq}\,\ketbra{01}{01}+\ketbra{10}{10}$, \mbox{$\bar{2}\,{\coloneqq}\,\ketbra{11}{11}\}$}, where $\bar{n}$ counts the number of received tokens, obtaining the correlation $P_{\token}$. Whenever the outcome $\bar{1}$ is obtained (so one token is received), the party who obtained it performs a second, dichotomic measurement given by $\{\ket{\bar{1}_0}\,{\coloneqq}\, c\ket{01}\,{+}\,s\ket{10}$, $\ket{\bar{1}_1}\,{\coloneqq}\, s\ket{01}\,{-}\,c\ket{10}\}$.

The three-outcome distribution $P_{\token}$ can be obtained from a classical strategy.
Indeed, since the operators $\{\bar{0}, \bar{1}, \bar{2}\}$ are diagonal in the token number basis, replacing the quantum sources by classical sources emitting the state $1/2 (\ketbra{10}{10}+\ketbra{01}{01})$ does not change $P_{\token}$ and provides a classical token-counting strategy.
Moreover, the authors of Refs.~\cite{Renou2020short,Renou2020long} prove the rigidity of $P_{\token}$, this is, they show that its observation self-tests the local variables and response functions in the triangle network.
The proof is based on the fact that the total number of tokens seen by all parties is always three: in particular, if a change in the value of source $\alpha$ produces a change in the number of tokens detected by Bob, then it must be compensated in the number of tokens detected by Charlie in order to keep the total sum invariant (since Alice does not have access to $\alpha$).

Hence, any hypothetical strategy simulating $P_\text{RGB4}$ induces a classical strategy for $P_{\token}$, which by rigidity must be the classical token-counting strategy outlined above. Therefore, in any hypothetical simulating strategy of $P_\text{RGB4}$, a new classical hidden variable can be introduced: the direction of each individual token.
For instance, postselecting on the case where all parties obtain $\bar{1}$, the tokens of the source must be sent incoherently in the clockwise ($t\,{=}\circlearrowright$) or in the anti-clockwise ($t\,{=}\circlearrowleft$) direction, as shown in Figure~\ref{fig:TriangleTC}.
Introducing the (unknown) joint probability distribution of this hidden variable with the parties output
\begin{equation}
  q(a,b,c,t)=\text{Pr}(a,b,c,t|\mathrm{all~parties~measure~\bar{1}~token}),
\end{equation}
it is possible to explicitly compute some of the marginals of $q(a,b,c,t)$, expressed in terms of the probability distribution $P_\text{RGB4}$.
By proving that those marginals are incompatible with any valid probability distribution for some specific choices of the operators $\{\bar{1}_0,\bar{1}_1\}$, the authors of Refs.~\cite{Renou2020short,Renou2020long} demonstrate that $P_\text{RGB4}$ does not admit a local model in the triangle.

Notice a key difference between classical and quantum token-counting strategies:
in the quantum strategy, when all parties measure $\bar{1}$ token, the global state is a coherent superposition of the tokens turning clockwise and anti-clockwise. This is, once all parties obtain one token, the global state of the sources is left in the state:
\begin{align*}
  \mathrm{Classical~token~counting:~}&\frac{1}{2}\left(\ketbra{\circlearrowleft}{\circlearrowleft}+\ketbra{\circlearrowright}{\circlearrowright}\right)\\
  \mathrm{Quantum~token~counting:~}&\frac{1}{\sqrt{2}}\left(\ket{\circlearrowleft}+\ket{\circlearrowright}\right).
\end{align*}
Hence, in the quantum case, one observes the creation of a global coherent superposition between all parties through local measurements in the network. This is the core feature that rigidity-based methods exploit, and what ultimately allows these methods to identify network nonlocality.

\paragraph*{Network nonlocality from token-counting rigidity:}
The method described above can be extended to a large class of networks beyond the triangle.
More precisely, a generic classical token-counting strategy involves sources distributing an arbitrary fixed number of tokens, and parties measuring the total number of tokens they receive, obtaining the token count distribution $P_{\token}$.
Such strategies are proven to be rigid in all networks where any pair of parties shares at most one source.
Corresponding quantum token-counting strategies are obtained by allowing superposition in the tokens sent by the sources, and measuring the origin of the tokens using entangled measurements.
Refs.~\cite{Renou2020short,Renou2020long} generalise the initial triangle example outlined above, which was first introduced in Ref.~\cite{Renou2019}, to prove the nonlocality of quantum token-counting distributions in the four-partite network where every pair of parties except $\{A_1, A_2\}$ share a bipartite source, and in all ring-type networks, where the every pair $A_j, A_{j+1}$ shares a bipartite source and $A_{N+1}=A_1$.

\paragraph*{Network nonlocality from color-matching rigidity:}
A second class of strategies is known to be rigid: these are known as the color-matching strategies.
In a quantum color-matching strategy, each source picks a color from a pre-established set, in superposition.
In the triangle network, the sources can for instance distribute the state $(\ket{00}+\ket{11}+\ket{22})/\sqrt{3}$, where $0,1,2$ represent colors.
A party first checks if all her neighbouring sources sent her the same color, in which case she outputs that color.
In the triangle, the measurement performed by each of the parties would correspond to rank-one projectors over $\{\ket{00},\ket{11},\ket{22}\}$, with an extra no-color-match output, corresponding to the projector over the remaining dimension-six vector space.
This gives the color-matching distribution $P_{\match}$.
At a second step, and in analogy with the procedure for token-counting, the no-color-match output can be refined by asking the parties to perform a second measurement using refined projectors to reveal information about the pattern of the received colors.
In the triangle, an example measurement would be $\{(\ket{01}\pm\ket{10})/\sqrt{2}, (\ket{02}\pm\ket{20})/\sqrt{2}, (\ket{12}\pm\ket{21})/\sqrt{2}\}$.

As in the token-counting case, $P_{\match}$ can be obtained from a classical color-matching strategy in which there is no superposition of colors in the states or measurements.
Using techniques based on Finner inequalities (which we describe in Section~\ref{subsec:FinnerIneq}), Refs.~\cite{Renou2020short,Renou2020long} prove color-matching strategies to be rigid in a number of networks, following the same procedure as for token-counting, just by replacing $P_{\token}$ with $P_{\match}$.
This allows to find network nonlocal quantum color-matching strategies in the four-partite network where $\{A_1, A_2, A_3\}$ and $\{A_1, A_2, A_4\}$ share each a tripartite source and $\{A_3, A_4\}$ share a bipartite source, and in all networks where every pair of parties shares a bipartite source.
Moreover, Refs.~\cite{Renou2020short,Renou2020long} establish a connection between network nonlocality and graph coloring, which is used to prove the nonlocality of quantum color-matching distributions in all networks where every party is connected to a different pair of sources and every possible pair of sources connects to a party.
For this, the parties are asked to check that two neighbouring sources have distinct colors.
When this is the case for all parties, it indicates a proper coloring of the graph of sources.
These measurements project the joint state of all sources to a coherent mixture of all possible proper colorings, which then leads to network nonlocality.

\subsection{Entropic characterisation}
\label{Entropy}
Network correlations can also be analysed in terms of the entropies of those variables that are jointly measurable in the network.
This approach, which was first employed in the bipartite Bell scenario in the late 1980s~\cite{Braunstein1988}, has been found useful in the study of network correlations for two main reasons.
Firstly, variable entropies are insensitive to their cardinalities.
In the analysis of correlations in networks, this implies that the conclusions derived from this approach hold independently of the number of outcomes.
Secondly, as will be shown later on, some constraints imposed by the network that are nonlinear in the elements of the probability distribution become linear in the entropic picture, enabling the use of numerical tools that cannot be leveraged in the direct study of compatible correlations.
The entropic approach to the study of correlations is reviewed in Ref.~\cite{Weilenmann2017}, covering not just network correlations but also correlations in more general scenarios.
While referring the reader to Ref.~\cite{Weilenmann2017} for more detailed discussions, we here survey the main ideas and advances concerning this approach in network nonlocality.

In the same way that the primary object in the probability picture is the set of (classical, quantum or theory-independent) distributions compatible with a network structure, the primary object in the entropic picture is the so-called entropy cone.
For a fixed number of random variables $n$, the entropy cone is the closure of the set of all vectors $v\in\mathbb{R}^{2^n-1}$ of the form
\begin{multline}
  v=\big(H(X_1), H(X_2), \dots, H(X_1 X_2), H(X_1 X_3), \\
  \dots, H(X_1 X_2 \dots X_n)\big),
\end{multline}
where $H$ is an entropy function.
In the traditional approach to the entropic analysis of causal structures, which assumes that all nodes in the structure are classical random variables, the entropy function used is the Shannon entropy, given by
\begin{equation}
  H(\mathcal{X})=-\sum_{x\in X}p(x)\log_2 p(x),
\end{equation}
where $p(x)$ is the distribution of events for a subset of variables $\mathcal{X}\,{\subset}\,\{X_1,\dots,X_n\}$ whose range is $X$.
Other entropies can be used alternatively, such as the R\'enyi entropies~\cite{Renyi1961} or the Tsallis entropies~\cite{vilasini2019tsallis}.
In particular, when one deals with random variables resulting from quantum measurements, a popular choice is the von Neumann entropy, $H(\mathcal{X})\,{=}\,-\Tr\left(\rho_\mathcal{X}\log\rho_\mathcal{X}\right)$.

The boundary of the entropy cone may be described by potentially infinitely many linear inequalities, and thus it is difficult to characterise in general.
This is the reason why outer approximations of the cone are used instead.
The most common approximation is the Shannon cone, which is the polyhedral cone characterised by three basic constraints, namely positivity (the uncertainty about a variable is a non-negative quantity), monotonicity (the uncertainty about a set of variables is not smaller than the uncertainty about any subset of it), and submodularity (the conditional mutual information is positive)\footnote{The corresponding mathematical formulations of these properties can be found, for instance, in \cite[Section 2(a)(i)]{Weilenmann2017} or in \cite[Appendix I]{Budroni2016}.}.
The Shannon cone coincides with the entropy cone only for $n\,{\leq}\,3$.
For $n\,{>}\,3$, the Shannon cone is an outer approximation of the entropy cone.
All those inequalities that describe the entropy cone but cannot be derived from the conditions above are called non-Shannon-type inequalities~\cite{Zhang1998entropy,Linden2005nonshannon,Dougherty2006nonshannon,dougherty2011nonshannon,Weilenmann2018}.

In order to deal with correlations in networks, the Shannon cone must be further constrained in order to account for the network structure.
In what follows, we review how this is done for local and quantum correlations.
A brief remark on the entropic analysis of more general correlations in networks is deferred to Section~\ref{sec:nonsignaling}.

\subsubsection{Local correlations}
\label{sec:centropy}
In the case of characterising local models in networks, the condition of locality implies that all variables are jointly measurable, and thus one can define a probability distribution over all events~\cite{Fine1982}.
Therefore, the corresponding Shannon cone can be built by imposing the Shannon constraints (recall, positivity, monotonicity, and submodularity) to all the possible subsets of variables in the network.
Then, the constraints implied by the network structure are imposed by the entropic analogue of the compatibility condition of a probability distribution with a network.
While in the probability picture compatibility was captured by Eq.~\eqref{Eqlocal}, in the entropic picture this notion is captured by
\begin{equation}
  \begin{split}
    I(\lambda_j:\lambda_k)&=0, \\
    I(A_j:A_k|\lambda_{i\rightarrow j,k})&=0,
  \end{split}
  \label{eq:compatibilityentropy}
\end{equation}
where $I(X\,{:}\,Y|Z)\,{=}\,H(X|Z)\,{+}\,H(Y|Z)\,{-}\,H(X,Y|Z)$ is known as the conditional mutual information, $\lambda$ denotes the sources, $A$ denotes the parties, and $\lambda_{i\rightarrow j,k}$ is a source that is connected to parties $A_j$ and $A_k$.
The first equation captures the independence of different sources, while the second establishes that all correlations between two parties are established exclusively via the sources that connect to them.
It is important to note that, while Eq.~\eqref{Eqlocal} gives rise to equality constraints that are polynomial in the elements of the probability distribution, Eq.~\eqref{eq:compatibilityentropy} is linear in the variables' entropies.
Indeed, the problem of whether a given entropy vector lies within the set defined by the Shannon cone and Eq.~\eqref{eq:compatibilityentropy} can be cast as a linear program, which can be efficiently solved using numerical methods~\cite{cvx1,cvx2,yalmip,mosek,gurobi}.
For example, the entropic constraint that imposes the independence of the sources in the bilocal scenario where Alice and Charlie can perform, respectively, $m_\text{A}$ and $m_\text{C}$ measurements is $H(A_1, \dots, A_{m_\text{A}}, C_1, \dots, C_{m_\text{C}}) = H(A_1, \dots, A_{m_\text{A}}) + H(C_1, \dots, C_{m_\text{C}})$~\cite{Chaves2012}.

In order to obtain constraints on the entropies of the variables an experimenter has access to, one must marginalise the entropic cone over the variables that represent the sources of shared randomness.
This procedure can be addressed with Fourier-Motzkin elimination~\cite{Fritz2013,chaves2014inferring}.
An explicit example of this procedure, that derives an entropic inequality in the triangle scenario, can be found in~\cite[Proof of Lemma 2.14]{Fritz2012}.
Another example, in the bilocal scenario, is given in Ref.~\cite{Budroni2016}.
This is the inequality
\begin{equation}
  H(A_0,C)\leq H(A_0,B) + H(C|A_1,B),
  \label{eq:centropy}
\end{equation}
where $A_0$ and $A_1$ are the random variables corresponding to the two possible measurements that Alice can choose to perform.
This inequality is framed in a scenario where Alice has a choice of measurement while Bob and Charlie do not.
While the distributions that arise in this scenario all admit a Bell-local model [recall Eq.~\eqref{BellLocal}], Ref.~\cite{Budroni2016} finds distributions that violate Eq.~\eqref{eq:centropy}.
As a final example, the entropic approach is used in Ref.~\cite{chaves2021causal} to analyse the freedom of choice loophole in the bipartite Bell scenario, by characterising the correlations compatible with the network that models an experiment where the loophole is present.

The fact that the entropic approach considers approximations of the entropic cone for $n\,{>}\,3$, even with the use of non-Shannon constraints, prevents it from identifying all the observations that can be produced by network local models.
Also, on a more fundamental perspective, the full probability distribution represents a larger amount of information than the variables' entropies.
While it is possible to identify nonlocal network correlations using the entropic approach (an example of this is the case of the GHZ distribution of Eq.~\eqref{eq:GHZ} violating \cite[Lemma 2.14]{Fritz2012}), it is not clear that an exact characterisation of the corresponding sets of correlations can be achieved by constraining the variables' entropies.
For example, the W distribution of Eq.~\eqref{eq:W} has not been identified as incompatible with a local model in the triangle using entropic techniques~\cite{Weilenmann2017}, while other methods such as inflation~\cite{wolfe2019inflation} have been capable of doing so.

\subsubsection{Quantum correlations}
\label{sec:qentropy}
In contrast with correlations admitting local models, quantum correlations are characterised by the fact that there exist sets of observables that cannot be jointly measured.
Examples of these sets are the set of different measurements performed by a same party, or the systems input to a measurement device and the variable representing the outcome of its measurement.
Therefore, when using the entropic approach to characterise quantum correlations in networks, one cannot set the basic Shannon constraints on the joint entropies of any set of variables, but only on the entropies of jointly measurable ones~\cite{Chaves2015}, for which a joint quantum state can be defined.
Moreover, in further analogy, the entropy function now corresponds to a quantum entropy, such as the von Neumann entropy. These new functions do not always satisfy the monotonicity constraint.
For the cases where monotonicity does not hold, a weaker notion is imposed \cite{Lieb1973weakmonotonicity,Linden2005nonshannon,Weilenmann2017}, which instead ensures the non-negativity of the sum of conditional entropies.

Finally, in order to relate the entropies of different jointly measurable sets, one makes use of data-processing inequalities, that capture the notion that the information content of a quantum system cannot be increased by means of local operations.
An example of this type of inequalities is the upper bound of the mutual information of two outcome variables by the mutual information of their parent quantum states.
More precisely, in the example of the triangle scenario of Figure~\ref{fig:triangle}, the data-processing inequalities would take the form $I(A\,{:}\,B)\,{\leq}\, I(\beta_\text{A} \gamma_\text{A}\,{:}\,\alpha_\text{B} \gamma_\text{B})$, and analogously for the pairs $\{A,C\}$ and $\{B,C\}$.

The quantum entropic approach is developed in Ref.~\cite{Chaves2015} and used to recover and generalise Information Causality, an information-theoretic principle satisfied by quantum mechanics~\cite{pawloski2009ic}.
In the same work, the quantum entropic approach is also applied to characterise the correlations achievable in the triangle scenario, observing that the entropic cones obtained by using just Shannon-type constraints is the same in the quantum and classical cases.
The classical cone has been since then further shrunk through the use of non-Shannon inequalities~\cite{Weilenmann2018}, and initial steps have been taken in the case of the quantum cone~\cite{Linden2005nonshannon}.

\section{Towards theory-independent network correlations}
\label{section4}

\begin{figure}[t]
  \includegraphics[width=0.9\columnwidth]{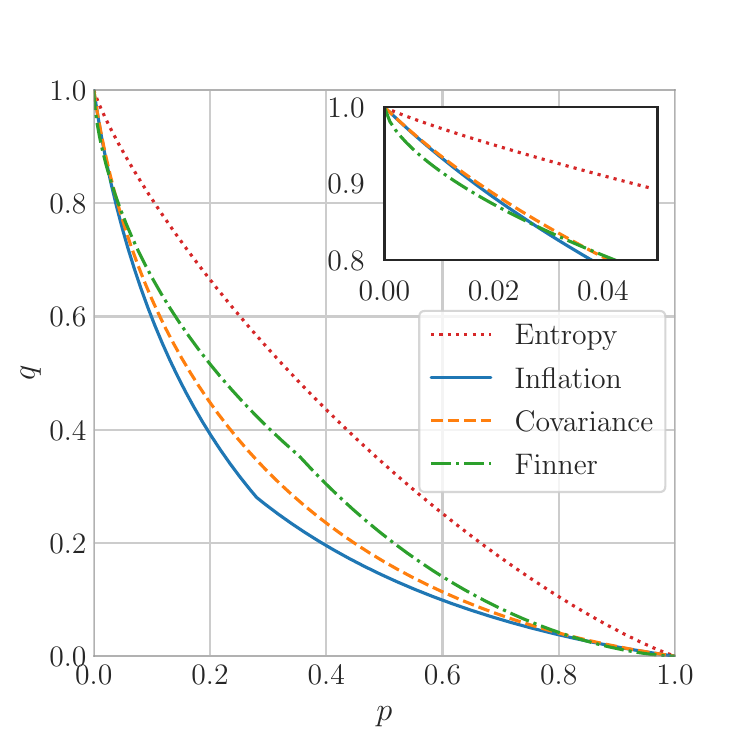}
  \caption{Application of all the methods discussed in Section~\ref{section4} for deducing necessary conditions for the compatibility of a family of probability distributions with the triangle network. The considered family is $P_{p,q}\,{\coloneqq}\,p[000]+q[111]+\frac{1-p-q}{6}([001]+[010]+[011]+[100]+[101]+[110])$, with $p\,{\geq}\,0$, $q\,{\geq}\,0$, $p\,{+}\,q\,{\leq}\,1$.
  The different curves denote the boundaries in the $p-q$ plane above which each method is capable of certifying that the distribution cannot be generated in the triangle scenario.
  The dotted red curve is implied by the entropic condition \eqref{entropicineq}. The solid blue curve is implied by the inflation-based arguments behind the condition \eqref{eq:nsi}. The dashed orange curve is implied by the covariance matrix approach described in Section~\ref{subsec:CovMatrix} (also analytically derived in~\cite[Figure 1]{Berg2020SDPCovarTestAllNetwork}). The dot-dashed green curve is given by the Finner inequality \eqref{eq:finnerexplicit}.
  The inset depicts a zoom-in on the region $p\,{\in}\,[0,0.05]$, where the boundaries illustrating the different methods intersect.
  }
  \label{fig:nosignaling}
\end{figure}

In all the situations studied thus far, we have phrased compatibility relative to a network with the sources sending systems described by a particular theory, be it classical [where we looked for compatibility with Eq.~\eqref{Eqlocal}] or quantum [where the corresponding model is Eq.~\eqref{eq:quantummodel}].
A very interesting avenue is removing as many assumptions as possible in the theory that describes the systems sent by the sources, leaving only those that are strictly necessary to have a reasonable notion of a system and transformations over it (see, for instance, Refs.~\cite{Barrett2007GPT,Henson2014,chiribella2009framework,Chiribella2010GPT}), and thus exploring the most general correlations that can be achieved in a network and the constraints on compatible correlations that arise exclusively from the network structure.
These correlations are the focus of this section.
There are several approaches to the matter, and those that we review below are compared through application to a simple problem illustrated in Figure~\ref{fig:nosignaling}\footnote{It must be noted, however, that the general performance of each method in arbitrary networks is still not fully understood.
Therefore, one should not be surprised to find that methods that seem to provide looser characterisations in Figure~\ref{fig:nosignaling} happen to be tighter when applied to different distributions or different networks.}.
On one hand, in Section~\ref{sec:nonsignaling} we consider general techniques that have been described throughout the review, which now are applied without any regard to the particular theory that describes the states and measurement devices, only focusing on the constraints imposed by the network structure.
On the other, in Sections \ref{subsec:CovMatrix} and \ref{subsec:FinnerIneq} we describe two interesting methods that, while originally tailored to be satisfied by network local models, hold also by network quantum correlations and in several generalised operational-probabilistic theories.

\subsection{Theory-independent network constraints}
\label{sec:nonsignaling}
Classical sources, represented by local variables, and quantum sources, represented by quantum states, may be viewed as particular instances of more general, theory-independent, sources.
The latters are only restricted by the no-signaling principle: operations on one subsystem cannot instantaneously relay information to another subsystem.
This principle stems from relativity and it is also the cornerstone behind the notion of distinct parties.
Thus, the no-signaling principle is the minimal requirement for a set of space-like separated observations to be compatible with a given network.
This naturally leads to the question of how one can determine the most general, theory-independent, correlations relevant in networks and how to understand the role of the no-signaling principle.

\subsubsection{Entropic approach}
One approach, which extends ideas encountered previously in Section~\ref{Entropy}, is based on using entropies of the observed correlations in order to capture theory-independent constraints. Entropic formulations of such correlations are, however, not straightforward because there exists no equivalent of the Shannon and von Neumann entropies defined for theory-independent systems. However, a characterisation of the entropic cones corresponding to theory-independent correlations in several networks is developed in Ref.~\cite{Budroni2016}. The entropy cone for a particular scenario is, in this case, the intersection of the Shannon cones (recall, characterised in Section~\ref{Entropy}) corresponding to every set of coexisting variables in the scenario. Since the independence of the sources implies linear constraints at the level of entropies, the relevant entropic characterisation is an intersection of the convex cone intersection of all Shannon cones (that imposes the standard no-signaling constraints) with a hyperplane (the linear constraints implied by the independence of the sources). Ref.~\cite{Budroni2016} completely characterises this set in the bilocal scenario with binary inputs and finds that there exist correlations that violate the entropic bilocal Bell inequality \eqref{eq:centropy}. Furthermore, in the triangle network without inputs, Ref.~\cite{Henson2014} proves that the inequality
\begin{equation}
\label{entropicineq}
I(A:B)+I(A:C)\leq H(A),
\end{equation}
initially proved in \cite{Fritz2012} to hold for correlations admitting triangle local models of the form of Eq.~\eqref{eq:localtriangle}, is in fact a theory-independent inequality, and any correlation generated in the triangle will satisfy it irrespective of the nature of the sources and measurement devices.
This inequality enables one to bound e.g.~the noise to which the GHZ distribution must be exposed in order to be compatible with the network structure (see Figure~\ref{fig:nosignaling}).

\subsubsection{Inflation}\label{sec:nsinflation}
\begin{figure}[t]
	\centering
	\includegraphics[width=0.8\columnwidth]{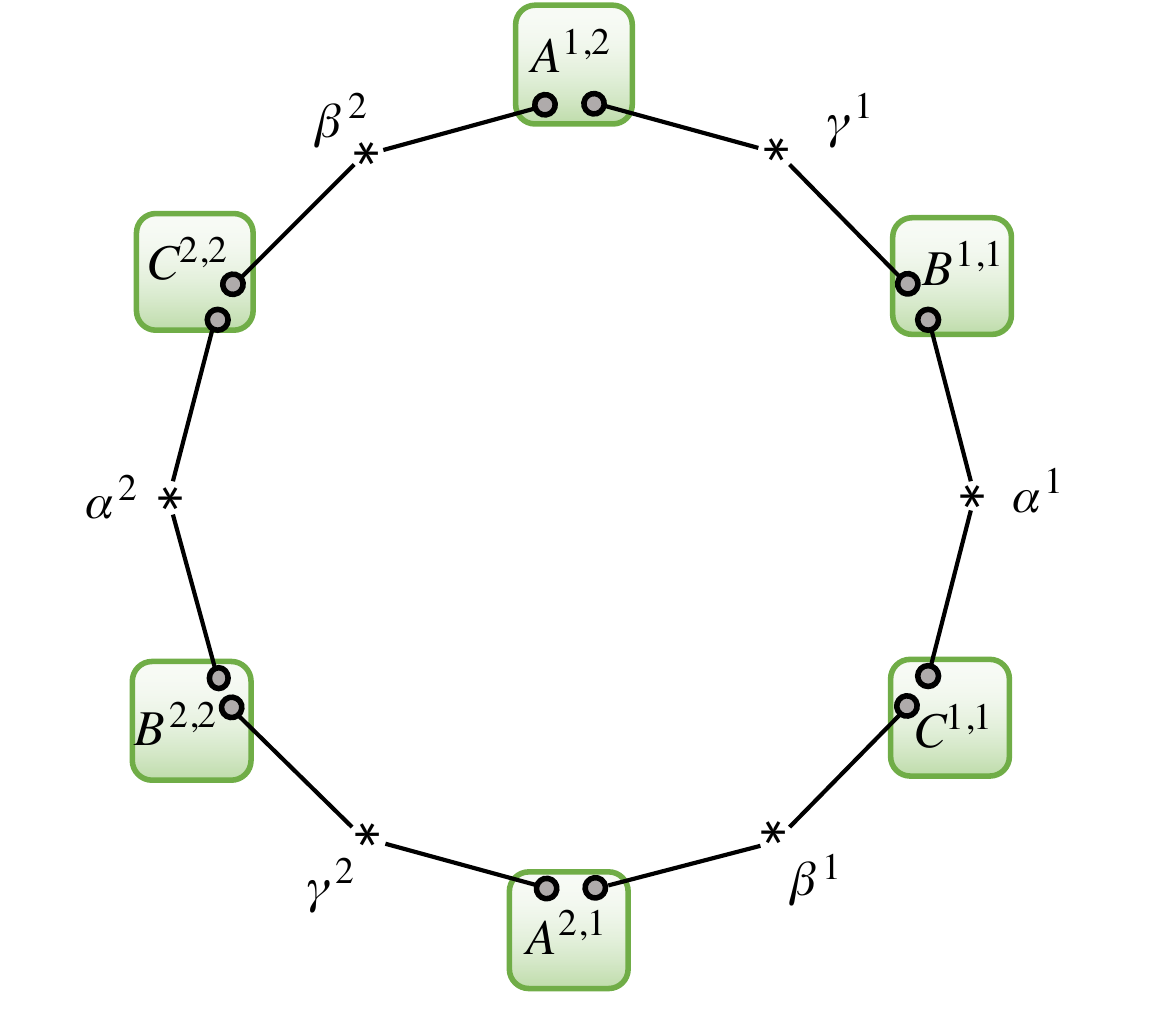}
	\caption{Example of non-fanout inflation: six-partite ring inflation of the triangle scenario (the arrows denoting inputs and outputs have been removed for clarity). In this inflation, the sources and measurement devices are exact copies of those in the original triangle (see Figure~\ref{fig:TriangleCInf}). In particular, the systems prepared by those sources are not cloned before being sent to the parties, in contrast with Figure~\ref{fig:webinflation}.
  Therefore, this inflation only captures theory-independent constraints on correlations compatible with the triangle scenario.}
	\label{fig:nonfanout}
\end{figure}

Another approach to theory-independent correlations in networks continues the idea of inflation, which was discussed for classical and quantum systems in Section~\ref{Inflation}.
A procedure for adapting the inflation technique to general correlations was already put forward in the original work \cite{wolfe2019inflation}.
Recall that the success of inflation for local correlations rests on the fact that classical information (such as a local variable) can be cloned and therefore several copies can be sent to different parties. Such cloning is not possible in generalised operational-probabilistic theories~\cite{Barrett2007GPT} (in fact, it is not possible already in quantum theory).
One can consider, however, inflations that do not require information cloning, which are known as non-fanout inflations.
An example is the so-called cut inflation, already depicted in Figure~\ref{fig:TriangleCInfInflated}.
Another one is illustrated in Figure~\ref{fig:nonfanout}, where the triangle network is inflated to a six-partite ring, and which contains the cut inflation upon marginalisation.
Due to the absence of cloned subsystems, these non-fanout inflations cannot directly constrain classical correlations, but they can be used to deduce constraints on any distribution arising in the original triangle scenario.
This is, one may deduce constraints implied by the network structure.
In fact, the proof of the validity of Eq.~\eqref{entropicineq} for theory-independent correlations compatible with the triangle is arguably centered around the cut inflation of Fig.~\ref{fig:TriangleCInfInflated}, before the formal concept of inflation was developed.
In practice, however, the constraints based on inflation arguments quickly become computationally demanding when one considers large networks or many copies.

Non-fanout inflations have been explicitly studied in the context of extending the role of the no-signaling principle to network scenarios. Ref.~\cite{Gisin2020} deduces no-signaling constraints in the triangle network when neither Alice, Bob nor Charlie have an input. Naively, one might believe that no-signaling is not relevant in the absence of inputs, but this is indeed not true in network scenarios due to the independence of the sources. Even in network scenarios where the parties do not have a choice of inputs, the outcomes of one party should be insensitive to whatever the remaining parties do, including the particular arrangement of the topology of their part of the network. This is the core of the notion of no-signaling presented in Ref.~\cite{Gisin2020}. For the case of binary outcomes, Ref.~\cite{Gisin2020} analyses the six-partite ring inflation of the triangle network in terms of the one-, two- and three-body correlators between Alice, Bob and Charlie\footnote{These are written $E_\text{A}$, $E_\text{B}$, $E_\text{C}$, $E_\text{AB}$, $E_\text{AC}$, $E_\text{BC}$, $E_\text{ABC}$. For instance, we have $E_\text{A}=\sum a^{1,2}\,p_\text{inf}(a^{1,2},b^{1,1},c^{1,1},a^{2,1},b^{2,2},c^{2,2})$ when all outcomes take values $\pm 1$, which due to the inflation constraints can be expressed equivalently as \mbox{$E_\text{A}=\sum a^{2,1} p_\text{inf}(a^{1,2},b^{1,1},c^{1,1},a^{2,1},b^{2,2},c^{2,2})$}.}. However, due to the six-partite ring structure effectively opening up the triangle network, the three-body correlator $E_\text{ABC}$ does not appear in the inflation. It is shown that no-signaling and independence  implies the following constraint on the one- and two-body correlators in the triangle network,
\begin{align}
  &\left(1+|E_\text{A}|+|E_\text{B}|+E_\text{AB}\right)^2 \notag\\
  &+\left(1+|E_\text{A}|+|E_\text{C}|+E_\text{AC}\right)^2\notag\\
  & +\left(1+|E_\text{B}|+|E_\text{C}|+E_\text{BC}\right)^2\notag \\
  & \leq 6\left(1+|E_\text{A}|\right)\left(1+|E_\text{B}|\right)\left(1+|E_\text{C}|\right).
  \label{eq:nsi}
\end{align}
In Figure~\ref{fig:nosignaling} this criterion is illustrated for the noisy GHZ distribution and found to give stronger constraints than the entropic inequality \eqref{entropicineq}. Notably, when the three one-body correlators vanish and the three two-body correlators are equal, the inequality can be saturated by wiring copies of a unique building block composed of a nonlocal source and a measurement device \cite{Bancal2021}. It is not yet known whether this building block can be realised in quantum theory.

In summary, theory-independent correlations in networks are strikingly different than in standard Bell scenarios. While the latter is characterised by a linear program [because, aside from positivity and normalisation, the only constraint to impose is no-signaling, which is linear in the probabilities, recall Eq.~\eqref{eq:nonsignaling}], currently the former is only partially characterised through a linear program e.g.~via analysis of entropies or inflations. This leaves an important open problem in whether it is possible to obtain simple and computable necessary and sufficient conditions that characterise the most general correlations achievable in networks.

\subsection{Covariance matrices}
\label{subsec:CovMatrix}
It is also possible to characterise network constraints through the analysis of the covariance matrix of the random variables that describe the parties' outcomes. This perspective, developed in Refs.~\cite{Kela2020SDPCovarTestClassicalNetwork,Berg2020SDPCovarTestAllNetwork,Kraft2020SDP,Beigi2021}, has the benefit of being notably simple.

\begin{figure}
  \centering
  \includegraphics[width=0.3\textwidth]{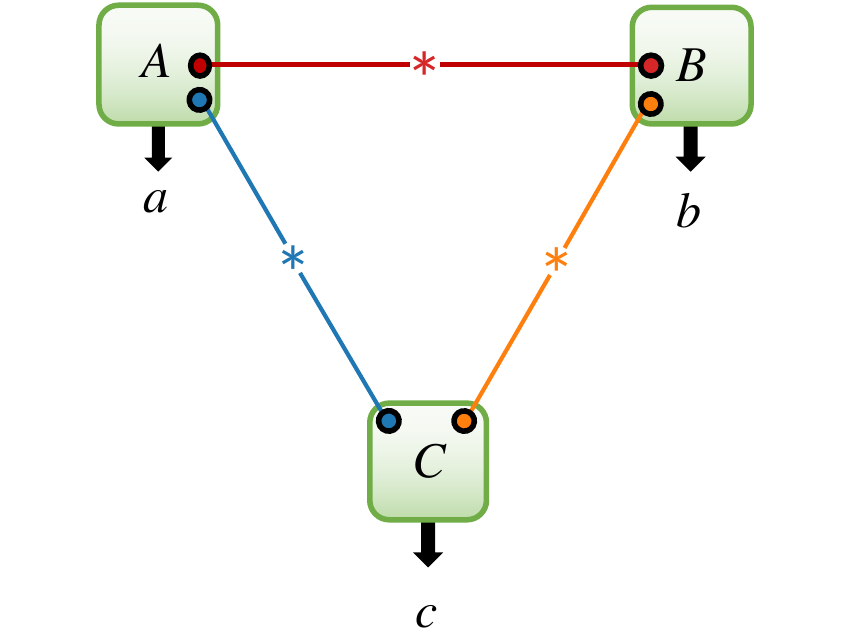}
  \\
  \begin{equation*}
    \Cov(\bv)=
    \begin{pmatrix}
      \squ[c1] & \squ[c1] & \\
      \squ[c1] & \squ[c1] & \\
       & & \squ[white]
    \end{pmatrix}
    +
    \begin{pmatrix}
      \squ[c2] & & \squ[c2] \\
       & \squ[white] & \\
      \squ[c2] & & \squ[c2]
    \end{pmatrix}
    +
    \begin{pmatrix}
      \squ[white] & & \\
      & \squ[c3] & \squ[c3] \\
      & \squ[c3] & \squ[c3]
    \end{pmatrix}
  \end{equation*}
  \caption{Example of semidefinite decomposition of the covariance matrix according to the network topology, in the triangle scenario without inputs. For distributions which can be generated in the triangle scenario (regardless of the nature of the sources), the corresponding covariance matrix decomposes into the block structure depicted in the bottom, where every block is a positive-semidefinite matrix. This method holds for any network, the decomposition being dictated by the network structure.}
  \label{covariance}
\end{figure}

As an illustration in the case of networks without inputs, consider the correlation $p(a_1, \dots, a_n)$.
In the covariance matrix approach, each party is associated a vector space, and each of the parties' outputs is mapped to an element of the corresponding vector space, i.e.~$a_j\,{\mapsto}\, \bv^{(i)}(a_j)$ for every party $i$ obtaining output $a_j$ upon measurement.
Thus, the final joint observation of outcomes $\bar{a}\,{=}\,(a_1,\dots,a_n)$ is mapped to a vector in the direct sum of the parties' spaces, \mbox{$\bv(\bar{a})\,{=}\,\bv^{(1)}(a_1)\oplus\dots\oplus\bv^{(n)}(a_n)$}.
When one considers the covariance matrix of this vector under the correlation,
\begin{equation}
  \begin{split}
    \Cov(\bv)_{i,j} =& \left\langle v_i v_j \right\rangle - \left\langle v_i \right\rangle \left\langle v_j \right\rangle \\
    =& \sum_{a_1,\dots,a_n}p(\bar{a})v_i(\bar{a}) v_j(\bar{a}) \\
    &-\sum_{a_1,\dots,a_n}p(\bar{a})v_i(\bar{a}) \sum_{a_1,\dots,a_n}p(\bar{a})v_j(\bar{a}),
  \end{split}
  \label{eq:CovMatrix}
\end{equation}
not only this matrix is positive semidefinite as any covariance matrix is, but Refs.~\cite{Kela2020SDPCovarTestClassicalNetwork,Berg2020SDPCovarTestAllNetwork} prove that, in the case of local and quantum realisations, if the distribution is compatible with a network, then it decomposes in a sum of positive-semidefinite blocks that is dictated by the network structure.
This is,
\begin{align}
  &\exists C_i =\Pi^{(i)} C_i \Pi^{(i)} \succeq 0 \nonumber \\
  &\st \Cov(\bv)= \sum_{i=1}^m C_i,
  \label{eq:CovMatrixDecompo}
\end{align}
where $\Pi^{(i)}$ are the projectors onto the subspaces corresponding to the parties that are connected to the source $i$ and $m$ is the total number of sources in the network.
Figure \ref{covariance} depicts the decomposition above in the case of the triangle scenario without inputs.
If such a decomposition does not exist, then the distribution cannot be generated in the network neither with local variables nor with quantum systems.

While this method uses distributions generated by parties that do not have a choice of input, Ref.~\cite{Berg2020SDPCovarTestAllNetwork} provides a generalisation to networks where the parties have input choices.
In such a situation, a decomposition analogous to Eq.~\eqref{eq:CovMatrixDecompo} still holds, that contains entries which cannot be computed directly from the probability distribution.
However, finding such a decomposition can still be formulated as a semidefinite programming problem, so the covariance matrix approach maintains its numerical simplicity.

As remarked by Ref.~\cite{Kraft2020SDP}, the decomposition~\eqref{eq:CovMatrixDecompo} can be directly deduced from block coherence theory, which, given a quantum state and an orthonormal basis, aims at characterising the minimal number of the basis elements needed to decompose the state.
This connection allows to directly solve the existence of such a decomposition in case of dichotomic measurements, by providing sufficient conditions for compatibility.
In Refs.~\cite{Kela2020SDPCovarTestClassicalNetwork,Berg2020SDPCovarTestAllNetwork,Kraft2020SDP}, the decomposition \eqref{eq:CovMatrixDecompo} is shown to hold in the restricted case of local and quantum realisations.
Ref.~\cite{Beigi2021} promotes it to a fully theory-independent network constraint.
The proof is restricted to the case of all networks where any pair of parties shares at most one source.
It is also restricted to the case of vectors $\bv^{(i)}$ of dimension one.
These allow to recover the standard formulation of the covariance matrix in Eq.~\eqref{eq:CovMatrix}, at the expense of providing
a weaker characterisation of the sets of theory-independent correlations than what is theoretically possible.
It first reformulates the decomposition~\eqref{eq:CovMatrixDecompo}, considering its dual cone in the set of positive matrices, which is characterised using quantum embezzlement concepts~\cite{vDam2003Embezzlement}.
Then, generalising the connection to block coherence theory established in Ref.~\cite{Kraft2020SDP}, it shows that this decomposition is a consequence of the positivity of the covariance matrices in the asymptotic non-fanout inflations of the networks.

\subsection{Limits from Finner inequalities}
\label{subsec:FinnerIneq}
Another method to determine whether a probability distribution can be generated in a particular network arises from the scrutiny of the problem under the perspective of pure probability theory.
Indeed, the Finner inequalities~\cite{Finner1992} give rise, when interpreted in the context of Bell nonlocality in networks, to necessary criteria for correlations to be compatible with a network.
These constraints have been proven to apply to local, quantum, and correlations arising in some generalised probabilistic theories in large families of networks.

Finner inequalities are a generalisation of the H\"older inequality~\cite{Rogers1888,Holder1889} (which bounds the expectation of a product of functions $f_j\,{:}\,\mathbb{R}\,{\mapsto}\,\mathbb{R}$ by the product of individual expectations), that are easy to interpret as necessary conditions that correlations compatible with a network must satisfy: Consider local stochastic post-processing functions $f_j\,{:}\,a_j\,{\mapsto}\, f_j(a_j)\,{\in}\,\mathbb{R}$ of the parties' outputs. Consider also a vector of weights $(\eta_1,\dots,\eta_m)$ satisfying $\forall i, \sum_{j\,{:}\, i\,{\rightarrow}\, j} \eta_j \,{\leq}\, 1$, where the sum is over all parties connected to source $i$. Then, the Finner inequalities are constraints on the joint expectation value of $\prod_j f_j$ that read
\begin{equation}
  \label{eq:FinnerIneq}
  \expec*{ \prod_j f_j} \leq \prod_j \expec*{f_j^{1/\eta_j}}^{\eta_j},
\end{equation}
where the averages are taken over the probability distribution under scrutiny.
In the case where there is just one source, one recovers the original Finner inequality by forgetting the post-processing functions and directly taking $a_j\,{=}\,\bar{\lambda}_j$.
For different networks the subsets of $(\eta_1,\dots,\eta_m)$ whose sum is bounded are different, giving rise to distinct Finner inequalities. Moreover, for a given network, there are infinitely many Finner inequalities, corresponding to all suitable choices of weights.
In certain cases, however, a finite collection of them provides a complete description of the space of allowed values.
An example of such cases can be found in \cite[Appendix D]{RenouFinner2019}.

The Finner inequalities were introduced in the context of probability theory and graph theory.
Their original proof~\cite{Finner1992}, when interpreted in the context of nonlocality in networks, demonstrates that they hold for any local distribution compatible with the network it applies to.
Beyond that, Ref.~\cite{RenouFinner2019} proves that the Finner inequalities also hold in networks composed of bipartite quantum sources, and in the triangle scenario where the sources are no-signaling boxes\footnote{No-signaling boxes are the building elements of the box-world generalised probabilistic theory~\cite{Barrett2007GPT} that generates stronger-than-quantum correlations~\cite{Popescu1994}.} and the parties are only allowed to do wirings of their systems.
It also provides an explicit example of one of these inequalities, namely
\begin{equation}
  p(a,b,c)\,{\leq}\,\sqrt{p(a)p(b)p(c)},
  \label{eq:finnerexplicit}
\end{equation}
which holds in the triangle scenario of Figure~\ref{fig:triangle} and is used in Ref.~\cite{Shukla2020} to analyse quantum networks obtained in the context of limited quantum resources.

Ref.~\cite{RenouFinner2019} conjectures that the Finner inequalities hold for any network and any no-signalling theory.
This conjecture is proved in Ref.~\cite{LuoFinner2021}, which also provides an explicit proof of the inequality $p(a,b,c)\,{\leq}\,p(a)^{1-\frac{1}{m}}p(b)^{\frac{1}{m}}p(c)^{1-\frac{1}{m}}$ holding in the bilocal scenario for any integer $m\,{\geq}\,2$.

Finner inequalities are naturally formulated in networks where the parties do not have a choice of measurement performed to their respective systems.
Nevertheless, there is a straightforward way of applying them in networks with inputs, by checking them for every possible choice of inputs.
The question of whether this approach is the only possible one, or in the contrary new inequalities can be devised for networks with inputs, remains open.

\section{Experiments}
\label{section5}
\label{secExperiments}
Quantum nonlocality in networks is now not just a theoretical concept, but also a subject of experiments. This constitutes a natural continuation of decades of experimental investigation of quantum nonlocality in standard Bell scenarios (for a review of experiments in standard Bell nonlocality see, e.g.,~Refs.~\cite{Genovese2005, Pan2012, BrunnerRMP}). Its interest lies not only in testing the predictions of quantum theory in more sophisticated scenarios, but also in the key role attributed to the experimental realisation of entangled measurements and for the future development of quantum information protocols in networks, such as the quantum internet \cite{Kimble2008, Wehner2018}. A program for quantum nonlocality in networks constitutes a formidable challenge, which ultimately includes the implementation of synchronised, high-quality, independent sources distributing entangled states over large distances, and scalable entangled measurements of several distinct systems. In addition, quantum nonlocality in networks faces a challenge inherited from standard Bell experiments, namely the closing of experimental loopholes, commonly associated to technical imperfections in Bell experiments, which can be exploited to reproduce quantum correlations in (network) local models\footnote{While the idea of nature conspiring against experimenters using seemingly contrived local models is arguably far-fetched, the practical importance of closing of such loopholes, beyond purely foundational considerations, becomes more evident in quantum information tasks such as random number generation and quantum key distribution which feature a natural notion of an adversary attempting to corrupt the protocol.}. Furthermore, a pivotal challenge is the justification of the fundamental assumption of network Bell experiments, namely the independence of the sources. Increasing confidence in this assumption is closely related to increasingly sophisticated realisations of synchronised, physically distinct, quantum sources. In what follows, we first discuss entanglement swapping experiments in the context of quantum nonlocality and then direct experimental tests of network nonlocality.

\subsection{Indirect tests: entanglement swapping and nonlocality}
Entanglement swapping is a key primitive for quantum nonlocality in networks. Indeed, the simplest entanglement swapping scenario is identical to the bilocal scenario (see Figure~\ref{fig:bilocal}): two independent sources each distribute an entangled pair of quantum systems  and by performing an entangled measurement on one share of each pair, the two remaining systems, which have never interacted, can become entangled \cite{Zukowski1993}. The paradigmatic measurement for entanglement swapping is the Bell state measurement \eqref{BSM}, i.e.~the projection of two qubits in a basis of maximally entangled states. The state of the \textit{a posteriori} entangled qubits (those held by Alice and Charlie) depends on the specific outcome of the Bell state measurement performed by Bob. By conditioning on this outcome, the swapped entanglement is said to be  event-ready. One can then attempt to use this entanglement (e.g.~as a quality benchmark) to violate a Bell inequality. Naturally, since standard local models are strictly stronger than network local models, it follows that a violation of a standard Bell inequality based on event-ready entanglement constitutes an indirect demonstration of network nonlocality. Therefore, such indirect tests significantly pre-date the research program aimed at understanding Bell nonlocality in networks.

In photonics-based setups, which are natural candidates due to the possibility of realising significant distances, Bell state measurements~\cite{Mantle1996} constitute a challenge. In fact, a complete Bell state measurement cannot be performed deterministically with linear optics~\cite{Lutkenhaus1999} (moreover, the efficiency cannot exceed $50\%$~\cite{Calsamiglia2001}) unless auxiliary degrees of freedom are made available~\cite{Kwiat1998}. Nevertheles, it is possible to implement partially resolved Bell state measurements, i.e.~measurements that only distinguish two of the four basis states, using linear optics without auxiliary degrees of freedom~\cite{Weinfurter1994, Braunstein1995}. Such a measurement was used to realise photonic entanglement swapping for the first time in 1998~\cite{Pan1998}. However, in this proof-of-principle experiment, the visibility of the event-ready entanglement was insufficient to enable a violation of the CHSH inequality. This limitation was overcome in 2001 when Ref.~\cite{Pan2001} demonstrated event-ready entanglement of a visibility well above the required threshold. A few years later, such visibilities were achieved with photons separated over a substantial distance (over 2 km) \cite{Riedmatten2005}. In 2002, entanglement swapping with high visibility was demonstrated in a different way, by using only two independent photons with each one being a physical carrier of entanglement between two of its degrees of freedom~\cite{Sciarrino2002}. Furthermore, entanglement swapping has also been demonstrated in continuous-variable systems~\cite{Jia2004, Takei2005}.

The first demonstration of nonlocality from event-ready entanglement was reported in Ref.~\cite{Jennewein2001} in 2001.
In this experiment, both entangled photon pairs were generated with a spontaneous parametric down-conversion (SPDC) process from the same laser pulse, which led to correlations in the phases of the two entangled photon pairs.
In 2006, this motivated a more faithful event-ready violation of the CHSH inequality which featured synchronised entanglement sources based on separate SPDC processes and separate (but optically dependent) lasers~\cite{Yang2006}.
Soon afterwards, Refs.~\cite{Kaltenbaek2006,Halder2007} implemented entanglement swapping with fully independent sources, but the event-ready entanglement was not strong enough to violate the CHSH inequality.
It is interesting note that the visibilities reported in Refs.~\cite{Kaltenbaek2006,Halder2007} would have been sufficient to violate the BRGP inequality \eqref{BRGP}, which was developed a few years later.
Subsequently, in 2009, Ref.~\cite{Kaltenbaek2009} reported on a stringent demonstration of nonlocality from event-ready entanglement with separate and independent sources.
Notably, taking another route to entanglement swapping, Ref.~\cite{Schmid2009} demonstrated a violation of the CHSH inequality via event-ready entanglement established through a probabilistically implemented complete Bell state measurement.
Furthermore, event-ready entanglement and nonlocality have been demonstrated in hybrid systems featuring one source of discrete-variable single-photon entanglement and one source of continuous-variable entanglement~\cite{Takeda2015,Guccione2020}.

Photonic entanglement swapping has also been demonstrated beyond the simplest network scenario. Ref.~\cite{Goebel2008} considered a four-party chain network involving three SPDC sources and performed two separate partial Bell state measurements in order to entangle two initially independent photons.
Also, a four-party star network has been realised in which three sources emit two-photon entangled states: three initially independent photons are projected onto a GHZ state in order to remotely prepare the remaining three photons in a GHZ state~\cite{Lu2009}.
In addition, entanglement swapping has been performed between two initially independent three-photon GHZ states~\cite{Su2016}.

Photons can also be employed as means for event-ready entanglement between two, initially independent, matter qubits.
To this end, one can consider sources of atom-photon entanglement in which the photon acts as a flying qubit~\cite{Blinov2004,Moehring2004,Volz2006}.
By subjecting the photons from two different sources to an entangled measurement, one obtains event-ready atom-atom entanglement with significant spatial separation.
This was demonstrated in 2007 and 2008 in Refs.~\cite{Moehring2007,Yuan2008}.
In Ref.~\cite{Moehring2007}, using a photonic partial Bell state measurement, two ions separated by one meter were entangled, albeit at a low success probability and with a visibility insufficient to violate the CHSH inequality.
The visibility of the event-ready entanglement was improved in 2008 and nonlocality was demonstrated with a high detection efficiency~\cite{Matsukevich2008}.
In 2012, event-ready atom-atom entanglement with a separation of 20 meters, as well as a demonstration of its nonlocality, was reported in Ref.~\cite{Hofmann2012}.
Notably, loophole-free violations of the CHSH inequality have been reported based on the implementation of entanglement swapping in photon-electron spin systems~\cite{HansonLoopholeFree2015} as well as in photon-atom systems~\cite{WeinfurterLoopholeFree}.
In contrast, doing away with photons and instead considering ion-trap systems, fully deterministic entanglement swapping and a violation of the CHSH inequality was reported in Ref.~\cite{Riebe2008}.
In 2019, high-fidelity deterministic entanglement swapping was demonstrated on superconducting qubits~\cite{Ning2019}.

\begin{table*}[t!]
	\begin{tabular}{c|c|c|c|c|c|c|c|c}
		Reference & Network                                                                  & System & Inequality                                                                                           & \begin{tabular}[c]{@{}c@{}}Type of \\  measurement\end{tabular}       & \begin{tabular}[c]{@{}c@{}}Independence\\ of sources\end{tabular}                              & LOC & DET & FOC \\ \hline \hline
 		Saunders et al. \cite{Saunders2017} & Bilocal                                                                  & Photons         & \cite{Branciard2012}                                                                                            & Partial BSM                                                          & \begin{tabular}[c]{@{}c@{}}Partly\end{tabular}  & Open                                                        & Open                                                         & Open                                                                 \\ \hline
		Carvacho et al. \cite{Carvacho2017} & Bilocal                                                                  & Photons         & \cite{Branciard2012}                                                                                            & Partial BSM                                                          & Assumed                                                                                        & Open                                                        & Open                                                         & Open                                                                 \\ \hline
		Sun et al. \cite{Sun2019} & Bilocal                                                                  & Photons         & \cite{Branciard2012}                                                                                            & Partial BSM                                                          & \begin{tabular}[c]{@{}c@{}}Strict\end{tabular} & Closed                                                      & Open                                                         & Closed                                                               \\ \hline
		Poderini et al. \cite{Poderini2020experimental} & \begin{tabular}[c]{@{}c@{}}Star\\ $N\,{\leq}\, 4$\end{tabular} & Photons         & \cite{Tavakoli2014}  \& \cite{Poderini2020experimental}  & Separable                                                            & Strict                                                                   & Open                                                        & Open                                                         & Open                                                                \\ \hline
		B\"aumer et al. \cite{Baumer2020} & \begin{tabular}[c]{@{}c@{}}Star\\ $N\,{\leq}\, 5$ \end{tabular} & Transmons       & \cite{Tavakoli2014} \& \cite{Tavakoli2020}                           & \begin{tabular}[c]{@{}c@{}} Full BSM \\ \& EJM\end{tabular} & Assumed                                                                                        & Open                                                        & Closed                                                       & Open
	\end{tabular}
	\caption{Summary of selected network nonlocality experiments. LOC, DET and FOC denote the locality, detection and freedom of choice loopholes, respectively.}\label{TableExperiments}
\end{table*}

\subsection{Direct tests: violating network Bell inequalities}
Direct experimental tests of network nonlocality, i.e.~experiments that implement protocols tailored for violating the constraints of network local models, is a recent and growing program. The first proof-of-principle photonics experiments were reported in 2017 by Saunders et al.~\cite{Saunders2017} and Carvacho et al.~\cite{Carvacho2017}, both focused on the bilocal scenario. In 2019, a more stringent violation of bilocal Bell inequalities was demonstrated, which closed several loopholes~\cite{Sun2019}. The bilocal Bell inequalities discussed in Section~\ref{sec:bilocality} have also been violated in quantum protocols without entangled measurements~\cite{Andreoli2017b}. In 2020, violations based on separable measurements were demonstrated for star networks with up to four branch observers~\cite{Poderini2020experimental}. Departing from photonic implementations, Ref.~\cite{Baumer2020} considered transmon quantum computers and star networks of up to five branch parties and demonstrated quantum correlations using fully entangled multi-qubit measurements. We now proceed to discuss these experiments one by one. Their key features are listed in Table~\ref{TableExperiments}.

Refs.~\cite{Saunders2017,Carvacho2017} report on bilocality experiments performed using polarisation qubits generated via SPDC sources. In the spirit of independent sources, two separate nonlinear crystals were used to generate the respective entangled pairs. However, in both experiments, the beams impinging on the respective crystals originate from the same laser. In the experiment of Saunders et al., this potential origin of correlation between the sources is addressed by using a quantum random number generator together with a phase shifter in order to give a random offset to the beam impinging on the second crystal. Both experiments use linear optics to violate the BRGP inequality~\eqref{BRGP}, achieving $\mathcal{S}_\text{BRGP}\,{=}\,1.25\,{\pm}\, 0.04$~\cite{Saunders2017} and $\mathcal{S}_\text{BRGP}\,{=}\,1.268\,{\pm}\, 0.014$~\cite{Carvacho2017}, respectively. The implemented protocol is based on the complete Bell state measurement. However, due to the limitations of standard linear optics~\cite{Loock2004}, the experiments actively exploit an assumption of the detected events constituting a fair sample of all de-facto events (the so-called detection loophole) in order to emulate this measurement by using several implementations of a partial Bell state measurement for entanglement swapping. Notably, Ref.~\cite{Saunders2017} also demonstrates a violation of a modified version of the BRGP inequality, which is tailored for a partial Bell state measurement. This protocol was implemented without active exploitation of the detection loophole.

Sun et al.~\cite{Sun2019} take a stringent approach to realising the bilocal scenario. In this experiment, the independence of the sources is imposed using two separate, synchronised, SPDC sources with their relative phases randomised by periodically shifting the laser from spontaneous emission to stimulated emission. Furthermore, the experiment implements fast switching, which allows for spacelike separation between the relevant parts of the experiment (state emissions, generation of input and performing the measurements). This allows for the elimination of the so-called locality loophole (the possibility of exchanging a signal between events) as well as the freedom of choice loophole (the possibility of correlating the inputs of Alice and Charlie when they are generated within the same lightcone). It does, however, leave the detection loophole open. The experiment also captures the spirit of a quantum network since the sources and parties are all subject to significant separations: the smallest distance between the parties and sources is roughly 100 meters. This experiment demonstrates a faithful violation of a a modified version of the BRGP inequality which is tailored for implementation with a partial Bell state measurement.

Recently, violations of network Bell inequalities have been demonstrated beyond the bilocal scenario. Poderini et al.~\cite{Poderini2020experimental} considered star networks with up to four branch parties, with each of the four bipartite sources of polarisation qubits being realised with independent lasers and SPDC processes, and parties and sources being separated by tens of meters. In contrast to the previous, this experiment is based on a separable measurement in the central node of the network. This considerably simplifies the implementation, as there is no need for interference in the central node between the photons originating from the different sources. Instead, the central node performs synchronised single-photon measurements implemented, for instance, using a large coincidence window. Therefore, this may  be viewed as a synchronisation of several independent standard Bell experiments. For star networks with two, three and four branch parties, the measured correlations were shown to violate a modified version of the star network Bell inequality~\eqref{StarIneq} based on binary outputs \cite{Tavakoli2014}. Furthermore, motivated by device-independent considerations, Ref.~\cite{Poderini2020experimental} also considered the case of adding more inputs for the branch parties. In addition, violations of bilocality based on separable measurements have also been demonstrated in Ref.~\cite{Andreoli2017b} without assuming a fixed reference frame between the involved parties. Furthermore, Ref.~\cite{Agresti2021} considered the bilocal network but without assuming the underlying network structure in its theoretical analysis, and performed two synchronised tests of the CHSH inequality in order to certify that the global state is a pair of singlets. The experiment performs the certification up to a global fidelity of $86.3\%$.

Network Bell experiments have also been performed  beyond optical setups, specifically in transmon quantum computers. In 2020, Ref.~\cite{Baumer2020} employed the publicly available, cloud-controlled IBM quantum computers~\cite{IBMexperience} to implement star network Bell experiments. Since all the qubits are located on a single quantum processor, this experiment does not follow the spirit of sizable separation between parties in Bell experiments. In this sense, it may be considered a simulation of network Bell experiments. Furthermore, the device's architecture is trusted, meaning that the independence of the sources is assumed. Nevertheless, Ref.~\cite{Baumer2020} evidences that the assumption of independent sources is accurate on the level of the measured probabilities. The experiment targets a violation of the star network Bell inequality \eqref{StarIneq} by implementing a deterministic, fully entangled, measurement for entanglement swapping in the central node, i.e.~a $2^m$-outcome Bell state measurement of $m$ qubits. Violations are reported for two, three, four and five branch parties. Furthermore, the same platform is used to implement the Elegant Joint Measurement \eqref{EJM} by realising a circuit, requiring three CNOT gates, proposed in Ref.~\cite{Tavakoli2020}. This is used both to violate  a bilocal Bell inequality and to implement the quantum protocol for the triangle network discussed in Section~\ref{subsec:OverviewTC}. However, it remains unclear whether the measured data violates triangle local models due to the lack of noise-robust analytical criteria for detecting the nonlocality of the distribution given in Eq.~\eqref{eq:elegant}. No postselection is made on the detections and therefore the nonlocality is not vulnerable to the detection loophole.

\section{Further topics and related areas}
\label{section6}
In this section we discuss additional topics in network nonlocality.
These include the current progress towards definitions of genuine nonlocality in networks, its application in device-independent quantum information processing protocols, its relationship to standard multipartite Bell nonlocality, and machine learning approaches to tackle problems in the area.
We also discuss the notion of entanglement in networks and the physics of more general causal structures, which are related to network nonlocality.

\subsection{Stronger notions of network nonlocality}
\label{genuine}
As discussed in Section \ref{subsec:OverviewReplaceInputs}, distinguishing disguised forms of network nonlocality from those genuine to the network structure on which they arise is central.
This challenging matter involves both the development of stronger, new notions of network nonlocality geared towards different aspects in which the standard definition proves itself unsatisfactory, and the development of new tools to explore their features and consequences.

Two conceptually different approaches have been introduced so far.
Ref.~\cite{PozasGenuine} takes a theory-independent approach to networks and considers as correlations of interest those which can be explained only if every source in a given network distributes systems that cannot be modeled by local variables.
Such correlations are termed fully network nonlocal.
Thus, if a distribution obtained on a given network can be modelled through at least one source admitting a local-variable description, it is not fully network nonlocal.
This definition is convenient in terms of numerical characterisation because it enables one to describe the desired correlations by considering hybrid models of inflation, in which one source is local and all other sources are general nonlocal resources \cite{wolfe2019inflation,Gisin2020}.
On the conceptual level, detection of full network nonlocality guarantees that no source of a network distributes classical systems.
Notably, adding more sources and parties that act classically, to a network that already displays full network nonlocality, causes the correlations in the expanded network to no longer be fully nonlocal.
Thus, it apropriately addresses e.g.~the Fritz example in the triangle (recall Section~\ref{subsec:OverviewReplaceInputs} and Appendix~\ref{AppendixDistributions}).
Another interesting consequence of this concept is that no known quantum violation of the BRGP inequality implies full network nonlocality, i.e.~the known violations can all be simulated in the bilocal scenario when one source  is described by a local variable.
In order to reveal full network nonlocality in quantum theory, concrete protocols based on i.a.~the Elegant Joint Measurement and the three-qubit Bell state measurement have been proposed \cite{PozasGenuine}.
However, full network nonlocality does not in principle (although it does for many concrete quantum examples) prevent wirings of standard Bell inequalities to be considered fully network nonlocal.
It can also be activated in a trivial manner, by grouping parties in a network to create a larger, single party.
This property is known as instability under composition, which notably also applies to genuine multipartite nonlocality~\cite{Svetlichny1987}, genuine multipartite entanglement~\cite{Seevinck2001multipartiteentanglement}, and robust genuine multipartite entanglement~\cite{Luo2021robust}.

A different, complementary, approach is proposed in Ref.~\cite{supicGenuine}.
Although in principle extendible also to theory-independent systems, the focus is on networks that are assumed to be governed by quantum theory.
The authors label a correlation arising in a given network as genuine network nonlocal if it cannot be reproduced by wiring standard Bell nonlocal distributions according to the network structure.
In other words, each source is considered to distribute a black box, and any distribution that cannot be obtained by classically processing the inputs and outputs to those boxes is considered to be genuinely network nonlocal.
Notably, this does not necessarily imply a need for a measurement with entangled eigenvectors due to the inability of simulating all joint product measurements by means of local operations and classical communication \cite{Bennett1999}.
For instance, genuine quantum network nonlocality apropriately addresses the Fritz distribution in the triangle and by construction also wirings of standard Bell experiments.
By employing self-testing techniques, Ref.~\cite{supicGenuine} proves that the correlation achieving the maximum known violation of the BRGP inequality is genuinely network nonlocal under the assumption of quantum theory.
However, noise-robust proofs of genuine network nonlocality, as well as reasonably versatile methods to characterise the non-genuine set of correlations, are still lacking.

Interestingly, neither of the two above discussed concepts are special cases of the other\footnote{Note that if genuine network nonlocality is defined for theory independent models, then it contains full network nonlocality as a special case.}.
For instance, two wired PR boxes in a bilocal scenario manifest full network nonlocality but not genuine network nonlocality (when quantum theory is not assumed).
Conversely, the largest known quantum violation of the BRGP inequality \eqref{BRGP} is not fully nonlocal but is genuinely nonlocal in the quantum bilocal scenario.
This is a clear example of the richness of the question, pointing to the intuition that new alternative notions of network nonlocality, complementary of these two, will probably be discovered in the future.

\subsection{Device-independent considerations}
\label{Device-independent}
Nonlocality is at the heart of the device-independent approach to quantum information science (see e.g.~Ref.~\cite{Pironio2016} for a short overview in the context of Bell scenarios).
In this approach, deductions are made based on experiments subject to minimal assumptions, most notably without any characterisation of the internal workings of the operating devices \cite{May98,Barrett05,Acin2007}.
This approach is naturally extended to network scenarios, but requires one to assume the independence of the sources. Notably, such independence cannot  be guaranteed in a device-independent manner.
Whereas source independence can be justified with increasing confidence by suitably adapting experimental setups (see Section~\ref{secExperiments}), it is ultimately an additional assumption.
This provides a distinction in network scenarios for device-independent protocols aiming to assert a property of nature and for device-independent protocols aiming to perform a quantum information task.
In the former, the independence of the sources is a reasonable assumption whereas in the latter, it is not far-fetched to consider a malicious party tampering with the sources in a coordinated manner, thus affecting their independence.

From another point of view, the assumption of independent sources has, in certain contexts, a status similar to the freedom of choice assumption in standard Bell scenarios.
For instance, in Section~\ref{subsec:OverviewReplaceInputs} (see Figure~\ref{fig:IsomorphismBellNonLocNetwNonloc}), we saw that correlations violating the CHSH Bell inequality can be identified with a nonlocal quantum distribution in the four-party chain network arising from parties without inputs.
This identification exchanges the freedom of choice assumption with the assumption of independent sources.

In the following we provide an overview of a few aspects where the device-independent framework and network nonlocality are combined.

\paragraph*{Quantum key distribution:}
The device-independent framework, when applied in quantum key distribution, allows two parties to extract a common key and guarantee its security by analysing the outcome statistics of a Bell-type experiment. It enables a high level of security which is relevant in view of increasingly sophisticated eavesdropping attacks \cite{Lydersen,Gisin2002,Xu2020,portmann2021security}. Generally, an eavesdropper can correlate a system of their possession with the systems measured by the parties in order to gain information about their outcomes. Such eavesdropping can be detected by testing the nonlocal correlations of the honest participants. Specific Bell inequalities are used for bounding the information leaked to the eavesdropper or, alternatively, the secure key rate (see e.g.~Refs.~\cite{Wigner,Braunstein,Barrett05,Acin2007,Masanes11,Vazirani14}).

In Bell scenarios, the eavesdropper is assumed to hold a (potentially stronger-than-quantum) system correlated with the source that distributes systems to the honest parties.
In network scenarios, this changes into independent systems that can each be correlated with different sources.
Ref.~\cite{lee2018di} introduced an intermediate device-independent scenario by trusting a subset of sources that are not correlated with a single system held by an eavesdropper.
For a chain network consisting of $n$ sources, Ref.~\cite{lee2018di} finds a monogamy relation between the amount of information that the eavesdropper can obtain about the outcomes of the extremal parties and the violation of the BRGP inequality that these parties can achieve: if the violation is larger, the leaked information is smaller.
Similar results were also proven to hold in star networks \cite{lee2018di}.
These results have been extended to general networks in Ref.~\cite{Luo2020a} by using the network Bell inequality derived in Ref.~\cite{Luo2018} as the quantifier of monogamy.
All these monogamy relations are generally conservative because of the assumptions of independence and that the eavesdroppers are allowed to manipulate stronger-than-quantum systems.
In the case of the chain network, Ref.~\cite{wolfe2019qinflation} strengthened the existing bounds by making the realistic assumption that the eavesdropper has access only to quantum systems.
Importantly, note that if all the independent sources are correlated by the eavesdropper, the quantum nonlocal correlations may be simulated by eavesdroppers with classical sources \cite{lee2018di}.
This stresses the pivotal role of the of the source independence assumption in device-independent protocols in networks~\cite{Luodevice2021}.

\paragraph*{Self-testing:} Another relevant device-independent consideration is self-testing \cite{MY04}, which aims to characterise an unknown quantum device based only on some observed correlation (see e.g.~Ref.~\cite{Supic2020} for a review).
In networks, it is possible to self-test entangled measurements~\cite{Bancal2018,Renou2018}.
In particular, the Bell state measurement can be certified in the bilocal scenario (recall Figure~\ref{fig:bilocal}), asking the central party to perform a complete Bell state measurement and the two remaining parties to perform the standard CHSH measurements.
This maximally violates four relabelings of the CHSH inequality.
A certification of the central party performing an entangled measurement is obtained for a violation of the CHSH inequality larger than $\mathcal{S}_\text{CHSH}\,{\approx}\,2.69$.
This protocol can be generalised to certifying the tilted Bell state measurement associated to four partially entangled two-qubit states as well as a measurement in a basis of GHZ states \cite{Renou2018}.
This protocol has been experimentally implemented in Ref.~\cite{Zhang2019}.

Self-testing has also been considered at the level of the physical theory itself.
This aims at identifying a task in networks where the optimal performance singles out quantum theory as the only possible physical model (up to reasonable degrees of freedom).
In standard Bell scenarios, no-signaling resources enable larger state spaces and therefore more general correlations than those of quantum theory.
However, enlarged state spaces typically come with restricted notions of joint measurements \cite{Barrett2007GPT}.
Since networks provide joint measurements with a pivotal role, it becomes interesting to ask whether they can be used to self-test quantum theory.
Recently, such arguments have been used in the bilocal scenario with the same protocol involving a central Bell state measurement and the maximal violation of four relabelled CHSH inequalities.
For this task, no generalised probabilistic theory performs better than quantum theory, and quantum theory outperforms many of them~\cite{Weilenmann2020PRL,Weilenmann2020PRA}.

\paragraph*{The role of complex numbers:} Complex numbers are essential for mathematical analysis. While some physical theories do not make use of complex numbers, others, such as electromagnetism, employ them as convenient, but in principle dispensible, tools for calculations. Similarly, complex numbers are highly useful in the standard Hilbert space formalism of quantum theory. However, it is unclear if they are truly necessary in the standard Hilbert space formalism or if it is possible to simulate all predictions of quantum theory using only real-valued Hilbert spaces. A partial answer is provided by the fact that the statistics of every quantum implementation of a Bell experiment can be simulated in another Bell experiment using a quantum model that only requires real Hilbert spaces \cite{Pal08,McKague09}. Such a simulation requires one to enlarge the Hilbert space by introducing ancillary real qubits; otherwise it cannot succeed \cite{Vertesi2010,Andersson2017,Smania2020,Smania2020b}. However, the situation changes when one examines network scenarios. Assuming that two sources are independent, it is possible to device-independently certify the need for complex numbers in the standard Hilbert space formalism of quantum theory~\cite{Mark2021}. This result is shown in a scenario reminiscent of the bilocal scenario; it involves three parties and two independent, bipartite quantum sources. In contrast to the bilocal scenario, all three parties are allowed to share a local variable. By Bob performing a Bell state measurement, Alice and Charlie can test a version of the CHSH inequality \cite{Bowles18} in order to device-independently assert the impossibility of a quantum simulation based on real numbers. This has been experimentally demonstrated in both superconducting circuits \cite{chen2021complex} and photonics \cite{li2021complex}.

\subsection{Relaxing source-independence}
\label{withoutsource-independent}
Quantum networks are present in many experimental implementations of Bell experiments~\cite{HansonLoopholeFree2015} and generation of entangled states~\cite{Duan2010,Sangouard2011,Rempe2012}. In these cases, the quantum network is the physical framework in which the quantum correlations are obtained. However, the theoretical analysis is performed by studying the generated correlations with respect to a single source of randomness being shared by all parties, this is, in the associated Bell scenario. In the following, we discuss a series of theoretical works in which the notion of a network is still present at the theoretical level, but the analysis of the correlations obtained is performed at the level of the corresponding Bell scenario. There are two different lines of research that do not use the source independence assumption. On one hand, several works studied how the nonlocality of an entangled state could be activated by distributing copies of the state in a quantum network. On the other hand, this framework has been employed in the analysis of multipartite nonlocal games.

Ref.~\cite{Sende2005} shows that the nonlocality of a Bell-local state can be activated by distributing many copies of it in a star network, using an event-ready protocol which postselects on the outcome of an entanglement swapping measurement. The same idea is used in Ref.~\cite{Cavalcanti2011} to activate nonlocality for all one-way distillable states. Event-ready protocols were first introduced to prove nonlocality of all multipartite entangled pure states in Ref.~\cite{Popescu92}. These can be used in chain networks (recall Figure~\ref{fig:chain}) to identify unknown nonlocal resources using erasure channels. Furthermore, it is also possible to activate nonlocality by allowing the black boxes held by the parties to admit quantum inputs~\cite{Luo2018b}. This latter work shows, additionally, that nonlocality is not an additive quantity.
Activation of nonlocality has also been analysed through the verification of genuine multipartite nonlocality~\cite{Svetlichny1987}. Ref.~\cite{Contreras2020} shows that genuine multipartite nonlocality can be activated in networks that exclusively contain bipartite sources. The main concept in the proof of Ref.~\cite{Contreras2020} is applying the lifting method of Ref.~\cite{Pironio05} for obtaining the boundary of the set of correlations in increasingly complex networks starting from inequalities for bipartite systems. Outside the context of activation of nonlocality, the approach of using new Bell inequalities with multiple inputs in order to verify the genuinely multipartite nonlocality of quantum networks is employed in Ref.~\cite{Luodevice2021} to analyse protocols in quantum key distribution, blind quantum computing, and quantum secret sharing. A third example has been demonstrated for the triangle network without inputs. Ref.~\cite{Supic2020independence} introduces a way of quantifying the amount of independence between sources, and investigates to what extent the independence assumption can be relaxed in order to demonstrate network nonlocality. By generalising Fritz's construction for disguising a bipartite Bell test in the triangle network~\cite{Fritz2012}, Ref.~\cite{Supic2020independence} shows that correlations incompatible with a local model in the triangle can be observed even for arbitrarily small levels of independence between the sources.

The second approach originates from the analysis of quantum games \cite{Brunner2013}. A nonlocal game is described as spatially separated players who interact with a referee. Assuming that the players cannot communicate with each other, they are required to answer questions in a coordinated manner. Nonlocal correlations are used to achieve a quantum advantage over similar experiments where the parties share classical resources. Generally, any nonlocal game naturally corresponds to a standard Bell inequality, although the converse is not true \cite{Silman2007}. As a natural extension, Ref.~\cite{Luo2019} considers nonlocal games in networks with bipartite sources, where sharing quantum states allows for larger winning probabilities than those achievable when the players share classical systems. The winning condition is consistent with the assumptions required for the derivation of the network Bell inequality of Ref.~\cite{Luo2018}. Thus, these games provide necessary and sufficient conditions for witnessing quantum networks consisting of bipartite sources.

\subsection{Machine learning in network nonlocality}
\label{sec:neural}
Machine learning provides tools to extract relevant features from data in an automated way. Its generality as a method has had the consequence that nowadays it is used in many fields of physics (see e.g.~the review \cite{MLphys}).
One of the central problems in network nonlocality, namely finding network local models for given correlations, is a computationally challenging but mathematically well-defined problem. Because of this, it is \textit{a priori} expected that machine learning techniques become useful in tackling it.
However, the heuristic nature of machine learning methods strongly contrasts with the mathematical rigour of proofs of nonlocality.

This has not prevented machine learning tools to be used in the study of network nonlocality.
Initial attempts followed the traditional approach in machine learning: using a neural network to approximate a function that is difficult to compute otherwise.
Ref.~\cite{Canabarro2019ML} uses this approach in the bilocal scenario to estimate the distance between the bipartite probabilities of the events of the extremal parties and the product of their corresponding single-party marginals.
This work shows that neural-network regression can lead to very accurate results, despite the fact of needing significant amounts of labeled data for training to be effective.

By using a neural-network-based parametrization of Eq.~\eqref{Eqlocal}, \cite{krivachy2019neural} exploits the deep learning toolbox  to find the parameters that minimise the distance of the corresponding model to a target distribution.
Such distance being made zero is a proof\footnote{Note that these proofs are not immediately rigorous due to the fact that machine precision is finite for all practical purposes.} that the target distribution admits a local variable model in the network.
However, the converse is not true: the failure in obtaining a vanishing distance does not necessarily imply that the distribution is nonlocal. It can instead  be a consequence of i.e.~insufficient expressiveness of the neural network architecture chosen or unsatisfactory parameter initialisation.

This construction is employed to address whether some interesting tripartite quantum distributions admit local models in the triangle network.
Interestingly, when applied to the RGB4 distribution given in Eq.~\eqref{eq:genuine}, the machine learning approach fails to find local models in a larger parameter range than the analytical arguments of Ref.~\cite{Renou2019} (recall also Section~\ref{subsec:TCandCM}).
Moreover, Ref.~\cite{krivachy2019neural} also reports the failure of finding a local model for the triangle distribution in Eq.~\eqref{eq:elegant}, providing an estimate on the visibility and the detection efficiency required to demonstrate its (potential) nonlocality.
These results support the conjectures of the nonlocality of those distributions, although formal proofs must yet be developed.
Finally, this approach is used in Ref.~\cite{abiuso2021singlephoton} to estimate parameter ranges and physical requirements of experimental realisations in the triangle network.

\subsection{Multipartite network entanglement}
\label{multipartiteentanglement}
The standard definition of genuine multipartite entanglement requires that a quantum state cannot be written as a convex mixture of states that are separable under a bipartition of the parties~\cite{Seevinck2001multipartiteentanglement}.
This notion, however, presents some arguably undesired features when considering entanglement in networks.
As shown in Ref.~\cite{navascues2020gnme}, sources capable of distributing bipartite entanglement can, by themselves, generate genuinely $k$-partite entangled states for any $k$.
Moreover, there are multipartite entangled states, such as cluster states \cite{Briegel2001} or graph states \cite{Hein}, that can be naturally defined through local operations on shares of several entangled sources.
Thus, parallel to the development of network Bell nonlocality, several works have proposed different ways of defining network entanglement.

On the one hand, Ref.~\cite{navascues2020gnme} proposes a definition whereby a quantum state is genuinely network $k$-entangled if it cannot be produced by applying local, completely positive and trace-preserving linear maps over several $(k\,{-}\,1)$-partite entangled states distributed among the parties, even with the aid of global shared randomness.
The simplest example is the state \mbox{$\ket{\psi}_\text{ABC}\,{=}\,\ket{\psi_1}_\text{AB${}_1$}\otimes\ket{\psi_2}_\text{B${}_2$C}$}, where $\ket{\psi_1}$ and $\ket{\psi_2}$ are bipartite entangled quantum states and system $B$ is composed by two subsystems, $B_1$ and $B_2$.
The state $\ket{\psi}_\text{ABC}$ is genuinely tripartite entangled per the standard definition, but only genuinely network 2-entangled. The definition provided by Ref.~\cite{navascues2020gnme} is particularly suitable for numerical characterisation, since the constraints that are derived from it are amenable to semidefinite programming. In Ref.~\cite{navascues2020gnme}, several witnesses of genuine network entanglement are presented, that make use of the quantum inflation technique discussed in Section~\ref{sec:qinflation} to derive analytical and numerical bounds to the fidelity of multipartite states that can be generated in network structures.

The same problem is considered in Ref.~\cite{kraft2020networkentanglement} for the specific case of the triangle network. In contrast with Ref.~\cite{navascues2020gnme}, in this work the parties are only allowed to perform local unitary transformations in the states that they receive, and classical randomness is shared not by the parties but by the sources.
The restrictions considered in Ref.~\cite{kraft2020networkentanglement} allow providing upper bounds to the dimension of the sources.
Moreover, Ref.~\cite{kraft2020networkentanglement} gives necessary conditions that are satisfied by all tripartite correlations that can be generated in the triangle network and proves that there are no three-qubit genuinely multipartite entangled states which, even when embedded in larger-dimensional systems, can be prepared in the triangle network
even when assisted by global shared randomness.

The same setup (parties are only allowed to perform unitary operations on their states) is characterised in Ref.~\cite{luo2020networkentanglement}.
It provides many witnesses of genuine multipartite network entanglement in terms of the fidelities with experimentally realisable states (e.g. the GHZ state, the W state \cite{Dur04}, or Dicke states \cite{Dicke}), and computes several visibility thresholds of mixed states that certify genuine network entanglement.
Furthermore, Ref.~\cite{luo2020networkentanglement} proves that any $n\,{\geq}\,3$-partite permutationally symmetric entangled pure state, and any $n\,{\geq}\,3$-partite entangled pure state composed of qubits and qutrits, are genuinely network entangled.

Quantum network entanglement has recently found use in the analysis of entanglement-based attacks to quantum bit commitment~\cite{Mayers97,LC97}. Bit commitment protocols aim to commit secret bits between two parties, with the certification that neither the committer nor the receiver can forge their observations. The no-go theorems of Refs.~\cite{Mayers97,LC97} forbid non-relativistic quantum bit commitment with single particles or bipartite entangled states. It is possible, however, to perform quantum bit commitment in relativistic settings~\cite{Kent1999,Kent2011}. Exploiting network entanglement, Ref.~\cite{Luo2020b} proposes an alternative protocol based on GHZ states, where one of the parties holds multiple shares of the state. Although a party may forge her operations by modifying the overall network to be the bilocal scenario (in Figure~\ref{fig:bilocal}), the fact that the works \cite{navascues2020gnme,kraft2020networkentanglement,luo2020networkentanglement} find certificates of incompatibility of quantum states with networks allow the parties to exclude this attack, thus making the protocol resistant to dishonest parties.

\subsection{Network-based genuinely multipartite nonlocality}
\label{multipartitenonlocality}
In analogy with the theory of multipartite entanglement, the traditional definition of genuinely multipartite nonlocality~\cite{Svetlichny1987} is based on the notion of bipartitions of the set of parties.
A correlation is genuinely multipartite nonlocal if and only if it cannot be written as a sum, over bipartitions of the set of parties, of convex combinations of distributions involving each of the subsets of the bipartition (see, for instance, \cite[Eq. 1]{Contreras2020} for its mathematical expression).
As in the case of genuinely multipartite entanglement, this notion includes distributions that confront the standard understanding of genuine multipartiteness: for instance, the tripartite distribution $p(a,b,c|x,y,z)\,{=}\,p_\text{CHSH}(a,b_1|x,y_1)p_\text{CHSH}(b_2,c|y_2,z)$, where $p_\text{CHSH}$ is the bipartite distribution that achieves the maximum quantum violation of the CHSH inequality (recall Section~\ref{secIntroBell}), is genuinely multipartite nonlocal according to this definition~\cite{Svetlichny1987}, yet it is easy to generate by considering two separate Bell tests and grouping one party of each of the tests in order to define the central party.
Moreover, this traditional definition is formulated under the assumption that parties can use local operations and classical communication freely, which is incompatible with the concept of nonlocality where spacelike separated parties cannot communicate with each other until the experiment is completed.

Refs.~\cite{coiteux2021short,coiteux2021long} introduced a new definition of genuinely multipartite nonlocality completely based in network notions, that parallels that of Ref.~\cite{navascues2020gnme} for the case of genuinely multipartite entanglement.
An $n$-partite distribution is genuinely network $k$-nonlocal if it cannot be produced by applying local operations over several $(k\,{-}\,1)$-partite nonlocal resources distributed among the parties even with the aid of global shared randomness, and thus it is genuinely network nonlocal if it cannot be generated by the use of $(n\,{-}\,1)$-partite nonlocal resources.

Based on this new definition, Refs.~\cite{coiteux2021short,coiteux2021long} provide several device-independent and noise-tolerant witnesses of genuine network nonlocality.
In particular, they show that the $n$-partite GHZ state and the tripartite W state (see, e.g., Ref.~\cite{coiteux2021long} for the definition of these states) can give rise to distributions that are genuinely network $n$-nonlocal, thus proving that only boundlessly multipartite nonlocal theories\footnote{These are the theories where, for any number of parties, there exist sources of multipartite states that cannot be simulated by combining sources that distribute states to fewer parties.} can explain all quantum correlations.
They furthermore suggest that an experimental device-independent proof of the genuinely tripartite character of nature is already feasible with current technology~\cite{Jennewein2014}.

\subsection{Bayesian causal inference and quantum causal models}
\label{sec:causalinference}
\begin{figure}[t!]
	\centering
	\includegraphics[width=0.5\columnwidth]{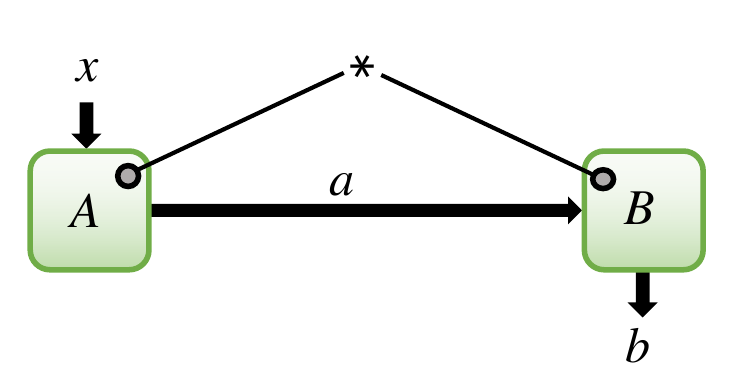}
	\caption{The instrumental scenario, depicting Alice and Bob receiving systems from a shared source. This structure is not a network, because the outcome of Alice acts as an input for Bob. In the particular case of this scenario, the question of whether a probability distribution $p(a,b|x)$ can be generated in it can be mapped to an analysis of the bipartite Bell scenario~\cite{Agresti2019instrumental,Himbeeck2019instrumental,gachechiladze2020qace}. However, this is not the general case in arbitrary causal structures.}
	\label{fig:instrumental}
\end{figure}

This review focuses on the correlations that can be generated in networks, where the parties' outcomes depend solely on their choice of input and the systems they receive from independent sources.
These are the fundamental properties that are captured by the network local and network quantum models of Eqs.~\eqref{Eqlocal} and \eqref{eq:quantummodel}.
It is natural to further wonder about the correlations that can be achieved in more general scenarios, where these conditions are not met.
A simple example of a scenario that is not a network is the so-called instrumental scenario depicted in Figure~\ref{fig:instrumental}, where the outcome of Alice serves as input for Bob's measurement.

The field of causal inference (see Ref.~\cite{pearl} for an introduction in the context of correlations admitting models based on local variables) deals with the task of understanding the correlations that can be generated in these general scenarios.
Causal scenarios\footnote{Causality here is defined in terms of Reichenbachs' principle, which states that if a correlation between two events, A and B, exists, then either A is a cause of B, B is a cause of A, or A and B are both caused by a third event, C, that occurs prior to A and B \cite{reichenbach}.} can be understood in terms of directed acyclic graphs.
In the language of directed acyclic graphs, network structures can be described as two-layer graphs where one layer of visible variables (corresponding to the parties' outcomes) is connected to both visible (denoting the input choices) and unobservable variables (that describe the sources) in another layer, and there are no intra-layer connections.
However, it is possible to envision many phenomena of relevance that cannot be captured by these two-layer graphs, such as the signaling of an outcome described above, or the controlled preparation of the systems sent by a source,\footnote{In the formalism of directed acyclic graphs, this is achieved by an observable variable connecting to an unobservable variable.} to name a few.

Causal inference benefits from the developments in the study of nonlocality in networks.
The clearest example is the classical inflation technique described in Section~\ref{sec:cinflation}, which gives asymptotically sufficient conditions for compatibility in any causal structure with just local variables~\cite{navascues2017convergence}.
Broadly, Ref.~\cite{navascues2017convergence} (and Ref.~\cite{wolfe2019qinflation} for the case of quantum variables) describes a series of steps that allow to reduce any causal structure and correlations generated in it to a network scenario and associated correlations that are dictated by those achievable in the original causal structure.
Then, any method for characterising correlations in that network can be applied in order to characterise the correlations of the original causal structure.
This procedure is exemplified in Ref.~\cite{wolfe2019qinflation} by bounding the quantum average causal effect~\cite{HollandACE} of a visible variable on another in a triangle scenario with signaling.
Also, the methods developed for the entropic characterisation of correlations can be readily applied to any causal structure, as described in Ref.~\cite{Weilenmann2017}, by properly enumerating its coexisting sets and their respective Shannon cones as described in Sections~\ref{Entropy} and \ref{sec:nonsignaling}.

Note that classical causal inference is an established field that has developed its own analysis techniques~\cite{pearl,shpitser2008Identification}.
Moreover, since the goal of classical causal inference can be understood to be finding models based in local variables in arbitrary directed acyclic graphs, the techniques developed within its formalism have found application in the certification of nonlocality in causal structures that are not networks in the sense of this review~\cite{chaves2018instrumental,agresti2020instrumental,gachechiladze2020qace}.
The maturity of classical causal inference motivates the generalisation of its framework to assess causal relations between quantum systems.
This is the effort of quantum causal modelling, which has since developed quantum generalisations of, for instance, Bayesian networks~\cite{Henson2014}, Reichenbach's common cause principle~\cite{Allen2017Qcausal}, $d$-separation~\cite{Pienaar2015dseparation}, the rules of do-calculus~\cite{barrett2019qcausal}, and causal discovery algorithms~\cite{giarmatzi2018discovery}.

\section{Conclusions and open problems}
\label{section8}

Nonlocality in networks is a research program that conceptually, technically and practically takes the decades-old field of Bell nonlocality beyond the traditional Einstein-Podolsky-Rosen scenario. This review article surveys it up to the time of writing. It appears sensible to claim that this rapidly developing topic presently finds itself at a point where several basic methods and tools for its systematic analysis are being established, but many elementary questions remain wide open. While we have mentioned many concrete open problems along the way, we conclude this article with a discussion and highlight of some of the most interesting and relevant lines of future research in network nonlocality.

\textbf{Genuine network nonlocality.---} Despite the advances outlined in Section~\ref{genuine}, many questions remain to be solved regarding this aspect. How can one formalise and analyse the intuitive idea that some forms of network nonlocality are more genuine than others? This is a key question which reasonably should not be expected to boil down to a single concept, but may rather motivate a systematic classification of conceptually different forms of network nonlocality. To what extent are nonlocal resources required to be distributed in a network in order to explain quantum nonlocality? When does network nonlocality reduce to building blocks of standard Bell nonlocality (as discussed in Section~\ref{subsec:OverviewReplaceInputs})? What is the role of entanglement swapping in more genuine notions of quantum network nonlocality? Important steps in this direction include developing efficient methods to identify disguised Bell nonlocality in networks and understanding the relationship between network nonlocality and standard multipartite nonlocality.

\textbf{Entangled measurements.---} Standard quantum nonlocality is powered by entangled states. However, quantum nonlocality in networks gives also entangled measurements (measurements in an entangled basis) a fundamental role. This motivates the development of entangled measurements, beyond the well-established Bell state measurement and the recent Elegant Joint Measurement, that possess conceptually or practically appealing properties. Are there new classes of entangled measurements that can reveal interesting features of quantum correlations in networks? How, and to what extent, can entangled measurements be viewed as resources in network nonlocality?

\textbf{Quantum versus no-signaling.---} In standard Bell scenarios, no-signaling correlations are sometimes called ``post-quantum'' or ``supra-quantum'' correlations, because they are strictly more powerful than quantum nonlocality. In networks, however, entangled measurements are brought to the forefront of the physical process. How does this impact the role and status of quantum theory in the broader landscape of no-signaling theories? How can we understand the counterpart of entanglement swapping in more general theories? Are there natural counterparts to the Popescu-Rohrlich nonlocal box in network scenarios, and what can they teach us? Is it possible to find experiments in which quantum theory is singled out as the unique no-signaling theory for explaining predicted correlations?

\textbf{Activation of nonlocality.---} Networks are known to reveal the nonlocality of some states that do not violate any Bell inequality, even without the crucial assumption of independent sources (see Section~\ref{withoutsource-independent}). It is therefore reasonably expected that networks with independent sources should be able to reveal such nonlocality in an even more powerful manner. However, there presently exists no example of using independent sources in a network to activate nonlocality in a way that cannot already be achieved in networks without assuming independent sources. Finding such an example, for instance based on the isotropic state \eqref{isotropic}, is an interesting problem with both fundamental and practical relevance. In a similar vein, given any entangled state, does there exist a network in which nonlocality can be detected if all sources emit one copy of the state?

\textbf{Mixed resources.---} This review, and indeed most research thus far, has largely focused on networks in which all sources are local, quantum or no-signaling, respectively. However, an interesting avenue of research concerns investigating the correlations that can arise in networks composed of sources described by different physical models. At a foundational level, this can shed light on the extent to which nonlocal resources are needed in a network in order to explain observed correlations. Do network nonlocal correlations truly exploit entanglement in all sources in the network in order to assert their nonlocality? This can also teach us about the relationship between different nonlocal theories in terms of studying their interplay on the level of correlations. Moreover, at a practical level, networks composed of both classical and quantum sources can find relevance in quantum information protocols for long-distance networks and allow to integrate natural assumptions such as the sources correlating via classical shared randomness.

\textbf{Device-independence.---} An interesting avenue is the exploration of network nonlocality as a phenomenon for powering device-independent quantum information protocols. Indeed, as discussed in Section~\ref{Device-independent}, initial steps in this direction have already been taken. Which tasks are natural to consider in a network scenario, as opposed to standard Bell scenarios? Do different networks enable qualitatively different protocols? Are there networks that support particularly well-performing protocols? What are the requirements for their implementation in relevant physical systems? More generally, what are the limits of device-independent inference in networks?

\textbf{Justifying the independence of the sources.---}
The source-independence assumption is critical in network nonlocality. Contrary to the no-signaling assumption, it cannot be justified via reference to a fundamental principle similar to space-like separation. In particular, shared randomness can be simulated in synchronised protocols, for instance by independent parties separately measuring some global physical quantity (e.g.~observing solar flare fluctuations). Hence, it is a potential loophole in any experiment or device-independent protocol based on network nonlocal correlations. How critical is this loophole?  Moreover, can the independence assumption be relaxed in some protocols, so that it is approximately or probabilistically satisfied? What are the best ways to experimentally justify the independence of the sources?

\textbf{Exterior- and interior-point methods.---} Developing methods for characterising the sets of local, quantum and no-signaling correlations in networks from both the interior and the exterior is a key challenge, since it allows to understand the ultimate limits of each theory. Many features of these methods are important to develop and improve upon. This includes which networks they apply to, which input/output scenarios in a given network can they deal with, how accurately can they bound the relevant correlations, and to what extent can they actually be computed with standard resources. While several such techniques have been discussed in this article, the need for powerful and practically useful methods, even in simple networks with a small number of inputs and outputs, remains central. In this line, an important aspect that remains open is to investigate how well the current techniques for bounding the correlation sets from the exterior characterise the sets of network correlations.

\textbf{Bilocal and triangle networks.---} Several concrete, yet elementary, questions lay open for the two smallest networks, namely the bilocal and the triangle scenarios. An interesting matter is to explore the geometry of the sets of local and quantum correlations in simple input/output scenarios in these networks. This may, e.g., enable the device-independent certification of entanglement swapping processes in the bilocal scenario. Furthermore, in the triangle network with no inputs and binary outputs, the conjecture that the local and quantum sets are identical remains open. Immediately related is the open problem of whether the set of triangle no-signaling correlations is equivalent to the quantum set. Another relevant open problem is to decide the triangle nonlocality of the highly symmetric quantum distribution, given in Eq.~\eqref{eq:elegant}, based on the Elegant Joint Measurement.

\textbf{Better network Bell inequalities.---} Many presently known network Bell inequalities ultimately rely on exploiting the core idea behind the BRGP bilocal Bell inequality. A natural challenge is to find qualitatively different and systematic ways of constructing network Bell inequalities. Particularly relevant are constructions that are not based on standard Bell inequalities and tailored to quantum protocols based on entanglement swapping. Indeed, from the standpoint of present knowledge, this is relevant for every non-trivial network (see Section~\ref{StarAndChainedNet}).

\textbf{Experimental considerations.---} Many fundamental aspects and technical challenges are relevant for experimental demonstrations and applications of network nonlocality. Perhaps most notably, an experiment that simultaneously closes the locality, fair sampling and freedom of choice loopholes is still missing. Moreover, none of the existing experiments go beyond the standard notion of network nonlocality; for instance, the network Bell inequality violations so far reported are insufficient for demonstrating stronger notions such as full network nonlocality. Furthermore, although several experiments have been performed in the bilocal scenario, no photonics experiment has thus far considered a different quantum network without resorting to wiring many standard violations of standard Bell inequalities. These more general networks may notably include multipartite sources, whose relevance in network nonlocality has yet not been demonstrated. Furthermore, a key challenge concerns the implementation and application of entangled measurements, in particular those that are different from a standard partial Bell state analyser. On this front, little is still known. These are important steps towards the long-term challenge of reliable and high-quality entanglement distribution and entanglement swapping in large-scale quantum networks.


\appendix

\section{Notable quantum distributions in the triangle network}\label{AppendixDistributions}
In this appendix we compile a number of relevant distributions generated in the triangle scenario by means of quantum resources.

\subsection{The Fritz distribution}
The Fritz distribution~\cite{Fritz2012} is the archetype example of a distribution that presents disguised Bell nonlocality.
It is built from the distribution $p(a,b|x,y)\,{=}\,\frac{1}{4}[1+\frac{1}{\sqrt{2}}(-1)^{a+b+xy}]$, which achieves the maximal quantum violation of the CHSH inequality in Eq.~\eqref{CHSH}, by transforming the bipartite Bell network into the triangle network as described in Section~\ref{subsec:OverviewReplaceInputs}.
The Fritz distribution has no inputs and four outputs per party, and is given by
\begin{equation}
  \begin{split}
    \frac{1}{32}(2\!+\!\sqrt{2})&=\!p(0,0,0)=p(1,1,0)=p(0,2,1)=p(1,3,1) \\
    &=\!p(2,0,2)=p(3,1,2)=p(2,3,3)=p(3,2,3), \\
    \frac{1}{32}(2\!-\!\sqrt{2})&=\!p(0,1,0)=p(1,0,0)=p(0,3,1)=p(1,2,1) \\
    &=\!p(2,1,2)=p(3,0,2)=p(2,2,3)=p(3,3,3),
  \end{split}
  \label{eq:fritz}
\end{equation}
with all probabilities not appearing above being equal to zero.

\subsection{The triangle distribution based on the Elegant Joint Measurement}
When each of the parties in the triangle scenario performs the Elegant Joint Measurement described in Eq.~\eqref{EJM} in the respective shares of singlets $\ket{\psi^-}$ sent by the sources, a particularly simple distribution without inputs and four outputs is generated.
It reads
\begin{equation}
  \begin{split}
    p(a=r,b=r,c=r)&=\frac{25}{256}\qquad \text{for } r=1,2,3,4,\\
    p(a=r,b=r,c=s)&=\frac{1}{256}\qquad \text{for } r\neq s, \\
    p(a=r,b=s,c=t)&=\frac{5}{256}\qquad \text{for } r\neq s\neq t\neq r,
  \end{split}
\label{eq:elegant}
\end{equation}
plus cyclic permutations of $(a,b,c)$. A number of interesting properties about this distribution are collected in Ref.~\cite{gisin2017elegant}.
Despite many efforts, neither a local model nor a proof of network nonlocality has been found for this distribution.

\subsection{The RGB4 distribution}
The RGB4 distribution is the first triangle nonlocal distribution that is, arguably, conceptually different from the Fritz distribution. It was originally derived in Ref.~\cite{Renou2019} by having the parties perform the same four-outcome measurement on their respective shares of states $\ket{\psi^+}$  distributed by the sources, and it is given by
\begin{equation}
  \begin{split}
    p(\bar{0},\bar{1}_1,\bar{2})=p(\bar{0},\bar{2},\bar{1}_0)&=\frac{c^2}{8}, \\
    p(\bar{0},\bar{1}_0,\bar{2})=p(\bar{0},\bar{2},\bar{1}_1)&=\frac{1-c^2}{8}, \\
    p(\bar{1}_i,\bar{1}_i,\bar{1}_i)&=\frac{[(1-c^2)^{3/2}+(-1)^i c^3]^2}{8}, \\
    p(\bar{1}_i,\bar{1}_i,\bar{1}_j)&=\frac{[c^2\sqrt{1-c^2}+(-1)^j c(1-c^2)]^2}{8},
  \end{split}
  \label{eq:genuine}
\end{equation}
plus cyclic permutations, all the remaining probabilities being $0$.
The distribution depends on a parameter $c\,{\in}\,[-1,1]$, and for values $0.785\,{\lesssim}\,c^2\,{<}\, 1$ it is known, using token-counting arguments (see Section~\ref{subsec:TCandCM}), not to admit a local model in the triangle network.

\acknowledgments
We are grateful to Cyril Branciard, Nicolas Gisin, Antonio Ac\'in, Nicolas Brunner, Elie Wolfe, Tristan Kraft and Emanuel-Cristian Boghiu for support and comments.
A.T.~is supported by the Swiss National Science Foundation and the Wenner-Gren Foundations.
A.P.-K.~is supported by the European Union's Horizon 2020 research and innovation programme-grant agreement No. 648913 and by the Spanish Ministry of Science and Innovation through the ``Severo Ochoa Programme for Centres of Excellence in R\&D'' (CEX2019-000904-S).
M.-X.L.~is supported by the National Natural Science Foundation of China (Nos. 61772437, 62172341) and Southern University of Science and Technology (No. SIQSE202105).
M.-O.R.~is supported by the Swiss National Fund Early Mobility Grant P2GEP2\_191444 and the grant PCI2021-122022-2B financed by MCIN/AEI/10.13039/501100011033 and by the European Union NextGenerationEU/PRTR, and acknowledges the Government of Spain (FIS2020-TRANQI and Severo Ochoa CEX2019-000910-S [MCIN/AEI/10.13039/501100011033]), Fundació Cellex, Fundació Mir-Puig, Generalitat de Catalunya (CERCA, AGAUR SGR 1381) and the ERC AdG CERQUTE.

\vspace{10mm}

\bibliographystyle{apsrev4-2-custom}
\bibliography{references}

\begin{thebibliography}{0}%
\makeatletter
\providecommand \@ifxundefined [1]{%
 \@ifx{#1\undefined}
}%
\providecommand \@ifnum [1]{%
 \ifnum #1\expandafter \@firstoftwo
 \else \expandafter \@secondoftwo
 \fi
}%
\providecommand \@ifx [1]{%
 \ifx #1\expandafter \@firstoftwo
 \else \expandafter \@secondoftwo
 \fi
}%
\providecommand \natexlab [1]{#1}%
\providecommand \enquote  [1]{``#1''}%
\providecommand \bibnamefont  [1]{#1}%
\providecommand \bibfnamefont [1]{#1}%
\providecommand \citenamefont [1]{#1}%
\providecommand \href@noop [0]{\@secondoftwo}%
\providecommand \href [0]{\begingroup \@sanitize@url \@href}%
\providecommand \@href[1]{\@@startlink{#1}\@@href}%
\providecommand \@@href[1]{\endgroup#1\@@endlink}%
\providecommand \@sanitize@url [0]{\catcode `\\12\catcode `\$12\catcode
  `\&12\catcode `\#12\catcode `\^12\catcode `\_12\catcode `\%12\relax}%
\providecommand \@@startlink[1]{}%
\providecommand \@@endlink[0]{}%
\providecommand \url  [0]{\begingroup\@sanitize@url \@url }%
\providecommand \@url [1]{\endgroup\@href {#1}{\urlprefix }}%
\providecommand \urlprefix  [0]{URL }%
\providecommand \Eprint [0]{\href }%
\providecommand \doibase [0]{https://doi.org/}%
\providecommand \selectlanguage [0]{\@gobble}%
\providecommand \bibinfo  [0]{\@secondoftwo}%
\providecommand \bibfield  [0]{\@secondoftwo}%
\providecommand \translation [1]{[#1]}%
\providecommand \BibitemOpen [0]{}%
\providecommand \bibitemStop [0]{}%
\providecommand \bibitemNoStop [0]{.\EOS\space}%
\providecommand \EOS [0]{\spacefactor3000\relax}%
\providecommand \BibitemShut  [1]{\csname bibitem#1\endcsname}%
\let\auto@bib@innerbib\@empty
\bibitem [{\citenamefont {Bell}(1964)}]{BellTheorem}%
  \BibitemOpen
  \bibfield  {author} {\bibinfo {author} {\bibfnamefont {J.~S.}\ \bibnamefont
  {Bell}},\ }\href {https://doi.org/10.1103/PhysicsPhysiqueFizika.1.195}
  {\bibfield  {journal} {\bibinfo  {journal} {Physics Physique Fizika}\
  }\textbf {\bibinfo {volume} {1}},\ \bibinfo {pages} {195} (\bibinfo {year}
  {1964})}\BibitemShut {NoStop}%
\bibitem [{\citenamefont {Genovese}(2005)}]{Genovese2005}%
  \BibitemOpen
  \bibfield  {author} {\bibinfo {author} {\bibfnamefont {M.}~\bibnamefont
  {Genovese}},\ }\href {https://doi.org/10.1016/j.physrep.2005.03.003}
  {\bibfield  {journal} {\bibinfo  {journal} {Phys. Rep.}\ }\textbf {\bibinfo
  {volume} {413}},\ \bibinfo {pages} {319} (\bibinfo {year}
  {2005})}\BibitemShut {NoStop}%
\bibitem [{\citenamefont {Brunner}\ \emph {et~al.}(2014)\citenamefont
  {Brunner}, \citenamefont {Cavalcanti}, \citenamefont {Pironio}, \citenamefont
  {Scarani},\ and\ \citenamefont {Wehner}}]{BrunnerRMP}%
  \BibitemOpen
  \bibfield  {author} {\bibinfo {author} {\bibfnamefont {N.}~\bibnamefont
  {Brunner}}, \bibinfo {author} {\bibfnamefont {D.}~\bibnamefont {Cavalcanti}},
  \bibinfo {author} {\bibfnamefont {S.}~\bibnamefont {Pironio}}, \bibinfo
  {author} {\bibfnamefont {V.}~\bibnamefont {Scarani}},\ and\ \bibinfo {author}
  {\bibfnamefont {S.}~\bibnamefont {Wehner}},\ }\href
  {https://doi.org/10.1103/RevModPhys.86.419} {\bibfield  {journal} {\bibinfo
  {journal} {Rev. Mod. Phys.}\ }\textbf {\bibinfo {volume} {86}},\ \bibinfo
  {pages} {419} (\bibinfo {year} {2014})}\BibitemShut {NoStop}%
\bibitem [{\citenamefont {Scarani}(2019)}]{ScaraniBook}%
  \BibitemOpen
  \bibfield  {author} {\bibinfo {author} {\bibfnamefont {V.}~\bibnamefont
  {Scarani}},\ }\href@noop {} {\emph {\bibinfo {title} {{Bell nonlocality}}}}\
  (\bibinfo  {publisher} {Oxford University Press},\ \bibinfo {year}
  {2019})\BibitemShut {NoStop}%
\bibitem [{\citenamefont {Einstein}\ \emph {et~al.}(1935)\citenamefont
  {Einstein}, \citenamefont {Podolsky},\ and\ \citenamefont {Rosen}}]{EPR1935}%
  \BibitemOpen
  \bibfield  {author} {\bibinfo {author} {\bibfnamefont {A.}~\bibnamefont
  {Einstein}}, \bibinfo {author} {\bibfnamefont {B.}~\bibnamefont {Podolsky}},\
  and\ \bibinfo {author} {\bibfnamefont {N.}~\bibnamefont {Rosen}},\ }\href
  {https://doi.org/10.1103/PhysRev.47.777} {\bibfield  {journal} {\bibinfo
  {journal} {Phys. Rev.}\ }\textbf {\bibinfo {volume} {47}},\ \bibinfo {pages}
  {777} (\bibinfo {year} {1935})}\BibitemShut {NoStop}%
\bibitem [{\citenamefont {Kimble}(2008)}]{Kimble2008}%
  \BibitemOpen
  \bibfield  {author} {\bibinfo {author} {\bibfnamefont {H.~J.}\ \bibnamefont
  {Kimble}},\ }\href {https://doi.org/10.1038/nature07127} {\bibfield
  {journal} {\bibinfo  {journal} {Nature}\ }\textbf {\bibinfo {volume} {453}},\
  \bibinfo {pages} {1023} (\bibinfo {year} {2008})}\BibitemShut {NoStop}%
\bibitem [{\citenamefont {Wehner}\ \emph {et~al.}(2018)\citenamefont {Wehner},
  \citenamefont {Elkouss},\ and\ \citenamefont {Hanson}}]{Wehner2018}%
  \BibitemOpen
  \bibfield  {author} {\bibinfo {author} {\bibfnamefont {S.}~\bibnamefont
  {Wehner}}, \bibinfo {author} {\bibfnamefont {D.}~\bibnamefont {Elkouss}},\
  and\ \bibinfo {author} {\bibfnamefont {R.}~\bibnamefont {Hanson}},\ }\href
  {https://science.sciencemag.org/content/362/6412/eaam9288} {\bibfield
  {journal} {\bibinfo  {journal} {Science}\ }\textbf {\bibinfo {volume}
  {362}},\ \bibinfo {pages} {6412} (\bibinfo {year} {2018})}\BibitemShut
  {NoStop}%
\bibitem [{\citenamefont {Kozlowski}\ and\ \citenamefont
  {Wehner}(2019)}]{Kozlowski2019}%
  \BibitemOpen
  \bibfield  {author} {\bibinfo {author} {\bibfnamefont {W.}~\bibnamefont
  {Kozlowski}}\ and\ \bibinfo {author} {\bibfnamefont {S.}~\bibnamefont
  {Wehner}},\ }in\ \href {https://doi.org/10.1145/3345312.3345497} {\emph
  {\bibinfo {booktitle} {Proceedings of the Sixth Annual ACM International
  Conference on Nanoscale Computing and Communication}}},\ \bibinfo {series and
  number} {NANOCOM '19}\ (\bibinfo  {publisher} {Association for Computing
  Machinery},\ \bibinfo {address} {New York, NY, USA},\ \bibinfo {year}
  {2019})\BibitemShut {NoStop}%
\bibitem [{\citenamefont {\ifmmode~\dot{Z}\else \.{Z}\fi{}ukowski}\ \emph
  {et~al.}(1993)\citenamefont {\ifmmode~\dot{Z}\else \.{Z}\fi{}ukowski},
  \citenamefont {Zeilinger}, \citenamefont {Horne},\ and\ \citenamefont
  {Ekert}}]{Zukowski1993}%
  \BibitemOpen
  \bibfield  {author} {\bibinfo {author} {\bibfnamefont {M.}~\bibnamefont
  {\ifmmode~\dot{Z}\else \.{Z}\fi{}ukowski}}, \bibinfo {author} {\bibfnamefont
  {A.}~\bibnamefont {Zeilinger}}, \bibinfo {author} {\bibfnamefont {M.~A.}\
  \bibnamefont {Horne}},\ and\ \bibinfo {author} {\bibfnamefont {A.~K.}\
  \bibnamefont {Ekert}},\ }\href {https://doi.org/10.1103/PhysRevLett.71.4287}
  {\bibfield  {journal} {\bibinfo  {journal} {Phys. Rev. Lett.}\ }\textbf
  {\bibinfo {volume} {71}},\ \bibinfo {pages} {4287} (\bibinfo {year}
  {1993})}\BibitemShut {NoStop}%
\bibitem [{\citenamefont {Fine}(1982)}]{Fine1982}%
  \BibitemOpen
  \bibfield  {author} {\bibinfo {author} {\bibfnamefont {A.}~\bibnamefont
  {Fine}},\ }\href {https://doi.org/10.1103/PhysRevLett.48.291} {\bibfield
  {journal} {\bibinfo  {journal} {Phys. Rev. Lett.}\ }\textbf {\bibinfo
  {volume} {48}},\ \bibinfo {pages} {291} (\bibinfo {year} {1982})}\BibitemShut
  {NoStop}%
\bibitem [{\citenamefont {Kaszlikowski}\ \emph {et~al.}(2000)\citenamefont
  {Kaszlikowski}, \citenamefont {Gnaci\ifmmode~\acute{n}\else \'{n}\fi{}ski},
  \citenamefont {\ifmmode~\dot{Z}\else \.{Z}\fi{}ukowski}, \citenamefont
  {Miklaszewski},\ and\ \citenamefont {Zeilinger}}]{Kaszlikowski2000}%
  \BibitemOpen
  \bibfield  {author} {\bibinfo {author} {\bibfnamefont {D.}~\bibnamefont
  {Kaszlikowski}}, \bibinfo {author} {\bibfnamefont {P.}~\bibnamefont
  {Gnaci\ifmmode~\acute{n}\else \'{n}\fi{}ski}}, \bibinfo {author}
  {\bibfnamefont {M.}~\bibnamefont {\ifmmode~\dot{Z}\else \.{Z}\fi{}ukowski}},
  \bibinfo {author} {\bibfnamefont {W.}~\bibnamefont {Miklaszewski}},\ and\
  \bibinfo {author} {\bibfnamefont {A.}~\bibnamefont {Zeilinger}},\ }\href
  {https://doi.org/10.1103/PhysRevLett.85.4418} {\bibfield  {journal} {\bibinfo
   {journal} {Phys. Rev. Lett.}\ }\textbf {\bibinfo {volume} {85}},\ \bibinfo
  {pages} {4418} (\bibinfo {year} {2000})}\BibitemShut {NoStop}%
\bibitem [{\citenamefont {\ifmmode~\dot{Z}\else \.{Z}\fi{}ukowski}\ \emph
  {et~al.}()\citenamefont {\ifmmode~\dot{Z}\else \.{Z}\fi{}ukowski},
  \citenamefont {Kaszlikowski}, \citenamefont {Baturo},\ and\ \citenamefont
  {Larsson}}]{Zukowski1999}%
  \BibitemOpen
  \bibfield  {author} {\bibinfo {author} {\bibfnamefont {M.}~\bibnamefont
  {\ifmmode~\dot{Z}\else \.{Z}\fi{}ukowski}}, \bibinfo {author} {\bibfnamefont
  {D.}~\bibnamefont {Kaszlikowski}}, \bibinfo {author} {\bibfnamefont
  {A.}~\bibnamefont {Baturo}},\ and\ \bibinfo {author} {\bibfnamefont
  {J.-{\AA}.}\ \bibnamefont {Larsson}},\ }\href@noop {} {\bibinfo {title}
  {{S}trengthening the {B}ell {T}heorem: conditions to falsify local realism in
  an experiment}},\ \Eprint {https://arxiv.org/abs/quant-ph/9910058}
  {arXiv:quant-ph/9910058} \BibitemShut {NoStop}%
\bibitem [{\citenamefont {Clauser}\ \emph {et~al.}(1969)\citenamefont
  {Clauser}, \citenamefont {Horne}, \citenamefont {Shimony},\ and\
  \citenamefont {Holt}}]{CHSH1969}%
  \BibitemOpen
  \bibfield  {author} {\bibinfo {author} {\bibfnamefont {J.~F.}\ \bibnamefont
  {Clauser}}, \bibinfo {author} {\bibfnamefont {M.~A.}\ \bibnamefont {Horne}},
  \bibinfo {author} {\bibfnamefont {A.}~\bibnamefont {Shimony}},\ and\ \bibinfo
  {author} {\bibfnamefont {R.~A.}\ \bibnamefont {Holt}},\ }\href
  {https://doi.org/10.1103/PhysRevLett.23.880} {\bibfield  {journal} {\bibinfo
  {journal} {Phys. Rev. Lett.}\ }\textbf {\bibinfo {volume} {23}},\ \bibinfo
  {pages} {880} (\bibinfo {year} {1969})}\BibitemShut {NoStop}%
\bibitem [{\citenamefont {Cirel'son}(1980)}]{Cirelson1980}%
  \BibitemOpen
  \bibfield  {author} {\bibinfo {author} {\bibfnamefont {B.~S.}\ \bibnamefont
  {Cirel'son}},\ }\href {https://doi.org/10.1007/BF00417500} {\bibfield
  {journal} {\bibinfo  {journal} {Lett. Math. Phys.}\ }\textbf {\bibinfo
  {volume} {4}},\ \bibinfo {pages} {93} (\bibinfo {year} {1980})}\BibitemShut
  {NoStop}%
\bibitem [{\citenamefont {Popescu}\ and\ \citenamefont
  {Rohrlich}(1994)}]{Popescu1994}%
  \BibitemOpen
  \bibfield  {author} {\bibinfo {author} {\bibfnamefont {S.}~\bibnamefont
  {Popescu}}\ and\ \bibinfo {author} {\bibfnamefont {D.}~\bibnamefont
  {Rohrlich}},\ }\href {https://doi.org/10.1007/BF02058098} {\bibfield
  {journal} {\bibinfo  {journal} {Found. Phys.}\ }\textbf {\bibinfo {volume}
  {24}},\ \bibinfo {pages} {379} (\bibinfo {year} {1994})}\BibitemShut
  {NoStop}%
\bibitem [{\citenamefont {Freedman}\ and\ \citenamefont
  {Clauser}(1972)}]{Freedman1972}%
  \BibitemOpen
  \bibfield  {author} {\bibinfo {author} {\bibfnamefont {S.~J.}\ \bibnamefont
  {Freedman}}\ and\ \bibinfo {author} {\bibfnamefont {J.~F.}\ \bibnamefont
  {Clauser}},\ }\href {https://doi.org/10.1103/PhysRevLett.28.938} {\bibfield
  {journal} {\bibinfo  {journal} {Phys. Rev. Lett.}\ }\textbf {\bibinfo
  {volume} {28}},\ \bibinfo {pages} {938} (\bibinfo {year} {1972})}\BibitemShut
  {NoStop}%
\bibitem [{\citenamefont {Aspect}\ \emph {et~al.}(1982)\citenamefont {Aspect},
  \citenamefont {Dalibard},\ and\ \citenamefont {Roger}}]{Aspect1982}%
  \BibitemOpen
  \bibfield  {author} {\bibinfo {author} {\bibfnamefont {A.}~\bibnamefont
  {Aspect}}, \bibinfo {author} {\bibfnamefont {J.}~\bibnamefont {Dalibard}},\
  and\ \bibinfo {author} {\bibfnamefont {G.}~\bibnamefont {Roger}},\ }\href
  {https://doi.org/10.1103/PhysRevLett.49.1804} {\bibfield  {journal} {\bibinfo
   {journal} {Phys. Rev. Lett.}\ }\textbf {\bibinfo {volume} {49}},\ \bibinfo
  {pages} {1804} (\bibinfo {year} {1982})}\BibitemShut {NoStop}%
\bibitem [{\citenamefont {Giustina}\ \emph {et~al.}(2015)\citenamefont
  {Giustina}, \citenamefont {Versteegh}, \citenamefont {Wengerowsky},
  \citenamefont {Handsteiner}, \citenamefont {Hochrainer}, \citenamefont
  {Phelan}, \citenamefont {Steinlechner}, \citenamefont {Kofler}, \citenamefont
  {Larsson}, \citenamefont {Abell\'an}, \citenamefont {Amaya}, \citenamefont
  {Pruneri}, \citenamefont {Mitchell}, \citenamefont {Beyer}, \citenamefont
  {Gerrits}, \citenamefont {Lita}, \citenamefont {Shalm}, \citenamefont {Nam},
  \citenamefont {Scheidl}, \citenamefont {Ursin}, \citenamefont {Wittmann},\
  and\ \citenamefont {Zeilinger}}]{ZeilingerLoopholeFree2015}%
  \BibitemOpen
  \bibfield  {author} {\bibinfo {author} {\bibfnamefont {M.}~\bibnamefont
  {Giustina}}, \bibinfo {author} {\bibfnamefont {M.~A.~M.}\ \bibnamefont
  {Versteegh}}, \bibinfo {author} {\bibfnamefont {S.}~\bibnamefont
  {Wengerowsky}}, \bibinfo {author} {\bibfnamefont {J.}~\bibnamefont
  {Handsteiner}}, \bibinfo {author} {\bibfnamefont {A.}~\bibnamefont
  {Hochrainer}}, \bibinfo {author} {\bibfnamefont {K.}~\bibnamefont {Phelan}},
  \bibinfo {author} {\bibfnamefont {F.}~\bibnamefont {Steinlechner}}, \bibinfo
  {author} {\bibfnamefont {J.}~\bibnamefont {Kofler}}, \bibinfo {author}
  {\bibfnamefont {J.-{\AA}.}\ \bibnamefont {Larsson}}, \bibinfo {author}
  {\bibfnamefont {C.}~\bibnamefont {Abell\'an}}, \bibinfo {author}
  {\bibfnamefont {W.}~\bibnamefont {Amaya}}, \bibinfo {author} {\bibfnamefont
  {V.}~\bibnamefont {Pruneri}}, \bibinfo {author} {\bibfnamefont {M.~W.}\
  \bibnamefont {Mitchell}}, \bibinfo {author} {\bibfnamefont {J.}~\bibnamefont
  {Beyer}}, \bibinfo {author} {\bibfnamefont {T.}~\bibnamefont {Gerrits}},
  \bibinfo {author} {\bibfnamefont {A.~E.}\ \bibnamefont {Lita}}, \bibinfo
  {author} {\bibfnamefont {L.~K.}\ \bibnamefont {Shalm}}, \bibinfo {author}
  {\bibfnamefont {S.~W.}\ \bibnamefont {Nam}}, \bibinfo {author} {\bibfnamefont
  {T.}~\bibnamefont {Scheidl}}, \bibinfo {author} {\bibfnamefont
  {R.}~\bibnamefont {Ursin}}, \bibinfo {author} {\bibfnamefont
  {B.}~\bibnamefont {Wittmann}},\ and\ \bibinfo {author} {\bibfnamefont
  {A.}~\bibnamefont {Zeilinger}},\ }\href
  {https://doi.org/10.1103/PhysRevLett.115.250401} {\bibfield  {journal}
  {\bibinfo  {journal} {Phys. Rev. Lett.}\ }\textbf {\bibinfo {volume} {115}},\
  \bibinfo {pages} {250401} (\bibinfo {year} {2015})}\BibitemShut {NoStop}%
\bibitem [{\citenamefont {Hensen}\ \emph {et~al.}(2015)\citenamefont {Hensen},
  \citenamefont {Bernien}, \citenamefont {Dréau}, \citenamefont {Reiserer},
  \citenamefont {Kalb}, \citenamefont {Blok}, \citenamefont {Ruitenberg},
  \citenamefont {Vermeulen}, \citenamefont {Schouten}, \citenamefont
  {Abellán}, \citenamefont {Amaya}, \citenamefont {Pruneri}, \citenamefont
  {Mitchell}, \citenamefont {Markham}, \citenamefont {Twitchen}, \citenamefont
  {Elkouss}, \citenamefont {Wehner}, \citenamefont {Taminiau},\ and\
  \citenamefont {Hanson}}]{HansonLoopholeFree2015}%
  \BibitemOpen
  \bibfield  {author} {\bibinfo {author} {\bibfnamefont {B.}~\bibnamefont
  {Hensen}}, \bibinfo {author} {\bibfnamefont {H.}~\bibnamefont {Bernien}},
  \bibinfo {author} {\bibfnamefont {A.~E.}\ \bibnamefont {Dréau}}, \bibinfo
  {author} {\bibfnamefont {A.}~\bibnamefont {Reiserer}}, \bibinfo {author}
  {\bibfnamefont {N.}~\bibnamefont {Kalb}}, \bibinfo {author} {\bibfnamefont
  {M.~S.}\ \bibnamefont {Blok}}, \bibinfo {author} {\bibfnamefont
  {J.}~\bibnamefont {Ruitenberg}}, \bibinfo {author} {\bibfnamefont {R.~F.~L.}\
  \bibnamefont {Vermeulen}}, \bibinfo {author} {\bibfnamefont {R.~N.}\
  \bibnamefont {Schouten}}, \bibinfo {author} {\bibfnamefont {C.}~\bibnamefont
  {Abellán}}, \bibinfo {author} {\bibfnamefont {W.}~\bibnamefont {Amaya}},
  \bibinfo {author} {\bibfnamefont {V.}~\bibnamefont {Pruneri}}, \bibinfo
  {author} {\bibfnamefont {M.~W.}\ \bibnamefont {Mitchell}}, \bibinfo {author}
  {\bibfnamefont {M.}~\bibnamefont {Markham}}, \bibinfo {author} {\bibfnamefont
  {D.~J.}\ \bibnamefont {Twitchen}}, \bibinfo {author} {\bibfnamefont
  {D.}~\bibnamefont {Elkouss}}, \bibinfo {author} {\bibfnamefont
  {S.}~\bibnamefont {Wehner}}, \bibinfo {author} {\bibfnamefont {T.~H.}\
  \bibnamefont {Taminiau}},\ and\ \bibinfo {author} {\bibfnamefont
  {R.}~\bibnamefont {Hanson}},\ }\href {https://doi.org/10.1038/nature15759}
  {\bibfield  {journal} {\bibinfo  {journal} {Nature}\ }\textbf {\bibinfo
  {volume} {526}},\ \bibinfo {pages} {682–686} (\bibinfo {year}
  {2015})}\BibitemShut {NoStop}%
\bibitem [{\citenamefont {Shalm}\ \emph {et~al.}(2015)\citenamefont {Shalm},
  \citenamefont {Meyer-Scott}, \citenamefont {Christensen}, \citenamefont
  {Bierhorst}, \citenamefont {Wayne}, \citenamefont {Stevens}, \citenamefont
  {Gerrits}, \citenamefont {Glancy}, \citenamefont {Hamel}, \citenamefont
  {Allman}, \citenamefont {Coakley}, \citenamefont {Dyer}, \citenamefont
  {Hodge}, \citenamefont {Lita}, \citenamefont {Verma}, \citenamefont
  {Lambrocco}, \citenamefont {Tortorici}, \citenamefont {Migdall},
  \citenamefont {Zhang}, \citenamefont {Kumor}, \citenamefont {Farr},
  \citenamefont {Marsili}, \citenamefont {Shaw}, \citenamefont {Stern},
  \citenamefont {Abell\'an}, \citenamefont {Amaya}, \citenamefont {Pruneri},
  \citenamefont {Jennewein}, \citenamefont {Mitchell}, \citenamefont {Kwiat},
  \citenamefont {Bienfang}, \citenamefont {Mirin}, \citenamefont {Knill},\ and\
  \citenamefont {Nam}}]{ShalmLoopholefree}%
  \BibitemOpen
  \bibfield  {author} {\bibinfo {author} {\bibfnamefont {L.~K.}\ \bibnamefont
  {Shalm}}, \bibinfo {author} {\bibfnamefont {E.}~\bibnamefont {Meyer-Scott}},
  \bibinfo {author} {\bibfnamefont {B.~G.}\ \bibnamefont {Christensen}},
  \bibinfo {author} {\bibfnamefont {P.}~\bibnamefont {Bierhorst}}, \bibinfo
  {author} {\bibfnamefont {M.~A.}\ \bibnamefont {Wayne}}, \bibinfo {author}
  {\bibfnamefont {M.~J.}\ \bibnamefont {Stevens}}, \bibinfo {author}
  {\bibfnamefont {T.}~\bibnamefont {Gerrits}}, \bibinfo {author} {\bibfnamefont
  {S.}~\bibnamefont {Glancy}}, \bibinfo {author} {\bibfnamefont {D.~R.}\
  \bibnamefont {Hamel}}, \bibinfo {author} {\bibfnamefont {M.~S.}\ \bibnamefont
  {Allman}}, \bibinfo {author} {\bibfnamefont {K.~J.}\ \bibnamefont {Coakley}},
  \bibinfo {author} {\bibfnamefont {S.~D.}\ \bibnamefont {Dyer}}, \bibinfo
  {author} {\bibfnamefont {C.}~\bibnamefont {Hodge}}, \bibinfo {author}
  {\bibfnamefont {A.~E.}\ \bibnamefont {Lita}}, \bibinfo {author}
  {\bibfnamefont {V.~B.}\ \bibnamefont {Verma}}, \bibinfo {author}
  {\bibfnamefont {C.}~\bibnamefont {Lambrocco}}, \bibinfo {author}
  {\bibfnamefont {E.}~\bibnamefont {Tortorici}}, \bibinfo {author}
  {\bibfnamefont {A.~L.}\ \bibnamefont {Migdall}}, \bibinfo {author}
  {\bibfnamefont {Y.}~\bibnamefont {Zhang}}, \bibinfo {author} {\bibfnamefont
  {D.~R.}\ \bibnamefont {Kumor}}, \bibinfo {author} {\bibfnamefont {W.~H.}\
  \bibnamefont {Farr}}, \bibinfo {author} {\bibfnamefont {F.}~\bibnamefont
  {Marsili}}, \bibinfo {author} {\bibfnamefont {M.~D.}\ \bibnamefont {Shaw}},
  \bibinfo {author} {\bibfnamefont {J.~A.}\ \bibnamefont {Stern}}, \bibinfo
  {author} {\bibfnamefont {C.}~\bibnamefont {Abell\'an}}, \bibinfo {author}
  {\bibfnamefont {W.}~\bibnamefont {Amaya}}, \bibinfo {author} {\bibfnamefont
  {V.}~\bibnamefont {Pruneri}}, \bibinfo {author} {\bibfnamefont
  {T.}~\bibnamefont {Jennewein}}, \bibinfo {author} {\bibfnamefont {M.~W.}\
  \bibnamefont {Mitchell}}, \bibinfo {author} {\bibfnamefont {P.~G.}\
  \bibnamefont {Kwiat}}, \bibinfo {author} {\bibfnamefont {J.~C.}\ \bibnamefont
  {Bienfang}}, \bibinfo {author} {\bibfnamefont {R.~P.}\ \bibnamefont {Mirin}},
  \bibinfo {author} {\bibfnamefont {E.}~\bibnamefont {Knill}},\ and\ \bibinfo
  {author} {\bibfnamefont {S.~W.}\ \bibnamefont {Nam}},\ }\href
  {https://doi.org/10.1103/PhysRevLett.115.250402} {\bibfield  {journal}
  {\bibinfo  {journal} {Phys. Rev. Lett.}\ }\textbf {\bibinfo {volume} {115}},\
  \bibinfo {pages} {250402} (\bibinfo {year} {2015})}\BibitemShut {NoStop}%
\bibitem [{\citenamefont {Rosenfeld}\ \emph {et~al.}(2017)\citenamefont
  {Rosenfeld}, \citenamefont {Burchardt}, \citenamefont {Garthoff},
  \citenamefont {Redeker}, \citenamefont {Ortegel}, \citenamefont {Rau},\ and\
  \citenamefont {Weinfurter}}]{WeinfurterLoopholeFree}%
  \BibitemOpen
  \bibfield  {author} {\bibinfo {author} {\bibfnamefont {W.}~\bibnamefont
  {Rosenfeld}}, \bibinfo {author} {\bibfnamefont {D.}~\bibnamefont
  {Burchardt}}, \bibinfo {author} {\bibfnamefont {R.}~\bibnamefont {Garthoff}},
  \bibinfo {author} {\bibfnamefont {K.}~\bibnamefont {Redeker}}, \bibinfo
  {author} {\bibfnamefont {N.}~\bibnamefont {Ortegel}}, \bibinfo {author}
  {\bibfnamefont {M.}~\bibnamefont {Rau}},\ and\ \bibinfo {author}
  {\bibfnamefont {H.}~\bibnamefont {Weinfurter}},\ }\href
  {https://doi.org/10.1103/PhysRevLett.119.010402} {\bibfield  {journal}
  {\bibinfo  {journal} {Phys. Rev. Lett.}\ }\textbf {\bibinfo {volume} {119}},\
  \bibinfo {pages} {010402} (\bibinfo {year} {2017})}\BibitemShut {NoStop}%
\bibitem [{\citenamefont {Svetlichny}(1987)}]{Svetlichny1987}%
  \BibitemOpen
  \bibfield  {author} {\bibinfo {author} {\bibfnamefont {G.}~\bibnamefont
  {Svetlichny}},\ }\href {https://doi.org/10.1103/PhysRevD.35.3066} {\bibfield
  {journal} {\bibinfo  {journal} {Phys. Rev. D}\ }\textbf {\bibinfo {volume}
  {35}},\ \bibinfo {pages} {3066} (\bibinfo {year} {1987})}\BibitemShut
  {NoStop}%
\bibitem [{\citenamefont {Fritz}(2016)}]{Fritz2016}%
  \BibitemOpen
  \bibfield  {author} {\bibinfo {author} {\bibfnamefont {T.}~\bibnamefont
  {Fritz}},\ }\href {https://doi.org/10.1007/s00220-015-2495-5} {\bibfield
  {journal} {\bibinfo  {journal} {Commun. Math. Phys.}\ }\textbf {\bibinfo
  {volume} {341}},\ \bibinfo {pages} {391} (\bibinfo {year}
  {2016})}\BibitemShut {NoStop}%
\bibitem [{\citenamefont {Rosset}\ \emph {et~al.}(2018)\citenamefont {Rosset},
  \citenamefont {Gisin},\ and\ \citenamefont {Wolfe}}]{Rosset2018}%
  \BibitemOpen
  \bibfield  {author} {\bibinfo {author} {\bibfnamefont {D.}~\bibnamefont
  {Rosset}}, \bibinfo {author} {\bibfnamefont {N.}~\bibnamefont {Gisin}},\ and\
  \bibinfo {author} {\bibfnamefont {E.}~\bibnamefont {Wolfe}},\ }\href
  {https://doi.org/10.26421/QIC18.11-12} {\bibfield  {journal} {\bibinfo
  {journal} {Quantum Inf. Comput.}\ }\textbf {\bibinfo {volume} {18}},\
  \bibinfo {pages} {0910} (\bibinfo {year} {2018})}\BibitemShut {NoStop}%
\bibitem [{\citenamefont {Branciard}\ \emph {et~al.}(2010)\citenamefont
  {Branciard}, \citenamefont {Gisin},\ and\ \citenamefont
  {Pironio}}]{Branciard2010}%
  \BibitemOpen
  \bibfield  {author} {\bibinfo {author} {\bibfnamefont {C.}~\bibnamefont
  {Branciard}}, \bibinfo {author} {\bibfnamefont {N.}~\bibnamefont {Gisin}},\
  and\ \bibinfo {author} {\bibfnamefont {S.}~\bibnamefont {Pironio}},\ }\href
  {https://doi.org/10.1103/PhysRevLett.104.170401} {\bibfield  {journal}
  {\bibinfo  {journal} {Phys. Rev. Lett.}\ }\textbf {\bibinfo {volume} {104}},\
  \bibinfo {pages} {170401} (\bibinfo {year} {2010})}\BibitemShut {NoStop}%
\bibitem [{\citenamefont {Fritz}(2012)}]{Fritz2012}%
  \BibitemOpen
  \bibfield  {author} {\bibinfo {author} {\bibfnamefont {T.}~\bibnamefont
  {Fritz}},\ }\href {https://doi.org/10.1088/1367-2630/14/10/103001} {\bibfield
   {journal} {\bibinfo  {journal} {New J. Phys.}\ }\textbf {\bibinfo {volume}
  {14}},\ \bibinfo {pages} {103001} (\bibinfo {year} {2012})}\BibitemShut
  {NoStop}%
\bibitem [{\citenamefont {Henson}\ \emph {et~al.}(2014)\citenamefont {Henson},
  \citenamefont {Lal},\ and\ \citenamefont {Pusey}}]{Henson2014}%
  \BibitemOpen
  \bibfield  {author} {\bibinfo {author} {\bibfnamefont {J.}~\bibnamefont
  {Henson}}, \bibinfo {author} {\bibfnamefont {R.}~\bibnamefont {Lal}},\ and\
  \bibinfo {author} {\bibfnamefont {M.~F.}\ \bibnamefont {Pusey}},\ }\href
  {https://doi.org/10.1088/1367-2630/16/11/113043} {\bibfield  {journal}
  {\bibinfo  {journal} {New J. Phys.}\ }\textbf {\bibinfo {volume} {16}},\
  \bibinfo {pages} {113043} (\bibinfo {year} {2014})}\BibitemShut {NoStop}%
\bibitem [{\citenamefont {Chiribella}\ \emph {et~al.}(2009)\citenamefont
  {Chiribella}, \citenamefont {D'Ariano},\ and\ \citenamefont
  {Perinotti}}]{chiribella2009framework}%
  \BibitemOpen
  \bibfield  {author} {\bibinfo {author} {\bibfnamefont {G.}~\bibnamefont
  {Chiribella}}, \bibinfo {author} {\bibfnamefont {G.~M.}\ \bibnamefont
  {D'Ariano}},\ and\ \bibinfo {author} {\bibfnamefont {P.}~\bibnamefont
  {Perinotti}},\ }\href {https://doi.org/10.1103/PhysRevA.80.022339} {\bibfield
   {journal} {\bibinfo  {journal} {Phys. Rev. A}\ }\textbf {\bibinfo {volume}
  {80}},\ \bibinfo {pages} {022339} (\bibinfo {year} {2009})}\BibitemShut
  {NoStop}%
\bibitem [{\citenamefont {Branciard}\ \emph {et~al.}(2012)\citenamefont
  {Branciard}, \citenamefont {Rosset}, \citenamefont {Gisin},\ and\
  \citenamefont {Pironio}}]{Branciard2012}%
  \BibitemOpen
  \bibfield  {author} {\bibinfo {author} {\bibfnamefont {C.}~\bibnamefont
  {Branciard}}, \bibinfo {author} {\bibfnamefont {D.}~\bibnamefont {Rosset}},
  \bibinfo {author} {\bibfnamefont {N.}~\bibnamefont {Gisin}},\ and\ \bibinfo
  {author} {\bibfnamefont {S.}~\bibnamefont {Pironio}},\ }\href
  {https://doi.org/10.1103/PhysRevA.85.032119} {\bibfield  {journal} {\bibinfo
  {journal} {Phys. Rev. A}\ }\textbf {\bibinfo {volume} {85}},\ \bibinfo
  {pages} {032119} (\bibinfo {year} {2012})}\BibitemShut {NoStop}%
\bibitem [{\citenamefont {Tavakoli}\ \emph {et~al.}(2021)\citenamefont
  {Tavakoli}, \citenamefont {Gisin},\ and\ \citenamefont
  {Branciard}}]{Tavakoli2020}%
  \BibitemOpen
  \bibfield  {author} {\bibinfo {author} {\bibfnamefont {A.}~\bibnamefont
  {Tavakoli}}, \bibinfo {author} {\bibfnamefont {N.}~\bibnamefont {Gisin}},\
  and\ \bibinfo {author} {\bibfnamefont {C.}~\bibnamefont {Branciard}},\ }\href
  {https://doi.org/10.1103/PhysRevLett.126.220401} {\bibfield  {journal}
  {\bibinfo  {journal} {Phys. Rev. Lett.}\ }\textbf {\bibinfo {volume} {126}},\
  \bibinfo {pages} {220401} (\bibinfo {year} {2021})}\BibitemShut {NoStop}%
\bibitem [{\citenamefont {Wolfe}\ \emph {et~al.}(2019)\citenamefont {Wolfe},
  \citenamefont {Spekkens},\ and\ \citenamefont {Fritz}}]{wolfe2019inflation}%
  \BibitemOpen
  \bibfield  {author} {\bibinfo {author} {\bibfnamefont {E.}~\bibnamefont
  {Wolfe}}, \bibinfo {author} {\bibfnamefont {R.~W.}\ \bibnamefont
  {Spekkens}},\ and\ \bibinfo {author} {\bibfnamefont {T.}~\bibnamefont
  {Fritz}},\ }\href {https://doi.org/10.1515/jci-2017-0020} {\bibfield
  {journal} {\bibinfo  {journal} {J. Causal Inference}\ }\textbf {\bibinfo
  {volume} {7}},\ \bibinfo {pages} {20170020} (\bibinfo {year}
  {2019})}\BibitemShut {NoStop}%
\bibitem [{\citenamefont {Tavakoli}\ \emph {et~al.}(2014)\citenamefont
  {Tavakoli}, \citenamefont {Skrzypczyk}, \citenamefont {Cavalcanti},\ and\
  \citenamefont {Ac\'{\i}n}}]{Tavakoli2014}%
  \BibitemOpen
  \bibfield  {author} {\bibinfo {author} {\bibfnamefont {A.}~\bibnamefont
  {Tavakoli}}, \bibinfo {author} {\bibfnamefont {P.}~\bibnamefont
  {Skrzypczyk}}, \bibinfo {author} {\bibfnamefont {D.}~\bibnamefont
  {Cavalcanti}},\ and\ \bibinfo {author} {\bibfnamefont {A.}~\bibnamefont
  {Ac\'{\i}n}},\ }\href {https://doi.org/10.1103/PhysRevA.90.062109} {\bibfield
   {journal} {\bibinfo  {journal} {Phys. Rev. A}\ }\textbf {\bibinfo {volume}
  {90}},\ \bibinfo {pages} {062109} (\bibinfo {year} {2014})}\BibitemShut
  {NoStop}%
\bibitem [{\citenamefont {Fraser}\ and\ \citenamefont
  {Wolfe}(2018)}]{Fraser2018}%
  \BibitemOpen
  \bibfield  {author} {\bibinfo {author} {\bibfnamefont {T.~C.}\ \bibnamefont
  {Fraser}}\ and\ \bibinfo {author} {\bibfnamefont {E.}~\bibnamefont {Wolfe}},\
  }\href {https://doi.org/10.1103/PhysRevA.98.022113} {\bibfield  {journal}
  {\bibinfo  {journal} {Phys. Rev. A}\ }\textbf {\bibinfo {volume} {98}},\
  \bibinfo {pages} {022113} (\bibinfo {year} {2018})}\BibitemShut {NoStop}%
\bibitem [{\citenamefont {Renou}\ \emph {et~al.}(2019)\citenamefont {Renou},
  \citenamefont {B\"aumer}, \citenamefont {Boreiri}, \citenamefont {Brunner},
  \citenamefont {Gisin},\ and\ \citenamefont {Beigi}}]{Renou2019}%
  \BibitemOpen
  \bibfield  {author} {\bibinfo {author} {\bibfnamefont {M.-O.}\ \bibnamefont
  {Renou}}, \bibinfo {author} {\bibfnamefont {E.}~\bibnamefont {B\"aumer}},
  \bibinfo {author} {\bibfnamefont {S.}~\bibnamefont {Boreiri}}, \bibinfo
  {author} {\bibfnamefont {N.}~\bibnamefont {Brunner}}, \bibinfo {author}
  {\bibfnamefont {N.}~\bibnamefont {Gisin}},\ and\ \bibinfo {author}
  {\bibfnamefont {S.}~\bibnamefont {Beigi}},\ }\href
  {https://doi.org/10.1103/PhysRevLett.123.140401} {\bibfield  {journal}
  {\bibinfo  {journal} {Phys. Rev. Lett.}\ }\textbf {\bibinfo {volume} {123}},\
  \bibinfo {pages} {140401} (\bibinfo {year} {2019})}\BibitemShut {NoStop}%
\bibitem [{\citenamefont {\ifmmode \check{S}\else
  \v{S}\fi{}upi\ifmmode~\acute{c}\else \'{c}\fi{}}\ \emph
  {et~al.}(2020)\citenamefont {\ifmmode \check{S}\else
  \v{S}\fi{}upi\ifmmode~\acute{c}\else \'{c}\fi{}}, \citenamefont {Bancal},\
  and\ \citenamefont {Brunner}}]{Supic2020independence}%
  \BibitemOpen
  \bibfield  {author} {\bibinfo {author} {\bibfnamefont {I.}~\bibnamefont
  {\ifmmode \check{S}\else \v{S}\fi{}upi\ifmmode~\acute{c}\else \'{c}\fi{}}},
  \bibinfo {author} {\bibfnamefont {J.-D.}\ \bibnamefont {Bancal}},\ and\
  \bibinfo {author} {\bibfnamefont {N.}~\bibnamefont {Brunner}},\ }\href
  {https://doi.org/10.1103/PhysRevLett.125.240403} {\bibfield  {journal}
  {\bibinfo  {journal} {Phys. Rev. Lett.}\ }\textbf {\bibinfo {volume} {125}},\
  \bibinfo {pages} {240403} (\bibinfo {year} {2020})}\BibitemShut {NoStop}%
\bibitem [{\citenamefont {Bennett}\ \emph {et~al.}(1999)\citenamefont
  {Bennett}, \citenamefont {DiVincenzo}, \citenamefont {Fuchs}, \citenamefont
  {Mor}, \citenamefont {Rains}, \citenamefont {Shor}, \citenamefont {Smolin},\
  and\ \citenamefont {Wootters}}]{Bennett1999}%
  \BibitemOpen
  \bibfield  {author} {\bibinfo {author} {\bibfnamefont {C.~H.}\ \bibnamefont
  {Bennett}}, \bibinfo {author} {\bibfnamefont {D.~P.}\ \bibnamefont
  {DiVincenzo}}, \bibinfo {author} {\bibfnamefont {C.~A.}\ \bibnamefont
  {Fuchs}}, \bibinfo {author} {\bibfnamefont {T.}~\bibnamefont {Mor}}, \bibinfo
  {author} {\bibfnamefont {E.}~\bibnamefont {Rains}}, \bibinfo {author}
  {\bibfnamefont {P.~W.}\ \bibnamefont {Shor}}, \bibinfo {author}
  {\bibfnamefont {J.~A.}\ \bibnamefont {Smolin}},\ and\ \bibinfo {author}
  {\bibfnamefont {W.~K.}\ \bibnamefont {Wootters}},\ }\href
  {https://doi.org/10.1103/PhysRevA.59.1070} {\bibfield  {journal} {\bibinfo
  {journal} {Phys. Rev. A}\ }\textbf {\bibinfo {volume} {59}},\ \bibinfo
  {pages} {1070} (\bibinfo {year} {1999})}\BibitemShut {NoStop}%
\bibitem [{\citenamefont {\ifmmode~\dot{Z}\else \.{Z}\fi{}ukowski}\ \emph
  {et~al.}(1995)\citenamefont {\ifmmode~\dot{Z}\else \.{Z}\fi{}ukowski},
  \citenamefont {Zeilinger},\ and\ \citenamefont {Weinfurter}}]{Zukowski1995}%
  \BibitemOpen
  \bibfield  {author} {\bibinfo {author} {\bibfnamefont {M.}~\bibnamefont
  {\ifmmode~\dot{Z}\else \.{Z}\fi{}ukowski}}, \bibinfo {author} {\bibfnamefont
  {A.}~\bibnamefont {Zeilinger}},\ and\ \bibinfo {author} {\bibfnamefont
  {H.}~\bibnamefont {Weinfurter}},\ }\href
  {https://doi.org/10.1111/j.1749-6632.1995.tb38959.x} {\bibfield  {journal}
  {\bibinfo  {journal} {Ann. New York Acad. Sci}\ }\textbf {\bibinfo {volume}
  {755}},\ \bibinfo {pages} {91} (\bibinfo {year} {1995})}\BibitemShut
  {NoStop}%
\bibitem [{\citenamefont {Bose}\ \emph {et~al.}(1999)\citenamefont {Bose},
  \citenamefont {Vedral},\ and\ \citenamefont {Knight}}]{Bose1999}%
  \BibitemOpen
  \bibfield  {author} {\bibinfo {author} {\bibfnamefont {S.}~\bibnamefont
  {Bose}}, \bibinfo {author} {\bibfnamefont {V.}~\bibnamefont {Vedral}},\ and\
  \bibinfo {author} {\bibfnamefont {P.~L.}\ \bibnamefont {Knight}},\ }\href
  {https://doi.org/10.1103/PhysRevA.60.194} {\bibfield  {journal} {\bibinfo
  {journal} {Phys. Rev. A}\ }\textbf {\bibinfo {volume} {60}},\ \bibinfo
  {pages} {194} (\bibinfo {year} {1999})}\BibitemShut {NoStop}%
\bibitem [{\citenamefont {Pan}\ \emph {et~al.}(1998)\citenamefont {Pan},
  \citenamefont {Bouwmeester}, \citenamefont {Weinfurter},\ and\ \citenamefont
  {Zeilinger}}]{Pan1998}%
  \BibitemOpen
  \bibfield  {author} {\bibinfo {author} {\bibfnamefont {J.-W.}\ \bibnamefont
  {Pan}}, \bibinfo {author} {\bibfnamefont {D.}~\bibnamefont {Bouwmeester}},
  \bibinfo {author} {\bibfnamefont {H.}~\bibnamefont {Weinfurter}},\ and\
  \bibinfo {author} {\bibfnamefont {A.}~\bibnamefont {Zeilinger}},\ }\href
  {https://doi.org/10.1103/PhysRevLett.80.3891} {\bibfield  {journal} {\bibinfo
   {journal} {Phys. Rev. Lett.}\ }\textbf {\bibinfo {volume} {80}},\ \bibinfo
  {pages} {3891} (\bibinfo {year} {1998})}\BibitemShut {NoStop}%
\bibitem [{\citenamefont {Jennewein}\ \emph {et~al.}(2001)\citenamefont
  {Jennewein}, \citenamefont {Weihs}, \citenamefont {Pan},\ and\ \citenamefont
  {Zeilinger}}]{Jennewein2001}%
  \BibitemOpen
  \bibfield  {author} {\bibinfo {author} {\bibfnamefont {T.}~\bibnamefont
  {Jennewein}}, \bibinfo {author} {\bibfnamefont {G.}~\bibnamefont {Weihs}},
  \bibinfo {author} {\bibfnamefont {J.-W.}\ \bibnamefont {Pan}},\ and\ \bibinfo
  {author} {\bibfnamefont {A.}~\bibnamefont {Zeilinger}},\ }\href
  {https://doi.org/10.1103/PhysRevLett.88.017903} {\bibfield  {journal}
  {\bibinfo  {journal} {Phys. Rev. Lett.}\ }\textbf {\bibinfo {volume} {88}},\
  \bibinfo {pages} {017903} (\bibinfo {year} {2001})}\BibitemShut {NoStop}%
\bibitem [{\citenamefont {Bennett}\ \emph {et~al.}(1993)\citenamefont
  {Bennett}, \citenamefont {Brassard}, \citenamefont {Cr\'epeau}, \citenamefont
  {Jozsa}, \citenamefont {Peres},\ and\ \citenamefont
  {Wootters}}]{Teleportation}%
  \BibitemOpen
  \bibfield  {author} {\bibinfo {author} {\bibfnamefont {C.~H.}\ \bibnamefont
  {Bennett}}, \bibinfo {author} {\bibfnamefont {G.}~\bibnamefont {Brassard}},
  \bibinfo {author} {\bibfnamefont {C.}~\bibnamefont {Cr\'epeau}}, \bibinfo
  {author} {\bibfnamefont {R.}~\bibnamefont {Jozsa}}, \bibinfo {author}
  {\bibfnamefont {A.}~\bibnamefont {Peres}},\ and\ \bibinfo {author}
  {\bibfnamefont {W.~K.}\ \bibnamefont {Wootters}},\ }\href
  {https://doi.org/10.1103/PhysRevLett.70.1895} {\bibfield  {journal} {\bibinfo
   {journal} {Phys. Rev. Lett.}\ }\textbf {\bibinfo {volume} {70}},\ \bibinfo
  {pages} {1895} (\bibinfo {year} {1993})}\BibitemShut {NoStop}%
\bibitem [{\citenamefont {Bennett}\ and\ \citenamefont
  {Wiesner}(1992)}]{Bennett1992}%
  \BibitemOpen
  \bibfield  {author} {\bibinfo {author} {\bibfnamefont {C.~H.}\ \bibnamefont
  {Bennett}}\ and\ \bibinfo {author} {\bibfnamefont {S.~J.}\ \bibnamefont
  {Wiesner}},\ }\href {https://doi.org/10.1103/PhysRevLett.69.2881} {\bibfield
  {journal} {\bibinfo  {journal} {Phys. Rev. Lett.}\ }\textbf {\bibinfo
  {volume} {69}},\ \bibinfo {pages} {2881} (\bibinfo {year}
  {1992})}\BibitemShut {NoStop}%
\bibitem [{\citenamefont {Saunders}\ \emph {et~al.}(2017)\citenamefont
  {Saunders}, \citenamefont {Bennet}, \citenamefont {Branciard},\ and\
  \citenamefont {Pryde}}]{Saunders2017}%
  \BibitemOpen
  \bibfield  {author} {\bibinfo {author} {\bibfnamefont {D.~J.}\ \bibnamefont
  {Saunders}}, \bibinfo {author} {\bibfnamefont {A.~J.}\ \bibnamefont
  {Bennet}}, \bibinfo {author} {\bibfnamefont {C.}~\bibnamefont {Branciard}},\
  and\ \bibinfo {author} {\bibfnamefont {G.~J.}\ \bibnamefont {Pryde}},\ }\href
  {https://doi.org/10.1126/sciadv.1602743} {\bibfield  {journal} {\bibinfo
  {journal} {Sci. Adv.}\ }\textbf {\bibinfo {volume} {3}},\ \bibinfo {pages}
  {e1602743} (\bibinfo {year} {2017})}\BibitemShut {NoStop}%
\bibitem [{\citenamefont {Gisin}\ \emph {et~al.}(2017)\citenamefont {Gisin},
  \citenamefont {Mei}, \citenamefont {Tavakoli}, \citenamefont {Renou},\ and\
  \citenamefont {Brunner}}]{Gisin2017}%
  \BibitemOpen
  \bibfield  {author} {\bibinfo {author} {\bibfnamefont {N.}~\bibnamefont
  {Gisin}}, \bibinfo {author} {\bibfnamefont {Q.}~\bibnamefont {Mei}}, \bibinfo
  {author} {\bibfnamefont {A.}~\bibnamefont {Tavakoli}}, \bibinfo {author}
  {\bibfnamefont {M.-O.}\ \bibnamefont {Renou}},\ and\ \bibinfo {author}
  {\bibfnamefont {N.}~\bibnamefont {Brunner}},\ }\href
  {https://doi.org/10.1103/PhysRevA.96.020304} {\bibfield  {journal} {\bibinfo
  {journal} {Phys. Rev. A}\ }\textbf {\bibinfo {volume} {96}},\ \bibinfo
  {pages} {020304} (\bibinfo {year} {2017})}\BibitemShut {NoStop}%
\bibitem [{\citenamefont {Andreoli}\ \emph {et~al.}(2017)\citenamefont
  {Andreoli}, \citenamefont {Carvacho}, \citenamefont {Santodonato},
  \citenamefont {Chaves},\ and\ \citenamefont {Sciarrino}}]{Andreoli2017}%
  \BibitemOpen
  \bibfield  {author} {\bibinfo {author} {\bibfnamefont {F.}~\bibnamefont
  {Andreoli}}, \bibinfo {author} {\bibfnamefont {G.}~\bibnamefont {Carvacho}},
  \bibinfo {author} {\bibfnamefont {L.}~\bibnamefont {Santodonato}}, \bibinfo
  {author} {\bibfnamefont {R.}~\bibnamefont {Chaves}},\ and\ \bibinfo {author}
  {\bibfnamefont {F.}~\bibnamefont {Sciarrino}},\ }\href
  {https://doi.org/10.1088/1367-2630/aa8b9b} {\bibfield  {journal} {\bibinfo
  {journal} {New J. Phys.}\ }\textbf {\bibinfo {volume} {19}},\ \bibinfo
  {pages} {113020} (\bibinfo {year} {2017})}\BibitemShut {NoStop}%
\bibitem [{\citenamefont {Horodecki}\ \emph {et~al.}(1995)\citenamefont
  {Horodecki}, \citenamefont {Horodecki},\ and\ \citenamefont
  {Horodecki}}]{Horodecki1995}%
  \BibitemOpen
  \bibfield  {author} {\bibinfo {author} {\bibfnamefont {R.}~\bibnamefont
  {Horodecki}}, \bibinfo {author} {\bibfnamefont {P.}~\bibnamefont
  {Horodecki}},\ and\ \bibinfo {author} {\bibfnamefont {M.}~\bibnamefont
  {Horodecki}},\ }\href {https://doi.org/10.1016/0375-9601(95)00214-N}
  {\bibfield  {journal} {\bibinfo  {journal} {Phys. Lett. A}\ }\textbf
  {\bibinfo {volume} {200}},\ \bibinfo {pages} {340} (\bibinfo {year}
  {1995})}\BibitemShut {NoStop}%
\bibitem [{\citenamefont {Mukherjee}\ \emph {et~al.}(2016)\citenamefont
  {Mukherjee}, \citenamefont {Paul},\ and\ \citenamefont
  {Sarkar}}]{Mukherjee2016}%
  \BibitemOpen
  \bibfield  {author} {\bibinfo {author} {\bibfnamefont {K.}~\bibnamefont
  {Mukherjee}}, \bibinfo {author} {\bibfnamefont {B.}~\bibnamefont {Paul}},\
  and\ \bibinfo {author} {\bibfnamefont {D.}~\bibnamefont {Sarkar}},\ }\href
  {https://doi.org/10.1007/s11128-016-1301-4} {\bibfield  {journal} {\bibinfo
  {journal} {Quantum Inf. Process.}\ }\textbf {\bibinfo {volume} {15}},\
  \bibinfo {pages} {2895} (\bibinfo {year} {2016})}\BibitemShut {NoStop}%
\bibitem [{\citenamefont {Ac\'{\i}n}\ \emph {et~al.}(2002)\citenamefont
  {Ac\'{\i}n}, \citenamefont {Durt}, \citenamefont {Gisin},\ and\ \citenamefont
  {Latorre}}]{Acin2002}%
  \BibitemOpen
  \bibfield  {author} {\bibinfo {author} {\bibfnamefont {A.}~\bibnamefont
  {Ac\'{\i}n}}, \bibinfo {author} {\bibfnamefont {T.}~\bibnamefont {Durt}},
  \bibinfo {author} {\bibfnamefont {N.}~\bibnamefont {Gisin}},\ and\ \bibinfo
  {author} {\bibfnamefont {J.~I.}\ \bibnamefont {Latorre}},\ }\href
  {https://doi.org/10.1103/PhysRevA.65.052325} {\bibfield  {journal} {\bibinfo
  {journal} {Phys. Rev. A}\ }\textbf {\bibinfo {volume} {65}},\ \bibinfo
  {pages} {052325} (\bibinfo {year} {2002})}\BibitemShut {NoStop}%
\bibitem [{\citenamefont {Maudlin}(1992)}]{Maudlin1992}%
  \BibitemOpen
  \bibfield  {author} {\bibinfo {author} {\bibfnamefont {T.}~\bibnamefont
  {Maudlin}},\ }\href {http://www.jstor.org/stable/192771} {\bibfield
  {journal} {\bibinfo  {journal} {PSA: Proceedings of the Biennial Meeting of
  the Philosophy of Science Association}\ }\textbf {\bibinfo {volume} {1992}},\
  \bibinfo {pages} {404} (\bibinfo {year} {1992})}\BibitemShut {NoStop}%
\bibitem [{\citenamefont {Brassard}\ \emph {et~al.}(1999)\citenamefont
  {Brassard}, \citenamefont {Cleve},\ and\ \citenamefont
  {Tapp}}]{Brassard1999}%
  \BibitemOpen
  \bibfield  {author} {\bibinfo {author} {\bibfnamefont {G.}~\bibnamefont
  {Brassard}}, \bibinfo {author} {\bibfnamefont {R.}~\bibnamefont {Cleve}},\
  and\ \bibinfo {author} {\bibfnamefont {A.}~\bibnamefont {Tapp}},\ }\href
  {https://doi.org/10.1103/PhysRevLett.83.1874} {\bibfield  {journal} {\bibinfo
   {journal} {Phys. Rev. Lett.}\ }\textbf {\bibinfo {volume} {83}},\ \bibinfo
  {pages} {1874} (\bibinfo {year} {1999})}\BibitemShut {NoStop}%
\bibitem [{\citenamefont {Steiner}(2000)}]{Steiner2000}%
  \BibitemOpen
  \bibfield  {author} {\bibinfo {author} {\bibfnamefont {M.}~\bibnamefont
  {Steiner}},\ }\href
  {https://doi.org/https://doi.org/10.1016/S0375-9601(00)00315-7} {\bibfield
  {journal} {\bibinfo  {journal} {Phys. Lett. A}\ }\textbf {\bibinfo {volume}
  {270}},\ \bibinfo {pages} {239 } (\bibinfo {year} {2000})}\BibitemShut
  {NoStop}%
\bibitem [{\citenamefont {Bacon}\ and\ \citenamefont
  {Toner}(2003)}]{Bacon2003}%
  \BibitemOpen
  \bibfield  {author} {\bibinfo {author} {\bibfnamefont {D.}~\bibnamefont
  {Bacon}}\ and\ \bibinfo {author} {\bibfnamefont {B.~F.}\ \bibnamefont
  {Toner}},\ }\href {https://doi.org/10.1103/PhysRevLett.90.157904} {\bibfield
  {journal} {\bibinfo  {journal} {Phys. Rev. Lett.}\ }\textbf {\bibinfo
  {volume} {90}},\ \bibinfo {pages} {157904} (\bibinfo {year}
  {2003})}\BibitemShut {NoStop}%
\bibitem [{\citenamefont {Toner}\ and\ \citenamefont
  {Bacon}(2003)}]{Toner2003}%
  \BibitemOpen
  \bibfield  {author} {\bibinfo {author} {\bibfnamefont {B.~F.}\ \bibnamefont
  {Toner}}\ and\ \bibinfo {author} {\bibfnamefont {D.}~\bibnamefont {Bacon}},\
  }\href {https://doi.org/10.1103/PhysRevLett.91.187904} {\bibfield  {journal}
  {\bibinfo  {journal} {Phys. Rev. Lett.}\ }\textbf {\bibinfo {volume} {91}},\
  \bibinfo {pages} {187904} (\bibinfo {year} {2003})}\BibitemShut {NoStop}%
\bibitem [{\citenamefont {Barrett}\ and\ \citenamefont
  {Gisin}(2011)}]{Barrett2011}%
  \BibitemOpen
  \bibfield  {author} {\bibinfo {author} {\bibfnamefont {J.}~\bibnamefont
  {Barrett}}\ and\ \bibinfo {author} {\bibfnamefont {N.}~\bibnamefont
  {Gisin}},\ }\href {https://doi.org/10.1103/PhysRevLett.106.100406} {\bibfield
   {journal} {\bibinfo  {journal} {Phys. Rev. Lett.}\ }\textbf {\bibinfo
  {volume} {106}},\ \bibinfo {pages} {100406} (\bibinfo {year}
  {2011})}\BibitemShut {NoStop}%
\bibitem [{\citenamefont {Branciard}\ \emph {et~al.}(2012)\citenamefont
  {Branciard}, \citenamefont {Brunner}, \citenamefont {Buhrman}, \citenamefont
  {Cleve}, \citenamefont {Gisin}, \citenamefont {Portmann}, \citenamefont
  {Rosset},\ and\ \citenamefont {Szegedy}}]{BranciardBrunner2012}%
  \BibitemOpen
  \bibfield  {author} {\bibinfo {author} {\bibfnamefont {C.}~\bibnamefont
  {Branciard}}, \bibinfo {author} {\bibfnamefont {N.}~\bibnamefont {Brunner}},
  \bibinfo {author} {\bibfnamefont {H.}~\bibnamefont {Buhrman}}, \bibinfo
  {author} {\bibfnamefont {R.}~\bibnamefont {Cleve}}, \bibinfo {author}
  {\bibfnamefont {N.}~\bibnamefont {Gisin}}, \bibinfo {author} {\bibfnamefont
  {S.}~\bibnamefont {Portmann}}, \bibinfo {author} {\bibfnamefont
  {D.}~\bibnamefont {Rosset}},\ and\ \bibinfo {author} {\bibfnamefont
  {M.}~\bibnamefont {Szegedy}},\ }\href
  {https://doi.org/10.1103/PhysRevLett.109.100401} {\bibfield  {journal}
  {\bibinfo  {journal} {Phys. Rev. Lett.}\ }\textbf {\bibinfo {volume} {109}},\
  \bibinfo {pages} {100401} (\bibinfo {year} {2012})}\BibitemShut {NoStop}%
\bibitem [{\citenamefont {Gisin}(2019)}]{Gisin2019}%
  \BibitemOpen
  \bibfield  {author} {\bibinfo {author} {\bibfnamefont {N.}~\bibnamefont
  {Gisin}},\ }\href {https://doi.org/10.3390/e21030325} {\bibfield  {journal}
  {\bibinfo  {journal} {Entropy}\ }\textbf {\bibinfo {volume} {21}},\ \bibinfo
  {pages} {325} (\bibinfo {year} {2019})}\BibitemShut {NoStop}%
\bibitem [{\citenamefont {Gisin}\ and\ \citenamefont
  {Popescu}(1999)}]{Gisin1999}%
  \BibitemOpen
  \bibfield  {author} {\bibinfo {author} {\bibfnamefont {N.}~\bibnamefont
  {Gisin}}\ and\ \bibinfo {author} {\bibfnamefont {S.}~\bibnamefont
  {Popescu}},\ }\href {https://doi.org/10.1103/PhysRevLett.83.432} {\bibfield
  {journal} {\bibinfo  {journal} {Phys. Rev. Lett.}\ }\textbf {\bibinfo
  {volume} {83}},\ \bibinfo {pages} {432} (\bibinfo {year} {1999})}\BibitemShut
  {NoStop}%
\bibitem [{\citenamefont {Czartowski}\ and\ \citenamefont
  {{\.{Z}}yczkowski}(2021)}]{Czartowski2020}%
  \BibitemOpen
  \bibfield  {author} {\bibinfo {author} {\bibfnamefont {J.}~\bibnamefont
  {Czartowski}}\ and\ \bibinfo {author} {\bibfnamefont {K.}~\bibnamefont
  {{\.{Z}}yczkowski}},\ }\href {https://doi.org/10.22331/q-2021-04-26-442}
  {\bibfield  {journal} {\bibinfo  {journal} {{Quantum}}\ }\textbf {\bibinfo
  {volume} {5}},\ \bibinfo {pages} {442} (\bibinfo {year} {2021})}\BibitemShut
  {NoStop}%
\bibitem [{\citenamefont {Renes}\ \emph {et~al.}(2004)\citenamefont {Renes},
  \citenamefont {Blume-Kohout}, \citenamefont {Scott},\ and\ \citenamefont
  {Caves}}]{Renes2004}%
  \BibitemOpen
  \bibfield  {author} {\bibinfo {author} {\bibfnamefont {J.~M.}\ \bibnamefont
  {Renes}}, \bibinfo {author} {\bibfnamefont {R.}~\bibnamefont {Blume-Kohout}},
  \bibinfo {author} {\bibfnamefont {A.~J.}\ \bibnamefont {Scott}},\ and\
  \bibinfo {author} {\bibfnamefont {C.~M.}\ \bibnamefont {Caves}},\ }\href
  {https://doi.org/10.1063/1.1737053} {\bibfield  {journal} {\bibinfo
  {journal} {J. Math. Phys.}\ }\textbf {\bibinfo {volume} {45}},\ \bibinfo
  {pages} {2171} (\bibinfo {year} {2004})}\BibitemShut {NoStop}%
\bibitem [{\citenamefont {Werner}(1989)}]{Werner1989}%
  \BibitemOpen
  \bibfield  {author} {\bibinfo {author} {\bibfnamefont {R.~F.}\ \bibnamefont
  {Werner}},\ }\href {https://doi.org/10.1103/PhysRevA.40.4277} {\bibfield
  {journal} {\bibinfo  {journal} {Phys. Rev. A}\ }\textbf {\bibinfo {volume}
  {40}},\ \bibinfo {pages} {4277} (\bibinfo {year} {1989})}\BibitemShut
  {NoStop}%
\bibitem [{\citenamefont {Renou}\ and\ \citenamefont
  {Beigi}(2022)}]{Renou2020short}%
  \BibitemOpen
  \bibfield  {author} {\bibinfo {author} {\bibfnamefont {M.-O.}\ \bibnamefont
  {Renou}}\ and\ \bibinfo {author} {\bibfnamefont {S.}~\bibnamefont {Beigi}},\
  }\href {https://doi.org/10.1103/PhysRevLett.128.060401} {\bibfield  {journal}
  {\bibinfo  {journal} {Phys. Rev. Lett.}\ }\textbf {\bibinfo {volume} {128}},\
  \bibinfo {pages} {060401} (\bibinfo {year} {2022})}\BibitemShut {NoStop}%
\bibitem [{\citenamefont {Renou}\ and\ \citenamefont
  {Beigi}(2022)}]{Renou2020long}%
  \BibitemOpen
  \bibfield  {author} {\bibinfo {author} {\bibfnamefont {M.-O.}\ \bibnamefont
  {Renou}}\ and\ \bibinfo {author} {\bibfnamefont {S.}~\bibnamefont {Beigi}},\
  }\href {https://doi.org/10.1103/PhysRevA.105.022408} {\bibfield  {journal}
  {\bibinfo  {journal} {Phys. Rev. A}\ }\textbf {\bibinfo {volume} {105}},\
  \bibinfo {pages} {022408} (\bibinfo {year} {2022})}\BibitemShut {NoStop}%
\bibitem [{\citenamefont {Abiuso}\ \emph {et~al.}(2022)\citenamefont {Abiuso},
  \citenamefont {Kriv\'achy}, \citenamefont {Boghiu}, \citenamefont {Renou},
  \citenamefont {Pozas-Kerstjens},\ and\ \citenamefont
  {Ac\'{\i}n}}]{abiuso2021singlephoton}%
  \BibitemOpen
  \bibfield  {author} {\bibinfo {author} {\bibfnamefont {P.}~\bibnamefont
  {Abiuso}}, \bibinfo {author} {\bibfnamefont {T.}~\bibnamefont {Kriv\'achy}},
  \bibinfo {author} {\bibfnamefont {E.-C.}\ \bibnamefont {Boghiu}}, \bibinfo
  {author} {\bibfnamefont {M.-O.}\ \bibnamefont {Renou}}, \bibinfo {author}
  {\bibfnamefont {A.}~\bibnamefont {Pozas-Kerstjens}},\ and\ \bibinfo {author}
  {\bibfnamefont {A.}~\bibnamefont {Ac\'{\i}n}},\ }\href
  {https://doi.org/10.1103/PhysRevResearch.4.L012041} {\bibfield  {journal}
  {\bibinfo  {journal} {Phys. Rev. Research}\ }\textbf {\bibinfo {volume}
  {4}},\ \bibinfo {pages} {L012041} (\bibinfo {year} {2022})}\BibitemShut
  {NoStop}%
\bibitem [{\citenamefont {Gisin}()}]{gisin2017elegant}%
  \BibitemOpen
  \bibfield  {author} {\bibinfo {author} {\bibfnamefont {N.}~\bibnamefont
  {Gisin}},\ }\href@noop {} {\bibinfo {title} {The {Elegant} joint quantum
  measurement and some conjectures about {N}-locality in the triangle and other
  configurations}},\ \Eprint {https://arxiv.org/abs/1708.05556}
  {arXiv:1708.05556} \BibitemShut {NoStop}%
\bibitem [{\citenamefont {Kriv\'achy}\ \emph {et~al.}(2020)\citenamefont
  {Kriv\'achy}, \citenamefont {Cai}, \citenamefont {Cavalcanti}, \citenamefont
  {Tavakoli}, \citenamefont {Gisin},\ and\ \citenamefont
  {Brunner}}]{krivachy2019neural}%
  \BibitemOpen
  \bibfield  {author} {\bibinfo {author} {\bibfnamefont {T.}~\bibnamefont
  {Kriv\'achy}}, \bibinfo {author} {\bibfnamefont {Y.}~\bibnamefont {Cai}},
  \bibinfo {author} {\bibfnamefont {D.}~\bibnamefont {Cavalcanti}}, \bibinfo
  {author} {\bibfnamefont {A.}~\bibnamefont {Tavakoli}}, \bibinfo {author}
  {\bibfnamefont {N.}~\bibnamefont {Gisin}},\ and\ \bibinfo {author}
  {\bibfnamefont {N.}~\bibnamefont {Brunner}},\ }\href
  {https://doi.org/10.1038/s41534-020-00305-x} {\bibfield  {journal} {\bibinfo
  {journal} {npj Quantum Inf.}\ }\textbf {\bibinfo {volume} {6}},\ \bibinfo
  {pages} {70} (\bibinfo {year} {2020})}\BibitemShut {NoStop}%
\bibitem [{\citenamefont {Steudel}\ and\ \citenamefont
  {Ay}(2015)}]{SteudelGHZ}%
  \BibitemOpen
  \bibfield  {author} {\bibinfo {author} {\bibfnamefont {B.}~\bibnamefont
  {Steudel}}\ and\ \bibinfo {author} {\bibfnamefont {N.}~\bibnamefont {Ay}},\
  }\href {https://doi.org/10.3390/e17042304} {\bibfield  {journal} {\bibinfo
  {journal} {Entropy}\ }\textbf {\bibinfo {volume} {17}},\ \bibinfo {pages}
  {2304} (\bibinfo {year} {2015})}\BibitemShut {NoStop}%
\bibitem [{\citenamefont {Wolfe}\ \emph {et~al.}(2021)\citenamefont {Wolfe},
  \citenamefont {Pozas-Kerstjens}, \citenamefont {Grinberg}, \citenamefont
  {Rosset}, \citenamefont {Ac\'{\i}n},\ and\ \citenamefont
  {Navascu\'es}}]{wolfe2019qinflation}%
  \BibitemOpen
  \bibfield  {author} {\bibinfo {author} {\bibfnamefont {E.}~\bibnamefont
  {Wolfe}}, \bibinfo {author} {\bibfnamefont {A.}~\bibnamefont
  {Pozas-Kerstjens}}, \bibinfo {author} {\bibfnamefont {M.}~\bibnamefont
  {Grinberg}}, \bibinfo {author} {\bibfnamefont {D.}~\bibnamefont {Rosset}},
  \bibinfo {author} {\bibfnamefont {A.}~\bibnamefont {Ac\'{\i}n}},\ and\
  \bibinfo {author} {\bibfnamefont {M.}~\bibnamefont {Navascu\'es}},\ }\href
  {https://doi.org/10.1103/PhysRevX.11.021043} {\bibfield  {journal} {\bibinfo
  {journal} {Phys. Rev. X}\ }\textbf {\bibinfo {volume} {11}},\ \bibinfo
  {pages} {021043} (\bibinfo {year} {2021})}\BibitemShut {NoStop}%
\bibitem [{\citenamefont {Mermin}(1990)}]{Mermin1990}%
  \BibitemOpen
  \bibfield  {author} {\bibinfo {author} {\bibfnamefont {N.~D.}\ \bibnamefont
  {Mermin}},\ }\href {https://doi.org/10.1103/PhysRevLett.65.1838} {\bibfield
  {journal} {\bibinfo  {journal} {Phys. Rev. Lett.}\ }\textbf {\bibinfo
  {volume} {65}},\ \bibinfo {pages} {1838} (\bibinfo {year}
  {1990})}\BibitemShut {NoStop}%
\bibitem [{\citenamefont {Gisin}\ \emph {et~al.}(2020)\citenamefont {Gisin},
  \citenamefont {Bancal}, \citenamefont {Cai}, \citenamefont {Remy},
  \citenamefont {Tavakoli}, \citenamefont {Zambrini~Cruzeiro}, \citenamefont
  {Popescu},\ and\ \citenamefont {Brunner}}]{Gisin2020}%
  \BibitemOpen
  \bibfield  {author} {\bibinfo {author} {\bibfnamefont {N.}~\bibnamefont
  {Gisin}}, \bibinfo {author} {\bibfnamefont {J.-D.}\ \bibnamefont {Bancal}},
  \bibinfo {author} {\bibfnamefont {Y.}~\bibnamefont {Cai}}, \bibinfo {author}
  {\bibfnamefont {P.}~\bibnamefont {Remy}}, \bibinfo {author} {\bibfnamefont
  {A.}~\bibnamefont {Tavakoli}}, \bibinfo {author} {\bibfnamefont
  {E.}~\bibnamefont {Zambrini~Cruzeiro}}, \bibinfo {author} {\bibfnamefont
  {S.}~\bibnamefont {Popescu}},\ and\ \bibinfo {author} {\bibfnamefont
  {N.}~\bibnamefont {Brunner}},\ }\href
  {https://doi.org/10.1038/s41467-020-16137-4} {\bibfield  {journal} {\bibinfo
  {journal} {Nat. Commun.}\ }\textbf {\bibinfo {volume} {11}},\ \bibinfo
  {pages} {2378} (\bibinfo {year} {2020})}\BibitemShut {NoStop}%
\bibitem [{\citenamefont {Luo}(2018)}]{Luo2018}%
  \BibitemOpen
  \bibfield  {author} {\bibinfo {author} {\bibfnamefont {M.-X.}\ \bibnamefont
  {Luo}},\ }\href {https://doi.org/10.1103/PhysRevLett.120.140402} {\bibfield
  {journal} {\bibinfo  {journal} {Phys. Rev. Lett.}\ }\textbf {\bibinfo
  {volume} {120}},\ \bibinfo {pages} {140402} (\bibinfo {year}
  {2018})}\BibitemShut {NoStop}%
\bibitem [{\citenamefont {Greenberger}\ \emph {et~al.}()\citenamefont
  {Greenberger}, \citenamefont {Horne},\ and\ \citenamefont
  {Zeilinger}}]{Greenberger1989}%
  \BibitemOpen
  \bibfield  {author} {\bibinfo {author} {\bibfnamefont {D.~M.}\ \bibnamefont
  {Greenberger}}, \bibinfo {author} {\bibfnamefont {M.~A.}\ \bibnamefont
  {Horne}},\ and\ \bibinfo {author} {\bibfnamefont {A.}~\bibnamefont
  {Zeilinger}},\ }\href@noop {} {\bibinfo {title} {{G}oing {B}eyond {B}ell's
  {T}heorem}},\ \bibinfo {howpublished} {in: {Bell}'s Theorem, Quantum Theory,
  and Conceptions of the Universe', M. Kafatos (Ed.), Kluwer, Dordrecht, 69-72
  (1989)},\ \Eprint {https://arxiv.org/abs/0712.0921} {arXiv:0712.0921}
  \BibitemShut {NoStop}%
\bibitem [{\citenamefont {Hsu}\ and\ \citenamefont {Chen}(2021)}]{Hsu2021}%
  \BibitemOpen
  \bibfield  {author} {\bibinfo {author} {\bibfnamefont {L.-Y.}\ \bibnamefont
  {Hsu}}\ and\ \bibinfo {author} {\bibfnamefont {C.-H.}\ \bibnamefont {Chen}},\
  }\href {https://doi.org/10.1103/PhysRevResearch.3.023139} {\bibfield
  {journal} {\bibinfo  {journal} {Phys. Rev. Research}\ }\textbf {\bibinfo
  {volume} {3}},\ \bibinfo {pages} {023139} (\bibinfo {year}
  {2021})}\BibitemShut {NoStop}%
\bibitem [{\citenamefont {Rosset}\ \emph {et~al.}(2016)\citenamefont {Rosset},
  \citenamefont {Branciard}, \citenamefont {Barnea}, \citenamefont {P\"utz},
  \citenamefont {Brunner},\ and\ \citenamefont {Gisin}}]{Rosset2016}%
  \BibitemOpen
  \bibfield  {author} {\bibinfo {author} {\bibfnamefont {D.}~\bibnamefont
  {Rosset}}, \bibinfo {author} {\bibfnamefont {C.}~\bibnamefont {Branciard}},
  \bibinfo {author} {\bibfnamefont {T.~J.}\ \bibnamefont {Barnea}}, \bibinfo
  {author} {\bibfnamefont {G.}~\bibnamefont {P\"utz}}, \bibinfo {author}
  {\bibfnamefont {N.}~\bibnamefont {Brunner}},\ and\ \bibinfo {author}
  {\bibfnamefont {N.}~\bibnamefont {Gisin}},\ }\href
  {https://doi.org/10.1103/PhysRevLett.116.010403} {\bibfield  {journal}
  {\bibinfo  {journal} {Phys. Rev. Lett.}\ }\textbf {\bibinfo {volume} {116}},\
  \bibinfo {pages} {010403} (\bibinfo {year} {2016})}\BibitemShut {NoStop}%
\bibitem [{\citenamefont {Tavakoli}(2016)}]{Tavakoli2016b}%
  \BibitemOpen
  \bibfield  {author} {\bibinfo {author} {\bibfnamefont {A.}~\bibnamefont
  {Tavakoli}},\ }\href {https://doi.org/10.1103/PhysRevA.93.030101} {\bibfield
  {journal} {\bibinfo  {journal} {Phys. Rev. A}\ }\textbf {\bibinfo {volume}
  {93}},\ \bibinfo {pages} {030101} (\bibinfo {year} {2016})}\BibitemShut
  {NoStop}%
\bibitem [{\citenamefont {Chaves}(2016)}]{Chaves2016}%
  \BibitemOpen
  \bibfield  {author} {\bibinfo {author} {\bibfnamefont {R.}~\bibnamefont
  {Chaves}},\ }\href {https://doi.org/10.1103/PhysRevLett.116.010402}
  {\bibfield  {journal} {\bibinfo  {journal} {Phys. Rev. Lett.}\ }\textbf
  {\bibinfo {volume} {116}},\ \bibinfo {pages} {010402} (\bibinfo {year}
  {2016})}\BibitemShut {NoStop}%
\bibitem [{\citenamefont {Budroni}\ and\ \citenamefont
  {Cabello}(2012)}]{Budroni2012}%
  \BibitemOpen
  \bibfield  {author} {\bibinfo {author} {\bibfnamefont {C.}~\bibnamefont
  {Budroni}}\ and\ \bibinfo {author} {\bibfnamefont {A.}~\bibnamefont
  {Cabello}},\ }\href {https://doi.org/10.1088/1751-8113/45/38/385304}
  {\bibfield  {journal} {\bibinfo  {journal} {J. Phys. A: Math. Theor.}\
  }\textbf {\bibinfo {volume} {45}},\ \bibinfo {pages} {385304} (\bibinfo
  {year} {2012})}\BibitemShut {NoStop}%
\bibitem [{\citenamefont {Fritz}\ and\ \citenamefont
  {Chaves}(2013)}]{Fritz2013}%
  \BibitemOpen
  \bibfield  {author} {\bibinfo {author} {\bibfnamefont {T.}~\bibnamefont
  {Fritz}}\ and\ \bibinfo {author} {\bibfnamefont {R.}~\bibnamefont {Chaves}},\
  }\href {https://doi.org/10.1109/TIT.2012.2222863} {\bibfield  {journal}
  {\bibinfo  {journal} {IEEE Trans. Inf. Theory}\ }\textbf {\bibinfo {volume}
  {59}},\ \bibinfo {pages} {803} (\bibinfo {year} {2013})}\BibitemShut
  {NoStop}%
\bibitem [{\citenamefont {Williams}(1986)}]{Williams1986}%
  \BibitemOpen
  \bibfield  {author} {\bibinfo {author} {\bibfnamefont {H.~P.}\ \bibnamefont
  {Williams}},\ }\href {http://www.jstor.org/stable/2322281} {\bibfield
  {journal} {\bibinfo  {journal} {Am. Math. Mon.}\ }\textbf {\bibinfo {volume}
  {93}},\ \bibinfo {pages} {681} (\bibinfo {year} {1986})}\BibitemShut
  {NoStop}%
\bibitem [{\citenamefont {Tavakoli}(2016)}]{Tavakoli2016}%
  \BibitemOpen
  \bibfield  {author} {\bibinfo {author} {\bibfnamefont {A.}~\bibnamefont
  {Tavakoli}},\ }\href {https://doi.org/10.1088/1751-8113/49/14/145304}
  {\bibfield  {journal} {\bibinfo  {journal} {J. Phys. A: Math. Theor.}\
  }\textbf {\bibinfo {volume} {49}},\ \bibinfo {pages} {145304} (\bibinfo
  {year} {2016})}\BibitemShut {NoStop}%
\bibitem [{\citenamefont {Tavakoli}\ \emph {et~al.}(2017)\citenamefont
  {Tavakoli}, \citenamefont {Renou}, \citenamefont {Gisin},\ and\ \citenamefont
  {Brunner}}]{Tavakoli2017}%
  \BibitemOpen
  \bibfield  {author} {\bibinfo {author} {\bibfnamefont {A.}~\bibnamefont
  {Tavakoli}}, \bibinfo {author} {\bibfnamefont {M.-O.}\ \bibnamefont {Renou}},
  \bibinfo {author} {\bibfnamefont {N.}~\bibnamefont {Gisin}},\ and\ \bibinfo
  {author} {\bibfnamefont {N.}~\bibnamefont {Brunner}},\ }\href
  {https://doi.org/10.1088/1367-2630/aa7673} {\bibfield  {journal} {\bibinfo
  {journal} {New J. Phys.}\ }\textbf {\bibinfo {volume} {19}},\ \bibinfo
  {pages} {073003} (\bibinfo {year} {2017})}\BibitemShut {NoStop}%
\bibitem [{\citenamefont {Augusiak}\ \emph {et~al.}(2014)\citenamefont
  {Augusiak}, \citenamefont {Demianowicz},\ and\ \citenamefont
  {Ac{\'{\i}}n}}]{Augusiak2014}%
  \BibitemOpen
  \bibfield  {author} {\bibinfo {author} {\bibfnamefont {R.}~\bibnamefont
  {Augusiak}}, \bibinfo {author} {\bibfnamefont {M.}~\bibnamefont
  {Demianowicz}},\ and\ \bibinfo {author} {\bibfnamefont {A.}~\bibnamefont
  {Ac{\'{\i}}n}},\ }\href {https://doi.org/10.1088/1751-8113/47/42/424002}
  {\bibfield  {journal} {\bibinfo  {journal} {J. Phys. A: Math. Theor.}\
  }\textbf {\bibinfo {volume} {47}},\ \bibinfo {pages} {424002} (\bibinfo
  {year} {2014})}\BibitemShut {NoStop}%
\bibitem [{\citenamefont {Sen(De)}\ \emph {et~al.}(2005)\citenamefont
  {Sen(De)}, \citenamefont {Sen}, \citenamefont {Brukner}, \citenamefont
  {Bu\ifmmode~\check{z}\else \v{z}\fi{}ek},\ and\ \citenamefont
  {\ifmmode~\dot{Z}\else \.{Z}\fi{}ukowski}}]{Sende2005}%
  \BibitemOpen
  \bibfield  {author} {\bibinfo {author} {\bibfnamefont {A.}~\bibnamefont
  {Sen(De)}}, \bibinfo {author} {\bibfnamefont {U.}~\bibnamefont {Sen}},
  \bibinfo {author} {\bibfnamefont {{\v C}.}~\bibnamefont {Brukner}}, \bibinfo
  {author} {\bibfnamefont {V.}~\bibnamefont {Bu\ifmmode~\check{z}\else
  \v{z}\fi{}ek}},\ and\ \bibinfo {author} {\bibfnamefont {M.}~\bibnamefont
  {\ifmmode~\dot{Z}\else \.{Z}\fi{}ukowski}},\ }\href
  {https://doi.org/10.1103/PhysRevA.72.042310} {\bibfield  {journal} {\bibinfo
  {journal} {Phys. Rev. A}\ }\textbf {\bibinfo {volume} {72}},\ \bibinfo
  {pages} {042310} (\bibinfo {year} {2005})}\BibitemShut {NoStop}%
\bibitem [{\citenamefont {Cavalcanti}\ \emph {et~al.}(2011)\citenamefont
  {Cavalcanti}, \citenamefont {Almeida}, \citenamefont {Scarani},\ and\
  \citenamefont {Ac{\'i}n}}]{Cavalcanti2011}%
  \BibitemOpen
  \bibfield  {author} {\bibinfo {author} {\bibfnamefont {D.}~\bibnamefont
  {Cavalcanti}}, \bibinfo {author} {\bibfnamefont {M.~L.}\ \bibnamefont
  {Almeida}}, \bibinfo {author} {\bibfnamefont {V.}~\bibnamefont {Scarani}},\
  and\ \bibinfo {author} {\bibfnamefont {A.}~\bibnamefont {Ac{\'i}n}},\ }\href
  {https://doi.org/10.1038/ncomms1193} {\bibfield  {journal} {\bibinfo
  {journal} {Nat. Commun.}\ }\textbf {\bibinfo {volume} {2}},\ \bibinfo {pages}
  {184} (\bibinfo {year} {2011})}\BibitemShut {NoStop}%
\bibitem [{\citenamefont {Briegel}\ \emph {et~al.}(1998)\citenamefont
  {Briegel}, \citenamefont {D\"ur}, \citenamefont {Cirac},\ and\ \citenamefont
  {Zoller}}]{Briegel1998}%
  \BibitemOpen
  \bibfield  {author} {\bibinfo {author} {\bibfnamefont {H.-J.}\ \bibnamefont
  {Briegel}}, \bibinfo {author} {\bibfnamefont {W.}~\bibnamefont {D\"ur}},
  \bibinfo {author} {\bibfnamefont {J.~I.}\ \bibnamefont {Cirac}},\ and\
  \bibinfo {author} {\bibfnamefont {P.}~\bibnamefont {Zoller}},\ }\href
  {https://doi.org/10.1103/PhysRevLett.81.5932} {\bibfield  {journal} {\bibinfo
   {journal} {Phys. Rev. Lett.}\ }\textbf {\bibinfo {volume} {81}},\ \bibinfo
  {pages} {5932} (\bibinfo {year} {1998})}\BibitemShut {NoStop}%
\bibitem [{\citenamefont {Sangouard}\ \emph {et~al.}(2011)\citenamefont
  {Sangouard}, \citenamefont {Simon}, \citenamefont {de~Riedmatten},\ and\
  \citenamefont {Gisin}}]{Sangouard2011}%
  \BibitemOpen
  \bibfield  {author} {\bibinfo {author} {\bibfnamefont {N.}~\bibnamefont
  {Sangouard}}, \bibinfo {author} {\bibfnamefont {C.}~\bibnamefont {Simon}},
  \bibinfo {author} {\bibfnamefont {H.}~\bibnamefont {de~Riedmatten}},\ and\
  \bibinfo {author} {\bibfnamefont {N.}~\bibnamefont {Gisin}},\ }\href
  {https://doi.org/10.1103/RevModPhys.83.33} {\bibfield  {journal} {\bibinfo
  {journal} {Rev. Mod. Phys.}\ }\textbf {\bibinfo {volume} {83}},\ \bibinfo
  {pages} {33} (\bibinfo {year} {2011})}\BibitemShut {NoStop}%
\bibitem [{\citenamefont {Mukherjee}\ \emph {et~al.}(2015)\citenamefont
  {Mukherjee}, \citenamefont {Paul},\ and\ \citenamefont
  {Sarkar}}]{Mukherjee2015}%
  \BibitemOpen
  \bibfield  {author} {\bibinfo {author} {\bibfnamefont {K.}~\bibnamefont
  {Mukherjee}}, \bibinfo {author} {\bibfnamefont {B.}~\bibnamefont {Paul}},\
  and\ \bibinfo {author} {\bibfnamefont {D.}~\bibnamefont {Sarkar}},\ }\href
  {https://doi.org/10.1007/s11128-015-0971-7} {\bibfield  {journal} {\bibinfo
  {journal} {Quantum Inf. Process.}\ }\textbf {\bibinfo {volume} {14}},\
  \bibinfo {pages} {2025} (\bibinfo {year} {2015})}\BibitemShut {NoStop}%
\bibitem [{\citenamefont {Goebel}\ \emph {et~al.}(2008)\citenamefont {Goebel},
  \citenamefont {Wagenknecht}, \citenamefont {Zhang}, \citenamefont {Chen},
  \citenamefont {Chen}, \citenamefont {Schmiedmayer},\ and\ \citenamefont
  {Pan}}]{Goebel2008}%
  \BibitemOpen
  \bibfield  {author} {\bibinfo {author} {\bibfnamefont {A.~M.}\ \bibnamefont
  {Goebel}}, \bibinfo {author} {\bibfnamefont {C.}~\bibnamefont {Wagenknecht}},
  \bibinfo {author} {\bibfnamefont {Q.}~\bibnamefont {Zhang}}, \bibinfo
  {author} {\bibfnamefont {Y.-A.}\ \bibnamefont {Chen}}, \bibinfo {author}
  {\bibfnamefont {K.}~\bibnamefont {Chen}}, \bibinfo {author} {\bibfnamefont
  {J.}~\bibnamefont {Schmiedmayer}},\ and\ \bibinfo {author} {\bibfnamefont
  {J.-W.}\ \bibnamefont {Pan}},\ }\href
  {https://doi.org/10.1103/PhysRevLett.101.080403} {\bibfield  {journal}
  {\bibinfo  {journal} {Phys. Rev. Lett.}\ }\textbf {\bibinfo {volume} {101}},\
  \bibinfo {pages} {080403} (\bibinfo {year} {2008})}\BibitemShut {NoStop}%
\bibitem [{\citenamefont {Mukherjee}\ \emph {et~al.}(2017)\citenamefont
  {Mukherjee}, \citenamefont {Paul},\ and\ \citenamefont
  {Sarkar}}]{Mukherjee2017}%
  \BibitemOpen
  \bibfield  {author} {\bibinfo {author} {\bibfnamefont {K.}~\bibnamefont
  {Mukherjee}}, \bibinfo {author} {\bibfnamefont {B.}~\bibnamefont {Paul}},\
  and\ \bibinfo {author} {\bibfnamefont {D.}~\bibnamefont {Sarkar}},\ }\href
  {https://doi.org/10.1103/PhysRevA.96.022103} {\bibfield  {journal} {\bibinfo
  {journal} {Phys. Rev. A}\ }\textbf {\bibinfo {volume} {96}},\ \bibinfo
  {pages} {022103} (\bibinfo {year} {2017})}\BibitemShut {NoStop}%
\bibitem [{\citenamefont {Mukherjee}\ \emph {et~al.}(2020)\citenamefont
  {Mukherjee}, \citenamefont {Paul},\ and\ \citenamefont
  {Roy}}]{Mukherjee2020}%
  \BibitemOpen
  \bibfield  {author} {\bibinfo {author} {\bibfnamefont {K.}~\bibnamefont
  {Mukherjee}}, \bibinfo {author} {\bibfnamefont {B.}~\bibnamefont {Paul}},\
  and\ \bibinfo {author} {\bibfnamefont {A.}~\bibnamefont {Roy}},\ }\href
  {https://doi.org/10.1103/PhysRevA.101.032328} {\bibfield  {journal} {\bibinfo
   {journal} {Phys. Rev. A}\ }\textbf {\bibinfo {volume} {101}},\ \bibinfo
  {pages} {032328} (\bibinfo {year} {2020})}\BibitemShut {NoStop}%
\bibitem [{\citenamefont {Boyd}\ and\ \citenamefont
  {Vandenberghe}(2004)}]{Boyd}%
  \BibitemOpen
  \bibfield  {author} {\bibinfo {author} {\bibfnamefont {S.}~\bibnamefont
  {Boyd}}\ and\ \bibinfo {author} {\bibfnamefont {L.}~\bibnamefont
  {Vandenberghe}},\ }\href {https://doi.org/10.1017/CBO9780511803161} {\emph
  {\bibinfo {title} {{Convex Optimization}}}}\ (\bibinfo  {publisher}
  {Cambridge University Press},\ \bibinfo {year} {2004})\BibitemShut {NoStop}%
\bibitem [{\citenamefont {Vandenberghe}\ and\ \citenamefont
  {Boyd}(1996)}]{Vandenberghe}%
  \BibitemOpen
  \bibfield  {author} {\bibinfo {author} {\bibfnamefont {L.}~\bibnamefont
  {Vandenberghe}}\ and\ \bibinfo {author} {\bibfnamefont {S.}~\bibnamefont
  {Boyd}},\ }\href {https://doi.org/10.1137/1038003} {\bibfield  {journal}
  {\bibinfo  {journal} {SIAM Rev.}\ }\textbf {\bibinfo {volume} {38}},\
  \bibinfo {pages} {49} (\bibinfo {year} {1996})}\BibitemShut {NoStop}%
\bibitem [{\citenamefont {Navascu\'es}\ \emph {et~al.}(2007)\citenamefont
  {Navascu\'es}, \citenamefont {Pironio},\ and\ \citenamefont
  {Ac\'{i}n}}]{navascues2007npa}%
  \BibitemOpen
  \bibfield  {author} {\bibinfo {author} {\bibfnamefont {M.}~\bibnamefont
  {Navascu\'es}}, \bibinfo {author} {\bibfnamefont {S.}~\bibnamefont
  {Pironio}},\ and\ \bibinfo {author} {\bibfnamefont {A.}~\bibnamefont
  {Ac\'{i}n}},\ }\href {https://link.aps.org/doi/10.1103/PhysRevLett.98.010401}
  {\bibfield  {journal} {\bibinfo  {journal} {Phys. Rev. Lett.}\ }\textbf
  {\bibinfo {volume} {98}},\ \bibinfo {pages} {010401} (\bibinfo {year}
  {2007})}\BibitemShut {NoStop}%
\bibitem [{\citenamefont {Navascu\'es}\ \emph {et~al.}(2008)\citenamefont
  {Navascu\'es}, \citenamefont {Pironio},\ and\ \citenamefont
  {Ac{\'i}n}}]{navascues2008npa2}%
  \BibitemOpen
  \bibfield  {author} {\bibinfo {author} {\bibfnamefont {M.}~\bibnamefont
  {Navascu\'es}}, \bibinfo {author} {\bibfnamefont {S.}~\bibnamefont
  {Pironio}},\ and\ \bibinfo {author} {\bibfnamefont {A.}~\bibnamefont
  {Ac{\'i}n}},\ }\href {https://doi.org/10.1088/1367-2630/10/7/073013}
  {\bibfield  {journal} {\bibinfo  {journal} {New J. Phys.}\ }\textbf {\bibinfo
  {volume} {10}},\ \bibinfo {pages} {073013} (\bibinfo {year}
  {2008})}\BibitemShut {NoStop}%
\bibitem [{\citenamefont {Navascu\'es}\ and\ \citenamefont
  {V\'ertesi}(2015)}]{Navascues2015}%
  \BibitemOpen
  \bibfield  {author} {\bibinfo {author} {\bibfnamefont {M.}~\bibnamefont
  {Navascu\'es}}\ and\ \bibinfo {author} {\bibfnamefont {T.}~\bibnamefont
  {V\'ertesi}},\ }\href {https://doi.org/10.1103/PhysRevLett.115.020501}
  {\bibfield  {journal} {\bibinfo  {journal} {Phys. Rev. Lett.}\ }\textbf
  {\bibinfo {volume} {115}},\ \bibinfo {pages} {020501} (\bibinfo {year}
  {2015})}\BibitemShut {NoStop}%
\bibitem [{\citenamefont {Kogias}\ \emph {et~al.}(2015)\citenamefont {Kogias},
  \citenamefont {Skrzypczyk}, \citenamefont {Cavalcanti}, \citenamefont
  {Ac\'{\i}n},\ and\ \citenamefont {Adesso}}]{Kogias2015}%
  \BibitemOpen
  \bibfield  {author} {\bibinfo {author} {\bibfnamefont {I.}~\bibnamefont
  {Kogias}}, \bibinfo {author} {\bibfnamefont {P.}~\bibnamefont {Skrzypczyk}},
  \bibinfo {author} {\bibfnamefont {D.}~\bibnamefont {Cavalcanti}}, \bibinfo
  {author} {\bibfnamefont {A.}~\bibnamefont {Ac\'{\i}n}},\ and\ \bibinfo
  {author} {\bibfnamefont {G.}~\bibnamefont {Adesso}},\ }\href
  {https://doi.org/10.1103/PhysRevLett.115.210401} {\bibfield  {journal}
  {\bibinfo  {journal} {Phys. Rev. Lett.}\ }\textbf {\bibinfo {volume} {115}},\
  \bibinfo {pages} {210401} (\bibinfo {year} {2015})}\BibitemShut {NoStop}%
\bibitem [{\citenamefont {Wang}\ \emph {et~al.}(2019)\citenamefont {Wang},
  \citenamefont {Primaatmaja}, \citenamefont {Lavie}, \citenamefont
  {Varvitsiotis},\ and\ \citenamefont {Lim}}]{Wang2019}%
  \BibitemOpen
  \bibfield  {author} {\bibinfo {author} {\bibfnamefont {Y.}~\bibnamefont
  {Wang}}, \bibinfo {author} {\bibfnamefont {I.~W.}\ \bibnamefont
  {Primaatmaja}}, \bibinfo {author} {\bibfnamefont {E.}~\bibnamefont {Lavie}},
  \bibinfo {author} {\bibfnamefont {A.}~\bibnamefont {Varvitsiotis}},\ and\
  \bibinfo {author} {\bibfnamefont {C.~C.~W.}\ \bibnamefont {Lim}},\ }\href
  {https://doi.org/10.1038/s41534-019-0133-3} {\bibfield  {journal} {\bibinfo
  {journal} {npj Quantum Inf.}\ }\textbf {\bibinfo {volume} {5}},\ \bibinfo
  {pages} {17} (\bibinfo {year} {2019})}\BibitemShut {NoStop}%
\bibitem [{\citenamefont {Bowles}\ \emph {et~al.}(2020)\citenamefont {Bowles},
  \citenamefont {Baccari},\ and\ \citenamefont
  {Salavrakos}}]{bowles2019sequential}%
  \BibitemOpen
  \bibfield  {author} {\bibinfo {author} {\bibfnamefont {J.}~\bibnamefont
  {Bowles}}, \bibinfo {author} {\bibfnamefont {F.}~\bibnamefont {Baccari}},\
  and\ \bibinfo {author} {\bibfnamefont {A.}~\bibnamefont {Salavrakos}},\
  }\href {https://doi.org/10.22331/q-2020-10-19-344} {\bibfield  {journal}
  {\bibinfo  {journal} {{Quantum}}\ }\textbf {\bibinfo {volume} {4}},\ \bibinfo
  {pages} {344} (\bibinfo {year} {2020})}\BibitemShut {NoStop}%
\bibitem [{\citenamefont {Tavakoli}\ \emph {et~al.}(2022)\citenamefont
  {Tavakoli}, \citenamefont {Zambrini~Cruzeiro}, \citenamefont {Woodhead},\
  and\ \citenamefont {Pironio}}]{Tavakoli2020b}%
  \BibitemOpen
  \bibfield  {author} {\bibinfo {author} {\bibfnamefont {A.}~\bibnamefont
  {Tavakoli}}, \bibinfo {author} {\bibfnamefont {E.}~\bibnamefont
  {Zambrini~Cruzeiro}}, \bibinfo {author} {\bibfnamefont {E.}~\bibnamefont
  {Woodhead}},\ and\ \bibinfo {author} {\bibfnamefont {S.}~\bibnamefont
  {Pironio}},\ }\href {https://doi.org/10.22331/q-2022-01-05-620} {\bibfield
  {journal} {\bibinfo  {journal} {{Quantum}}\ }\textbf {\bibinfo {volume}
  {6}},\ \bibinfo {pages} {620} (\bibinfo {year} {2022})}\BibitemShut {NoStop}%
\bibitem [{\citenamefont {Tavakoli}\ \emph {et~al.}(2021)\citenamefont
  {Tavakoli}, \citenamefont {Cruzeiro}, \citenamefont {Uola},\ and\
  \citenamefont {Abbott}}]{Tavakoli2020c}%
  \BibitemOpen
  \bibfield  {author} {\bibinfo {author} {\bibfnamefont {A.}~\bibnamefont
  {Tavakoli}}, \bibinfo {author} {\bibfnamefont {E.~Z.}\ \bibnamefont
  {Cruzeiro}}, \bibinfo {author} {\bibfnamefont {R.}~\bibnamefont {Uola}},\
  and\ \bibinfo {author} {\bibfnamefont {A.~A.}\ \bibnamefont {Abbott}},\
  }\href {https://doi.org/10.1103/PRXQuantum.2.020334} {\bibfield  {journal}
  {\bibinfo  {journal} {PRX Quantum}\ }\textbf {\bibinfo {volume} {2}},\
  \bibinfo {pages} {020334} (\bibinfo {year} {2021})}\BibitemShut {NoStop}%
\bibitem [{\citenamefont {Chaturvedi}\ \emph {et~al.}(2021)\citenamefont
  {Chaturvedi}, \citenamefont {Farkas},\ and\ \citenamefont
  {Wright}}]{Chaturvedi2020}%
  \BibitemOpen
  \bibfield  {author} {\bibinfo {author} {\bibfnamefont {A.}~\bibnamefont
  {Chaturvedi}}, \bibinfo {author} {\bibfnamefont {M.}~\bibnamefont {Farkas}},\
  and\ \bibinfo {author} {\bibfnamefont {V.~J.}\ \bibnamefont {Wright}},\
  }\href {https://doi.org/10.22331/q-2021-06-29-484} {\bibfield  {journal}
  {\bibinfo  {journal} {{Quantum}}\ }\textbf {\bibinfo {volume} {5}},\ \bibinfo
  {pages} {484} (\bibinfo {year} {2021})}\BibitemShut {NoStop}%
\bibitem [{\citenamefont {Jee}\ \emph {et~al.}(2021)\citenamefont {Jee},
  \citenamefont {Sparaciari}, \citenamefont {Fawzi},\ and\ \citenamefont
  {Berta}}]{Jee2020}%
  \BibitemOpen
  \bibfield  {author} {\bibinfo {author} {\bibfnamefont {H.~H.}\ \bibnamefont
  {Jee}}, \bibinfo {author} {\bibfnamefont {C.}~\bibnamefont {Sparaciari}},
  \bibinfo {author} {\bibfnamefont {O.}~\bibnamefont {Fawzi}},\ and\ \bibinfo
  {author} {\bibfnamefont {M.}~\bibnamefont {Berta}},\ }in\ \href
  {https://doi.org/10.4230/LIPIcs.ICALP.2021.82} {\emph {\bibinfo {booktitle}
  {48th International Colloquium on Automata, Languages, and Programming (ICALP
  2021)}}},\ \bibinfo {series} {Leibniz International Proceedings in
  Informatics (LIPIcs)}, Vol.\ \bibinfo {volume} {198},\ \bibinfo {editor}
  {edited by\ \bibinfo {editor} {\bibfnamefont {N.}~\bibnamefont {Bansal}},
  \bibinfo {editor} {\bibfnamefont {E.}~\bibnamefont {Merelli}},\ and\ \bibinfo
  {editor} {\bibfnamefont {J.}~\bibnamefont {Worrell}}}\ (\bibinfo  {publisher}
  {Schloss Dagstuhl -- Leibniz-Zentrum f{\"u}r Informatik},\ \bibinfo {address}
  {Dagstuhl, Germany},\ \bibinfo {year} {2021})\ pp.\ \bibinfo {pages}
  {82:1--82:20}\BibitemShut {NoStop}%
\bibitem [{\citenamefont {Pironio}\ \emph {et~al.}(2010)\citenamefont
  {Pironio}, \citenamefont {Navascu\'es},\ and\ \citenamefont
  {Ac{\'i}n}}]{npo}%
  \BibitemOpen
  \bibfield  {author} {\bibinfo {author} {\bibfnamefont {S.}~\bibnamefont
  {Pironio}}, \bibinfo {author} {\bibfnamefont {M.}~\bibnamefont
  {Navascu\'es}},\ and\ \bibinfo {author} {\bibfnamefont {A.}~\bibnamefont
  {Ac{\'i}n}},\ }\href {https://doi.org/10.1137/090760155} {\bibfield
  {journal} {\bibinfo  {journal} {SIAM J. Optim.}\ }\textbf {\bibinfo {volume}
  {20}},\ \bibinfo {pages} {2157} (\bibinfo {year} {2010})}\BibitemShut
  {NoStop}%
\bibitem [{\citenamefont {Slofstra}(2019)}]{Slofstra}%
  \BibitemOpen
  \bibfield  {author} {\bibinfo {author} {\bibfnamefont {W.}~\bibnamefont
  {Slofstra}},\ }\href {https://doi.org/10.1017/fmp.2018.3} {\bibfield
  {journal} {\bibinfo  {journal} {For. Math., Pi}\ }\textbf {\bibinfo {volume}
  {7}},\ \bibinfo {pages} {e1} (\bibinfo {year} {2019})}\BibitemShut {NoStop}%
\bibitem [{\citenamefont {{Ji}}\ \emph {et~al.}()\citenamefont {{Ji}},
  \citenamefont {{Natarajan}}, \citenamefont {{Vidick}}, \citenamefont
  {{Wright}},\ and\ \citenamefont {{Yuen}}}]{Ji2020Connes}%
  \BibitemOpen
  \bibfield  {author} {\bibinfo {author} {\bibfnamefont {Z.}~\bibnamefont
  {{Ji}}}, \bibinfo {author} {\bibfnamefont {A.}~\bibnamefont {{Natarajan}}},
  \bibinfo {author} {\bibfnamefont {T.}~\bibnamefont {{Vidick}}}, \bibinfo
  {author} {\bibfnamefont {J.}~\bibnamefont {{Wright}}},\ and\ \bibinfo
  {author} {\bibfnamefont {H.}~\bibnamefont {{Yuen}}},\ }\href@noop {}
  {\bibinfo {title} {{MIP*=RE}}},\ \Eprint {https://arxiv.org/abs/2001.04383}
  {arXiv:2001.04383} \BibitemShut {NoStop}%
\bibitem [{\citenamefont {Parrilo}(2003)}]{parrilo}%
  \BibitemOpen
  \bibfield  {author} {\bibinfo {author} {\bibfnamefont {P.~A.}\ \bibnamefont
  {Parrilo}},\ }\href {https://doi.org/10.1007/s10107-003-0387-5} {\bibfield
  {journal} {\bibinfo  {journal} {Math. Program.}\ }\textbf {\bibinfo {volume}
  {96}},\ \bibinfo {pages} {293} (\bibinfo {year} {2003})}\BibitemShut
  {NoStop}%
\bibitem [{\citenamefont {Anjos}\ and\ \citenamefont
  {Lasserre}(2011)}]{LasserreBook}%
  \BibitemOpen
  \bibfield  {author} {\bibinfo {author} {\bibfnamefont {M.~F.}\ \bibnamefont
  {Anjos}}\ and\ \bibinfo {author} {\bibfnamefont {J.~B.}\ \bibnamefont
  {Lasserre}},\ }\href {https://doi.org/10.1007/978-1-4614-0769-0} {\emph
  {\bibinfo {title} {Handbook on Semidefinite, Conic and Polynomial
  Optimization}}}\ (\bibinfo  {publisher} {Springer},\ \bibinfo {year}
  {2011})\BibitemShut {NoStop}%
\bibitem [{\citenamefont {Grant}\ and\ \citenamefont {Boyd}(2014)}]{cvx1}%
  \BibitemOpen
  \bibfield  {author} {\bibinfo {author} {\bibfnamefont {M.}~\bibnamefont
  {Grant}}\ and\ \bibinfo {author} {\bibfnamefont {S.}~\bibnamefont {Boyd}},\
  }\href {http://cvxr.com/cvx} {\emph {\bibinfo {title} {{CVX}: {MATLAB}
  Software for Disciplined Convex Programming, version 2.1}}} (\bibinfo {year}
  {2014})\BibitemShut {NoStop}%
\bibitem [{\citenamefont {Grant}\ and\ \citenamefont {Boyd}(2008)}]{cvx2}%
  \BibitemOpen
  \bibfield  {author} {\bibinfo {author} {\bibfnamefont {M.}~\bibnamefont
  {Grant}}\ and\ \bibinfo {author} {\bibfnamefont {S.}~\bibnamefont {Boyd}},\
  }in\ \href {http://stanford.edu/~boyd/graph_dcp.html} {\emph {\bibinfo
  {booktitle} {Recent Advances in Learning and Control}}},\ \bibinfo {series
  and number} {Lecture Notes in Control and Information Sciences},\ \bibinfo
  {editor} {edited by\ \bibinfo {editor} {\bibfnamefont {V.}~\bibnamefont
  {Blondel}}, \bibinfo {editor} {\bibfnamefont {S.}~\bibnamefont {Boyd}},\ and\
  \bibinfo {editor} {\bibfnamefont {H.}~\bibnamefont {Kimura}}}\ (\bibinfo
  {publisher} {Springer-Verlag Limited},\ \bibinfo {year} {2008})\ pp.\
  \bibinfo {pages} {95--110}\BibitemShut {NoStop}%
\bibitem [{\citenamefont {{L\"{o}fberg}}(2004)}]{yalmip}%
  \BibitemOpen
  \bibfield  {author} {\bibinfo {author} {\bibfnamefont {J.}~\bibnamefont
  {{L\"{o}fberg}}},\ }in\ \href {https://doi.org/10.1109/CACSD.2004.1393890}
  {\emph {\bibinfo {booktitle} {2004 IEEE International Conference on Robotics
  and Automation (IEEE Cat. No.04CH37508)}}}\ (\bibinfo {year} {2004})\ pp.\
  \bibinfo {pages} {284--289}\BibitemShut {NoStop}%
\bibitem [{\citenamefont {Wittek}(2015)}]{ncpol2sdpa}%
  \BibitemOpen
  \bibfield  {author} {\bibinfo {author} {\bibfnamefont {P.}~\bibnamefont
  {Wittek}},\ }\href {https://doi.org/10.1145/2699464} {\bibfield  {journal}
  {\bibinfo  {journal} {ACM Trans. Math. Softw.}\ }\textbf {\bibinfo {volume}
  {41}},\ \bibinfo {pages} {21} (\bibinfo {year} {2015})}\BibitemShut {NoStop}%
\bibitem [{\citenamefont {Sagnol}\ and\ \citenamefont
  {Stahlberg}(2020)}]{picos}%
  \BibitemOpen
  \bibfield  {author} {\bibinfo {author} {\bibfnamefont {G.}~\bibnamefont
  {Sagnol}}\ and\ \bibinfo {author} {\bibfnamefont {M.}~\bibnamefont
  {Stahlberg}},\ }\href {https://picos-api.gitlab.io/picos/} {\emph {\bibinfo
  {title} {{PICOS}: A {Python} interface to conic optimization solvers}}}
  (\bibinfo {year} {2020})\BibitemShut {NoStop}%
\bibitem [{\citenamefont {{MOSEK ApS}}(2019)}]{mosek}%
  \BibitemOpen
  \bibfield  {author} {\bibinfo {author} {\bibnamefont {{MOSEK ApS}}},\ }\href
  {http://docs.mosek.com/9.0/toolbox/index.html} {\emph {\bibinfo {title} {The
  {MOSEK} optimization toolbox for MATLAB manual. Version 9.0.}}} (\bibinfo
  {year} {2019})\BibitemShut {NoStop}%
\bibitem [{\citenamefont {Yamashita}\ \emph {et~al.}(2011)\citenamefont
  {Yamashita}, \citenamefont {Fujisawa}, \citenamefont {Fukuda}, \citenamefont
  {Kobayashi}, \citenamefont {Nakta},\ and\ \citenamefont {Nakata}}]{sdpa}%
  \BibitemOpen
  \bibfield  {author} {\bibinfo {author} {\bibfnamefont {M.}~\bibnamefont
  {Yamashita}}, \bibinfo {author} {\bibfnamefont {K.}~\bibnamefont {Fujisawa}},
  \bibinfo {author} {\bibfnamefont {M.}~\bibnamefont {Fukuda}}, \bibinfo
  {author} {\bibfnamefont {K.}~\bibnamefont {Kobayashi}}, \bibinfo {author}
  {\bibfnamefont {K.}~\bibnamefont {Nakta}},\ and\ \bibinfo {author}
  {\bibfnamefont {M.}~\bibnamefont {Nakata}},\ }in\ \href
  {http://sdpa.sourceforge.net/} {\emph {\bibinfo {booktitle} {Handbook on
  Semidefinite, Cone and Polynomial Optimization: Theory, Algorithms, Software
  and Applications}}},\ \bibinfo {editor} {edited by\ \bibinfo {editor}
  {\bibfnamefont {M.~F.}\ \bibnamefont {Anjos}}\ and\ \bibinfo {editor}
  {\bibfnamefont {J.~B.}\ \bibnamefont {Lasserre}}}\ (\bibinfo  {publisher}
  {Springer-Verlag Limited},\ \bibinfo {year} {2011})\ Chap.~\bibinfo {chapter}
  {24}, pp.\ \bibinfo {pages} {687--714}\BibitemShut {NoStop}%
\bibitem [{\citenamefont {{Gurobi Optimization, LLC}}(2020)}]{gurobi}%
  \BibitemOpen
  \bibfield  {author} {\bibinfo {author} {\bibnamefont {{Gurobi Optimization,
  LLC}}},\ }\href {http://www.gurobi.com} {\emph {\bibinfo {title} {Gurobi
  Optimizer Reference Manual}}} (\bibinfo {year} {2020})\BibitemShut {NoStop}%
\bibitem [{\citenamefont {Pozas-Kerstjens}\ \emph {et~al.}(2019)\citenamefont
  {Pozas-Kerstjens}, \citenamefont {Rabelo}, \citenamefont {Rudnicki},
  \citenamefont {Chaves}, \citenamefont {Cavalcanti}, \citenamefont
  {Navascu\'es},\ and\ \citenamefont {Ac\'{\i}n}}]{Pozas2019}%
  \BibitemOpen
  \bibfield  {author} {\bibinfo {author} {\bibfnamefont {A.}~\bibnamefont
  {Pozas-Kerstjens}}, \bibinfo {author} {\bibfnamefont {R.}~\bibnamefont
  {Rabelo}}, \bibinfo {author} {\bibfnamefont {{\L}.}~\bibnamefont {Rudnicki}},
  \bibinfo {author} {\bibfnamefont {R.}~\bibnamefont {Chaves}}, \bibinfo
  {author} {\bibfnamefont {D.}~\bibnamefont {Cavalcanti}}, \bibinfo {author}
  {\bibfnamefont {M.}~\bibnamefont {Navascu\'es}},\ and\ \bibinfo {author}
  {\bibfnamefont {A.}~\bibnamefont {Ac\'{\i}n}},\ }\href
  {https://doi.org/10.1103/PhysRevLett.123.140503} {\bibfield  {journal}
  {\bibinfo  {journal} {Phys. Rev. Lett.}\ }\textbf {\bibinfo {volume} {123}},\
  \bibinfo {pages} {140503} (\bibinfo {year} {2019})}\BibitemShut {NoStop}%
\bibitem [{\citenamefont {Pozas-Kerstjens}(2019)}]{AlexThesis}%
  \BibitemOpen
  \bibfield  {author} {\bibinfo {author} {\bibfnamefont {A.}~\bibnamefont
  {Pozas-Kerstjens}},\ }\emph {\bibinfo {title} {Quantum information outside
  quantum information}},\ \href {http://hdl.handle.net/10803/667696} {Ph.D.
  thesis} (\bibinfo {year} {2019})\BibitemShut {NoStop}%
\bibitem [{\citenamefont {Pozas-Kerstjens}(2019)}]{scalarextension}%
  \BibitemOpen
  \bibfield  {author} {\bibinfo {author} {\bibfnamefont {A.}~\bibnamefont
  {Pozas-Kerstjens}},\ }\href {https://doi.org/10.5281/zenodo.2648405}
  {\bibfield  {journal} {\bibinfo  {journal} {Zenodo}\ }\textbf {\bibinfo
  {volume} {2648405}} (\bibinfo {year} {2019})}\BibitemShut {NoStop}%
\bibitem [{\citenamefont {Klep}\ \emph {et~al.}(2021)\citenamefont {Klep},
  \citenamefont {Magron},\ and\ \citenamefont
  {Volčič}}]{klep2020optimization}%
  \BibitemOpen
  \bibfield  {author} {\bibinfo {author} {\bibfnamefont {I.}~\bibnamefont
  {Klep}}, \bibinfo {author} {\bibfnamefont {V.}~\bibnamefont {Magron}},\ and\
  \bibinfo {author} {\bibfnamefont {J.}~\bibnamefont {Volčič}},\ }\href
  {https://doi.org/10.1007/s00023-021-01095-4} {\bibfield  {journal} {\bibinfo
  {journal} {Ann. Henri Poincaré}\ }\textbf {\bibinfo {volume} {23}},\
  \bibinfo {pages} {67} (\bibinfo {year} {2021})}\BibitemShut {NoStop}%
\bibitem [{\citenamefont {ver Steeg}\ and\ \citenamefont
  {Galstyan}(2011)}]{Steeg2011}%
  \BibitemOpen
  \bibfield  {author} {\bibinfo {author} {\bibfnamefont {G.}~\bibnamefont {ver
  Steeg}}\ and\ \bibinfo {author} {\bibfnamefont {A.}~\bibnamefont
  {Galstyan}},\ }in\ \href {https://arxiv.org/abs/1106.1636} {\emph {\bibinfo
  {booktitle} {Proceedings of the Twenty-Seventh Conference on Uncertainty in
  Artificial Intelligence}}},\ \bibinfo {series and number} {UAI'11}\ (\bibinfo
   {publisher} {AUAI Press},\ \bibinfo {address} {Arlington, Virginia, USA},\
  \bibinfo {year} {2011})\ p.\ \bibinfo {pages} {717–726}\BibitemShut
  {NoStop}%
\bibitem [{\citenamefont {Eberhard}(1993)}]{eberhard1993detection}%
  \BibitemOpen
  \bibfield  {author} {\bibinfo {author} {\bibfnamefont {P.~H.}\ \bibnamefont
  {Eberhard}},\ }\href {https://doi.org/10.1103/PhysRevA.47.R747} {\bibfield
  {journal} {\bibinfo  {journal} {Phys. Rev. A}\ }\textbf {\bibinfo {volume}
  {47}},\ \bibinfo {pages} {R747} (\bibinfo {year} {1993})}\BibitemShut
  {NoStop}%
\bibitem [{\citenamefont {Massar}\ and\ \citenamefont
  {Pironio}(2003)}]{massar2003detection}%
  \BibitemOpen
  \bibfield  {author} {\bibinfo {author} {\bibfnamefont {S.}~\bibnamefont
  {Massar}}\ and\ \bibinfo {author} {\bibfnamefont {S.}~\bibnamefont
  {Pironio}},\ }\href {https://doi.org/10.1103/PhysRevA.68.062109} {\bibfield
  {journal} {\bibinfo  {journal} {Phys. Rev. A}\ }\textbf {\bibinfo {volume}
  {68}},\ \bibinfo {pages} {062109} (\bibinfo {year} {2003})}\BibitemShut
  {NoStop}%
\bibitem [{\citenamefont {Navascu\'{e}s}\ and\ \citenamefont
  {Wolfe}(2020)}]{navascues2017convergence}%
  \BibitemOpen
  \bibfield  {author} {\bibinfo {author} {\bibfnamefont {M.}~\bibnamefont
  {Navascu\'{e}s}}\ and\ \bibinfo {author} {\bibfnamefont {E.}~\bibnamefont
  {Wolfe}},\ }\href {https://doi.org/10.1515/jci-2018-0008} {\bibfield
  {journal} {\bibinfo  {journal} {J. Causal Inference}\ }\textbf {\bibinfo
  {volume} {8}},\ \bibinfo {pages} {70 } (\bibinfo {year} {2020})}\BibitemShut
  {NoStop}%
\bibitem [{\citenamefont {Scarani}\ \emph {et~al.}(2005)\citenamefont
  {Scarani}, \citenamefont {Iblisdir}, \citenamefont {Gisin},\ and\
  \citenamefont {Ac\'{\i}n}}]{cloning}%
  \BibitemOpen
  \bibfield  {author} {\bibinfo {author} {\bibfnamefont {V.}~\bibnamefont
  {Scarani}}, \bibinfo {author} {\bibfnamefont {S.}~\bibnamefont {Iblisdir}},
  \bibinfo {author} {\bibfnamefont {N.}~\bibnamefont {Gisin}},\ and\ \bibinfo
  {author} {\bibfnamefont {A.}~\bibnamefont {Ac\'{\i}n}},\ }\href
  {https://doi.org/10.1103/RevModPhys.77.1225} {\bibfield  {journal} {\bibinfo
  {journal} {Rev. Mod. Phys.}\ }\textbf {\bibinfo {volume} {77}},\ \bibinfo
  {pages} {1225} (\bibinfo {year} {2005})}\BibitemShut {NoStop}%
\bibitem [{\citenamefont {Ligthart}\ \emph {et~al.}()\citenamefont {Ligthart},
  \citenamefont {Gachechiladze},\ and\ \citenamefont {Gross}}]{ligthart2021}%
  \BibitemOpen
  \bibfield  {author} {\bibinfo {author} {\bibfnamefont {L.~T.}\ \bibnamefont
  {Ligthart}}, \bibinfo {author} {\bibfnamefont {M.}~\bibnamefont
  {Gachechiladze}},\ and\ \bibinfo {author} {\bibfnamefont {D.}~\bibnamefont
  {Gross}},\ }\href@noop {} {\bibinfo {title} {A convergent inflation hierarchy
  for quantum causal structures}},\ \Eprint {https://arxiv.org/abs/2110.14659}
  {arXiv:2110.14659} \BibitemShut {NoStop}%
\bibitem [{\citenamefont {Braunstein}\ and\ \citenamefont
  {Caves}(1988)}]{Braunstein1988}%
  \BibitemOpen
  \bibfield  {author} {\bibinfo {author} {\bibfnamefont {S.~L.}\ \bibnamefont
  {Braunstein}}\ and\ \bibinfo {author} {\bibfnamefont {C.~M.}\ \bibnamefont
  {Caves}},\ }\href {https://doi.org/10.1103/PhysRevLett.61.662} {\bibfield
  {journal} {\bibinfo  {journal} {Phys. Rev. Lett.}\ }\textbf {\bibinfo
  {volume} {61}},\ \bibinfo {pages} {662} (\bibinfo {year} {1988})}\BibitemShut
  {NoStop}%
\bibitem [{\citenamefont {Weilenmann}\ and\ \citenamefont
  {Colbeck}(2017)}]{Weilenmann2017}%
  \BibitemOpen
  \bibfield  {author} {\bibinfo {author} {\bibfnamefont {M.}~\bibnamefont
  {Weilenmann}}\ and\ \bibinfo {author} {\bibfnamefont {R.}~\bibnamefont
  {Colbeck}},\ }\href {https://doi.org/10.1098/rspa.2017.0483} {\bibfield
  {journal} {\bibinfo  {journal} {Proc. R. Soc. A}\ }\textbf {\bibinfo {volume}
  {473}},\ \bibinfo {pages} {20170483} (\bibinfo {year} {2017})}\BibitemShut
  {NoStop}%
\bibitem [{\citenamefont {R\'enyi}(1961)}]{Renyi1961}%
  \BibitemOpen
  \bibfield  {author} {\bibinfo {author} {\bibfnamefont {A.}~\bibnamefont
  {R\'enyi}},\ }in\ \href {https://projecteuclid.org/euclid.bsmsp/1200512181}
  {\emph {\bibinfo {booktitle} {Proceedings of the Fourth Berkeley Symposium on
  Mathematical Statistics and Probability, Volume 1: Contributions to the
  Theory of Statistics}}}\ (\bibinfo  {publisher} {University of California
  Press},\ \bibinfo {address} {Berkeley, Calif.},\ \bibinfo {year} {1961})\
  pp.\ \bibinfo {pages} {547--561}\BibitemShut {NoStop}%
\bibitem [{\citenamefont {Vilasini}\ and\ \citenamefont
  {Colbeck}(2019)}]{vilasini2019tsallis}%
  \BibitemOpen
  \bibfield  {author} {\bibinfo {author} {\bibfnamefont {V.}~\bibnamefont
  {Vilasini}}\ and\ \bibinfo {author} {\bibfnamefont {R.}~\bibnamefont
  {Colbeck}},\ }\href {https://doi.org/10.1103/PhysRevA.100.062108} {\bibfield
  {journal} {\bibinfo  {journal} {Phys. Rev. A}\ }\textbf {\bibinfo {volume}
  {100}},\ \bibinfo {pages} {062108} (\bibinfo {year} {2019})}\BibitemShut
  {NoStop}%
\bibitem [{\citenamefont {Chaves}\ and\ \citenamefont
  {Budroni}(2016)}]{Budroni2016}%
  \BibitemOpen
  \bibfield  {author} {\bibinfo {author} {\bibfnamefont {R.}~\bibnamefont
  {Chaves}}\ and\ \bibinfo {author} {\bibfnamefont {C.}~\bibnamefont
  {Budroni}},\ }\href {https://doi.org/10.1103/PhysRevLett.116.240501}
  {\bibfield  {journal} {\bibinfo  {journal} {Phys. Rev. Lett.}\ }\textbf
  {\bibinfo {volume} {116}},\ \bibinfo {pages} {240501} (\bibinfo {year}
  {2016})}\BibitemShut {NoStop}%
\bibitem [{\citenamefont {Zhang}\ and\ \citenamefont
  {Yeung}(1998)}]{Zhang1998entropy}%
  \BibitemOpen
  \bibfield  {author} {\bibinfo {author} {\bibfnamefont {Z.}~\bibnamefont
  {Zhang}}\ and\ \bibinfo {author} {\bibfnamefont {R.~W.}\ \bibnamefont
  {Yeung}},\ }\href {https://doi.org/10.1109/18.681320} {\bibfield  {journal}
  {\bibinfo  {journal} {IEEE Trans. Inf. Theory}\ }\textbf {\bibinfo {volume}
  {44}},\ \bibinfo {pages} {1440} (\bibinfo {year} {1998})}\BibitemShut
  {NoStop}%
\bibitem [{\citenamefont {Linden}\ and\ \citenamefont
  {Winter}(2005)}]{Linden2005nonshannon}%
  \BibitemOpen
  \bibfield  {author} {\bibinfo {author} {\bibfnamefont {N.}~\bibnamefont
  {Linden}}\ and\ \bibinfo {author} {\bibfnamefont {A.}~\bibnamefont
  {Winter}},\ }\href {https://doi.org/10.1007/s00220-005-1361-2} {\bibfield
  {journal} {\bibinfo  {journal} {Commun. Math. Phys.}\ }\textbf {\bibinfo
  {volume} {259}},\ \bibinfo {pages} {129} (\bibinfo {year}
  {2005})}\BibitemShut {NoStop}%
\bibitem [{\citenamefont {{Dougherty}}\ \emph {et~al.}(2006)\citenamefont
  {{Dougherty}}, \citenamefont {{Freiling}},\ and\ \citenamefont
  {{Zeger}}}]{Dougherty2006nonshannon}%
  \BibitemOpen
  \bibfield  {author} {\bibinfo {author} {\bibfnamefont {R.}~\bibnamefont
  {{Dougherty}}}, \bibinfo {author} {\bibfnamefont {C.}~\bibnamefont
  {{Freiling}}},\ and\ \bibinfo {author} {\bibfnamefont {K.}~\bibnamefont
  {{Zeger}}},\ }in\ \href {https://doi.org/10.1109/ISIT.2006.261840} {\emph
  {\bibinfo {booktitle} {2006 IEEE International Symposium on Information
  Theory}}}\ (\bibinfo {year} {2006})\ pp.\ \bibinfo {pages}
  {233--236}\BibitemShut {NoStop}%
\bibitem [{\citenamefont {Dougherty}\ \emph {et~al.}()\citenamefont
  {Dougherty}, \citenamefont {Freiling},\ and\ \citenamefont
  {Zeger}}]{dougherty2011nonshannon}%
  \BibitemOpen
  \bibfield  {author} {\bibinfo {author} {\bibfnamefont {R.}~\bibnamefont
  {Dougherty}}, \bibinfo {author} {\bibfnamefont {C.}~\bibnamefont
  {Freiling}},\ and\ \bibinfo {author} {\bibfnamefont {K.}~\bibnamefont
  {Zeger}},\ }\href@noop {} {\bibinfo {title} {Non-{Shannon} information
  inequalities in four random variables}},\ \Eprint
  {https://arxiv.org/abs/1104.3602} {arXiv:1104.3602} \BibitemShut {NoStop}%
\bibitem [{\citenamefont {Weilenmann}\ and\ \citenamefont
  {Colbeck}(2018)}]{Weilenmann2018}%
  \BibitemOpen
  \bibfield  {author} {\bibinfo {author} {\bibfnamefont {M.}~\bibnamefont
  {Weilenmann}}\ and\ \bibinfo {author} {\bibfnamefont {R.}~\bibnamefont
  {Colbeck}},\ }\href {https://doi.org/10.22331/q-2018-03-14-57} {\bibfield
  {journal} {\bibinfo  {journal} {{Quantum}}\ }\textbf {\bibinfo {volume}
  {2}},\ \bibinfo {pages} {57} (\bibinfo {year} {2018})}\BibitemShut {NoStop}%
\bibitem [{\citenamefont {Chaves}\ and\ \citenamefont
  {Fritz}(2012)}]{Chaves2012}%
  \BibitemOpen
  \bibfield  {author} {\bibinfo {author} {\bibfnamefont {R.}~\bibnamefont
  {Chaves}}\ and\ \bibinfo {author} {\bibfnamefont {T.}~\bibnamefont {Fritz}},\
  }\href {https://doi.org/10.1103/PhysRevA.85.032113} {\bibfield  {journal}
  {\bibinfo  {journal} {Phys. Rev. A}\ }\textbf {\bibinfo {volume} {85}},\
  \bibinfo {pages} {032113} (\bibinfo {year} {2012})}\BibitemShut {NoStop}%
\bibitem [{\citenamefont {Chaves}\ \emph {et~al.}(2014)\citenamefont {Chaves},
  \citenamefont {Luft}, \citenamefont {Maciel}, \citenamefont {Gross},
  \citenamefont {Janzing},\ and\ \citenamefont
  {Sch{\"o}lkopf}}]{chaves2014inferring}%
  \BibitemOpen
  \bibfield  {author} {\bibinfo {author} {\bibfnamefont {R.}~\bibnamefont
  {Chaves}}, \bibinfo {author} {\bibfnamefont {L.}~\bibnamefont {Luft}},
  \bibinfo {author} {\bibfnamefont {T.~O.}\ \bibnamefont {Maciel}}, \bibinfo
  {author} {\bibfnamefont {D.}~\bibnamefont {Gross}}, \bibinfo {author}
  {\bibfnamefont {D.}~\bibnamefont {Janzing}},\ and\ \bibinfo {author}
  {\bibfnamefont {B.}~\bibnamefont {Sch{\"o}lkopf}},\ }in\ \href
  {https://arxiv.org/abs/1407.2256} {\emph {\bibinfo {booktitle} {Proc. of the
  30th Conference on Uncertainty in Artificial Intelligence}}}\ (\bibinfo
  {organization} {AUAI},\ \bibinfo {year} {2014})\ pp.\ \bibinfo {pages}
  {112--121}\BibitemShut {NoStop}%
\bibitem [{\citenamefont {Chaves}\ \emph {et~al.}(2021)\citenamefont {Chaves},
  \citenamefont {Moreno}, \citenamefont {Polino}, \citenamefont {Poderini},
  \citenamefont {Agresti}, \citenamefont {Suprano}, \citenamefont {Barros},
  \citenamefont {Carvacho}, \citenamefont {Wolfe}, \citenamefont {Canabarro},
  \citenamefont {Spekkens},\ and\ \citenamefont
  {Sciarrino}}]{chaves2021causal}%
  \BibitemOpen
  \bibfield  {author} {\bibinfo {author} {\bibfnamefont {R.}~\bibnamefont
  {Chaves}}, \bibinfo {author} {\bibfnamefont {G.}~\bibnamefont {Moreno}},
  \bibinfo {author} {\bibfnamefont {E.}~\bibnamefont {Polino}}, \bibinfo
  {author} {\bibfnamefont {D.}~\bibnamefont {Poderini}}, \bibinfo {author}
  {\bibfnamefont {I.}~\bibnamefont {Agresti}}, \bibinfo {author} {\bibfnamefont
  {A.}~\bibnamefont {Suprano}}, \bibinfo {author} {\bibfnamefont {M.~R.}\
  \bibnamefont {Barros}}, \bibinfo {author} {\bibfnamefont {G.}~\bibnamefont
  {Carvacho}}, \bibinfo {author} {\bibfnamefont {E.}~\bibnamefont {Wolfe}},
  \bibinfo {author} {\bibfnamefont {A.}~\bibnamefont {Canabarro}}, \bibinfo
  {author} {\bibfnamefont {R.~W.}\ \bibnamefont {Spekkens}},\ and\ \bibinfo
  {author} {\bibfnamefont {F.}~\bibnamefont {Sciarrino}},\ }\href
  {https://doi.org/10.1103/PRXQuantum.2.040323} {\bibfield  {journal} {\bibinfo
   {journal} {PRX Quantum}\ }\textbf {\bibinfo {volume} {2}},\ \bibinfo {pages}
  {040323} (\bibinfo {year} {2021})}\BibitemShut {NoStop}%
\bibitem [{\citenamefont {Chaves}\ \emph {et~al.}(2015)\citenamefont {Chaves},
  \citenamefont {Majenz},\ and\ \citenamefont {Gross}}]{Chaves2015}%
  \BibitemOpen
  \bibfield  {author} {\bibinfo {author} {\bibfnamefont {R.}~\bibnamefont
  {Chaves}}, \bibinfo {author} {\bibfnamefont {C.}~\bibnamefont {Majenz}},\
  and\ \bibinfo {author} {\bibfnamefont {D.}~\bibnamefont {Gross}},\ }\href
  {https://doi.org/10.1038/ncomms6766} {\bibfield  {journal} {\bibinfo
  {journal} {Nat. Commun.}\ }\textbf {\bibinfo {volume} {6}},\ \bibinfo {pages}
  {5766} (\bibinfo {year} {2015})}\BibitemShut {NoStop}%
\bibitem [{\citenamefont {Lieb}\ and\ \citenamefont
  {Ruskai}(1973)}]{Lieb1973weakmonotonicity}%
  \BibitemOpen
  \bibfield  {author} {\bibinfo {author} {\bibfnamefont {E.~H.}\ \bibnamefont
  {Lieb}}\ and\ \bibinfo {author} {\bibfnamefont {M.~B.}\ \bibnamefont
  {Ruskai}},\ }\href {https://doi.org/10.1063/1.1666274} {\bibfield  {journal}
  {\bibinfo  {journal} {J. Math. Phys.}\ }\textbf {\bibinfo {volume} {14}},\
  \bibinfo {pages} {1938} (\bibinfo {year} {1973})}\BibitemShut {NoStop}%
\bibitem [{\citenamefont {Paw\l{}owski}\ \emph {et~al.}(2009)\citenamefont
  {Paw\l{}owski}, \citenamefont {Paterek}, \citenamefont {Kaszlikowski},
  \citenamefont {Scarani}, \citenamefont {Winter},\ and\ \citenamefont
  {\.Zukowski}}]{pawloski2009ic}%
  \BibitemOpen
  \bibfield  {author} {\bibinfo {author} {\bibfnamefont {M.}~\bibnamefont
  {Paw\l{}owski}}, \bibinfo {author} {\bibfnamefont {T.}~\bibnamefont
  {Paterek}}, \bibinfo {author} {\bibfnamefont {D.}~\bibnamefont
  {Kaszlikowski}}, \bibinfo {author} {\bibfnamefont {V.}~\bibnamefont
  {Scarani}}, \bibinfo {author} {\bibfnamefont {A.}~\bibnamefont {Winter}},\
  and\ \bibinfo {author} {\bibfnamefont {M.}~\bibnamefont {\.Zukowski}},\
  }\href {https://doi.org/10.1038/nature08400} {\bibfield  {journal} {\bibinfo
  {journal} {Nature}\ }\textbf {\bibinfo {volume} {461}},\ \bibinfo {pages}
  {1101} (\bibinfo {year} {2009})}\BibitemShut {NoStop}%
\bibitem [{\citenamefont {\AA{}berg}\ \emph {et~al.}(2020)\citenamefont
  {\AA{}berg}, \citenamefont {Nery}, \citenamefont {Duarte},\ and\
  \citenamefont {Chaves}}]{Berg2020SDPCovarTestAllNetwork}%
  \BibitemOpen
  \bibfield  {author} {\bibinfo {author} {\bibfnamefont {J.}~\bibnamefont
  {\AA{}berg}}, \bibinfo {author} {\bibfnamefont {R.}~\bibnamefont {Nery}},
  \bibinfo {author} {\bibfnamefont {C.}~\bibnamefont {Duarte}},\ and\ \bibinfo
  {author} {\bibfnamefont {R.}~\bibnamefont {Chaves}},\ }\href
  {https://doi.org/10.1103/PhysRevLett.125.110505} {\bibfield  {journal}
  {\bibinfo  {journal} {Phys. Rev. Lett.}\ }\textbf {\bibinfo {volume} {125}},\
  \bibinfo {pages} {110505} (\bibinfo {year} {2020})}\BibitemShut {NoStop}%
\bibitem [{\citenamefont {Barrett}(2007)}]{Barrett2007GPT}%
  \BibitemOpen
  \bibfield  {author} {\bibinfo {author} {\bibfnamefont {J.}~\bibnamefont
  {Barrett}},\ }\href {https://link.aps.org/doi/10.1103/PhysRevA.75.032304}
  {\bibfield  {journal} {\bibinfo  {journal} {Phys. Rev. A}\ }\textbf {\bibinfo
  {volume} {75}},\ \bibinfo {pages} {032304} (\bibinfo {year}
  {2007})}\BibitemShut {NoStop}%
\bibitem [{\citenamefont {Chiribella}\ \emph {et~al.}(2010)\citenamefont
  {Chiribella}, \citenamefont {D'Ariano},\ and\ \citenamefont
  {Perinotti}}]{Chiribella2010GPT}%
  \BibitemOpen
  \bibfield  {author} {\bibinfo {author} {\bibfnamefont {G.}~\bibnamefont
  {Chiribella}}, \bibinfo {author} {\bibfnamefont {G.~M.}\ \bibnamefont
  {D'Ariano}},\ and\ \bibinfo {author} {\bibfnamefont {P.}~\bibnamefont
  {Perinotti}},\ }\href {https://doi.org/10.1103/PhysRevA.81.062348} {\bibfield
   {journal} {\bibinfo  {journal} {Phys. Rev. A}\ }\textbf {\bibinfo {volume}
  {81}},\ \bibinfo {pages} {062348} (\bibinfo {year} {2010})}\BibitemShut
  {NoStop}%
\bibitem [{\citenamefont {Bancal}\ and\ \citenamefont
  {Gisin}(2021)}]{Bancal2021}%
  \BibitemOpen
  \bibfield  {author} {\bibinfo {author} {\bibfnamefont {J.-D.}\ \bibnamefont
  {Bancal}}\ and\ \bibinfo {author} {\bibfnamefont {N.}~\bibnamefont {Gisin}},\
  }\href {https://doi.org/10.1103/PhysRevA.104.052212} {\bibfield  {journal}
  {\bibinfo  {journal} {Phys. Rev. A}\ }\textbf {\bibinfo {volume} {104}},\
  \bibinfo {pages} {052212} (\bibinfo {year} {2021})}\BibitemShut {NoStop}%
\bibitem [{\citenamefont {Kela}\ \emph {et~al.}(2020)\citenamefont {Kela},
  \citenamefont {{Von Prillwitz}}, \citenamefont {{\AA{}berg}}, \citenamefont
  {{Chaves}},\ and\ \citenamefont
  {{Gross}}}]{Kela2020SDPCovarTestClassicalNetwork}%
  \BibitemOpen
  \bibfield  {author} {\bibinfo {author} {\bibfnamefont {A.}~\bibnamefont
  {Kela}}, \bibinfo {author} {\bibfnamefont {K.}~\bibnamefont {{Von
  Prillwitz}}}, \bibinfo {author} {\bibfnamefont {J.}~\bibnamefont
  {{\AA{}berg}}}, \bibinfo {author} {\bibfnamefont {R.}~\bibnamefont
  {{Chaves}}},\ and\ \bibinfo {author} {\bibfnamefont {D.}~\bibnamefont
  {{Gross}}},\ }\href {https://doi.org/10.1109/TIT.2019.2935755} {\bibfield
  {journal} {\bibinfo  {journal} {IEEE Trans. Inf. Theory}\ }\textbf {\bibinfo
  {volume} {66}},\ \bibinfo {pages} {339} (\bibinfo {year} {2020})}\BibitemShut
  {NoStop}%
\bibitem [{\citenamefont {Kraft}\ \emph {et~al.}(2021)\citenamefont {Kraft},
  \citenamefont {Spee}, \citenamefont {Yu},\ and\ \citenamefont
  {G\"uhne}}]{Kraft2020SDP}%
  \BibitemOpen
  \bibfield  {author} {\bibinfo {author} {\bibfnamefont {T.}~\bibnamefont
  {Kraft}}, \bibinfo {author} {\bibfnamefont {C.}~\bibnamefont {Spee}},
  \bibinfo {author} {\bibfnamefont {X.-D.}\ \bibnamefont {Yu}},\ and\ \bibinfo
  {author} {\bibfnamefont {O.}~\bibnamefont {G\"uhne}},\ }\href
  {https://doi.org/10.1103/PhysRevA.103.052405} {\bibfield  {journal} {\bibinfo
   {journal} {Phys. Rev. A}\ }\textbf {\bibinfo {volume} {103}},\ \bibinfo
  {pages} {052405} (\bibinfo {year} {2021})}\BibitemShut {NoStop}%
\bibitem [{\citenamefont {Beigi}\ and\ \citenamefont
  {Renou}(2022)}]{Beigi2021}%
  \BibitemOpen
  \bibfield  {author} {\bibinfo {author} {\bibfnamefont {S.}~\bibnamefont
  {Beigi}}\ and\ \bibinfo {author} {\bibfnamefont {M.-O.}\ \bibnamefont
  {Renou}},\ }\href {https://doi.org/10.1109/TIT.2021.3119651} {\bibfield
  {journal} {\bibinfo  {journal} {IEEE Trans. Inf. Theory}\ }\textbf {\bibinfo
  {volume} {68}},\ \bibinfo {pages} {384} (\bibinfo {year} {2022})}\BibitemShut
  {NoStop}%
\bibitem [{\citenamefont {van Dam}\ and\ \citenamefont
  {Hayden}(2003)}]{vDam2003Embezzlement}%
  \BibitemOpen
  \bibfield  {author} {\bibinfo {author} {\bibfnamefont {W.}~\bibnamefont {van
  Dam}}\ and\ \bibinfo {author} {\bibfnamefont {P.}~\bibnamefont {Hayden}},\
  }\href {https://doi.org/10.1103/PhysRevA.67.060302} {\bibfield  {journal}
  {\bibinfo  {journal} {Phys. Rev. A}\ }\textbf {\bibinfo {volume} {67}},\
  \bibinfo {pages} {060302} (\bibinfo {year} {2003})}\BibitemShut {NoStop}%
\bibitem [{\citenamefont {Finner}(1992)}]{Finner1992}%
  \BibitemOpen
  \bibfield  {author} {\bibinfo {author} {\bibfnamefont {H.}~\bibnamefont
  {Finner}},\ }\href {https://doi.org/10.1214/aop/1176989534} {\bibfield
  {journal} {\bibinfo  {journal} {Ann. Probab.}\ }\textbf {\bibinfo {volume}
  {20}},\ \bibinfo {pages} {1893} (\bibinfo {year} {1992})}\BibitemShut
  {NoStop}%
\bibitem [{\citenamefont {Rogers}(1888)}]{Rogers1888}%
  \BibitemOpen
  \bibfield  {author} {\bibinfo {author} {\bibfnamefont {L.~J.}\ \bibnamefont
  {Rogers}},\ }\href
  {https://archive.org/stream/messengermathem01unkngoog#page/n183/mode/1up}
  {\bibfield  {journal} {\bibinfo  {journal} {Messenger Math. New Ser.}\
  }\textbf {\bibinfo {volume} {XVII}},\ \bibinfo {pages} {145} (\bibinfo {year}
  {1888})}\BibitemShut {NoStop}%
\bibitem [{\citenamefont {H{\"o}lder}(1889)}]{Holder1889}%
  \BibitemOpen
  \bibfield  {author} {\bibinfo {author} {\bibfnamefont {O.}~\bibnamefont
  {H{\"o}lder}},\ }\href
  {http://resolver.sub.uni-goettingen.de/purl?GDZPPN00252421X} {\bibfield
  {journal} {\bibinfo  {journal} {Nachr. K{\"o}nigl. Ges. Wiss.
  Georg-Augusts-Univ.}\ }\textbf {\bibinfo {volume} {1889}},\ \bibinfo {pages}
  {38} (\bibinfo {year} {1889})}\BibitemShut {NoStop}%
\bibitem [{\citenamefont {Renou}\ \emph {et~al.}(2019)\citenamefont {Renou},
  \citenamefont {Wang}, \citenamefont {Boreiri}, \citenamefont {Beigi},
  \citenamefont {Gisin},\ and\ \citenamefont {Brunner}}]{RenouFinner2019}%
  \BibitemOpen
  \bibfield  {author} {\bibinfo {author} {\bibfnamefont {M.-O.}\ \bibnamefont
  {Renou}}, \bibinfo {author} {\bibfnamefont {Y.}~\bibnamefont {Wang}},
  \bibinfo {author} {\bibfnamefont {S.}~\bibnamefont {Boreiri}}, \bibinfo
  {author} {\bibfnamefont {S.}~\bibnamefont {Beigi}}, \bibinfo {author}
  {\bibfnamefont {N.}~\bibnamefont {Gisin}},\ and\ \bibinfo {author}
  {\bibfnamefont {N.}~\bibnamefont {Brunner}},\ }\href
  {https://doi.org/10.1103/PhysRevLett.123.070403} {\bibfield  {journal}
  {\bibinfo  {journal} {Phys. Rev. Lett.}\ }\textbf {\bibinfo {volume} {123}},\
  \bibinfo {pages} {070403} (\bibinfo {year} {2019})}\BibitemShut {NoStop}%
\bibitem [{\citenamefont {Shukla}\ \emph {et~al.}()\citenamefont {Shukla},
  \citenamefont {Huang}, \citenamefont {Chakrabarty},\ and\ \citenamefont
  {Wu}}]{Shukla2020}%
  \BibitemOpen
  \bibfield  {author} {\bibinfo {author} {\bibfnamefont {M.~K.}\ \bibnamefont
  {Shukla}}, \bibinfo {author} {\bibfnamefont {M.}~\bibnamefont {Huang}},
  \bibinfo {author} {\bibfnamefont {I.}~\bibnamefont {Chakrabarty}},\ and\
  \bibinfo {author} {\bibfnamefont {J.}~\bibnamefont {Wu}},\ }\href@noop {}
  {\bibinfo {title} {Three party quantum networks created by quantum
  cloning}},\ \Eprint {https://arxiv.org/abs/2011.07554} {arXiv:2011.07554}
  \BibitemShut {NoStop}%
\bibitem [{\citenamefont {Luo}()}]{LuoFinner2021}%
  \BibitemOpen
  \bibfield  {author} {\bibinfo {author} {\bibfnamefont {M.-X.}\ \bibnamefont
  {Luo}},\ }\href@noop {} {\bibinfo {title} {Network configuration theory for
  all networks}},\ \Eprint {https://arxiv.org/abs/2107.05846}
  {arXiv:2107.05846} \BibitemShut {NoStop}%
\bibitem [{\citenamefont {Pan}\ \emph {et~al.}(2012)\citenamefont {Pan},
  \citenamefont {Chen}, \citenamefont {Lu}, \citenamefont {Weinfurter},
  \citenamefont {Zeilinger},\ and\ \citenamefont {\ifmmode~\dot{Z}\else
  \.{Z}\fi{}ukowski}}]{Pan2012}%
  \BibitemOpen
  \bibfield  {author} {\bibinfo {author} {\bibfnamefont {J.-W.}\ \bibnamefont
  {Pan}}, \bibinfo {author} {\bibfnamefont {Z.-B.}\ \bibnamefont {Chen}},
  \bibinfo {author} {\bibfnamefont {C.-Y.}\ \bibnamefont {Lu}}, \bibinfo
  {author} {\bibfnamefont {H.}~\bibnamefont {Weinfurter}}, \bibinfo {author}
  {\bibfnamefont {A.}~\bibnamefont {Zeilinger}},\ and\ \bibinfo {author}
  {\bibfnamefont {M.}~\bibnamefont {\ifmmode~\dot{Z}\else \.{Z}\fi{}ukowski}},\
  }\href {https://doi.org/10.1103/RevModPhys.84.777} {\bibfield  {journal}
  {\bibinfo  {journal} {Rev. Mod. Phys.}\ }\textbf {\bibinfo {volume} {84}},\
  \bibinfo {pages} {777} (\bibinfo {year} {2012})}\BibitemShut {NoStop}%
\bibitem [{\citenamefont {Mattle}\ \emph {et~al.}(1996)\citenamefont {Mattle},
  \citenamefont {Weinfurter}, \citenamefont {Kwiat},\ and\ \citenamefont
  {Zeilinger}}]{Mantle1996}%
  \BibitemOpen
  \bibfield  {author} {\bibinfo {author} {\bibfnamefont {K.}~\bibnamefont
  {Mattle}}, \bibinfo {author} {\bibfnamefont {H.}~\bibnamefont {Weinfurter}},
  \bibinfo {author} {\bibfnamefont {P.~G.}\ \bibnamefont {Kwiat}},\ and\
  \bibinfo {author} {\bibfnamefont {A.}~\bibnamefont {Zeilinger}},\ }\href
  {https://doi.org/10.1103/PhysRevLett.76.4656} {\bibfield  {journal} {\bibinfo
   {journal} {Phys. Rev. Lett.}\ }\textbf {\bibinfo {volume} {76}},\ \bibinfo
  {pages} {4656} (\bibinfo {year} {1996})}\BibitemShut {NoStop}%
\bibitem [{\citenamefont {L\"utkenhaus}\ \emph {et~al.}(1999)\citenamefont
  {L\"utkenhaus}, \citenamefont {Calsamiglia},\ and\ \citenamefont
  {Suominen}}]{Lutkenhaus1999}%
  \BibitemOpen
  \bibfield  {author} {\bibinfo {author} {\bibfnamefont {N.}~\bibnamefont
  {L\"utkenhaus}}, \bibinfo {author} {\bibfnamefont {J.}~\bibnamefont
  {Calsamiglia}},\ and\ \bibinfo {author} {\bibfnamefont {K.-A.}\ \bibnamefont
  {Suominen}},\ }\href {https://doi.org/10.1103/PhysRevA.59.3295} {\bibfield
  {journal} {\bibinfo  {journal} {Phys. Rev. A}\ }\textbf {\bibinfo {volume}
  {59}},\ \bibinfo {pages} {3295} (\bibinfo {year} {1999})}\BibitemShut
  {NoStop}%
\bibitem [{\citenamefont {Calsamiglia}\ and\ \citenamefont
  {L{\"u}tkenhaus}(2001)}]{Calsamiglia2001}%
  \BibitemOpen
  \bibfield  {author} {\bibinfo {author} {\bibfnamefont {J.}~\bibnamefont
  {Calsamiglia}}\ and\ \bibinfo {author} {\bibfnamefont {N.}~\bibnamefont
  {L{\"u}tkenhaus}},\ }\href {https://doi.org/10.1007/s003400000484} {\bibfield
   {journal} {\bibinfo  {journal} {Appl. Phys. B}\ }\textbf {\bibinfo {volume}
  {72}},\ \bibinfo {pages} {67} (\bibinfo {year} {2001})}\BibitemShut {NoStop}%
\bibitem [{\citenamefont {Kwiat}\ and\ \citenamefont
  {Weinfurter}(1998)}]{Kwiat1998}%
  \BibitemOpen
  \bibfield  {author} {\bibinfo {author} {\bibfnamefont {P.~G.}\ \bibnamefont
  {Kwiat}}\ and\ \bibinfo {author} {\bibfnamefont {H.}~\bibnamefont
  {Weinfurter}},\ }\href {https://doi.org/10.1103/PhysRevA.58.R2623} {\bibfield
   {journal} {\bibinfo  {journal} {Phys. Rev. A}\ }\textbf {\bibinfo {volume}
  {58}},\ \bibinfo {pages} {R2623} (\bibinfo {year} {1998})}\BibitemShut
  {NoStop}%
\bibitem [{\citenamefont {Weinfurter}(1994)}]{Weinfurter1994}%
  \BibitemOpen
  \bibfield  {author} {\bibinfo {author} {\bibfnamefont {H.}~\bibnamefont
  {Weinfurter}},\ }\href {https://doi.org/10.1209/0295-5075/25/8/001}
  {\bibfield  {journal} {\bibinfo  {journal} {EPL}\ }\textbf {\bibinfo {volume}
  {25}},\ \bibinfo {pages} {559} (\bibinfo {year} {1994})}\BibitemShut
  {NoStop}%
\bibitem [{\citenamefont {Braunstein}\ and\ \citenamefont
  {Mann}(1995)}]{Braunstein1995}%
  \BibitemOpen
  \bibfield  {author} {\bibinfo {author} {\bibfnamefont {S.~L.}\ \bibnamefont
  {Braunstein}}\ and\ \bibinfo {author} {\bibfnamefont {A.}~\bibnamefont
  {Mann}},\ }\href {https://doi.org/10.1103/PhysRevA.51.R1727} {\bibfield
  {journal} {\bibinfo  {journal} {Phys. Rev. A}\ }\textbf {\bibinfo {volume}
  {51}},\ \bibinfo {pages} {R1727} (\bibinfo {year} {1995})}\BibitemShut
  {NoStop}%
\bibitem [{\citenamefont {Pan}\ \emph {et~al.}(2001)\citenamefont {Pan},
  \citenamefont {Daniell}, \citenamefont {Gasparoni}, \citenamefont {Weihs},\
  and\ \citenamefont {Zeilinger}}]{Pan2001}%
  \BibitemOpen
  \bibfield  {author} {\bibinfo {author} {\bibfnamefont {J.-W.}\ \bibnamefont
  {Pan}}, \bibinfo {author} {\bibfnamefont {M.}~\bibnamefont {Daniell}},
  \bibinfo {author} {\bibfnamefont {S.}~\bibnamefont {Gasparoni}}, \bibinfo
  {author} {\bibfnamefont {G.}~\bibnamefont {Weihs}},\ and\ \bibinfo {author}
  {\bibfnamefont {A.}~\bibnamefont {Zeilinger}},\ }\href
  {https://doi.org/10.1103/PhysRevLett.86.4435} {\bibfield  {journal} {\bibinfo
   {journal} {Phys. Rev. Lett.}\ }\textbf {\bibinfo {volume} {86}},\ \bibinfo
  {pages} {4435} (\bibinfo {year} {2001})}\BibitemShut {NoStop}%
\bibitem [{\citenamefont {de~Riedmatten}\ \emph {et~al.}(2005)\citenamefont
  {de~Riedmatten}, \citenamefont {Marcikic}, \citenamefont {van Houwelingen},
  \citenamefont {Tittel}, \citenamefont {Zbinden},\ and\ \citenamefont
  {Gisin}}]{Riedmatten2005}%
  \BibitemOpen
  \bibfield  {author} {\bibinfo {author} {\bibfnamefont {H.}~\bibnamefont
  {de~Riedmatten}}, \bibinfo {author} {\bibfnamefont {I.}~\bibnamefont
  {Marcikic}}, \bibinfo {author} {\bibfnamefont {J.~A.~W.}\ \bibnamefont {van
  Houwelingen}}, \bibinfo {author} {\bibfnamefont {W.}~\bibnamefont {Tittel}},
  \bibinfo {author} {\bibfnamefont {H.}~\bibnamefont {Zbinden}},\ and\ \bibinfo
  {author} {\bibfnamefont {N.}~\bibnamefont {Gisin}},\ }\href
  {https://doi.org/10.1103/PhysRevA.71.050302} {\bibfield  {journal} {\bibinfo
  {journal} {Phys. Rev. A}\ }\textbf {\bibinfo {volume} {71}},\ \bibinfo
  {pages} {050302} (\bibinfo {year} {2005})}\BibitemShut {NoStop}%
\bibitem [{\citenamefont {Sciarrino}\ \emph {et~al.}(2002)\citenamefont
  {Sciarrino}, \citenamefont {Lombardi}, \citenamefont {Milani},\ and\
  \citenamefont {De~Martini}}]{Sciarrino2002}%
  \BibitemOpen
  \bibfield  {author} {\bibinfo {author} {\bibfnamefont {F.}~\bibnamefont
  {Sciarrino}}, \bibinfo {author} {\bibfnamefont {E.}~\bibnamefont {Lombardi}},
  \bibinfo {author} {\bibfnamefont {G.}~\bibnamefont {Milani}},\ and\ \bibinfo
  {author} {\bibfnamefont {F.}~\bibnamefont {De~Martini}},\ }\href
  {https://doi.org/10.1103/PhysRevA.66.024309} {\bibfield  {journal} {\bibinfo
  {journal} {Phys. Rev. A}\ }\textbf {\bibinfo {volume} {66}},\ \bibinfo
  {pages} {024309} (\bibinfo {year} {2002})}\BibitemShut {NoStop}%
\bibitem [{\citenamefont {Jia}\ \emph {et~al.}(2004)\citenamefont {Jia},
  \citenamefont {Su}, \citenamefont {Pan}, \citenamefont {Gao}, \citenamefont
  {Xie},\ and\ \citenamefont {Peng}}]{Jia2004}%
  \BibitemOpen
  \bibfield  {author} {\bibinfo {author} {\bibfnamefont {X.}~\bibnamefont
  {Jia}}, \bibinfo {author} {\bibfnamefont {X.}~\bibnamefont {Su}}, \bibinfo
  {author} {\bibfnamefont {Q.}~\bibnamefont {Pan}}, \bibinfo {author}
  {\bibfnamefont {J.}~\bibnamefont {Gao}}, \bibinfo {author} {\bibfnamefont
  {C.}~\bibnamefont {Xie}},\ and\ \bibinfo {author} {\bibfnamefont
  {K.}~\bibnamefont {Peng}},\ }\href
  {https://doi.org/10.1103/PhysRevLett.93.250503} {\bibfield  {journal}
  {\bibinfo  {journal} {Phys. Rev. Lett.}\ }\textbf {\bibinfo {volume} {93}},\
  \bibinfo {pages} {250503} (\bibinfo {year} {2004})}\BibitemShut {NoStop}%
\bibitem [{\citenamefont {Takei}\ \emph {et~al.}(2005)\citenamefont {Takei},
  \citenamefont {Yonezawa}, \citenamefont {Aoki},\ and\ \citenamefont
  {Furusawa}}]{Takei2005}%
  \BibitemOpen
  \bibfield  {author} {\bibinfo {author} {\bibfnamefont {N.}~\bibnamefont
  {Takei}}, \bibinfo {author} {\bibfnamefont {H.}~\bibnamefont {Yonezawa}},
  \bibinfo {author} {\bibfnamefont {T.}~\bibnamefont {Aoki}},\ and\ \bibinfo
  {author} {\bibfnamefont {A.}~\bibnamefont {Furusawa}},\ }\href
  {https://doi.org/10.1103/PhysRevLett.94.220502} {\bibfield  {journal}
  {\bibinfo  {journal} {Phys. Rev. Lett.}\ }\textbf {\bibinfo {volume} {94}},\
  \bibinfo {pages} {220502} (\bibinfo {year} {2005})}\BibitemShut {NoStop}%
\bibitem [{\citenamefont {Yang}\ \emph {et~al.}(2006)\citenamefont {Yang},
  \citenamefont {Zhang}, \citenamefont {Chen}, \citenamefont {Lu},
  \citenamefont {Yin}, \citenamefont {Pan}, \citenamefont {Wei}, \citenamefont
  {Tian},\ and\ \citenamefont {Zhang}}]{Yang2006}%
  \BibitemOpen
  \bibfield  {author} {\bibinfo {author} {\bibfnamefont {T.}~\bibnamefont
  {Yang}}, \bibinfo {author} {\bibfnamefont {Q.}~\bibnamefont {Zhang}},
  \bibinfo {author} {\bibfnamefont {T.-Y.}\ \bibnamefont {Chen}}, \bibinfo
  {author} {\bibfnamefont {S.}~\bibnamefont {Lu}}, \bibinfo {author}
  {\bibfnamefont {J.}~\bibnamefont {Yin}}, \bibinfo {author} {\bibfnamefont
  {J.-W.}\ \bibnamefont {Pan}}, \bibinfo {author} {\bibfnamefont {Z.-Y.}\
  \bibnamefont {Wei}}, \bibinfo {author} {\bibfnamefont {J.-R.}\ \bibnamefont
  {Tian}},\ and\ \bibinfo {author} {\bibfnamefont {J.}~\bibnamefont {Zhang}},\
  }\href {https://doi.org/10.1103/PhysRevLett.96.110501} {\bibfield  {journal}
  {\bibinfo  {journal} {Phys. Rev. Lett.}\ }\textbf {\bibinfo {volume} {96}},\
  \bibinfo {pages} {110501} (\bibinfo {year} {2006})}\BibitemShut {NoStop}%
\bibitem [{\citenamefont {Kaltenbaek}\ \emph {et~al.}(2006)\citenamefont
  {Kaltenbaek}, \citenamefont {Blauensteiner}, \citenamefont
  {\ifmmode~\dot{Z}\else \.{Z}\fi{}ukowski}, \citenamefont {Aspelmeyer},\ and\
  \citenamefont {Zeilinger}}]{Kaltenbaek2006}%
  \BibitemOpen
  \bibfield  {author} {\bibinfo {author} {\bibfnamefont {R.}~\bibnamefont
  {Kaltenbaek}}, \bibinfo {author} {\bibfnamefont {B.}~\bibnamefont
  {Blauensteiner}}, \bibinfo {author} {\bibfnamefont {M.}~\bibnamefont
  {\ifmmode~\dot{Z}\else \.{Z}\fi{}ukowski}}, \bibinfo {author} {\bibfnamefont
  {M.}~\bibnamefont {Aspelmeyer}},\ and\ \bibinfo {author} {\bibfnamefont
  {A.}~\bibnamefont {Zeilinger}},\ }\href
  {https://doi.org/10.1103/PhysRevLett.96.240502} {\bibfield  {journal}
  {\bibinfo  {journal} {Phys. Rev. Lett.}\ }\textbf {\bibinfo {volume} {96}},\
  \bibinfo {pages} {240502} (\bibinfo {year} {2006})}\BibitemShut {NoStop}%
\bibitem [{\citenamefont {Halder}\ \emph {et~al.}(2007)\citenamefont {Halder},
  \citenamefont {Beveratos}, \citenamefont {Gisin}, \citenamefont {Scarani},
  \citenamefont {Simon},\ and\ \citenamefont {Zbinden}}]{Halder2007}%
  \BibitemOpen
  \bibfield  {author} {\bibinfo {author} {\bibfnamefont {M.}~\bibnamefont
  {Halder}}, \bibinfo {author} {\bibfnamefont {A.}~\bibnamefont {Beveratos}},
  \bibinfo {author} {\bibfnamefont {N.}~\bibnamefont {Gisin}}, \bibinfo
  {author} {\bibfnamefont {V.}~\bibnamefont {Scarani}}, \bibinfo {author}
  {\bibfnamefont {C.}~\bibnamefont {Simon}},\ and\ \bibinfo {author}
  {\bibfnamefont {H.}~\bibnamefont {Zbinden}},\ }\href
  {https://doi.org/10.1038/nphys700} {\bibfield  {journal} {\bibinfo  {journal}
  {Nat. Phys.}\ }\textbf {\bibinfo {volume} {3}},\ \bibinfo {pages} {692}
  (\bibinfo {year} {2007})}\BibitemShut {NoStop}%
\bibitem [{\citenamefont {Kaltenbaek}\ \emph {et~al.}(2009)\citenamefont
  {Kaltenbaek}, \citenamefont {Prevedel}, \citenamefont {Aspelmeyer},\ and\
  \citenamefont {Zeilinger}}]{Kaltenbaek2009}%
  \BibitemOpen
  \bibfield  {author} {\bibinfo {author} {\bibfnamefont {R.}~\bibnamefont
  {Kaltenbaek}}, \bibinfo {author} {\bibfnamefont {R.}~\bibnamefont
  {Prevedel}}, \bibinfo {author} {\bibfnamefont {M.}~\bibnamefont
  {Aspelmeyer}},\ and\ \bibinfo {author} {\bibfnamefont {A.}~\bibnamefont
  {Zeilinger}},\ }\href {https://doi.org/10.1103/PhysRevA.79.040302} {\bibfield
   {journal} {\bibinfo  {journal} {Phys. Rev. A}\ }\textbf {\bibinfo {volume}
  {79}},\ \bibinfo {pages} {040302} (\bibinfo {year} {2009})}\BibitemShut
  {NoStop}%
\bibitem [{\citenamefont {Schmid}\ \emph {et~al.}(2009)\citenamefont {Schmid},
  \citenamefont {Kiesel}, \citenamefont {Weber}, \citenamefont {Ursin},
  \citenamefont {Zeilinger},\ and\ \citenamefont {Weinfurter}}]{Schmid2009}%
  \BibitemOpen
  \bibfield  {author} {\bibinfo {author} {\bibfnamefont {C.}~\bibnamefont
  {Schmid}}, \bibinfo {author} {\bibfnamefont {N.}~\bibnamefont {Kiesel}},
  \bibinfo {author} {\bibfnamefont {U.~K.}\ \bibnamefont {Weber}}, \bibinfo
  {author} {\bibfnamefont {R.}~\bibnamefont {Ursin}}, \bibinfo {author}
  {\bibfnamefont {A.}~\bibnamefont {Zeilinger}},\ and\ \bibinfo {author}
  {\bibfnamefont {H.}~\bibnamefont {Weinfurter}},\ }\href
  {https://doi.org/10.1088/0031-8949/11/3/033008} {\bibfield  {journal}
  {\bibinfo  {journal} {New J. Phys.}\ }\textbf {\bibinfo {volume} {11}},\
  \bibinfo {pages} {033008} (\bibinfo {year} {2009})}\BibitemShut {NoStop}%
\bibitem [{\citenamefont {Takeda}\ \emph {et~al.}(2015)\citenamefont {Takeda},
  \citenamefont {Fuwa}, \citenamefont {van Loock},\ and\ \citenamefont
  {Furusawa}}]{Takeda2015}%
  \BibitemOpen
  \bibfield  {author} {\bibinfo {author} {\bibfnamefont {S.}~\bibnamefont
  {Takeda}}, \bibinfo {author} {\bibfnamefont {M.}~\bibnamefont {Fuwa}},
  \bibinfo {author} {\bibfnamefont {P.}~\bibnamefont {van Loock}},\ and\
  \bibinfo {author} {\bibfnamefont {A.}~\bibnamefont {Furusawa}},\ }\href
  {https://doi.org/10.1103/PhysRevLett.114.100501} {\bibfield  {journal}
  {\bibinfo  {journal} {Phys. Rev. Lett.}\ }\textbf {\bibinfo {volume} {114}},\
  \bibinfo {pages} {100501} (\bibinfo {year} {2015})}\BibitemShut {NoStop}%
\bibitem [{\citenamefont {Guccione}\ \emph {et~al.}(2020)\citenamefont
  {Guccione}, \citenamefont {Darras}, \citenamefont {Le~Jeannic}, \citenamefont
  {Verma}, \citenamefont {Nam}, \citenamefont {Cavaill{\`e}s},\ and\
  \citenamefont {Laurat}}]{Guccione2020}%
  \BibitemOpen
  \bibfield  {author} {\bibinfo {author} {\bibfnamefont {G.}~\bibnamefont
  {Guccione}}, \bibinfo {author} {\bibfnamefont {T.}~\bibnamefont {Darras}},
  \bibinfo {author} {\bibfnamefont {H.}~\bibnamefont {Le~Jeannic}}, \bibinfo
  {author} {\bibfnamefont {V.~B.}\ \bibnamefont {Verma}}, \bibinfo {author}
  {\bibfnamefont {S.~W.}\ \bibnamefont {Nam}}, \bibinfo {author} {\bibfnamefont
  {A.}~\bibnamefont {Cavaill{\`e}s}},\ and\ \bibinfo {author} {\bibfnamefont
  {J.}~\bibnamefont {Laurat}},\ }\href
  {https://advances.sciencemag.org/content/6/22/eaba4508} {\bibfield  {journal}
  {\bibinfo  {journal} {Sci. Adv.}\ }\textbf {\bibinfo {volume} {6}},\ \bibinfo
  {pages} {eaba4508} (\bibinfo {year} {2020})}\BibitemShut {NoStop}%
\bibitem [{\citenamefont {Lu}\ \emph {et~al.}(2009)\citenamefont {Lu},
  \citenamefont {Yang},\ and\ \citenamefont {Pan}}]{Lu2009}%
  \BibitemOpen
  \bibfield  {author} {\bibinfo {author} {\bibfnamefont {C.-Y.}\ \bibnamefont
  {Lu}}, \bibinfo {author} {\bibfnamefont {T.}~\bibnamefont {Yang}},\ and\
  \bibinfo {author} {\bibfnamefont {J.-W.}\ \bibnamefont {Pan}},\ }\href
  {https://doi.org/10.1103/PhysRevLett.103.020501} {\bibfield  {journal}
  {\bibinfo  {journal} {Phys. Rev. Lett.}\ }\textbf {\bibinfo {volume} {103}},\
  \bibinfo {pages} {020501} (\bibinfo {year} {2009})}\BibitemShut {NoStop}%
\bibitem [{\citenamefont {Su}\ \emph {et~al.}(2016)\citenamefont {Su},
  \citenamefont {Tian}, \citenamefont {Deng}, \citenamefont {Li}, \citenamefont
  {Xie},\ and\ \citenamefont {Peng}}]{Su2016}%
  \BibitemOpen
  \bibfield  {author} {\bibinfo {author} {\bibfnamefont {X.}~\bibnamefont
  {Su}}, \bibinfo {author} {\bibfnamefont {C.}~\bibnamefont {Tian}}, \bibinfo
  {author} {\bibfnamefont {X.}~\bibnamefont {Deng}}, \bibinfo {author}
  {\bibfnamefont {Q.}~\bibnamefont {Li}}, \bibinfo {author} {\bibfnamefont
  {C.}~\bibnamefont {Xie}},\ and\ \bibinfo {author} {\bibfnamefont
  {K.}~\bibnamefont {Peng}},\ }\href
  {https://doi.org/10.1103/PhysRevLett.117.240503} {\bibfield  {journal}
  {\bibinfo  {journal} {Phys. Rev. Lett.}\ }\textbf {\bibinfo {volume} {117}},\
  \bibinfo {pages} {240503} (\bibinfo {year} {2016})}\BibitemShut {NoStop}%
\bibitem [{\citenamefont {Blinov}\ \emph {et~al.}(2004)\citenamefont {Blinov},
  \citenamefont {Moehring}, \citenamefont {Duan},\ and\ \citenamefont
  {Monroe}}]{Blinov2004}%
  \BibitemOpen
  \bibfield  {author} {\bibinfo {author} {\bibfnamefont {B.~B.}\ \bibnamefont
  {Blinov}}, \bibinfo {author} {\bibfnamefont {D.~L.}\ \bibnamefont
  {Moehring}}, \bibinfo {author} {\bibfnamefont {L.}~\bibnamefont {Duan}},\
  and\ \bibinfo {author} {\bibfnamefont {C.}~\bibnamefont {Monroe}},\ }\href
  {https://doi.org/10.1038/nature02377} {\bibfield  {journal} {\bibinfo
  {journal} {Nature}\ }\textbf {\bibinfo {volume} {428}},\ \bibinfo {pages}
  {153} (\bibinfo {year} {2004})}\BibitemShut {NoStop}%
\bibitem [{\citenamefont {Moehring}\ \emph {et~al.}(2004)\citenamefont
  {Moehring}, \citenamefont {Madsen}, \citenamefont {Blinov},\ and\
  \citenamefont {Monroe}}]{Moehring2004}%
  \BibitemOpen
  \bibfield  {author} {\bibinfo {author} {\bibfnamefont {D.~L.}\ \bibnamefont
  {Moehring}}, \bibinfo {author} {\bibfnamefont {M.~J.}\ \bibnamefont
  {Madsen}}, \bibinfo {author} {\bibfnamefont {B.~B.}\ \bibnamefont {Blinov}},\
  and\ \bibinfo {author} {\bibfnamefont {C.}~\bibnamefont {Monroe}},\ }\href
  {https://doi.org/10.1103/PhysRevLett.93.090410} {\bibfield  {journal}
  {\bibinfo  {journal} {Phys. Rev. Lett.}\ }\textbf {\bibinfo {volume} {93}},\
  \bibinfo {pages} {090410} (\bibinfo {year} {2004})}\BibitemShut {NoStop}%
\bibitem [{\citenamefont {Volz}\ \emph {et~al.}(2006)\citenamefont {Volz},
  \citenamefont {Weber}, \citenamefont {Schlenk}, \citenamefont {Rosenfeld},
  \citenamefont {Vrana}, \citenamefont {Saucke}, \citenamefont {Kurtsiefer},\
  and\ \citenamefont {Weinfurter}}]{Volz2006}%
  \BibitemOpen
  \bibfield  {author} {\bibinfo {author} {\bibfnamefont {J.}~\bibnamefont
  {Volz}}, \bibinfo {author} {\bibfnamefont {M.}~\bibnamefont {Weber}},
  \bibinfo {author} {\bibfnamefont {D.}~\bibnamefont {Schlenk}}, \bibinfo
  {author} {\bibfnamefont {W.}~\bibnamefont {Rosenfeld}}, \bibinfo {author}
  {\bibfnamefont {J.}~\bibnamefont {Vrana}}, \bibinfo {author} {\bibfnamefont
  {K.}~\bibnamefont {Saucke}}, \bibinfo {author} {\bibfnamefont
  {C.}~\bibnamefont {Kurtsiefer}},\ and\ \bibinfo {author} {\bibfnamefont
  {H.}~\bibnamefont {Weinfurter}},\ }\href
  {https://doi.org/10.1103/PhysRevLett.96.030404} {\bibfield  {journal}
  {\bibinfo  {journal} {Phys. Rev. Lett.}\ }\textbf {\bibinfo {volume} {96}},\
  \bibinfo {pages} {030404} (\bibinfo {year} {2006})}\BibitemShut {NoStop}%
\bibitem [{\citenamefont {Moehring}\ \emph {et~al.}(2007)\citenamefont
  {Moehring}, \citenamefont {Maunz}, \citenamefont {Olmschenk}, \citenamefont
  {Younge}, \citenamefont {Matsukevich}, \citenamefont {Duan},\ and\
  \citenamefont {Monroe}}]{Moehring2007}%
  \BibitemOpen
  \bibfield  {author} {\bibinfo {author} {\bibfnamefont {D.~L.}\ \bibnamefont
  {Moehring}}, \bibinfo {author} {\bibfnamefont {P.}~\bibnamefont {Maunz}},
  \bibinfo {author} {\bibfnamefont {S.}~\bibnamefont {Olmschenk}}, \bibinfo
  {author} {\bibfnamefont {K.~C.}\ \bibnamefont {Younge}}, \bibinfo {author}
  {\bibfnamefont {D.~N.}\ \bibnamefont {Matsukevich}}, \bibinfo {author}
  {\bibfnamefont {L.}~\bibnamefont {Duan}},\ and\ \bibinfo {author}
  {\bibfnamefont {C.}~\bibnamefont {Monroe}},\ }\href
  {https://doi.org/10.1038/nature06118} {\bibfield  {journal} {\bibinfo
  {journal} {Nature}\ }\textbf {\bibinfo {volume} {449}},\ \bibinfo {pages}
  {68} (\bibinfo {year} {2007})}\BibitemShut {NoStop}%
\bibitem [{\citenamefont {Yuan}\ \emph {et~al.}(2008)\citenamefont {Yuan},
  \citenamefont {Chen}, \citenamefont {Zhao}, \citenamefont {Chen},
  \citenamefont {Schmiedmayer},\ and\ \citenamefont {Pan}}]{Yuan2008}%
  \BibitemOpen
  \bibfield  {author} {\bibinfo {author} {\bibfnamefont {Z.-S.}\ \bibnamefont
  {Yuan}}, \bibinfo {author} {\bibfnamefont {Y.-A.}\ \bibnamefont {Chen}},
  \bibinfo {author} {\bibfnamefont {B.}~\bibnamefont {Zhao}}, \bibinfo {author}
  {\bibfnamefont {S.}~\bibnamefont {Chen}}, \bibinfo {author} {\bibfnamefont
  {J.}~\bibnamefont {Schmiedmayer}},\ and\ \bibinfo {author} {\bibfnamefont
  {J.-W.}\ \bibnamefont {Pan}},\ }\href {https://doi.org/10.1038/nature07241}
  {\bibfield  {journal} {\bibinfo  {journal} {Nature}\ }\textbf {\bibinfo
  {volume} {454}},\ \bibinfo {pages} {1098} (\bibinfo {year}
  {2008})}\BibitemShut {NoStop}%
\bibitem [{\citenamefont {Matsukevich}\ \emph {et~al.}(2008)\citenamefont
  {Matsukevich}, \citenamefont {Maunz}, \citenamefont {Moehring}, \citenamefont
  {Olmschenk},\ and\ \citenamefont {Monroe}}]{Matsukevich2008}%
  \BibitemOpen
  \bibfield  {author} {\bibinfo {author} {\bibfnamefont {D.~N.}\ \bibnamefont
  {Matsukevich}}, \bibinfo {author} {\bibfnamefont {P.}~\bibnamefont {Maunz}},
  \bibinfo {author} {\bibfnamefont {D.~L.}\ \bibnamefont {Moehring}}, \bibinfo
  {author} {\bibfnamefont {S.}~\bibnamefont {Olmschenk}},\ and\ \bibinfo
  {author} {\bibfnamefont {C.}~\bibnamefont {Monroe}},\ }\href
  {https://doi.org/10.1103/PhysRevLett.100.150404} {\bibfield  {journal}
  {\bibinfo  {journal} {Phys. Rev. Lett.}\ }\textbf {\bibinfo {volume} {100}},\
  \bibinfo {pages} {150404} (\bibinfo {year} {2008})}\BibitemShut {NoStop}%
\bibitem [{\citenamefont {Hofmann}\ \emph {et~al.}(2012)\citenamefont
  {Hofmann}, \citenamefont {Krug}, \citenamefont {Ortegel}, \citenamefont
  {G{\'e}rard}, \citenamefont {Weber}, \citenamefont {Rosenfeld},\ and\
  \citenamefont {Weinfurter}}]{Hofmann2012}%
  \BibitemOpen
  \bibfield  {author} {\bibinfo {author} {\bibfnamefont {J.}~\bibnamefont
  {Hofmann}}, \bibinfo {author} {\bibfnamefont {M.}~\bibnamefont {Krug}},
  \bibinfo {author} {\bibfnamefont {N.}~\bibnamefont {Ortegel}}, \bibinfo
  {author} {\bibfnamefont {L.}~\bibnamefont {G{\'e}rard}}, \bibinfo {author}
  {\bibfnamefont {M.}~\bibnamefont {Weber}}, \bibinfo {author} {\bibfnamefont
  {W.}~\bibnamefont {Rosenfeld}},\ and\ \bibinfo {author} {\bibfnamefont
  {H.}~\bibnamefont {Weinfurter}},\ }\href
  {https://doi.org/10.1126/science.1221856} {\bibfield  {journal} {\bibinfo
  {journal} {Science}\ }\textbf {\bibinfo {volume} {337}},\ \bibinfo {pages}
  {72} (\bibinfo {year} {2012})}\BibitemShut {NoStop}%
\bibitem [{\citenamefont {Riebe}\ \emph {et~al.}(2008)\citenamefont {Riebe},
  \citenamefont {Monz}, \citenamefont {Kim}, \citenamefont {Villar},
  \citenamefont {Schindler}, \citenamefont {Chwalla}, \citenamefont
  {Hennrich},\ and\ \citenamefont {Blatt}}]{Riebe2008}%
  \BibitemOpen
  \bibfield  {author} {\bibinfo {author} {\bibfnamefont {M.}~\bibnamefont
  {Riebe}}, \bibinfo {author} {\bibfnamefont {T.}~\bibnamefont {Monz}},
  \bibinfo {author} {\bibfnamefont {K.}~\bibnamefont {Kim}}, \bibinfo {author}
  {\bibfnamefont {A.~S.}\ \bibnamefont {Villar}}, \bibinfo {author}
  {\bibfnamefont {P.}~\bibnamefont {Schindler}}, \bibinfo {author}
  {\bibfnamefont {M.}~\bibnamefont {Chwalla}}, \bibinfo {author} {\bibfnamefont
  {M.}~\bibnamefont {Hennrich}},\ and\ \bibinfo {author} {\bibfnamefont
  {R.}~\bibnamefont {Blatt}},\ }\href {https://doi.org/10.1038/nphys1107}
  {\bibfield  {journal} {\bibinfo  {journal} {Nat. Phys.}\ }\textbf {\bibinfo
  {volume} {4}},\ \bibinfo {pages} {839} (\bibinfo {year} {2008})}\BibitemShut
  {NoStop}%
\bibitem [{\citenamefont {Ning}\ \emph {et~al.}(2019)\citenamefont {Ning},
  \citenamefont {Huang}, \citenamefont {Han}, \citenamefont {Li}, \citenamefont
  {Deng}, \citenamefont {Yang}, \citenamefont {Zhong}, \citenamefont {Xia},
  \citenamefont {Xu}, \citenamefont {Zheng},\ and\ \citenamefont
  {Zheng}}]{Ning2019}%
  \BibitemOpen
  \bibfield  {author} {\bibinfo {author} {\bibfnamefont {W.}~\bibnamefont
  {Ning}}, \bibinfo {author} {\bibfnamefont {X.-J.}\ \bibnamefont {Huang}},
  \bibinfo {author} {\bibfnamefont {P.-R.}\ \bibnamefont {Han}}, \bibinfo
  {author} {\bibfnamefont {H.}~\bibnamefont {Li}}, \bibinfo {author}
  {\bibfnamefont {H.}~\bibnamefont {Deng}}, \bibinfo {author} {\bibfnamefont
  {Z.-B.}\ \bibnamefont {Yang}}, \bibinfo {author} {\bibfnamefont {Z.-R.}\
  \bibnamefont {Zhong}}, \bibinfo {author} {\bibfnamefont {Y.}~\bibnamefont
  {Xia}}, \bibinfo {author} {\bibfnamefont {K.}~\bibnamefont {Xu}}, \bibinfo
  {author} {\bibfnamefont {D.}~\bibnamefont {Zheng}},\ and\ \bibinfo {author}
  {\bibfnamefont {S.-B.}\ \bibnamefont {Zheng}},\ }\href
  {https://doi.org/10.1103/PhysRevLett.123.060502} {\bibfield  {journal}
  {\bibinfo  {journal} {Phys. Rev. Lett.}\ }\textbf {\bibinfo {volume} {123}},\
  \bibinfo {pages} {060502} (\bibinfo {year} {2019})}\BibitemShut {NoStop}%
\bibitem [{\citenamefont {Carvacho}\ \emph {et~al.}(2017)\citenamefont
  {Carvacho}, \citenamefont {Andreoli}, \citenamefont {Santodonato},
  \citenamefont {Bentivegna}, \citenamefont {Chaves},\ and\ \citenamefont
  {Sciarrino}}]{Carvacho2017}%
  \BibitemOpen
  \bibfield  {author} {\bibinfo {author} {\bibfnamefont {G.}~\bibnamefont
  {Carvacho}}, \bibinfo {author} {\bibfnamefont {F.}~\bibnamefont {Andreoli}},
  \bibinfo {author} {\bibfnamefont {L.}~\bibnamefont {Santodonato}}, \bibinfo
  {author} {\bibfnamefont {M.}~\bibnamefont {Bentivegna}}, \bibinfo {author}
  {\bibfnamefont {R.}~\bibnamefont {Chaves}},\ and\ \bibinfo {author}
  {\bibfnamefont {F.}~\bibnamefont {Sciarrino}},\ }\href
  {https://doi.org/10.1038/ncomms14775} {\bibfield  {journal} {\bibinfo
  {journal} {Nat. Commun.}\ }\textbf {\bibinfo {volume} {8}},\ \bibinfo {pages}
  {14775} (\bibinfo {year} {2017})}\BibitemShut {NoStop}%
\bibitem [{\citenamefont {Sun}\ \emph {et~al.}(2019)\citenamefont {Sun},
  \citenamefont {Jiang}, \citenamefont {Bai}, \citenamefont {Zhang},
  \citenamefont {Li}, \citenamefont {Jiang}, \citenamefont {Zhang},
  \citenamefont {You}, \citenamefont {Chen}, \citenamefont {Wang},
  \citenamefont {Zhang}, \citenamefont {Fan},\ and\ \citenamefont
  {Pan}}]{Sun2019}%
  \BibitemOpen
  \bibfield  {author} {\bibinfo {author} {\bibfnamefont {Q.-C.}\ \bibnamefont
  {Sun}}, \bibinfo {author} {\bibfnamefont {Y.-F.}\ \bibnamefont {Jiang}},
  \bibinfo {author} {\bibfnamefont {B.}~\bibnamefont {Bai}}, \bibinfo {author}
  {\bibfnamefont {W.}~\bibnamefont {Zhang}}, \bibinfo {author} {\bibfnamefont
  {H.}~\bibnamefont {Li}}, \bibinfo {author} {\bibfnamefont {X.}~\bibnamefont
  {Jiang}}, \bibinfo {author} {\bibfnamefont {J.}~\bibnamefont {Zhang}},
  \bibinfo {author} {\bibfnamefont {L.}~\bibnamefont {You}}, \bibinfo {author}
  {\bibfnamefont {X.}~\bibnamefont {Chen}}, \bibinfo {author} {\bibfnamefont
  {Z.}~\bibnamefont {Wang}}, \bibinfo {author} {\bibfnamefont {Q.}~\bibnamefont
  {Zhang}}, \bibinfo {author} {\bibfnamefont {J.}~\bibnamefont {Fan}},\ and\
  \bibinfo {author} {\bibfnamefont {J.-W.}\ \bibnamefont {Pan}},\ }\href
  {https://doi.org/10.1038/s41566-019-0502-7} {\bibfield  {journal} {\bibinfo
  {journal} {Nat. Photonics}\ }\textbf {\bibinfo {volume} {13}},\ \bibinfo
  {pages} {687} (\bibinfo {year} {2019})}\BibitemShut {NoStop}%
\bibitem [{\citenamefont {Poderini}\ \emph {et~al.}(2020)\citenamefont
  {Poderini}, \citenamefont {Agresti}, \citenamefont {Marchese}, \citenamefont
  {Polino}, \citenamefont {Giordani}, \citenamefont {Suprano}, \citenamefont
  {Valeri}, \citenamefont {Milani}, \citenamefont {Spagnolo}, \citenamefont
  {Carvacho}, \citenamefont {Chaves},\ and\ \citenamefont
  {Sciarrino}}]{Poderini2020experimental}%
  \BibitemOpen
  \bibfield  {author} {\bibinfo {author} {\bibfnamefont {D.}~\bibnamefont
  {Poderini}}, \bibinfo {author} {\bibfnamefont {I.}~\bibnamefont {Agresti}},
  \bibinfo {author} {\bibfnamefont {G.}~\bibnamefont {Marchese}}, \bibinfo
  {author} {\bibfnamefont {E.}~\bibnamefont {Polino}}, \bibinfo {author}
  {\bibfnamefont {T.}~\bibnamefont {Giordani}}, \bibinfo {author}
  {\bibfnamefont {A.}~\bibnamefont {Suprano}}, \bibinfo {author} {\bibfnamefont
  {M.}~\bibnamefont {Valeri}}, \bibinfo {author} {\bibfnamefont
  {G.}~\bibnamefont {Milani}}, \bibinfo {author} {\bibfnamefont
  {N.}~\bibnamefont {Spagnolo}}, \bibinfo {author} {\bibfnamefont
  {G.}~\bibnamefont {Carvacho}}, \bibinfo {author} {\bibfnamefont
  {R.}~\bibnamefont {Chaves}},\ and\ \bibinfo {author} {\bibfnamefont
  {F.}~\bibnamefont {Sciarrino}},\ }\href
  {https://doi.org/10.1038/s41467-020-16189-6} {\bibfield  {journal} {\bibinfo
  {journal} {Nat. Commun.}\ }\textbf {\bibinfo {volume} {11}},\ \bibinfo
  {pages} {2467} (\bibinfo {year} {2020})}\BibitemShut {NoStop}%
\bibitem [{\citenamefont {Bäumer}\ \emph {et~al.}(2021)\citenamefont
  {Bäumer}, \citenamefont {Gisin},\ and\ \citenamefont
  {Tavakoli}}]{Baumer2020}%
  \BibitemOpen
  \bibfield  {author} {\bibinfo {author} {\bibfnamefont {E.}~\bibnamefont
  {Bäumer}}, \bibinfo {author} {\bibfnamefont {N.}~\bibnamefont {Gisin}},\
  and\ \bibinfo {author} {\bibfnamefont {A.}~\bibnamefont {Tavakoli}},\ }\href
  {https://doi.org/10.1038/s41534-021-00450-x} {\bibfield  {journal} {\bibinfo
  {journal} {npj Quantum Inf.}\ }\textbf {\bibinfo {volume} {7}},\ \bibinfo
  {pages} {117} (\bibinfo {year} {2021})}\BibitemShut {NoStop}%
\bibitem [{\citenamefont {Andreoli}\ \emph {et~al.}(2017)\citenamefont
  {Andreoli}, \citenamefont {Carvacho}, \citenamefont {Santodonato},
  \citenamefont {Bentivegna}, \citenamefont {Chaves},\ and\ \citenamefont
  {Sciarrino}}]{Andreoli2017b}%
  \BibitemOpen
  \bibfield  {author} {\bibinfo {author} {\bibfnamefont {F.}~\bibnamefont
  {Andreoli}}, \bibinfo {author} {\bibfnamefont {G.}~\bibnamefont {Carvacho}},
  \bibinfo {author} {\bibfnamefont {L.}~\bibnamefont {Santodonato}}, \bibinfo
  {author} {\bibfnamefont {M.}~\bibnamefont {Bentivegna}}, \bibinfo {author}
  {\bibfnamefont {R.}~\bibnamefont {Chaves}},\ and\ \bibinfo {author}
  {\bibfnamefont {F.}~\bibnamefont {Sciarrino}},\ }\href
  {https://doi.org/10.1103/PhysRevA.95.062315} {\bibfield  {journal} {\bibinfo
  {journal} {Phys. Rev. A}\ }\textbf {\bibinfo {volume} {95}},\ \bibinfo
  {pages} {062315} (\bibinfo {year} {2017})}\BibitemShut {NoStop}%
\bibitem [{\citenamefont {van Loock}\ and\ \citenamefont
  {L\"utkenhaus}(2004)}]{Loock2004}%
  \BibitemOpen
  \bibfield  {author} {\bibinfo {author} {\bibfnamefont {P.}~\bibnamefont {van
  Loock}}\ and\ \bibinfo {author} {\bibfnamefont {N.}~\bibnamefont
  {L\"utkenhaus}},\ }\href {https://doi.org/10.1103/PhysRevA.69.012302}
  {\bibfield  {journal} {\bibinfo  {journal} {Phys. Rev. A}\ }\textbf {\bibinfo
  {volume} {69}},\ \bibinfo {pages} {012302} (\bibinfo {year}
  {2004})}\BibitemShut {NoStop}%
\bibitem [{\citenamefont {Agresti}\ \emph {et~al.}(2021)\citenamefont
  {Agresti}, \citenamefont {Polacchi}, \citenamefont {Poderini}, \citenamefont
  {Polino}, \citenamefont {Suprano}, \citenamefont {\ifmmode \check{S}\else
  \v{S}\fi{}upi\ifmmode~\acute{c}\else \'{c}\fi{}}, \citenamefont {Bowles},
  \citenamefont {Carvacho}, \citenamefont {Cavalcanti},\ and\ \citenamefont
  {Sciarrino}}]{Agresti2021}%
  \BibitemOpen
  \bibfield  {author} {\bibinfo {author} {\bibfnamefont {I.}~\bibnamefont
  {Agresti}}, \bibinfo {author} {\bibfnamefont {B.}~\bibnamefont {Polacchi}},
  \bibinfo {author} {\bibfnamefont {D.}~\bibnamefont {Poderini}}, \bibinfo
  {author} {\bibfnamefont {E.}~\bibnamefont {Polino}}, \bibinfo {author}
  {\bibfnamefont {A.}~\bibnamefont {Suprano}}, \bibinfo {author} {\bibfnamefont
  {I.}~\bibnamefont {\ifmmode \check{S}\else
  \v{S}\fi{}upi\ifmmode~\acute{c}\else \'{c}\fi{}}}, \bibinfo {author}
  {\bibfnamefont {J.}~\bibnamefont {Bowles}}, \bibinfo {author} {\bibfnamefont
  {G.}~\bibnamefont {Carvacho}}, \bibinfo {author} {\bibfnamefont
  {D.}~\bibnamefont {Cavalcanti}},\ and\ \bibinfo {author} {\bibfnamefont
  {F.}~\bibnamefont {Sciarrino}},\ }\href
  {https://doi.org/10.1103/PRXQuantum.2.020346} {\bibfield  {journal} {\bibinfo
   {journal} {PRX Quantum}\ }\textbf {\bibinfo {volume} {2}},\ \bibinfo {pages}
  {020346} (\bibinfo {year} {2021})}\BibitemShut {NoStop}%
\bibitem [{\citenamefont {{IBM Quantum Experience}}()}]{IBMexperience}%
  \BibitemOpen
  \bibfield  {author} {\bibinfo {author} {\bibnamefont {{IBM Quantum
  Experience}}},\ }\href@noop {} {}\bibinfo {howpublished}
  {\url{http://www.research.ibm.com/ibm-q/}}\BibitemShut {NoStop}%
\bibitem [{\citenamefont {Pozas-Kerstjens}\ \emph {et~al.}(2022)\citenamefont
  {Pozas-Kerstjens}, \citenamefont {Gisin},\ and\ \citenamefont
  {Tavakoli}}]{PozasGenuine}%
  \BibitemOpen
  \bibfield  {author} {\bibinfo {author} {\bibfnamefont {A.}~\bibnamefont
  {Pozas-Kerstjens}}, \bibinfo {author} {\bibfnamefont {N.}~\bibnamefont
  {Gisin}},\ and\ \bibinfo {author} {\bibfnamefont {A.}~\bibnamefont
  {Tavakoli}},\ }\href {https://doi.org/10.1103/PhysRevLett.128.010403}
  {\bibfield  {journal} {\bibinfo  {journal} {Phys. Rev. Lett.}\ }\textbf
  {\bibinfo {volume} {128}},\ \bibinfo {pages} {010403} (\bibinfo {year}
  {2022})}\BibitemShut {NoStop}%
\bibitem [{\citenamefont {Seevinck}\ and\ \citenamefont
  {Uffink}(2001)}]{Seevinck2001multipartiteentanglement}%
  \BibitemOpen
  \bibfield  {author} {\bibinfo {author} {\bibfnamefont {M.}~\bibnamefont
  {Seevinck}}\ and\ \bibinfo {author} {\bibfnamefont {J.}~\bibnamefont
  {Uffink}},\ }\href {https://doi.org/10.1103/PhysRevA.65.012107} {\bibfield
  {journal} {\bibinfo  {journal} {Phys. Rev. A}\ }\textbf {\bibinfo {volume}
  {65}},\ \bibinfo {pages} {012107} (\bibinfo {year} {2001})}\BibitemShut
  {NoStop}%
\bibitem [{\citenamefont {Luo}\ and\ \citenamefont
  {Fei}(2021)}]{Luo2021robust}%
  \BibitemOpen
  \bibfield  {author} {\bibinfo {author} {\bibfnamefont {M.-X.}\ \bibnamefont
  {Luo}}\ and\ \bibinfo {author} {\bibfnamefont {S.-M.}\ \bibnamefont {Fei}},\
  }\href {https://doi.org/10.1103/PhysRevResearch.3.043120} {\bibfield
  {journal} {\bibinfo  {journal} {Phys. Rev. Research}\ }\textbf {\bibinfo
  {volume} {3}},\ \bibinfo {pages} {043120} (\bibinfo {year}
  {2021})}\BibitemShut {NoStop}%
\bibitem [{\citenamefont {\ifmmode \check{S}\else
  \v{S}\fi{}upi\ifmmode~\acute{c}\else \'{c}\fi{}}\ \emph
  {et~al.}(2022)\citenamefont {\ifmmode \check{S}\else
  \v{S}\fi{}upi\ifmmode~\acute{c}\else \'{c}\fi{}}, \citenamefont {Bancal},
  \citenamefont {Cai},\ and\ \citenamefont {Brunner}}]{supicGenuine}%
  \BibitemOpen
  \bibfield  {author} {\bibinfo {author} {\bibfnamefont {I.}~\bibnamefont
  {\ifmmode \check{S}\else \v{S}\fi{}upi\ifmmode~\acute{c}\else \'{c}\fi{}}},
  \bibinfo {author} {\bibfnamefont {J.-D.}\ \bibnamefont {Bancal}}, \bibinfo
  {author} {\bibfnamefont {Y.}~\bibnamefont {Cai}},\ and\ \bibinfo {author}
  {\bibfnamefont {N.}~\bibnamefont {Brunner}},\ }\href
  {https://doi.org/10.1103/PhysRevA.105.022206} {\bibfield  {journal} {\bibinfo
   {journal} {Phys. Rev. A}\ }\textbf {\bibinfo {volume} {105}},\ \bibinfo
  {pages} {022206} (\bibinfo {year} {2022})}\BibitemShut {NoStop}%
\bibitem [{\citenamefont {Pironio}\ \emph {et~al.}(2016)\citenamefont
  {Pironio}, \citenamefont {Scarani},\ and\ \citenamefont
  {Vidick}}]{Pironio2016}%
  \BibitemOpen
  \bibfield  {author} {\bibinfo {author} {\bibfnamefont {S.}~\bibnamefont
  {Pironio}}, \bibinfo {author} {\bibfnamefont {V.}~\bibnamefont {Scarani}},\
  and\ \bibinfo {author} {\bibfnamefont {T.}~\bibnamefont {Vidick}},\ }\href
  {https://doi.org/10.1088/1367-2630/18/10/100202} {\bibfield  {journal}
  {\bibinfo  {journal} {New J. Phys.}\ }\textbf {\bibinfo {volume} {18}},\
  \bibinfo {pages} {100202} (\bibinfo {year} {2016})}\BibitemShut {NoStop}%
\bibitem [{\citenamefont {Mayers}\ and\ \citenamefont {Yao}(1998)}]{May98}%
  \BibitemOpen
  \bibfield  {author} {\bibinfo {author} {\bibfnamefont {D.}~\bibnamefont
  {Mayers}}\ and\ \bibinfo {author} {\bibfnamefont {A.}~\bibnamefont {Yao}},\
  }in\ \href {https://doi.org/10.1109/SFCS.1998.743501} {\emph {\bibinfo
  {booktitle} {Proceedings 39th Annual Symposium on Foundations of Computer
  Science (FOCS)}}}\ (\bibinfo {year} {1998})\ pp.\ \bibinfo {pages}
  {503--509}\BibitemShut {NoStop}%
\bibitem [{\citenamefont {Barrett}\ \emph {et~al.}(2005)\citenamefont
  {Barrett}, \citenamefont {Hardy},\ and\ \citenamefont {Kent}}]{Barrett05}%
  \BibitemOpen
  \bibfield  {author} {\bibinfo {author} {\bibfnamefont {J.}~\bibnamefont
  {Barrett}}, \bibinfo {author} {\bibfnamefont {L.}~\bibnamefont {Hardy}},\
  and\ \bibinfo {author} {\bibfnamefont {A.}~\bibnamefont {Kent}},\ }\href
  {https://doi.org/10.1103/PhysRevLett.95.010503} {\bibfield  {journal}
  {\bibinfo  {journal} {Phys. Rev. Lett.}\ }\textbf {\bibinfo {volume} {95}},\
  \bibinfo {pages} {010503} (\bibinfo {year} {2005})}\BibitemShut {NoStop}%
\bibitem [{\citenamefont {Ac\'{\i}n}\ \emph {et~al.}(2007)\citenamefont
  {Ac\'{\i}n}, \citenamefont {Brunner}, \citenamefont {Gisin}, \citenamefont
  {Massar}, \citenamefont {Pironio},\ and\ \citenamefont {Scarani}}]{Acin2007}%
  \BibitemOpen
  \bibfield  {author} {\bibinfo {author} {\bibfnamefont {A.}~\bibnamefont
  {Ac\'{\i}n}}, \bibinfo {author} {\bibfnamefont {N.}~\bibnamefont {Brunner}},
  \bibinfo {author} {\bibfnamefont {N.}~\bibnamefont {Gisin}}, \bibinfo
  {author} {\bibfnamefont {S.}~\bibnamefont {Massar}}, \bibinfo {author}
  {\bibfnamefont {S.}~\bibnamefont {Pironio}},\ and\ \bibinfo {author}
  {\bibfnamefont {V.}~\bibnamefont {Scarani}},\ }\href
  {https://doi.org/10.1103/PhysRevLett.98.230501} {\bibfield  {journal}
  {\bibinfo  {journal} {Phys. Rev. Lett.}\ }\textbf {\bibinfo {volume} {98}},\
  \bibinfo {pages} {230501} (\bibinfo {year} {2007})}\BibitemShut {NoStop}%
\bibitem [{\citenamefont {Lydersen}\ \emph {et~al.}(2010)\citenamefont
  {Lydersen}, \citenamefont {Wiechers}, \citenamefont {Wittmann}, \citenamefont
  {Elser}, \citenamefont {Skaar},\ and\ \citenamefont {Makarov}}]{Lydersen}%
  \BibitemOpen
  \bibfield  {author} {\bibinfo {author} {\bibfnamefont {L.}~\bibnamefont
  {Lydersen}}, \bibinfo {author} {\bibfnamefont {C.}~\bibnamefont {Wiechers}},
  \bibinfo {author} {\bibfnamefont {C.}~\bibnamefont {Wittmann}}, \bibinfo
  {author} {\bibfnamefont {D.}~\bibnamefont {Elser}}, \bibinfo {author}
  {\bibfnamefont {J.}~\bibnamefont {Skaar}},\ and\ \bibinfo {author}
  {\bibfnamefont {V.}~\bibnamefont {Makarov}},\ }\href
  {https://doi.org/10.1038/nphoton.2010.214} {\bibfield  {journal} {\bibinfo
  {journal} {Nat. Photonics}\ }\textbf {\bibinfo {volume} {4}} (\bibinfo {year}
  {2010})}\BibitemShut {NoStop}%
\bibitem [{\citenamefont {Gisin}\ \emph {et~al.}(2002)\citenamefont {Gisin},
  \citenamefont {Ribordy}, \citenamefont {Tittel},\ and\ \citenamefont
  {Zbinden}}]{Gisin2002}%
  \BibitemOpen
  \bibfield  {author} {\bibinfo {author} {\bibfnamefont {N.}~\bibnamefont
  {Gisin}}, \bibinfo {author} {\bibfnamefont {G.}~\bibnamefont {Ribordy}},
  \bibinfo {author} {\bibfnamefont {W.}~\bibnamefont {Tittel}},\ and\ \bibinfo
  {author} {\bibfnamefont {H.}~\bibnamefont {Zbinden}},\ }\href
  {https://doi.org/10.1103/RevModPhys.74.145} {\bibfield  {journal} {\bibinfo
  {journal} {Rev. Mod. Phys.}\ }\textbf {\bibinfo {volume} {74}},\ \bibinfo
  {pages} {145} (\bibinfo {year} {2002})}\BibitemShut {NoStop}%
\bibitem [{\citenamefont {Xu}\ \emph {et~al.}(2020)\citenamefont {Xu},
  \citenamefont {Ma}, \citenamefont {Zhang}, \citenamefont {Lo},\ and\
  \citenamefont {Pan}}]{Xu2020}%
  \BibitemOpen
  \bibfield  {author} {\bibinfo {author} {\bibfnamefont {F.}~\bibnamefont
  {Xu}}, \bibinfo {author} {\bibfnamefont {X.}~\bibnamefont {Ma}}, \bibinfo
  {author} {\bibfnamefont {Q.}~\bibnamefont {Zhang}}, \bibinfo {author}
  {\bibfnamefont {H.-K.}\ \bibnamefont {Lo}},\ and\ \bibinfo {author}
  {\bibfnamefont {J.-W.}\ \bibnamefont {Pan}},\ }\href
  {https://doi.org/10.1103/RevModPhys.92.025002} {\bibfield  {journal}
  {\bibinfo  {journal} {Rev. Mod. Phys.}\ }\textbf {\bibinfo {volume} {92}},\
  \bibinfo {pages} {025002} (\bibinfo {year} {2020})}\BibitemShut {NoStop}%
\bibitem [{\citenamefont {Portmann}\ and\ \citenamefont
  {Renner}()}]{portmann2021security}%
  \BibitemOpen
  \bibfield  {author} {\bibinfo {author} {\bibfnamefont {C.}~\bibnamefont
  {Portmann}}\ and\ \bibinfo {author} {\bibfnamefont {R.}~\bibnamefont
  {Renner}},\ }\href@noop {} {\bibinfo {title} {Security in quantum
  cryptography}},\ \Eprint {https://arxiv.org/abs/2102.00021}
  {arXiv:2102.00021} \BibitemShut {NoStop}%
\bibitem [{\citenamefont {Wigner}(1970)}]{Wigner}%
  \BibitemOpen
  \bibfield  {author} {\bibinfo {author} {\bibfnamefont {E.~P.}\ \bibnamefont
  {Wigner}},\ }\href {https://doi.org/10.1119/1.1976526} {\bibfield  {journal}
  {\bibinfo  {journal} {Am. J. Phys.}\ }\textbf {\bibinfo {volume} {38}},\
  \bibinfo {pages} {1005} (\bibinfo {year} {1970})}\BibitemShut {NoStop}%
\bibitem [{\citenamefont {Braunstein}\ and\ \citenamefont
  {Caves}(1990)}]{Braunstein}%
  \BibitemOpen
  \bibfield  {author} {\bibinfo {author} {\bibfnamefont {S.~L.}\ \bibnamefont
  {Braunstein}}\ and\ \bibinfo {author} {\bibfnamefont {C.~M.}\ \bibnamefont
  {Caves}},\ }\href
  {https://doi.org/https://doi.org/10.1016/0003-4916(90)90339-P} {\bibfield
  {journal} {\bibinfo  {journal} {Ann. Phys. (N.Y.)}\ }\textbf {\bibinfo
  {volume} {202}},\ \bibinfo {pages} {22} (\bibinfo {year} {1990})}\BibitemShut
  {NoStop}%
\bibitem [{\citenamefont {Masanes}\ \emph {et~al.}(2011)\citenamefont
  {Masanes}, \citenamefont {Pironio},\ and\ \citenamefont
  {Ac{\'i}n}}]{Masanes11}%
  \BibitemOpen
  \bibfield  {author} {\bibinfo {author} {\bibfnamefont {L.}~\bibnamefont
  {Masanes}}, \bibinfo {author} {\bibfnamefont {S.}~\bibnamefont {Pironio}},\
  and\ \bibinfo {author} {\bibfnamefont {A.}~\bibnamefont {Ac{\'i}n}},\ }\href
  {https://doi.org/10.1038/ncomms1244} {\bibfield  {journal} {\bibinfo
  {journal} {Nat. Commun.}\ }\textbf {\bibinfo {volume} {2}},\ \bibinfo {pages}
  {238} (\bibinfo {year} {2011})}\BibitemShut {NoStop}%
\bibitem [{\citenamefont {Vazirani}\ and\ \citenamefont
  {Vidick}(2014)}]{Vazirani14}%
  \BibitemOpen
  \bibfield  {author} {\bibinfo {author} {\bibfnamefont {U.}~\bibnamefont
  {Vazirani}}\ and\ \bibinfo {author} {\bibfnamefont {T.}~\bibnamefont
  {Vidick}},\ }\href {https://doi.org/10.1103/PhysRevLett.113.140501}
  {\bibfield  {journal} {\bibinfo  {journal} {Phys. Rev. Lett.}\ }\textbf
  {\bibinfo {volume} {113}},\ \bibinfo {pages} {140501} (\bibinfo {year}
  {2014})}\BibitemShut {NoStop}%
\bibitem [{\citenamefont {Lee}\ and\ \citenamefont {Hoban}(2018)}]{lee2018di}%
  \BibitemOpen
  \bibfield  {author} {\bibinfo {author} {\bibfnamefont {C.~M.}\ \bibnamefont
  {Lee}}\ and\ \bibinfo {author} {\bibfnamefont {M.~J.}\ \bibnamefont
  {Hoban}},\ }\href {https://doi.org/10.1103/PhysRevLett.120.020504} {\bibfield
   {journal} {\bibinfo  {journal} {Phys. Rev. Lett.}\ }\textbf {\bibinfo
  {volume} {120}},\ \bibinfo {pages} {020504} (\bibinfo {year}
  {2018})}\BibitemShut {NoStop}%
\bibitem [{\citenamefont {Luo}(2020)}]{Luo2020a}%
  \BibitemOpen
  \bibfield  {author} {\bibinfo {author} {\bibfnamefont {M.-X.}\ \bibnamefont
  {Luo}},\ }\href {https://doi.org/10.1103/PhysRevA.101.062317} {\bibfield
  {journal} {\bibinfo  {journal} {Phys. Rev. A}\ }\textbf {\bibinfo {volume}
  {101}},\ \bibinfo {pages} {062317} (\bibinfo {year} {2020})}\BibitemShut
  {NoStop}%
\bibitem [{\citenamefont {Luo}()}]{Luodevice2021}%
  \BibitemOpen
  \bibfield  {author} {\bibinfo {author} {\bibfnamefont {M.-X.}\ \bibnamefont
  {Luo}},\ }\href@noop {} {\bibinfo {title} {Fully device-independent model on
  quantum networks}},\ \Eprint {https://arxiv.org/abs/2106.15840}
  {arXiv:2106.15840} \BibitemShut {NoStop}%
\bibitem [{\citenamefont {Mayers}\ and\ \citenamefont {Yao}(2004)}]{MY04}%
  \BibitemOpen
  \bibfield  {author} {\bibinfo {author} {\bibfnamefont {D.}~\bibnamefont
  {Mayers}}\ and\ \bibinfo {author} {\bibfnamefont {A.}~\bibnamefont {Yao}},\
  }\href {http://dl.acm.org/citation.cfm?id=2011827.2011830} {\bibfield
  {journal} {\bibinfo  {journal} {Quantum Inf. Comput.}\ }\textbf {\bibinfo
  {volume} {4}},\ \bibinfo {pages} {273} (\bibinfo {year} {2004})}\BibitemShut
  {NoStop}%
\bibitem [{\citenamefont {\ifmmode \check{S}\else
  \v{S}\fi{}upi\ifmmode~\acute{c}\else \'{c}\fi{}}\ and\ \citenamefont
  {Bowles}(2020)}]{Supic2020}%
  \BibitemOpen
  \bibfield  {author} {\bibinfo {author} {\bibfnamefont {I.}~\bibnamefont
  {\ifmmode \check{S}\else \v{S}\fi{}upi\ifmmode~\acute{c}\else \'{c}\fi{}}}\
  and\ \bibinfo {author} {\bibfnamefont {J.}~\bibnamefont {Bowles}},\ }\href
  {https://doi.org/https://doi.org/10.22331/q-2020-09-30-337} {\bibfield
  {journal} {\bibinfo  {journal} {Quantum}\ }\textbf {\bibinfo {volume} {4}},\
  \bibinfo {pages} {337} (\bibinfo {year} {2020})}\BibitemShut {NoStop}%
\bibitem [{\citenamefont {Bancal}\ \emph {et~al.}(2018)\citenamefont {Bancal},
  \citenamefont {Sangouard},\ and\ \citenamefont {Sekatski}}]{Bancal2018}%
  \BibitemOpen
  \bibfield  {author} {\bibinfo {author} {\bibfnamefont {J.-D.}\ \bibnamefont
  {Bancal}}, \bibinfo {author} {\bibfnamefont {N.}~\bibnamefont {Sangouard}},\
  and\ \bibinfo {author} {\bibfnamefont {P.}~\bibnamefont {Sekatski}},\ }\href
  {https://doi.org/10.1103/PhysRevLett.121.250506} {\bibfield  {journal}
  {\bibinfo  {journal} {Phys. Rev. Lett.}\ }\textbf {\bibinfo {volume} {121}},\
  \bibinfo {pages} {250506} (\bibinfo {year} {2018})}\BibitemShut {NoStop}%
\bibitem [{\citenamefont {Renou}\ \emph {et~al.}(2018)\citenamefont {Renou},
  \citenamefont {Kaniewski},\ and\ \citenamefont {Brunner}}]{Renou2018}%
  \BibitemOpen
  \bibfield  {author} {\bibinfo {author} {\bibfnamefont {M.-O.}\ \bibnamefont
  {Renou}}, \bibinfo {author} {\bibfnamefont {J.}~\bibnamefont {Kaniewski}},\
  and\ \bibinfo {author} {\bibfnamefont {N.}~\bibnamefont {Brunner}},\ }\href
  {https://doi.org/10.1103/PhysRevLett.121.250507} {\bibfield  {journal}
  {\bibinfo  {journal} {Phys. Rev. Lett.}\ }\textbf {\bibinfo {volume} {121}},\
  \bibinfo {pages} {250507} (\bibinfo {year} {2018})}\BibitemShut {NoStop}%
\bibitem [{\citenamefont {Zhang}\ \emph {et~al.}(2019)\citenamefont {Zhang},
  \citenamefont {Chen}, \citenamefont {Peng}, \citenamefont {Ye}, \citenamefont
  {Yin}, \citenamefont {Xu}, \citenamefont {Xu}, \citenamefont {Li},\ and\
  \citenamefont {Guo}}]{Zhang2019}%
  \BibitemOpen
  \bibfield  {author} {\bibinfo {author} {\bibfnamefont {W.-H.}\ \bibnamefont
  {Zhang}}, \bibinfo {author} {\bibfnamefont {G.}~\bibnamefont {Chen}},
  \bibinfo {author} {\bibfnamefont {X.-X.}\ \bibnamefont {Peng}}, \bibinfo
  {author} {\bibfnamefont {X.-J.}\ \bibnamefont {Ye}}, \bibinfo {author}
  {\bibfnamefont {P.}~\bibnamefont {Yin}}, \bibinfo {author} {\bibfnamefont
  {X.-Y.}\ \bibnamefont {Xu}}, \bibinfo {author} {\bibfnamefont {J.-S.}\
  \bibnamefont {Xu}}, \bibinfo {author} {\bibfnamefont {C.-F.}\ \bibnamefont
  {Li}},\ and\ \bibinfo {author} {\bibfnamefont {G.-C.}\ \bibnamefont {Guo}},\
  }\href {https://doi.org/10.1103/PhysRevLett.122.090402} {\bibfield  {journal}
  {\bibinfo  {journal} {Phys. Rev. Lett.}\ }\textbf {\bibinfo {volume} {122}},\
  \bibinfo {pages} {90402} (\bibinfo {year} {2019})}\BibitemShut {NoStop}%
\bibitem [{\citenamefont {Weilenmann}\ and\ \citenamefont
  {Colbeck}(2020)}]{Weilenmann2020PRL}%
  \BibitemOpen
  \bibfield  {author} {\bibinfo {author} {\bibfnamefont {M.}~\bibnamefont
  {Weilenmann}}\ and\ \bibinfo {author} {\bibfnamefont {R.}~\bibnamefont
  {Colbeck}},\ }\href {https://doi.org/10.1103/PhysRevLett.125.060406}
  {\bibfield  {journal} {\bibinfo  {journal} {Phys. Rev. Lett.}\ }\textbf
  {\bibinfo {volume} {125}},\ \bibinfo {pages} {060406} (\bibinfo {year}
  {2020})}\BibitemShut {NoStop}%
\bibitem [{\citenamefont {Weilenmann}\ and\ \citenamefont
  {Colbeck}(2020)}]{Weilenmann2020PRA}%
  \BibitemOpen
  \bibfield  {author} {\bibinfo {author} {\bibfnamefont {M.}~\bibnamefont
  {Weilenmann}}\ and\ \bibinfo {author} {\bibfnamefont {R.}~\bibnamefont
  {Colbeck}},\ }\href {https://doi.org/10.1103/PhysRevA.102.022203} {\bibfield
  {journal} {\bibinfo  {journal} {Phys. Rev. A}\ }\textbf {\bibinfo {volume}
  {102}},\ \bibinfo {pages} {022203} (\bibinfo {year} {2020})}\BibitemShut
  {NoStop}%
\bibitem [{\citenamefont {P\'al}\ and\ \citenamefont
  {V\'ertesi}(2008)}]{Pal08}%
  \BibitemOpen
  \bibfield  {author} {\bibinfo {author} {\bibfnamefont {K.~F.}\ \bibnamefont
  {P\'al}}\ and\ \bibinfo {author} {\bibfnamefont {T.}~\bibnamefont
  {V\'ertesi}},\ }\href {https://doi.org/10.1103/PhysRevA.77.042105} {\bibfield
   {journal} {\bibinfo  {journal} {Phys. Rev. A}\ }\textbf {\bibinfo {volume}
  {77}},\ \bibinfo {pages} {042105} (\bibinfo {year} {2008})}\BibitemShut
  {NoStop}%
\bibitem [{\citenamefont {McKague}\ \emph {et~al.}(2009)\citenamefont
  {McKague}, \citenamefont {Mosca},\ and\ \citenamefont {Gisin}}]{McKague09}%
  \BibitemOpen
  \bibfield  {author} {\bibinfo {author} {\bibfnamefont {M.}~\bibnamefont
  {McKague}}, \bibinfo {author} {\bibfnamefont {M.}~\bibnamefont {Mosca}},\
  and\ \bibinfo {author} {\bibfnamefont {N.}~\bibnamefont {Gisin}},\ }\href
  {https://doi.org/10.1103/PhysRevLett.102.020505} {\bibfield  {journal}
  {\bibinfo  {journal} {Phys. Rev. Lett.}\ }\textbf {\bibinfo {volume} {102}},\
  \bibinfo {pages} {020505} (\bibinfo {year} {2009})}\BibitemShut {NoStop}%
\bibitem [{\citenamefont {V\'ertesi}\ and\ \citenamefont
  {Bene}(2010)}]{Vertesi2010}%
  \BibitemOpen
  \bibfield  {author} {\bibinfo {author} {\bibfnamefont {T.}~\bibnamefont
  {V\'ertesi}}\ and\ \bibinfo {author} {\bibfnamefont {E.}~\bibnamefont
  {Bene}},\ }\href {https://doi.org/10.1103/PhysRevA.82.062115} {\bibfield
  {journal} {\bibinfo  {journal} {Phys. Rev. A}\ }\textbf {\bibinfo {volume}
  {82}},\ \bibinfo {pages} {062115} (\bibinfo {year} {2010})}\BibitemShut
  {NoStop}%
\bibitem [{\citenamefont {Andersson}\ \emph {et~al.}(2017)\citenamefont
  {Andersson}, \citenamefont {Badzi\k{a}g}, \citenamefont {Bengtsson},
  \citenamefont {Dumitru},\ and\ \citenamefont {Cabello}}]{Andersson2017}%
  \BibitemOpen
  \bibfield  {author} {\bibinfo {author} {\bibfnamefont {O.}~\bibnamefont
  {Andersson}}, \bibinfo {author} {\bibfnamefont {P.}~\bibnamefont
  {Badzi\k{a}g}}, \bibinfo {author} {\bibfnamefont {I.}~\bibnamefont
  {Bengtsson}}, \bibinfo {author} {\bibfnamefont {I.}~\bibnamefont {Dumitru}},\
  and\ \bibinfo {author} {\bibfnamefont {A.}~\bibnamefont {Cabello}},\ }\href
  {https://doi.org/10.1103/PhysRevA.96.032119} {\bibfield  {journal} {\bibinfo
  {journal} {Phys. Rev. A}\ }\textbf {\bibinfo {volume} {96}},\ \bibinfo
  {pages} {032119} (\bibinfo {year} {2017})}\BibitemShut {NoStop}%
\bibitem [{\citenamefont {Tavakoli}\ \emph {et~al.}(2020)\citenamefont
  {Tavakoli}, \citenamefont {Smania}, \citenamefont {V{\'e}rtesi},
  \citenamefont {Brunner},\ and\ \citenamefont {Bourennane}}]{Smania2020}%
  \BibitemOpen
  \bibfield  {author} {\bibinfo {author} {\bibfnamefont {A.}~\bibnamefont
  {Tavakoli}}, \bibinfo {author} {\bibfnamefont {M.}~\bibnamefont {Smania}},
  \bibinfo {author} {\bibfnamefont {T.}~\bibnamefont {V{\'e}rtesi}}, \bibinfo
  {author} {\bibfnamefont {N.}~\bibnamefont {Brunner}},\ and\ \bibinfo {author}
  {\bibfnamefont {M.}~\bibnamefont {Bourennane}},\ }\href
  {https://doi.org/10.1126/sciadv.aaw6664} {\bibfield  {journal} {\bibinfo
  {journal} {Sci. Adv.}\ }\textbf {\bibinfo {volume} {6}},\ \bibinfo {pages}
  {aaw6664} (\bibinfo {year} {2020})}\BibitemShut {NoStop}%
\bibitem [{\citenamefont {Smania}\ \emph {et~al.}(2020)\citenamefont {Smania},
  \citenamefont {Mironowicz}, \citenamefont {Nawareg}, \citenamefont
  {Paw{\l}owski}, \citenamefont {Cabello},\ and\ \citenamefont
  {Bourennane}}]{Smania2020b}%
  \BibitemOpen
  \bibfield  {author} {\bibinfo {author} {\bibfnamefont {M.}~\bibnamefont
  {Smania}}, \bibinfo {author} {\bibfnamefont {P.}~\bibnamefont {Mironowicz}},
  \bibinfo {author} {\bibfnamefont {M.}~\bibnamefont {Nawareg}}, \bibinfo
  {author} {\bibfnamefont {M.}~\bibnamefont {Paw{\l}owski}}, \bibinfo {author}
  {\bibfnamefont {A.}~\bibnamefont {Cabello}},\ and\ \bibinfo {author}
  {\bibfnamefont {M.}~\bibnamefont {Bourennane}},\ }\href
  {https://doi.org/10.1364/OPTICA.377959} {\bibfield  {journal} {\bibinfo
  {journal} {Optica}\ }\textbf {\bibinfo {volume} {7}},\ \bibinfo {pages} {123}
  (\bibinfo {year} {2020})}\BibitemShut {NoStop}%
\bibitem [{\citenamefont {Renou}\ \emph {et~al.}(2021)\citenamefont {Renou},
  \citenamefont {Trillo}, \citenamefont {Weilenmann}, \citenamefont {Thinh},
  \citenamefont {Tavakoli}, \citenamefont {Gisin}, \citenamefont {Ac\'in},\
  and\ \citenamefont {Navascu\'es}}]{Mark2021}%
  \BibitemOpen
  \bibfield  {author} {\bibinfo {author} {\bibfnamefont {M.-O.}\ \bibnamefont
  {Renou}}, \bibinfo {author} {\bibfnamefont {D.}~\bibnamefont {Trillo}},
  \bibinfo {author} {\bibfnamefont {M.}~\bibnamefont {Weilenmann}}, \bibinfo
  {author} {\bibfnamefont {L.~P.}\ \bibnamefont {Thinh}}, \bibinfo {author}
  {\bibfnamefont {A.}~\bibnamefont {Tavakoli}}, \bibinfo {author}
  {\bibfnamefont {N.}~\bibnamefont {Gisin}}, \bibinfo {author} {\bibfnamefont
  {A.}~\bibnamefont {Ac\'in}},\ and\ \bibinfo {author} {\bibfnamefont
  {M.}~\bibnamefont {Navascu\'es}},\ }\href
  {https://doi.org/10.1038/s41586-021-04160-4} {\bibfield  {journal} {\bibinfo
  {journal} {Nature}\ }\textbf {\bibinfo {volume} {600}},\ \bibinfo {pages}
  {625} (\bibinfo {year} {2021})}\BibitemShut {NoStop}%
\bibitem [{\citenamefont {Bowles}\ \emph {et~al.}(2018)\citenamefont {Bowles},
  \citenamefont {\ifmmode \check{S}\else \v{S}\fi{}upi\ifmmode~\acute{c}\else
  \'{c}\fi{}}, \citenamefont {Cavalcanti},\ and\ \citenamefont
  {Ac\'{\i}n}}]{Bowles18}%
  \BibitemOpen
  \bibfield  {author} {\bibinfo {author} {\bibfnamefont {J.}~\bibnamefont
  {Bowles}}, \bibinfo {author} {\bibfnamefont {I.}~\bibnamefont {\ifmmode
  \check{S}\else \v{S}\fi{}upi\ifmmode~\acute{c}\else \'{c}\fi{}}}, \bibinfo
  {author} {\bibfnamefont {D.}~\bibnamefont {Cavalcanti}},\ and\ \bibinfo
  {author} {\bibfnamefont {A.}~\bibnamefont {Ac\'{\i}n}},\ }\href
  {https://doi.org/10.1103/PhysRevA.98.042336} {\bibfield  {journal} {\bibinfo
  {journal} {Phys. Rev. A}\ }\textbf {\bibinfo {volume} {98}},\ \bibinfo
  {pages} {042336} (\bibinfo {year} {2018})}\BibitemShut {NoStop}%
\bibitem [{\citenamefont {Chen}\ \emph {et~al.}(2022)\citenamefont {Chen},
  \citenamefont {Wang}, \citenamefont {Liu}, \citenamefont {Wang},
  \citenamefont {Ying}, \citenamefont {Shang}, \citenamefont {Wu},
  \citenamefont {Gong}, \citenamefont {Deng}, \citenamefont {Liang},
  \citenamefont {Zhang}, \citenamefont {Peng}, \citenamefont {Zhu},
  \citenamefont {Cabello}, \citenamefont {Lu},\ and\ \citenamefont
  {Pan}}]{chen2021complex}%
  \BibitemOpen
  \bibfield  {author} {\bibinfo {author} {\bibfnamefont {M.-C.}\ \bibnamefont
  {Chen}}, \bibinfo {author} {\bibfnamefont {C.}~\bibnamefont {Wang}}, \bibinfo
  {author} {\bibfnamefont {F.-M.}\ \bibnamefont {Liu}}, \bibinfo {author}
  {\bibfnamefont {J.-W.}\ \bibnamefont {Wang}}, \bibinfo {author}
  {\bibfnamefont {C.}~\bibnamefont {Ying}}, \bibinfo {author} {\bibfnamefont
  {Z.-X.}\ \bibnamefont {Shang}}, \bibinfo {author} {\bibfnamefont
  {Y.}~\bibnamefont {Wu}}, \bibinfo {author} {\bibfnamefont {M.}~\bibnamefont
  {Gong}}, \bibinfo {author} {\bibfnamefont {H.}~\bibnamefont {Deng}}, \bibinfo
  {author} {\bibfnamefont {F.-T.}\ \bibnamefont {Liang}}, \bibinfo {author}
  {\bibfnamefont {Q.}~\bibnamefont {Zhang}}, \bibinfo {author} {\bibfnamefont
  {C.-Z.}\ \bibnamefont {Peng}}, \bibinfo {author} {\bibfnamefont
  {X.}~\bibnamefont {Zhu}}, \bibinfo {author} {\bibfnamefont {A.}~\bibnamefont
  {Cabello}}, \bibinfo {author} {\bibfnamefont {C.-Y.}\ \bibnamefont {Lu}},\
  and\ \bibinfo {author} {\bibfnamefont {J.-W.}\ \bibnamefont {Pan}},\ }\href
  {https://doi.org/10.1103/PhysRevLett.128.040403} {\bibfield  {journal}
  {\bibinfo  {journal} {Phys. Rev. Lett.}\ }\textbf {\bibinfo {volume} {128}},\
  \bibinfo {pages} {040403} (\bibinfo {year} {2022})}\BibitemShut {NoStop}%
\bibitem [{\citenamefont {Li}\ \emph {et~al.}(2022)\citenamefont {Li},
  \citenamefont {Mao}, \citenamefont {Weilenmann}, \citenamefont {Tavakoli},
  \citenamefont {Chen}, \citenamefont {Feng}, \citenamefont {Yang},
  \citenamefont {Renou}, \citenamefont {Trillo}, \citenamefont {Le},
  \citenamefont {Gisin}, \citenamefont {Ac\'{\i}n}, \citenamefont
  {Navascu\'es}, \citenamefont {Wang},\ and\ \citenamefont
  {Fan}}]{li2021complex}%
  \BibitemOpen
  \bibfield  {author} {\bibinfo {author} {\bibfnamefont {Z.-D.}\ \bibnamefont
  {Li}}, \bibinfo {author} {\bibfnamefont {Y.-L.}\ \bibnamefont {Mao}},
  \bibinfo {author} {\bibfnamefont {M.}~\bibnamefont {Weilenmann}}, \bibinfo
  {author} {\bibfnamefont {A.}~\bibnamefont {Tavakoli}}, \bibinfo {author}
  {\bibfnamefont {H.}~\bibnamefont {Chen}}, \bibinfo {author} {\bibfnamefont
  {L.}~\bibnamefont {Feng}}, \bibinfo {author} {\bibfnamefont {S.-J.}\
  \bibnamefont {Yang}}, \bibinfo {author} {\bibfnamefont {M.-O.}\ \bibnamefont
  {Renou}}, \bibinfo {author} {\bibfnamefont {D.}~\bibnamefont {Trillo}},
  \bibinfo {author} {\bibfnamefont {T.~P.}\ \bibnamefont {Le}}, \bibinfo
  {author} {\bibfnamefont {N.}~\bibnamefont {Gisin}}, \bibinfo {author}
  {\bibfnamefont {A.}~\bibnamefont {Ac\'{\i}n}}, \bibinfo {author}
  {\bibfnamefont {M.}~\bibnamefont {Navascu\'es}}, \bibinfo {author}
  {\bibfnamefont {Z.}~\bibnamefont {Wang}},\ and\ \bibinfo {author}
  {\bibfnamefont {J.}~\bibnamefont {Fan}},\ }\href
  {https://doi.org/10.1103/PhysRevLett.128.040402} {\bibfield  {journal}
  {\bibinfo  {journal} {Phys. Rev. Lett.}\ }\textbf {\bibinfo {volume} {128}},\
  \bibinfo {pages} {040402} (\bibinfo {year} {2022})}\BibitemShut {NoStop}%
\bibitem [{\citenamefont {Duan}\ and\ \citenamefont {Monroe}(2010)}]{Duan2010}%
  \BibitemOpen
  \bibfield  {author} {\bibinfo {author} {\bibfnamefont {L.-M.}\ \bibnamefont
  {Duan}}\ and\ \bibinfo {author} {\bibfnamefont {C.}~\bibnamefont {Monroe}},\
  }\href {https://doi.org/10.1103/RevModPhys.82.1209} {\bibfield  {journal}
  {\bibinfo  {journal} {Rev. Mod. Phys.}\ }\textbf {\bibinfo {volume} {82}},\
  \bibinfo {pages} {1209} (\bibinfo {year} {2010})}\BibitemShut {NoStop}%
\bibitem [{\citenamefont {Ritter}\ \emph {et~al.}(2012)\citenamefont {Ritter},
  \citenamefont {Nölleke}, \citenamefont {Hahn}, \citenamefont {Reiserer},
  \citenamefont {Neuzner}, \citenamefont {Uphoff}, \citenamefont {Mücke},
  \citenamefont {Figueroa}, \citenamefont {Bochmann},\ and\ \citenamefont
  {Rempe}}]{Rempe2012}%
  \BibitemOpen
  \bibfield  {author} {\bibinfo {author} {\bibfnamefont {S.}~\bibnamefont
  {Ritter}}, \bibinfo {author} {\bibfnamefont {C.}~\bibnamefont {Nölleke}},
  \bibinfo {author} {\bibfnamefont {C.}~\bibnamefont {Hahn}}, \bibinfo {author}
  {\bibfnamefont {A.}~\bibnamefont {Reiserer}}, \bibinfo {author}
  {\bibfnamefont {A.}~\bibnamefont {Neuzner}}, \bibinfo {author} {\bibfnamefont
  {M.}~\bibnamefont {Uphoff}}, \bibinfo {author} {\bibfnamefont
  {M.}~\bibnamefont {Mücke}}, \bibinfo {author} {\bibfnamefont
  {E.}~\bibnamefont {Figueroa}}, \bibinfo {author} {\bibfnamefont
  {J.}~\bibnamefont {Bochmann}},\ and\ \bibinfo {author} {\bibfnamefont
  {G.}~\bibnamefont {Rempe}},\ }\href {https://doi.org/10.1038/nature11023}
  {\bibfield  {journal} {\bibinfo  {journal} {Nature}\ }\textbf {\bibinfo
  {volume} {484}},\ \bibinfo {pages} {195} (\bibinfo {year}
  {2012})}\BibitemShut {NoStop}%
\bibitem [{\citenamefont {Popescu}\ and\ \citenamefont
  {Rohrlich}(1992)}]{Popescu92}%
  \BibitemOpen
  \bibfield  {author} {\bibinfo {author} {\bibfnamefont {S.}~\bibnamefont
  {Popescu}}\ and\ \bibinfo {author} {\bibfnamefont {D.}~\bibnamefont
  {Rohrlich}},\ }\href {https://doi.org/10.1016/0375-9601(92)90711-T}
  {\bibfield  {journal} {\bibinfo  {journal} {Phys. Lett. A}\ }\textbf
  {\bibinfo {volume} {166}},\ \bibinfo {pages} {293 } (\bibinfo {year}
  {1992})}\BibitemShut {NoStop}%
\bibitem [{\citenamefont {Luo}(2018)}]{Luo2018b}%
  \BibitemOpen
  \bibfield  {author} {\bibinfo {author} {\bibfnamefont {M.-X.}\ \bibnamefont
  {Luo}},\ }\href {https://doi.org/10.1103/PhysRevA.98.042317} {\bibfield
  {journal} {\bibinfo  {journal} {Phys. Rev. A}\ }\textbf {\bibinfo {volume}
  {98}},\ \bibinfo {pages} {042317} (\bibinfo {year} {2018})}\BibitemShut
  {NoStop}%
\bibitem [{\citenamefont {Contreras-Tejada}\ \emph {et~al.}(2021)\citenamefont
  {Contreras-Tejada}, \citenamefont {Palazuelos},\ and\ \citenamefont
  {de~Vicente}}]{Contreras2020}%
  \BibitemOpen
  \bibfield  {author} {\bibinfo {author} {\bibfnamefont {P.}~\bibnamefont
  {Contreras-Tejada}}, \bibinfo {author} {\bibfnamefont {C.}~\bibnamefont
  {Palazuelos}},\ and\ \bibinfo {author} {\bibfnamefont {J.~I.}\ \bibnamefont
  {de~Vicente}},\ }\href {https://doi.org/10.1103/PhysRevLett.126.040501}
  {\bibfield  {journal} {\bibinfo  {journal} {Phys. Rev. Lett.}\ }\textbf
  {\bibinfo {volume} {126}},\ \bibinfo {pages} {040501} (\bibinfo {year}
  {2021})}\BibitemShut {NoStop}%
\bibitem [{\citenamefont {Pironio}(2005)}]{Pironio05}%
  \BibitemOpen
  \bibfield  {author} {\bibinfo {author} {\bibfnamefont {S.}~\bibnamefont
  {Pironio}},\ }\href {https://doi.org/10.1063/1.1928727} {\bibfield  {journal}
  {\bibinfo  {journal} {J. Math. Phys.}\ }\textbf {\bibinfo {volume} {46}},\
  \bibinfo {pages} {062112} (\bibinfo {year} {2005})}\BibitemShut {NoStop}%
\bibitem [{\citenamefont {Brunner}\ and\ \citenamefont
  {Linden}(2013)}]{Brunner2013}%
  \BibitemOpen
  \bibfield  {author} {\bibinfo {author} {\bibfnamefont {N.}~\bibnamefont
  {Brunner}}\ and\ \bibinfo {author} {\bibfnamefont {N.}~\bibnamefont
  {Linden}},\ }\href {https://doi.org/10.1038/ncomms3057} {\bibfield  {journal}
  {\bibinfo  {journal} {Nat. Commun.}\ }\textbf {\bibinfo {volume} {4}},\
  \bibinfo {pages} {2057} (\bibinfo {year} {2013})}\BibitemShut {NoStop}%
\bibitem [{\citenamefont {Silman}\ \emph {et~al.}(2008)\citenamefont {Silman},
  \citenamefont {Machnes},\ and\ \citenamefont {Aharon}}]{Silman2007}%
  \BibitemOpen
  \bibfield  {author} {\bibinfo {author} {\bibfnamefont {J.}~\bibnamefont
  {Silman}}, \bibinfo {author} {\bibfnamefont {S.}~\bibnamefont {Machnes}},\
  and\ \bibinfo {author} {\bibfnamefont {N.}~\bibnamefont {Aharon}},\ }\href
  {https://doi.org/https://doi.org/10.1016/j.physleta.2008.03.001} {\bibfield
  {journal} {\bibinfo  {journal} {Phys. Lett. A}\ }\textbf {\bibinfo {volume}
  {372}},\ \bibinfo {pages} {3796 } (\bibinfo {year} {2008})}\BibitemShut
  {NoStop}%
\bibitem [{\citenamefont {Luo}(2019)}]{Luo2019}%
  \BibitemOpen
  \bibfield  {author} {\bibinfo {author} {\bibfnamefont {M.-X.}\ \bibnamefont
  {Luo}},\ }\href {https://doi.org/10.1038/s41534-019-0203-6} {\bibfield
  {journal} {\bibinfo  {journal} {npj Quantum Inf.}\ }\textbf {\bibinfo
  {volume} {5}},\ \bibinfo {pages} {91} (\bibinfo {year} {2019})}\BibitemShut
  {NoStop}%
\bibitem [{\citenamefont {Carleo}\ \emph {et~al.}(2019)\citenamefont {Carleo},
  \citenamefont {Cirac}, \citenamefont {Cranmer}, \citenamefont {Daudet},
  \citenamefont {Schuld}, \citenamefont {Tishby}, \citenamefont
  {Vogt-Maranto},\ and\ \citenamefont {Zdeborov\'a}}]{MLphys}%
  \BibitemOpen
  \bibfield  {author} {\bibinfo {author} {\bibfnamefont {G.}~\bibnamefont
  {Carleo}}, \bibinfo {author} {\bibfnamefont {I.}~\bibnamefont {Cirac}},
  \bibinfo {author} {\bibfnamefont {K.}~\bibnamefont {Cranmer}}, \bibinfo
  {author} {\bibfnamefont {L.}~\bibnamefont {Daudet}}, \bibinfo {author}
  {\bibfnamefont {M.}~\bibnamefont {Schuld}}, \bibinfo {author} {\bibfnamefont
  {N.}~\bibnamefont {Tishby}}, \bibinfo {author} {\bibfnamefont
  {L.}~\bibnamefont {Vogt-Maranto}},\ and\ \bibinfo {author} {\bibfnamefont
  {L.}~\bibnamefont {Zdeborov\'a}},\ }\href
  {https://doi.org/10.1103/RevModPhys.91.045002} {\bibfield  {journal}
  {\bibinfo  {journal} {Rev. Mod. Phys.}\ }\textbf {\bibinfo {volume} {91}},\
  \bibinfo {pages} {045002} (\bibinfo {year} {2019})}\BibitemShut {NoStop}%
\bibitem [{\citenamefont {Canabarro}\ \emph {et~al.}(2019)\citenamefont
  {Canabarro}, \citenamefont {Brito},\ and\ \citenamefont
  {Chaves}}]{Canabarro2019ML}%
  \BibitemOpen
  \bibfield  {author} {\bibinfo {author} {\bibfnamefont {A.}~\bibnamefont
  {Canabarro}}, \bibinfo {author} {\bibfnamefont {S.}~\bibnamefont {Brito}},\
  and\ \bibinfo {author} {\bibfnamefont {R.}~\bibnamefont {Chaves}},\ }\href
  {https://doi.org/10.1103/PhysRevLett.122.200401} {\bibfield  {journal}
  {\bibinfo  {journal} {Phys. Rev. Lett.}\ }\textbf {\bibinfo {volume} {122}},\
  \bibinfo {pages} {200401} (\bibinfo {year} {2019})}\BibitemShut {NoStop}%
\bibitem [{\citenamefont {Navascu\'es}\ \emph {et~al.}(2020)\citenamefont
  {Navascu\'es}, \citenamefont {Wolfe}, \citenamefont {Rosset},\ and\
  \citenamefont {Pozas-Kerstjens}}]{navascues2020gnme}%
  \BibitemOpen
  \bibfield  {author} {\bibinfo {author} {\bibfnamefont {M.}~\bibnamefont
  {Navascu\'es}}, \bibinfo {author} {\bibfnamefont {E.}~\bibnamefont {Wolfe}},
  \bibinfo {author} {\bibfnamefont {D.}~\bibnamefont {Rosset}},\ and\ \bibinfo
  {author} {\bibfnamefont {A.}~\bibnamefont {Pozas-Kerstjens}},\ }\href
  {https://doi.org/10.1103/PhysRevLett.125.240505} {\bibfield  {journal}
  {\bibinfo  {journal} {Phys. Rev. Lett.}\ }\textbf {\bibinfo {volume} {125}},\
  \bibinfo {pages} {240505} (\bibinfo {year} {2020})}\BibitemShut {NoStop}%
\bibitem [{\citenamefont {Briegel}\ and\ \citenamefont
  {Raussendorf}(2001)}]{Briegel2001}%
  \BibitemOpen
  \bibfield  {author} {\bibinfo {author} {\bibfnamefont {H.~J.}\ \bibnamefont
  {Briegel}}\ and\ \bibinfo {author} {\bibfnamefont {R.}~\bibnamefont
  {Raussendorf}},\ }\href {https://doi.org/10.1103/PhysRevLett.86.910}
  {\bibfield  {journal} {\bibinfo  {journal} {Phys. Rev. Lett.}\ }\textbf
  {\bibinfo {volume} {86}},\ \bibinfo {pages} {910} (\bibinfo {year}
  {2001})}\BibitemShut {NoStop}%
\bibitem [{\citenamefont {Hein}\ \emph {et~al.}(2004)\citenamefont {Hein},
  \citenamefont {Eisert},\ and\ \citenamefont {Briegel}}]{Hein}%
  \BibitemOpen
  \bibfield  {author} {\bibinfo {author} {\bibfnamefont {M.}~\bibnamefont
  {Hein}}, \bibinfo {author} {\bibfnamefont {J.}~\bibnamefont {Eisert}},\ and\
  \bibinfo {author} {\bibfnamefont {H.~J.}\ \bibnamefont {Briegel}},\ }\href
  {https://doi.org/10.1103/PhysRevA.69.062311} {\bibfield  {journal} {\bibinfo
  {journal} {Phys. Rev. A}\ }\textbf {\bibinfo {volume} {69}},\ \bibinfo
  {pages} {062311} (\bibinfo {year} {2004})}\BibitemShut {NoStop}%
\bibitem [{\citenamefont {Kraft}\ \emph {et~al.}(2021)\citenamefont {Kraft},
  \citenamefont {Designolle}, \citenamefont {Ritz}, \citenamefont {Brunner},
  \citenamefont {G\"uhne},\ and\ \citenamefont
  {Huber}}]{kraft2020networkentanglement}%
  \BibitemOpen
  \bibfield  {author} {\bibinfo {author} {\bibfnamefont {T.}~\bibnamefont
  {Kraft}}, \bibinfo {author} {\bibfnamefont {S.}~\bibnamefont {Designolle}},
  \bibinfo {author} {\bibfnamefont {C.}~\bibnamefont {Ritz}}, \bibinfo {author}
  {\bibfnamefont {N.}~\bibnamefont {Brunner}}, \bibinfo {author} {\bibfnamefont
  {O.}~\bibnamefont {G\"uhne}},\ and\ \bibinfo {author} {\bibfnamefont
  {M.}~\bibnamefont {Huber}},\ }\href
  {https://doi.org/10.1103/PhysRevA.103.L060401} {\bibfield  {journal}
  {\bibinfo  {journal} {Phys. Rev. A}\ }\textbf {\bibinfo {volume} {103}},\
  \bibinfo {pages} {L060401} (\bibinfo {year} {2021})}\BibitemShut {NoStop}%
\bibitem [{\citenamefont {Luo}(2021)}]{luo2020networkentanglement}%
  \BibitemOpen
  \bibfield  {author} {\bibinfo {author} {\bibfnamefont {M.-X.}\ \bibnamefont
  {Luo}},\ }\href {https://doi.org/10.1002/qute.202000123} {\bibfield
  {journal} {\bibinfo  {journal} {Adv. Quantum Technol.}\ }\textbf {\bibinfo
  {volume} {4}},\ \bibinfo {pages} {2000123} (\bibinfo {year}
  {2021})}\BibitemShut {NoStop}%
\bibitem [{\citenamefont {D\"ur}\ \emph {et~al.}(2000)\citenamefont {D\"ur},
  \citenamefont {Vidal},\ and\ \citenamefont {Cirac}}]{Dur04}%
  \BibitemOpen
  \bibfield  {author} {\bibinfo {author} {\bibfnamefont {W.}~\bibnamefont
  {D\"ur}}, \bibinfo {author} {\bibfnamefont {G.}~\bibnamefont {Vidal}},\ and\
  \bibinfo {author} {\bibfnamefont {J.~I.}\ \bibnamefont {Cirac}},\ }\href
  {https://doi.org/10.1103/PhysRevA.62.062314} {\bibfield  {journal} {\bibinfo
  {journal} {Phys. Rev. A}\ }\textbf {\bibinfo {volume} {62}},\ \bibinfo
  {pages} {062314} (\bibinfo {year} {2000})}\BibitemShut {NoStop}%
\bibitem [{\citenamefont {T\'{o}th}(2007)}]{Dicke}%
  \BibitemOpen
  \bibfield  {author} {\bibinfo {author} {\bibfnamefont {G.}~\bibnamefont
  {T\'{o}th}},\ }\href {https://doi.org/10.1364/JOSAB.24.000275} {\bibfield
  {journal} {\bibinfo  {journal} {J. Opt. Soc. Am. B}\ }\textbf {\bibinfo
  {volume} {24}},\ \bibinfo {pages} {275} (\bibinfo {year} {2007})}\BibitemShut
  {NoStop}%
\bibitem [{\citenamefont {Mayers}(1997)}]{Mayers97}%
  \BibitemOpen
  \bibfield  {author} {\bibinfo {author} {\bibfnamefont {D.}~\bibnamefont
  {Mayers}},\ }\href {https://doi.org/10.1103/PhysRevLett.78.3414} {\bibfield
  {journal} {\bibinfo  {journal} {Phys. Rev. Lett.}\ }\textbf {\bibinfo
  {volume} {78}},\ \bibinfo {pages} {3414} (\bibinfo {year}
  {1997})}\BibitemShut {NoStop}%
\bibitem [{\citenamefont {Lo}\ and\ \citenamefont {Chau}(1997)}]{LC97}%
  \BibitemOpen
  \bibfield  {author} {\bibinfo {author} {\bibfnamefont {H.-K.}\ \bibnamefont
  {Lo}}\ and\ \bibinfo {author} {\bibfnamefont {H.~F.}\ \bibnamefont {Chau}},\
  }\href {https://doi.org/10.1103/PhysRevLett.78.3410} {\bibfield  {journal}
  {\bibinfo  {journal} {Phys. Rev. Lett.}\ }\textbf {\bibinfo {volume} {78}},\
  \bibinfo {pages} {3410} (\bibinfo {year} {1997})}\BibitemShut {NoStop}%
\bibitem [{\citenamefont {Kent}(1999)}]{Kent1999}%
  \BibitemOpen
  \bibfield  {author} {\bibinfo {author} {\bibfnamefont {A.}~\bibnamefont
  {Kent}},\ }\href {https://doi.org/10.1103/PhysRevLett.83.1447} {\bibfield
  {journal} {\bibinfo  {journal} {Phys. Rev. Lett.}\ }\textbf {\bibinfo
  {volume} {83}},\ \bibinfo {pages} {1447} (\bibinfo {year}
  {1999})}\BibitemShut {NoStop}%
\bibitem [{\citenamefont {Kent}(2011)}]{Kent2011}%
  \BibitemOpen
  \bibfield  {author} {\bibinfo {author} {\bibfnamefont {A.}~\bibnamefont
  {Kent}},\ }\href {https://doi.org/10.1088/1367-2630/13/11/113015} {\bibfield
  {journal} {\bibinfo  {journal} {New J. Phys.}\ }\textbf {\bibinfo {volume}
  {13}},\ \bibinfo {pages} {113015} (\bibinfo {year} {2011})}\BibitemShut
  {NoStop}%
\bibitem [{\citenamefont {Luo}\ and\ \citenamefont {Wang}(2020)}]{Luo2020b}%
  \BibitemOpen
  \bibfield  {author} {\bibinfo {author} {\bibfnamefont {M.-X.}\ \bibnamefont
  {Luo}}\ and\ \bibinfo {author} {\bibfnamefont {X.}~\bibnamefont {Wang}},\
  }\href {https://eprint.iacr.org/2020/1077} {\bibinfo {title} {Unconditionally
  secure quantum bit commitment: Revised}},\ \bibinfo {howpublished}
  {Cryptology ePrint Archive, Report 2020/1077} (\bibinfo {year}
  {2020})\BibitemShut {NoStop}%
\bibitem [{\citenamefont {Coiteux-Roy}\ \emph {et~al.}(2021)\citenamefont
  {Coiteux-Roy}, \citenamefont {Wolfe},\ and\ \citenamefont
  {Renou}}]{coiteux2021short}%
  \BibitemOpen
  \bibfield  {author} {\bibinfo {author} {\bibfnamefont {X.}~\bibnamefont
  {Coiteux-Roy}}, \bibinfo {author} {\bibfnamefont {E.}~\bibnamefont {Wolfe}},\
  and\ \bibinfo {author} {\bibfnamefont {M.-O.}\ \bibnamefont {Renou}},\ }\href
  {https://doi.org/10.1103/PhysRevLett.127.200401} {\bibfield  {journal}
  {\bibinfo  {journal} {Phys. Rev. Lett.}\ }\textbf {\bibinfo {volume} {127}},\
  \bibinfo {pages} {200401} (\bibinfo {year} {2021})}\BibitemShut {NoStop}%
\bibitem [{\citenamefont {Coiteux-Roy}\ \emph {et~al.}(2021)\citenamefont
  {Coiteux-Roy}, \citenamefont {Wolfe},\ and\ \citenamefont
  {Renou}}]{coiteux2021long}%
  \BibitemOpen
  \bibfield  {author} {\bibinfo {author} {\bibfnamefont {X.}~\bibnamefont
  {Coiteux-Roy}}, \bibinfo {author} {\bibfnamefont {E.}~\bibnamefont {Wolfe}},\
  and\ \bibinfo {author} {\bibfnamefont {M.-O.}\ \bibnamefont {Renou}},\ }\href
  {https://doi.org/10.1103/PhysRevA.104.052207} {\bibfield  {journal} {\bibinfo
   {journal} {Phys. Rev. A}\ }\textbf {\bibinfo {volume} {104}},\ \bibinfo
  {pages} {052207} (\bibinfo {year} {2021})}\BibitemShut {NoStop}%
\bibitem [{\citenamefont {Hamel}\ \emph {et~al.}(2014)\citenamefont {Hamel},
  \citenamefont {Shalm}, \citenamefont {H\"ubel}, \citenamefont {Miller},
  \citenamefont {Verma}, \citenamefont {Mirin}, \citenamefont {Woo~Nam},
  \citenamefont {Resch},\ and\ \citenamefont {Jennewein}}]{Jennewein2014}%
  \BibitemOpen
  \bibfield  {author} {\bibinfo {author} {\bibfnamefont {D.~R.}\ \bibnamefont
  {Hamel}}, \bibinfo {author} {\bibfnamefont {L.~K.}\ \bibnamefont {Shalm}},
  \bibinfo {author} {\bibfnamefont {H.}~\bibnamefont {H\"ubel}}, \bibinfo
  {author} {\bibfnamefont {A.~J.}\ \bibnamefont {Miller}}, \bibinfo {author}
  {\bibfnamefont {V.~B.}\ \bibnamefont {Verma}}, \bibinfo {author}
  {\bibfnamefont {R.~P.}\ \bibnamefont {Mirin}}, \bibinfo {author}
  {\bibfnamefont {S.}~\bibnamefont {Woo~Nam}}, \bibinfo {author} {\bibfnamefont
  {K.~J.}\ \bibnamefont {Resch}},\ and\ \bibinfo {author} {\bibfnamefont
  {T.}~\bibnamefont {Jennewein}},\ }\href
  {https://doi.org/10.1038/nphoton.2014.218} {\bibfield  {journal} {\bibinfo
  {journal} {Nat. Photonics}\ }\textbf {\bibinfo {volume} {8}},\ \bibinfo
  {pages} {801} (\bibinfo {year} {2014})}\BibitemShut {NoStop}%
\bibitem [{\citenamefont {Agresti}\ \emph {et~al.}(2019)\citenamefont
  {Agresti}, \citenamefont {Carvacho}, \citenamefont {Poderini}, \citenamefont
  {Aolita}, \citenamefont {Chaves},\ and\ \citenamefont
  {Sciarrino}}]{Agresti2019instrumental}%
  \BibitemOpen
  \bibfield  {author} {\bibinfo {author} {\bibfnamefont {I.}~\bibnamefont
  {Agresti}}, \bibinfo {author} {\bibfnamefont {G.}~\bibnamefont {Carvacho}},
  \bibinfo {author} {\bibfnamefont {D.}~\bibnamefont {Poderini}}, \bibinfo
  {author} {\bibfnamefont {L.}~\bibnamefont {Aolita}}, \bibinfo {author}
  {\bibfnamefont {R.}~\bibnamefont {Chaves}},\ and\ \bibinfo {author}
  {\bibfnamefont {F.}~\bibnamefont {Sciarrino}},\ }\href
  {https://doi.org/10.3390/proceedings2019012027} {\bibfield  {journal}
  {\bibinfo  {journal} {Proc. MDPI AG}\ }\textbf {\bibinfo {volume} {12}},\
  \bibinfo {pages} {27} (\bibinfo {year} {2019})}\BibitemShut {NoStop}%
\bibitem [{\citenamefont {van Himbeeck}\ \emph {et~al.}(2019)\citenamefont {van
  Himbeeck}, \citenamefont {Bohr~Brask}, \citenamefont {Pironio}, \citenamefont
  {Ramanathan}, \citenamefont {Sainz},\ and\ \citenamefont
  {Wolfe}}]{Himbeeck2019instrumental}%
  \BibitemOpen
  \bibfield  {author} {\bibinfo {author} {\bibfnamefont {T.}~\bibnamefont {van
  Himbeeck}}, \bibinfo {author} {\bibfnamefont {J.}~\bibnamefont {Bohr~Brask}},
  \bibinfo {author} {\bibfnamefont {S.}~\bibnamefont {Pironio}}, \bibinfo
  {author} {\bibfnamefont {R.}~\bibnamefont {Ramanathan}}, \bibinfo {author}
  {\bibfnamefont {A.~B.}\ \bibnamefont {Sainz}},\ and\ \bibinfo {author}
  {\bibfnamefont {E.}~\bibnamefont {Wolfe}},\ }\href
  {https://doi.org/10.22331/q-2019-09-16-186} {\bibfield  {journal} {\bibinfo
  {journal} {{Quantum}}\ }\textbf {\bibinfo {volume} {3}},\ \bibinfo {pages}
  {186} (\bibinfo {year} {2019})}\BibitemShut {NoStop}%
\bibitem [{\citenamefont {Gachechiladze}\ \emph {et~al.}(2020)\citenamefont
  {Gachechiladze}, \citenamefont {Miklin},\ and\ \citenamefont
  {Chaves}}]{gachechiladze2020qace}%
  \BibitemOpen
  \bibfield  {author} {\bibinfo {author} {\bibfnamefont {M.}~\bibnamefont
  {Gachechiladze}}, \bibinfo {author} {\bibfnamefont {N.}~\bibnamefont
  {Miklin}},\ and\ \bibinfo {author} {\bibfnamefont {R.}~\bibnamefont
  {Chaves}},\ }\href {https://doi.org/10.1103/PhysRevLett.125.230401}
  {\bibfield  {journal} {\bibinfo  {journal} {Phys. Rev. Lett.}\ }\textbf
  {\bibinfo {volume} {125}},\ \bibinfo {pages} {230401} (\bibinfo {year}
  {2020})}\BibitemShut {NoStop}%
\bibitem [{\citenamefont {Pearl}(2009)}]{pearl}%
  \BibitemOpen
  \bibfield  {author} {\bibinfo {author} {\bibfnamefont {J.}~\bibnamefont
  {Pearl}},\ }\href {https://doi.org/10.1017/CBO9780511803161} {\emph {\bibinfo
  {title} {{Causality: Models, Reasoning, and Inference}}}}\ (\bibinfo
  {publisher} {Cambridge University Press},\ \bibinfo {year}
  {2009})\BibitemShut {NoStop}%
\bibitem [{\citenamefont {Reichenbach}\ and\ \citenamefont
  {Reichenbach}(1991)}]{reichenbach}%
  \BibitemOpen
  \bibfield  {author} {\bibinfo {author} {\bibfnamefont {H.}~\bibnamefont
  {Reichenbach}}\ and\ \bibinfo {author} {\bibfnamefont {M.}~\bibnamefont
  {Reichenbach}},\ }\href@noop {} {\emph {\bibinfo {title} {The Direction of
  Time}}}\ (\bibinfo  {publisher} {University of California Press},\ \bibinfo
  {year} {1991})\BibitemShut {NoStop}%
\bibitem [{\citenamefont {Holland}(1986)}]{HollandACE}%
  \BibitemOpen
  \bibfield  {author} {\bibinfo {author} {\bibfnamefont {P.~W.}\ \bibnamefont
  {Holland}},\ }\href {https://doi.org/10.1080/01621459.1986.10478354}
  {\bibfield  {journal} {\bibinfo  {journal} {J. Am. Stat. Assoc.}\ }\textbf
  {\bibinfo {volume} {81}},\ \bibinfo {pages} {945} (\bibinfo {year}
  {1986})}\BibitemShut {NoStop}%
\bibitem [{\citenamefont {Shpitser}\ and\ \citenamefont
  {Pearl}(2008)}]{shpitser2008Identification}%
  \BibitemOpen
  \bibfield  {author} {\bibinfo {author} {\bibfnamefont {I.}~\bibnamefont
  {Shpitser}}\ and\ \bibinfo {author} {\bibfnamefont {J.}~\bibnamefont
  {Pearl}},\ }\href
  {http://www.jmlr.org/papers/volume9/shpitser08a/shpitser08a.pdf} {\bibfield
  {journal} {\bibinfo  {journal} {J. Mach. Learn. Res.}\ }\textbf {\bibinfo
  {volume} {9}},\ \bibinfo {pages} {1941} (\bibinfo {year} {2008})}\BibitemShut
  {NoStop}%
\bibitem [{\citenamefont {Chaves}\ \emph {et~al.}(2018)\citenamefont {Chaves},
  \citenamefont {Carvacho}, \citenamefont {Agresti}, \citenamefont {Di~Giulio},
  \citenamefont {Aolita}, \citenamefont {Giacomini},\ and\ \citenamefont
  {Sciarrino}}]{chaves2018instrumental}%
  \BibitemOpen
  \bibfield  {author} {\bibinfo {author} {\bibfnamefont {R.}~\bibnamefont
  {Chaves}}, \bibinfo {author} {\bibfnamefont {G.}~\bibnamefont {Carvacho}},
  \bibinfo {author} {\bibfnamefont {I.}~\bibnamefont {Agresti}}, \bibinfo
  {author} {\bibfnamefont {V.}~\bibnamefont {Di~Giulio}}, \bibinfo {author}
  {\bibfnamefont {L.}~\bibnamefont {Aolita}}, \bibinfo {author} {\bibfnamefont
  {S.}~\bibnamefont {Giacomini}},\ and\ \bibinfo {author} {\bibfnamefont
  {F.}~\bibnamefont {Sciarrino}},\ }\href
  {https://doi.org/10.1038/s41567-017-0008-5} {\bibfield  {journal} {\bibinfo
  {journal} {Nat. Phys.}\ }\textbf {\bibinfo {volume} {14}},\ \bibinfo {pages}
  {291} (\bibinfo {year} {2018})}\BibitemShut {NoStop}%
\bibitem [{\citenamefont {Agresti}\ \emph {et~al.}(2020)\citenamefont
  {Agresti}, \citenamefont {Poderini}, \citenamefont {Guerini}, \citenamefont
  {Mancusi}, \citenamefont {Carvacho}, \citenamefont {Aolita}, \citenamefont
  {Cavalcanti}, \citenamefont {Chaves},\ and\ \citenamefont
  {Sciarrino}}]{agresti2020instrumental}%
  \BibitemOpen
  \bibfield  {author} {\bibinfo {author} {\bibfnamefont {I.}~\bibnamefont
  {Agresti}}, \bibinfo {author} {\bibfnamefont {D.}~\bibnamefont {Poderini}},
  \bibinfo {author} {\bibfnamefont {L.}~\bibnamefont {Guerini}}, \bibinfo
  {author} {\bibfnamefont {M.}~\bibnamefont {Mancusi}}, \bibinfo {author}
  {\bibfnamefont {G.}~\bibnamefont {Carvacho}}, \bibinfo {author}
  {\bibfnamefont {L.}~\bibnamefont {Aolita}}, \bibinfo {author} {\bibfnamefont
  {D.}~\bibnamefont {Cavalcanti}}, \bibinfo {author} {\bibfnamefont
  {R.}~\bibnamefont {Chaves}},\ and\ \bibinfo {author} {\bibfnamefont
  {F.}~\bibnamefont {Sciarrino}},\ }\href
  {https://doi.org/10.1038/s42005-020-0375-6} {\bibfield  {journal} {\bibinfo
  {journal} {Commun. Phys.}\ }\textbf {\bibinfo {volume} {3}},\ \bibinfo
  {pages} {110} (\bibinfo {year} {2020})}\BibitemShut {NoStop}%
\bibitem [{\citenamefont {Allen}\ \emph {et~al.}(2017)\citenamefont {Allen},
  \citenamefont {Barrett}, \citenamefont {Horsman}, \citenamefont {Lee},\ and\
  \citenamefont {Spekkens}}]{Allen2017Qcausal}%
  \BibitemOpen
  \bibfield  {author} {\bibinfo {author} {\bibfnamefont {J.-M.~A.}\
  \bibnamefont {Allen}}, \bibinfo {author} {\bibfnamefont {J.}~\bibnamefont
  {Barrett}}, \bibinfo {author} {\bibfnamefont {D.~C.}\ \bibnamefont
  {Horsman}}, \bibinfo {author} {\bibfnamefont {C.~M.}\ \bibnamefont {Lee}},\
  and\ \bibinfo {author} {\bibfnamefont {R.~W.}\ \bibnamefont {Spekkens}},\
  }\href {https://doi.org/10.1103/PhysRevX.7.031021} {\bibfield  {journal}
  {\bibinfo  {journal} {Phys. Rev. X}\ }\textbf {\bibinfo {volume} {7}},\
  \bibinfo {pages} {031021} (\bibinfo {year} {2017})}\BibitemShut {NoStop}%
\bibitem [{\citenamefont {Pienaar}\ and\ \citenamefont
  {Brukner}(2015)}]{Pienaar2015dseparation}%
  \BibitemOpen
  \bibfield  {author} {\bibinfo {author} {\bibfnamefont {J.}~\bibnamefont
  {Pienaar}}\ and\ \bibinfo {author} {\bibfnamefont {{\v{C}}.}~\bibnamefont
  {Brukner}},\ }\href {https://doi.org/10.1088/1367-2630/17/7/073020}
  {\bibfield  {journal} {\bibinfo  {journal} {New J. Phys.}\ }\textbf {\bibinfo
  {volume} {17}},\ \bibinfo {pages} {073020} (\bibinfo {year}
  {2015})}\BibitemShut {NoStop}%
\bibitem [{\citenamefont {Barrett}\ \emph {et~al.}()\citenamefont {Barrett},
  \citenamefont {Lorenz},\ and\ \citenamefont {Oreshkov}}]{barrett2019qcausal}%
  \BibitemOpen
  \bibfield  {author} {\bibinfo {author} {\bibfnamefont {J.}~\bibnamefont
  {Barrett}}, \bibinfo {author} {\bibfnamefont {R.}~\bibnamefont {Lorenz}},\
  and\ \bibinfo {author} {\bibfnamefont {O.}~\bibnamefont {Oreshkov}},\
  }\href@noop {} {\bibinfo {title} {Quantum causal models}},\ \Eprint
  {https://arxiv.org/abs/1906.10726} {arXiv:1906.10726} \BibitemShut {NoStop}%
\bibitem [{\citenamefont {Giarmatzi}\ and\ \citenamefont
  {Costa}(2018)}]{giarmatzi2018discovery}%
  \BibitemOpen
  \bibfield  {author} {\bibinfo {author} {\bibfnamefont {C.}~\bibnamefont
  {Giarmatzi}}\ and\ \bibinfo {author} {\bibfnamefont {F.}~\bibnamefont
  {Costa}},\ }\href {https://doi.org/10.1038/s41534-018-0062-6} {\bibfield
  {journal} {\bibinfo  {journal} {npj Quantum Inf.}\ }\textbf {\bibinfo
  {volume} {4}} (\bibinfo {year} {2018})}\BibitemShut {NoStop}%
\end{thebibliography}%

\end{document}